\documentclass[11pt,a4paper,twoside,onehalfspacing,fleqn]{report}
\usepackage{natbib}
\usepackage{fixltx2e}
\usepackage{textcomp}
\usepackage{graphicx}	% Including figure files
\usepackage{amsmath}	% Advanced maths commands
\usepackage{amssymb}	% Extra maths symbols
\usepackage{verbatim}
\usepackage[T1]{fontenc}
\usepackage{ae,aecompl}
\usepackage{enumerate}
\usepackage[font=small,format=plain,labelfont=bf,up]{caption}
\usepackage{multirow}
\usepackage{longtable}
\usepackage{mathtools}
\usepackage{lscape}
\usepackage{natbib}
\usepackage{txfonts}
\usepackage{subfig}
\usepackage{slashed}
\usepackage{epsfig}
\usepackage{color}
\usepackage{float}
\usepackage{subfig}
\usepackage{parskip, setspace, fullpage}
\usepackage{appendix}
\usepackage{rotating}
\usepackage[a4paper,left=1in,right=1in,top=1in,bindingoffset=0.7in,bottom=1.2in]{geometry}
\usepackage{hyperref}
\hypersetup{}
\usepackage{bm}
\usepackage[frozencache]{minted}
\usepackage{makeidx}
\usepackage{epigraph}

\newcommand{\rxte}{{\textit{RXTE}}}

\newcommand{\ms}{$M_{\odot}$}

\newcommand{\la}{\,\rlap{\raise 0.5ex\hbox{$<$}}{\lower 1.0ex\hbox{$\sim$}}\,}
\newcommand\fd{\hbox{$.\!\!^{\reset@font\romn d}$}}
\newcommand\fh{\hbox{$.\!\!^{\reset@font\romn h}$}}
\newcommand\fm{\hbox{$.\!\!^{\reset@font\romn m}$}}
\newcommand\fs{\hbox{$.\!\!^{\reset@font\romn s}$}}

\newcommand\fp{\hbox{$.\!\!^{\reset@font\reset@font\scriptscriptstyle\romn p}$}}

\newcommand{\ga}{\,\rlap{\raise 0.5ex\hbox{$>$}}{\lower 1.0ex\hbox{$\sim$}}\,}
\newcommand{\spcu}{\,cts\,s$^{-1}$\,PCU$^{-1}$}
\newcommand{\ergf}{\,ergs~s$^{-1}$~cm$^{-2}$}

\newcommand{\dv}{\bm{\nabla}}
\newcommand{\indexalpha}{\index{Variability class!Ga@$\alpha$}}
\newcommand{\indexbeta}{\index{Variability class!Gb@$\beta$}}
\newcommand{\indexgamma}{\index{Variability class!Gc@$\gamma$}}
\newcommand{\indexdelta}{\index{Variability class!Gd@$\delta$}}

\newcommand{\indextheta}{\index{Variability class!Gh@$\theta$}}

\newcommand{\indexkappa}{\index{Variability class!Gj@$\kappa$}}
\newcommand{\indexlambda}{\index{Variability class!Gk@$\lambda$}}
\newcommand{\indexmu}{\index{Variability class!Gl@$\mu$}}
\newcommand{\indexnu}{\index{Variability class!Gm@$\nu$}}

\newcommand{\indexrho}{\index{Variability class!Gq@$\rho$}}

\newcommand{\indexphi}{\index{Variability class!Gu@$\phi$}}
\newcommand{\indexchi}{\index{Variability class!Gv@$\chi$}}

\newcommand{\indexomega}{\index{Variability class!Gx@$\omega$}}
\newcommand{\indexi}{\index{Variability class!I1@I}}
\newcommand{\indexii}{\index{Variability class!I2@II}}
\newcommand{\indexiii}{\index{Variability class!I3@III}}
\newcommand{\indexiv}{\index{Variability class!I4@IV}}
\newcommand{\indexv}{\index{Variability class!I5@V}}
\newcommand{\indexvi}{\index{Variability class!I6@VI}}
\newcommand{\indexvii}{\index{Variability class!I7@VII}}
\newcommand{\indexviii}{\index{Variability class!I8@VIII}}
\newcommand{\indexix}{\index{Variability class!I9@IX}}
\newcommand{\indexrxte}{\index{RXTE@\textit{RXTE}}}
 \newcommand{\indexpca}{\index{RXTE@\textit{RXTE}!PCA}}
 \newcommand{\indexhexte}{\index{RXTE@\textit{RXTE}!HEXTE}}
 \newcommand{\indexasm}{\index{RXTE@\textit{RXTE}!ASM}}
 \newcommand{\indexsto}{\index{RXTE@\textit{RXTE}!PCA!Standard1 data@\texttt{Standard1} data}}
 \newcommand{\indexstt}{\index{RXTE@\textit{RXTE}!PCA!Standard1 data@\texttt{Standard2} data}}
 \newcommand{\indexgx}{\index{RXTE@\textit{RXTE}!PCA!GoodXenon data@\texttt{GoodXenon} data}}
\newcommand{\indexsuzaku}{\index{Suzaku@\textit{Suzaku}}}
 \newcommand{\indexxrs}{\index{Suzaku@\textit{Suzaku}!XRS}}
 \newcommand{\indexxis}{\index{Suzaku@\textit{Suzaku}!XIS}}
 \newcommand{\indexhxd}{\index{Suzaku@\textit{Suzaku}!HXD}}
\newcommand{\indexchandra}{\index{Chandra@\textit{Chandra}}}
 \newcommand{\indexacis}{\index{Chandra@\textit{Chandra}!ACIS}}
 \newcommand{\indexhrc}{\index{Chandra@\textit{Chandra}!HRC}}
\newcommand{\indexintegral}{\index{Integral@\textit{INTEGRAL}}}
 \newcommand{\indexibis}{\index{Integral@\textit{INTEGRAL}!IBIS}}
 \newcommand{\indexspi}{\index{Integral@\textit{INTEGRAL}!SPI}}
 \newcommand{\indexjemx}{\index{Integral@\textit{INTEGRAL}!JEM-X}}
 \newcommand{\indexomc}{\index{Integral@\textit{INTEGRAL}!OMC}}
\newcommand{\indexnustar}{\index{NuSTAR@\textit{NuSTAR}}}
\newcommand{\indexswift}{\index{Swift@\textit{Swift}}}
 \newcommand{\indexuvot}{\index{Swift@\textit{Swift}!UVOT}}
 \newcommand{\indexbat}{\index{Swift@\textit{Swift}!BAT}}
 \newcommand{\indexxrt}{\index{Swift@\textit{Swift}!XRT}}
\newcommand{\indexxmm}{\index{XMM-Newton@\textit{XMM-Newton}}}
 \newcommand{\indexepic}{\index{XMM-Newton@\textit{XMM-Newton}!EPIC}}
 \newcommand{\indexom}{\index{XMM-Newton@\textit{XMM-Newton}!OM}}
 \newcommand{\indexrgs}{\index{XMM-Newton@\textit{XMM-Newton}!RGS}}
\newcommand{\thesistitle}{Recurrent Instability in LMXB Accretion Disks: How Strange is GRS 1915+105?}
\newcommand{\indexq}{\index{q-value@$q$-value}}

\makeindex

\begin{document}
\pagenumbering{gobble}
\begin{titlepage} % Suppresses displaying the page number on the title page and the subsequent page counts as page 1
	\newcommand{\HRule}{\rule{\linewidth}{0.5mm}} % Defines a new command for horizontal lines, change thickness here
	
	\center % Centre everything on the page
	
	%------------------------------------------------
	%	Headings
	%------------------------------------------------
	
	\textsc{\LARGE University of Southampton}\\[1.5cm] % Main heading such as the name of your university/college
	
	\textsc{\Large Faculty of Engineering and Physical Sciences}\\[0.5cm] % Major heading such as course name
	
	\textsc{\underline{\large Department of Physics and Astronomy}}\\[0.5cm] % Minor heading such as course title
	
	%------------------------------------------------
	%	Title
	%------------------------------------------------
	
	\HRule\\[0.4cm]
	
	{\huge\bfseries \thesistitle}\\[0.4cm] % Title of your document
	
	\HRule\\[1.5cm]
	
	%------------------------------------------------
	%	Author(s)
	%------------------------------------------------
	
	%\begin{minipage}{0.4\textwidth}
	%	\begin{flushleft}
	%		\large
	%		\textit{Author}\\
	%		B.J. \textsc{J.M.C. Court} % Your name
	%	\end{flushleft}
	%\end{minipage}
	%~
	%\begin{minipage}{0.4\textwidth}
	%	\begin{flushright}
	%		\large
	%		\textit{Supervisor}\\
	%		Dr. Diego \textsc{Altamirano} % Supervisor's name
	%	\end{flushright}
	%\end{minipage}
	
	% If you don't want a supervisor, uncomment the two lines below and comment the code above
	{\large\textit{Author}}\\
	\textbf{James Matthew Christopher \textsc{Court}}\\ % Your name
	ORCID ID \href{https://orcid.org/0000-0002-0873-926X}{0000-0002-0873-926X}\\[1.5cm]
	Thesis for the degree of\\\underline{Doctor of Philosophy}
	
	%------------------------------------------------
	%	Date
	%------------------------------------------------
	
	\vfill\vfill\vfill % Position the date 3/4 down the remaining page
	
	{\large\today} % Date, change the \today to a set date if you want to be precise
	
	%------------------------------------------------
	%	Logo
	%------------------------------------------------
	
	%\vfill\vfill
	%\includegraphics[width=0.2\textwidth]{placeholder.jpg}\\[1cm] % Include a department/university logo - this will require the graphicx package
	 
	%----------------------------------------------------------------------------------------
	
	\vfill % Push the date up 1/4 of the remaining page
	
\end{titlepage}

\thispagestyle{empty}

\cleardoublepage

\chapter*{Abstract}

\par Low Mass X-Ray Binaries (LMXBs) are systems in which a black hole or neutron star accretes matter from a stellar binary companion.  The accreted matter forms a disk of material around the compact object, known as an accretion disk.  The X-ray properties of LMXBs show strong variability over timescales ranging from milliseconds to decades.  Many of these types of variability are tied to the extreme environment of the inner accretion disk, and hence an understanding of this behaviour is key to understanding how matter behaves in such an environment.  GRS 1915+105 and MXB 1730-335 (also known as the Rapid Burster) are two LMXBs which show particularly unusual variability.  GRS 1915+105 shows a large number of distinct `classes' of second-to-minute scale variability, consisting of repeated patterns of dips and flares.  The Rapid Burster on the other hand shows `Type II X-ray Bursts'; second-to-minute scale increases in X-ray intensity with a sudden onset and a slower decay.  For many years both of these objects were thought to be unique amongst all known LMXBs.  More recently, two new objects, IGR J17091-3624 and GRO J1744-28 (also known as the Bursting Pulsar) have been shown to display similar behaviour to those seen in GRS 1915+105 and the Rapid Burster respectively.

\par In this thesis, I first present a new framework with which to classify variability seen in IGR J17091-3624.  Using my set of independent variability classes constructed for IGR J17091-3624, I perform a study of the similarities and differences between this source and GRS 1915+105 to better constrain their underlying physics.  In GRS 1915, hard X-ray emission lags soft X-ray emission in all variability classes; in IGR J17091, I find that the sign of this lag is different in variability classes.  Additionally, while GRS 1915+105 accretes at close to its Eddington Limit, I find that IGR J17091-3624 accretes at only $\sim5$--33\% of its Eddington Limit.  With these results I rule out any models which require near-Eddington accretion or hard corona reacting to the disk.  I also perform a study of the variability seen in the Bursting Pulsar.  I find that the flaring behaviour in the Bursting Pulsar is significantly more complex than in the Rapid Burster, consisting of at least 4 separate phenomena which may have separate physical origins.  One of these phenomena, `Structured Bursting', consists of patterns of flares and dips which are similar to those seen in GRS 1915+105 and IGR J17091-3624.  I compare these two types of variability and discuss the possibility that they are caused by the same physical instability.  I also present the alternative hypothesis that Structured Bursting is a manifestation of `hiccup' variability; a bimodal flickering of the accretion rate seen in systems approaching the `propeller' regime.

\thispagestyle{empty}

\cleardoublepage
\pagenumbering{roman}
    \setcounter{page}{1}

\tableofcontents

\cleardoublepage

\listoffigures
\addcontentsline{toc}{chapter}{List of Figures}

\cleardoublepage

\listoftables
\addcontentsline{toc}{chapter}{List of Tables}

\raggedright

\cleardoublepage
%
%\include{preface}
%
%\cleardoublepage

\chapter*{Dedication}

\addcontentsline{toc}{chapter}{Dedication}

\par To the memory of my brother Christopher, whose name will forever appear alongside my own.

\cleardoublepage

\chapter*{Acknowledgements}

\addcontentsline{toc}{chapter}{Acknowledgements}

\par This work was made possible by financial support from Science and Technology Facility Council (STFC) and the Royal Astronomical Society (RAS).
\par I would like to express sincere gratitude to my supervisor Dr. Diego Altamirano, referred to in this thesis as \textsf{D.A.}.  Without his experience, \textit{incredible} patience and willingness to push me to improve myself, this work would not have been possible.  I also thank Dr. Phil Uttley and Dr. Michael Childress for their roles as examiners during my PhD viva.
\par I would like to thank Professor Tomaso Belloni, Professor Ranjeev Misra and Dr Andrea Sanna for hosting me at their respective institutes at various times in my studies.  I would also like to acknowledge the co-authors on papers I have produced during this PhD:
\begin{itemize}
\item Professor Tomaso Belloni, referred to in this thesis as \textsf{T.B.}, for many insightful discussions regarding IGR J17091-3624 and assistance with producing Figures \ref{fig:global_ob}, \ref{fig:ob_evo1} and \ref{fig:ob_evo2}.
\item Dr Andrea Sanna, referred to in this thesis as \textsf{A.S.}, for contributing pulsar pulse analysis to the results presented in Chapters \ref{ch:BPbig} and \ref{ch:BPletter}.
\item Arianna Albayati, referred to in this thesis as \textsf{A.A.}, for performing the first round of analysis on the bursts in the Bursting Pulsar, and conceiving of the four classes presented in Chapter \ref{ch:BPbig}.
\item Professor Kazutaka Yamaoka, referred to in this thesis as \textsf{K.Y.}, for assisting in the reduction of \textit{Suzaku} data for work presented in Chapters \ref{ch:IGR} and \ref{ch:BPbig}.
\item Dr. Margarita Pereyra and Dr. Nathalie Degenaar, referred to in this thesis as \textsf{M.P.} and \textsf{N.D.} respectively, for assisting in the reduction of \textit{Chandra} data for work presented in Chapters \ref{ch:IGR} and \ref{ch:BPbig}.
\item Dr. Chris Boon, referred to in this thesis as \textsf{C.B.}, for assisting in the reduction and analysis of \textit{INTEGRAL} presented in Chapter \ref{ch:IGR}.
\item Dr. Adam Hill, for assisting in the reduction of \textit{Fermi} data.
\item Toyah Overton, referred to in this thesis as \textsf{T.O.}, for performing hardness-intensity analysis of the bursts in the Bursting Pulsar presented in Chapter \ref{ch:BPbig}.
\item Professor Rudy Wijnands, Professor Christian Knigge, Dr. Mayukh Pahari and Professor Omer Blaes for useful discussions and comments.
\end{itemize}
\par I also thank other members of the Southampton astronomy group for their support, including Professor Poshak Gandhi, Professor Tony Bird, Dr. Matt Middleton, Dr. Charlotte Angus, Marta Venanzi and Simon Harris (for Knowing How To Make Computers Do Things).  I would also like to thank my undergraduate tutor, Professor Steve King.
\par On a more personal note, I would like to thank my family for their unwavering support during this at-times arduous task.  I would like to thank Jacob Blamey, Rory Brown, Simon Duncan, Mahesh Herath, Sam Jones, David Williams, Ryan Wood \& Paul Wright for helping me to survive the Master's Degree that enabled me to get to this point.  I thank the new friends I have made during my time in the time in the department, including (but not limited to) Pip Grylls, Steven Browett, Bella Boulderstone, Dr. John Coxon, Michael Johnson (who is the worst), Lisa Kelsey, Sam Mangham, Pete Boorman \& Dr. Aarran Shaw.  All of you have helped me immensely, whether you realise it or not.
\par I thank my high school physics teacher Colin Piper for his enthusiasm which cemented my place on this path through academia.  I thank my undergraduate tutor Professor Steve King for helping me get up to speed with courses I had missed during a difficult second year.  And I thank my student mentor Susannah Wettone for 8 years of helping me cope with an at-times tremendously difficult studentship.
\par I also thank the staff of Titchfield Haven National Nature Reserve, Lymington and Keyhaven Marshes Local Nature Reserve, and Farlington Marshes Wildlife Reserve, for maintaining these beautiful places and giving me somewhere calm to visit at the end of stressful weeks.
\par Finally I thank my mother and father for their undying support throughout this entire process.  My father for believing in me, for his constant pushing for me to succeed, and for spending an entire day sat in a car in insect-infested Delaware to feed my birdwatching habit.  My mother for her care and patience, for years of driving to Southampton on a weekly basis to make sure everything was okay and for always being a 2 hour train ride away with a roast dinner and a chicken \& chorizo.

\cleardoublepage

\chapter*{Declaration of Authorship}

\addcontentsline{toc}{chapter}{Declaration of Authorship}

I, James Matthew Christopher Court, declare that this thesis entitled \textit{\thesistitle} and the work presented herein are my own and has been generated by me as the result of my own original research. I confirm that:

\begin{itemize}
	\item{This work was done wholly or mainly while in candidature for a research degree at this University;}
	\item{Where any part of this thesis has previously been submitted for a degree or any other qualification at this University or any other institution, this has been clearly stated;}
	\item{Where I have consulted the published work of others, this is always clearly attributed;}
	\item{Where I have quoted from the work of others, the source is always given. With the exception of such quotations, this thesis is entirely my own work;}
	\item{I have acknowledged all main sources of help;}
	\item{Where the thesis is based on work done by myself jointly with others, I have made clear exactly what was done by others and what I have contributed myself;}
	\item{Parts of this work have been published as:}
\begin{itemize}
		\item{Chapter \ref{ch:IGR}: \textit{	
	An atlas of exotic variability in IGR J17091-3624: a comparison with GRS 1915+105}, 2017 MNRAS 468 4748-4771}, hereafter \citet{IGR}.
		\item{Chapter \ref{ch:BPbig}: \textit{The Evolution of X-ray Bursts in the "Bursting Pulsar" GRO J1744-28}, 2018 MNRAS 481 2273-2239}, hereafter \citet{BPpaper}.
		\item{Chapter \ref{ch:BPletter}: \textit{The Bursting Pulsar GRO J1744-28: the slowest transitional pulsar?}, 2018 MNRASL 477 L106-L110}, hereafter \citet{BPletter}.
\end{itemize}
\end{itemize}

Signed: .............................................

Date: ...............................................
\cleardoublepage
\newpage{   }

\setcounter{chapter}{0}
	\clearpage
    \setcounter{page}{0}
\pagenumbering{arabic}
\setcounter{page}{1}

\chapter{Introduction}

\epigraph{\textit{Light thinks it travels faster than anything but it is wrong. No matter how fast light travels, it finds the darkness has always got there first, and is waiting for it.}}{Terry Pratchett -- \textit{Reaper Man}}

\vspace{1cm}
\par\noindent In this thesis, I discuss the physics of matter in close proximity to neutron stars\index{Neutron star} and black holes\index{Black hole}.  These astrophysical entities, collectively referred to as `compact objects'\index{Compact object}, are the densest objects known to exist in our universe, and are formed in the death throes of massive stars.
\par When a star with a mass between $\sim8$--10\,$M_\odot$\footnote{1\,$M_\odot\approx2\times10^{30}$\,kg, or one times the mass of our Sun.} (e.g. \citealp{Bildsten_NS}) runs out of nuclear fuel in its core, it is no longer able to support its own weight and collapses inwards.  This collapse generates a shockwave which disrupts the star, resulting in most of the star being ejected in an event known as a supernova\index{Supernova}.  The core of the star survives this disruption and continues collapsing.  The core of a massive star is supported by electron degeneracy pressure; a pressure caused by the fact that no two fermions can occupy the same quantum state \citep{Pauli_Exclusion}.  However, during the collapse of the core in a supernova, even electron degeneracy pressure cannot support the star; when the core has a mass greater than 1.4\,M$_\odot$ (the Chandrasekhar Limit, \citealp{Chandrasekhar_Mass}), electrons merge with protons via inverse $\beta$-decay, forming an object supported mostly by neutron degeneracy pressure.  The resulting `neutron star'\index{Neutron star} is an extremely dense object, with a mass of several M$_\odot$ compressed into a sphere with a radius of $\sim20$\,km.  Additionally as the core collapses, it spins up to conserve angular momentum until it is rotating at a rate of $\sim100$\,Hz.  The extreme gravitational field in the proximity of such a strong object results in a region of space which is strongly affected by the effects predicted by general relativity \citep{Einstein_GR}\index{General relativity}.  The extremely rapid rotation of neutron stars, and the associated high-velocity electron and proton populations present in their cores (e.g. \citealp{Alpar_Impure}), can result in magnetic fields as strong as $\sim10^{15}$\,G\footnote{1 Gauss, or 1\,G, is equal to 10$^{-4}$ Tesla, where Tesla is the SI-derived unit of magnetic field strength.} \citep{Woltjer_NSB,Gold_Pulsar,Kaspi_Magnetar}.  For a collapsing star with a mass of greater than $\sim10$\,M$\odot$, the end product is even more extreme.  The core of such a star can become so dense during a supernova that even neutron degeneracy pressure cannot support it, and instead it collapses into a black hole\index{Black hole}; a region of space with such a strong gravitational field that no information can escape it.
\par Unfortunately, compact objects\index{Compact object} are inherently faint objects.  In fact, an isolated black hole\index{Black hole} is theoretically only visible via the effects its gravitational well has on the light from stars located behind it.  As such, observational research into these objects tends to focus one of two types of system: Active Galactic Nuclei (AGN) and X-Ray Binaries (XRBs)\index{X-ray binary}.  In both of these types of system a compact object gravitationally attracts matter from its surrounding enviroment, a process known as `accretion'\index{Accretion}.  The act of matter falling into such a steep gravitational well causes large amounts of energy to be released; as such, these systems shine brightly in high-energy regions of the electromagnetic spectrum such as the X-rays and $\gamma$-rays.
\par AGN\index{Active galactic nucleus} contain supermassive black holes\index{Black hole}\index{Black hole!Supermassive black hole} with masses upwards of $\sim10^6$\,$M_\odot$ (e.g. \citealp{Miyoshi_SMBH}).  These black holes are believed to be present at the centre of all large galaxies but many, such as Sagittarius A$^\star$ in our Milky Way, are currently dormant and not significantly accreting\index{Accretion} \citep{LyndenBell_Quasar,Schodel_SagA}.  AGN are the brightest persistent sources of electromagnetic radiation in the universe, and they launch powerful `jets' of matter out to distances of many kiloparsec (kpc\footnote{$1$\,kpc$ =1000$\,parsec $\approx3\times10^{19}$\,m.  A parsec is the distance of an object that shows a parallax of 1'' (1 arcsecond, or $\frac{1}{3600}$ of a degree) against background objects when viewed from opposing points along the orbit of the Earth.}).  AGN have been implicated as having an important role in the development of their host galaxies via a process known as AGN feedback, in which mechanical and electromagnetic power from the AGN is `fed back' into its host galaxy and influences its evolution.
\par Active Galactic Nuclei are very distant systems.  Because of the large size of these objects, they also only evolve over timescales of thousands of years.  These facts make studying some of the properties of matter in a relativistic regime difficult to determine by only observing AGN.  Thankfully, there exists a population of bright, accreting\index{Accretion} compact objects\index{Compact object} much closer to home: XRBs\index{X-ray binary}.

\section{Anatomy of an X-Ray Binary}

\index{LMXB|see {X-ray binary, Low mass}}
\index{IMXB|see {X-ray binary, Intermediate mass}}
\index{HMXB|see {X-ray binary, High mass}}
\index{XRB|see {X-ray binary}}
\par In this thesis, I will be focusing on XRBs\index{X-ray binary}.  These systems are physically much smaller than AGN\index{Active galactic nucleus}\index{AGN|see {Active galactic nucleus}}, with compact objects\index{Compact object} no more massive than $\sim20\,M_\odot$, but in many ways they can be more extreme.  The gravitational tidal forces close to the compact object are greater than in AGN and, due to their small size, XRBs can evolve rapidly over timescales of seconds or less.
\par An XRB is a system containing a compact object\index{Compact object}\footnote{A black hole\index{Black hole} or a neutron star\index{Neutron star}.  Similar systems with a white dwarf\index{White dwarf} as their compact object are referred to as Cataclysmic Variables (CVs)\index{Cataclysmic variable}\index{CV|see {Cataclysmic variable}}.} and a main sequence or giant companion star\index{Companion star}\index{Stellar companion|see {Companion star}}.  By various processes, matter is lost from the companion star and transferred onto the compact object.  In order to conserve angular momentum\index{Angular momentum}, matter cannot simply fall onto the compact object; instead this matter spirals inwards, forming a large disk of material.  Frictional forces in the inner portions heat this `accretion disk'\index{Accretion disk} to extreme temperatures $\gtrsim1$\,keV\footnote{1\,keV$=1000$\,eV$=1.6\times10^{-16}$\,J .  1\,eV (electron-Volt) is the amount of energy an electron gains by crossing a potential difference of 1\,V.  Although this is a unit of energy, it is often used in high-energy physics to denote temperature by describing the energy at which the emission of a black body at that temperature is peaked.  1\,keV corresponds to a temperature of $\sim1.16\times10^7$\,K}.  In some XRBs, so much X-ray radiation is released in this process that the pressure from photons\index{Radiation pressure}, which is negligible but non-zero under standard conditions, becomes important to describe the equation of state of the disk.

\subsection{Types of X-Ray Binaries: High and Low-Mass}

\par XRBs are divided into two broad categories depending on the mass of the companion star\index{Companion star} and, in turn, the predominant mechanism responsible from transferring matter from the star to the compact object\index{Compact object}.  High Mass X-ray Binaries (HMXBs)\index{X-ray binary!High mass} have a companion star with a mass $\gtrsim10\,M_\odot$.  High mass stars tend to be unstable, and these objects can eject large quantities of matter in a stellar wind.  In a HMXB, part of this stellar wind is gravitationally captured by the compact object and feeds the accreting\index{Accretion} compact object.
\par Low Mass\index{X-ray binary!Low mass} and Intermediate Mass\index{X-ray binary!Intermediate mass} X-Ray Binaries (LMXBs/IMXBs), systems in which the mass $M$ of the companion star is $M\lesssim1\,M_\odot$ and 1M$_\odot\gtrsim M\lesssim10$M$_\odot$ repsectively, accrete\index{Accretion} matter in a different way.  Each object in an astrophysical binary system has a Roche Lobe\index{Roche lobe}: a teardrop-shaped region of space in which it is gravitationally dominant.  Inside the Roche Lobe matter is gravitationally bound to the central star, while matter outside of the lobe is free to escape.
\par Under some circumstances, it is possible for a star to become larger than its Roche lobe.  This can happen in two main ways:
\begin{enumerate}
\item The radius of the binary orbit decreases, shrinking the Roche Lobe of each object.
\item The radius of the star increases.  This can happen, for example, when the star evolves from the Main Sequence onto the Giant branch.
\end{enumerate}
In either scenario, a portion of the star ends up within the Roche lobe\index{Roche lobe} of the compact object\index{Compact object}.  This matter is free to spiral onto the compact object, forming the accretion disk\index{Accretion disk} (e.g. \citealp{Lewin_SSRev}).

\subsection{Components of a Low Mass X-Ray Binary}

\par
\begin{figure}
    \includegraphics[width=\columnwidth, trim = 0mm 0mm 0mm 0mm]{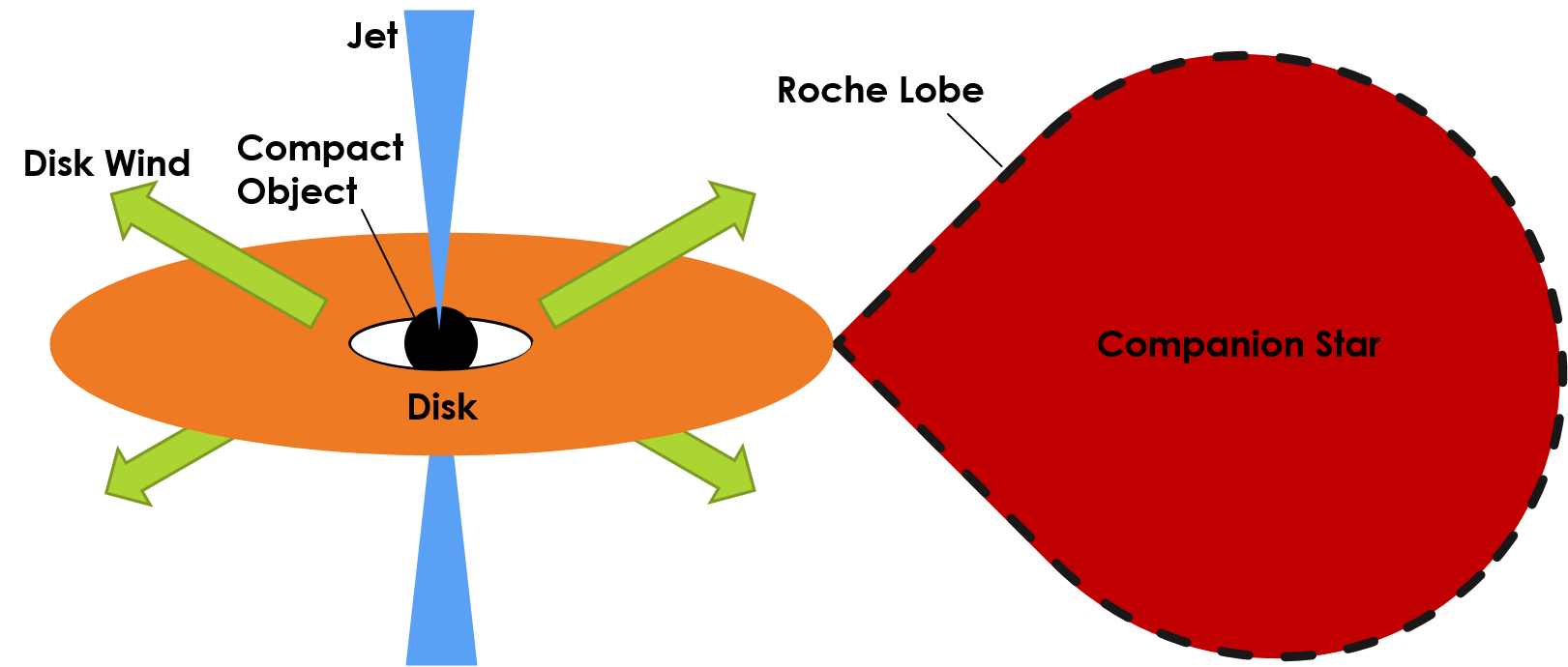}
    \captionsetup{singlelinecheck=off}
    \caption[A cartoon illustrating the basic geometry of a simple X-ray binary.]{A cartoon illustrating the basic geometry of a simple low mass X-ray binary\index{X-ray binary!Low mass}.  Not shown is the non-thermal corona\index{Corona} of material which can be inferred from spectroscopy, as the geometry of this feature is disputed.  Diagram not to scale.}
   \label{fig:xrbcartoon}
\end{figure}

\begin{figure}
   \centering
    \includegraphics[width=0.6\columnwidth, trim = 0.1mm 0.2mm 0.1mm 0.2mm, clip]{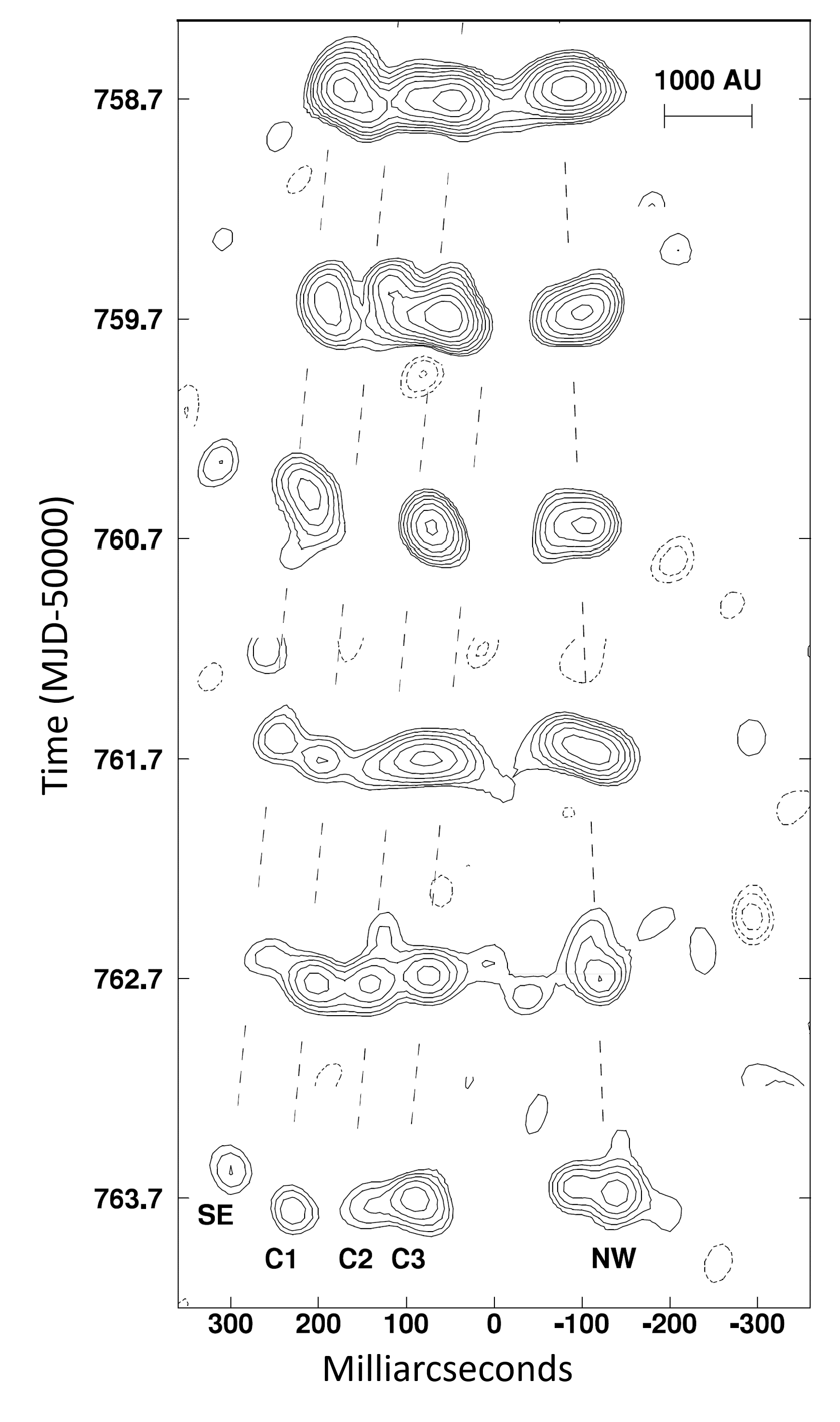}
    \captionsetup{singlelinecheck=off}
    \caption[A series of 5\,GHz radio images from \citet{Fender_1915} showing a jet being launched from the LMXB GRS 1915+105.]{A series of 5\,GHz radio images, advancing in time from top to bottom, showing a two-lobed jet of material flowing away from GRS 1915+105\index{GRS 1915+105} (at 0 milliarcseconds) at speeds approaching $c$.  Figure adapted from \citet{Fender_1915}.}
   \label{fig:jet}
\end{figure}

 As well as the accretion disk\index{Accretion disk}, there are several additional features present in a typical X-Ray Binary; I show a schematic of an LMXB in Figure \ref{fig:xrbcartoon}.   Radio observations of nearby XRBs (e.g. \citealp{Mirabel_Microquasar,Geldzahler_Jet}) have shown that these systems can show axial jets\index{Jet} of material similar to those seen in AGN; in Figure \ref{fig:jet} I show a radio image from \citet{Fender_1915} showing a jet being launched from the LMXB GRS 1915+105\index{GRS 1915+105}. These jets can eject matter at velocities approaching the speed of light $c$ (e.g. \citealp{Mirabel_Microquasar}).

\par
\begin{figure}
   \centering
    \includegraphics[width=\columnwidth, trim = 10mm 0mm 10mm 10mm, clip]{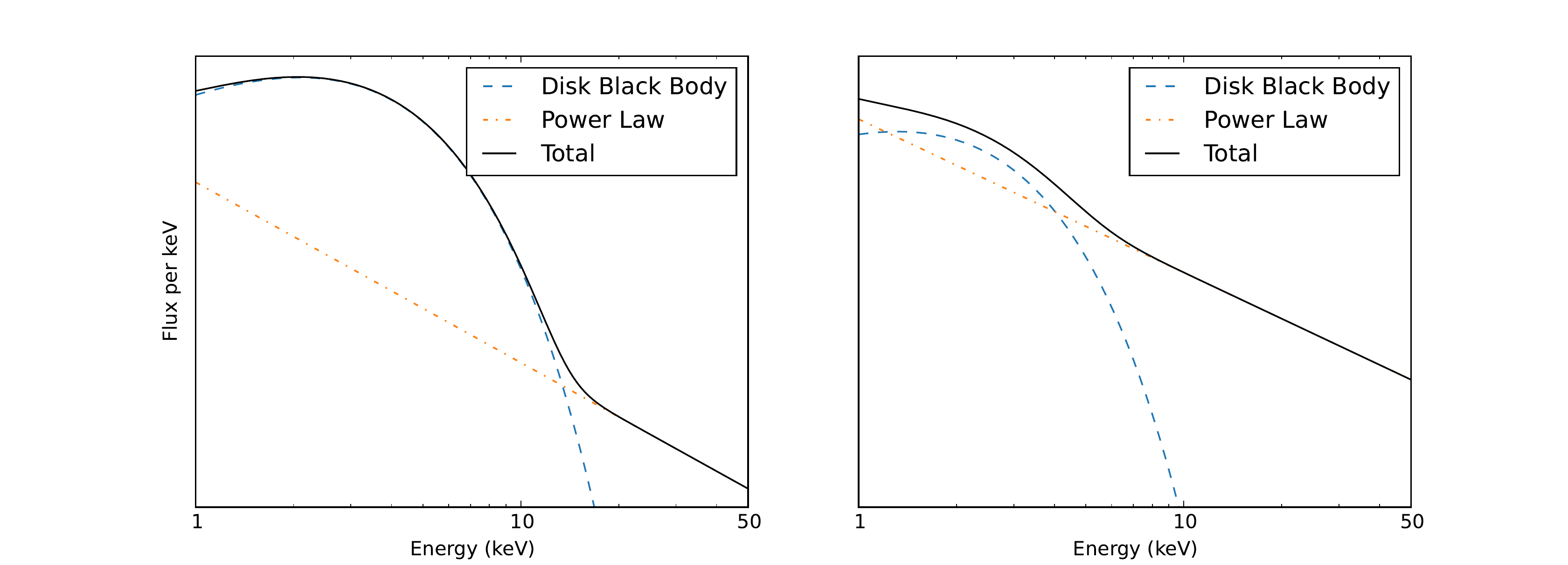}
    \captionsetup{singlelinecheck=off}
    \caption[Two simulated, simplified spectra of an LMXB, showing the two main components visible in X-ray: the accretion disk and the corona.]{Two simulated, simplified spectrum of an LMXB, showing the two main components visible in X-ray: the accretion disk\index{Accretion disk} (blue) and the corona\index{Corona} (orange).  The disk is generally modelled as a disk black body, a sum of black bodies at different temperatures corresponding to different annuli in the disk (e.g. \citealp{Mitsuda_diskbb}), while the corona is modeled as a power law.  The left panel shows a typical spectrum of an LMXB in the high/soft state, while the right panel shows a typical spectrum of an LMXB in the low/hard state: see Section \ref{sec:states} for a discussion of accretion states.  These spectra are based on spectral fits to the LMXB MXB 1658-298, performed by \citet{Sharma_vals}.}
   \label{fig:toyspec}
\end{figure}

 X-ray spectral studies of LMXBs find that, in addition to a black-body\footnote{The radiative power per unit frequency of a black body at temperature $T$ is given by \[F_T(\nu)=\frac{N\nu^3}{e^\frac{h\nu}{k_BT}-1}\] for some constant $N$ \citep{Planck}.  $k_B$ is the Boltzmann Constant, and $c$ is the speed of light in a vacuum.} like accretion disk\index{Accretion disk}, the systems must each contain a non-thermal `corona'\index{Corona} component.  The corona is a region of non-thermal electrons somewhere in the vicinity of the compact object\index{Compact object}, and it emits X-rays via Compton upscattering\index{Compton scattering}.  In this process, photons emitted from the disk collide with energetic electrons in the corona.  The photons, on average, gain energy from these collisions and are scattered back into space; some in the direction of observers on the Earth.  This leads to a characteristic power-law\footnote{A power-law distribution is any distribution with the functional form $f(x)=cx^k$ for some constants $c$ and $k$.} energy distribution signature at high energies, which can be seen in the spectra of LMXBs.  As I show in the simulated LMXB energy spectrum in Figure \ref{fig:toyspec}, the emission from the corona tends to dominate above energies of $\sim10$\,keV.
\par Models of the geometry of the coronal region have evolved over the years.  While the corona has been historically treated as if it was a single point fixed above the centre of the disk\index{Accretion disk} (the so-called `Lamp Post' model, e.g. \citealp{Rozanska_Lamppost}), more recent models tend to treat it either as an optically thin\footnote{An optically thin medium is defined as a medium in which an average photon interacts $<1$ times while passing through.} flow of material onto the compact object\index{Compact object} or equate it with the base of the radio jet\index{Jet} (e.g. \citealp{Skipper_CoronaGeo}).

%\begin{figure}
%   \centering
%    \includegraphics[width=0.7\columnwidth, trim = 10mm 0mm 10mm 10mm, clip]{images/toy_spec.eps}
%    \captionsetup{singlelinecheck=off}
%    \caption[A simulated, simplified spectrum of an LMXB in the high/soft state, showing the two main components visible in X-ray: the accretion disk and the corona.]{A simulated, simplified spectrum of an LMXB in the high/soft state  (see Section \ref{sec:states}), showing the two main components visible in X-ray: the accretion disk (blue) and the corona (orange).  The disk is generally modelled as a disk black body, a sum of black bodies at different temperatures corresponding to different annuli in the disk (e.g. \citealp{Mitsuda_diskbb}), while the corona is modeled as a power law.  Spectrum based on spectral fits to the LMXB MXB 1658-298, performed by \citet{Sharma_vals}.}
%   \label{fig:toyspec}
%\end{figure}

\par Another important component of an X-ray binary\index{X-ray binary} is the disk wind\index{Wind}\index{Disk wind|see {Wind}} \citep{vanParadijs_Wind}.  Due to the high temperatures and pressures in the inner part of the accretion disk\index{Accretion disk}, matter on the surface of the disk can obtain enough energy to escape the gravitational well of the compact object\index{Compact object}.  This matter is ejected from the system in large-scale, high velocity winds.  Studies of the spectral lines present in these winds have shown that they can have speeds approaching the speed of light (e.g. \citealp{Ponti_Wind,Degenaar_BPSpec}).

\subsubsection{Neutron Star X-ray Binaries}

\label{sec:NSintro}

\par The geometry of an X-ray binary is somewhat more complicated when the compact object\index{Compact object} is a neutron star\index{Neutron star}.  Unlike black holes, neutron stars are in general highly-magnetised systems, and the introduction of a large, strong magnetic field\index{Magnetic field} to an XRB has implications for the geometry of the accretion\index{Accretion} flow.  At some radius in the inner accretion disk\index{Accretion disk}, it is possible that the pressure\index{Magnetic pressure} exerted by this magnetic field becomes dominant over the gas\index{Ram pressure} and photon pressures\index{Radiation pressure}\index{Photon pressure|see {Radiation pressure}}.  At this point, ionised material becomes `frozen-in' to the magnetic field lines, and is only able to freely move along them.  Due to the extreme temperatures present in the inner portion of the accretion disk\index{Accretion disk}, the vast majority of material in this region is ionised.  This acts to disrupt the flow of material in the inner part of the accretion disk, and matter is funneled along field lines and onto the poles of the neutron star.  This causes the poles of the neutron star to become extremely hot.  As the neutron star spins, it appears to pulse as seen by an external observer due to the highly radiating magnetic poles coming in and out of view.  These objects are referred to as accreting X-ray pulsars\index{Pulsar}.
\par In addition to the effects of the magnetic field, there is another significant difference between neutron star\index{Neutron star} and black hole\index{Black hole} binaries.  Black holes are surrounded by an event horizon\index{Event horizon} from which no light can emerge, therefore there can be no direct emission from the compact object in a black hole X-ray binary.  Neutron stars on the other hand have a visible surface.  As such the surface of the neutron star itself, and any phenomena that take place there, can in principle be seen.
\par One of the most spectacular events that can occur on the surface of a neutron star\index{Neutron star} is a Type I X-ray burst \citep{Grindlay_TypeI}\index{X-ray burst!Type I}.  These occur when matter accreted\index{Accretion} onto the surface of the neutron star reaches a critical temperature and density ($\sim2.2\times10^9$\,K and $3\times10^6$\,g\,cm$^{-3}$, \citealp{Joss_TypeI}), and nuclear fusion is triggered.  This results in a flash of energy, which causes a runaway thermonuclear\index{Thermonuclear burning} explosion across most or all of the neutron star surface.  Type I bursts appear in data as a sudden increase in X-ray flux (1--2 orders of magnitude), followed by an power-law decay as the neutron star surface cools.  As Type I bursts are distinctive features which require a surface on which to occur, they are often used as a diagnostic tool to identify an unknown compact object\index{Compact object} as a neutron star.

\section{Low Mass X-Ray Binary Behaviour}

\par LMXBs\index{X-ray binary!Low mass} are not static systems, and most show variations in their luminosities over timescales of milliseconds to years.  Broadly speaking, LMXBs can be divided into persistent systems\index{Persistent source} and transient\index{Transient source} systems.  Persistent systems have always observed to be bright since their discovery, implying a high rate\index{Accretion rate} of accretion\index{Accretion} at all times.  In some objects, this bright, high-accretion rate state has persisted for $\geq20$ years (e.g. GRS 1915+105\index{GRS 1915+105}, \citealp{Deegan_1915}).
\par Transient\index{Transient source} LMXBs\index{X-ray binary!Low mass} have a somewhat more complicated life cycle.  These objects spend most of their time in a `quiescent' state\index{Quiescence}, during which they are faint in X-rays and only a relatively small amount of material is being accreted\index{Accretion}.  However these objects also undergo `outbursts'\index{Outburst}, during which their luminosity increases by many orders of magnitude for a period of days to years (e.g. \citealp{Frank_Outbursts}).  The frequency of these outbursts varies widely between sources, ranging from one every month or so to one every few decades or longer.

\label{sec:states}

\par
\begin{figure}
   \centering
    \includegraphics[width=\columnwidth, trim = 10mm 13mm 10mm 15mm, clip]{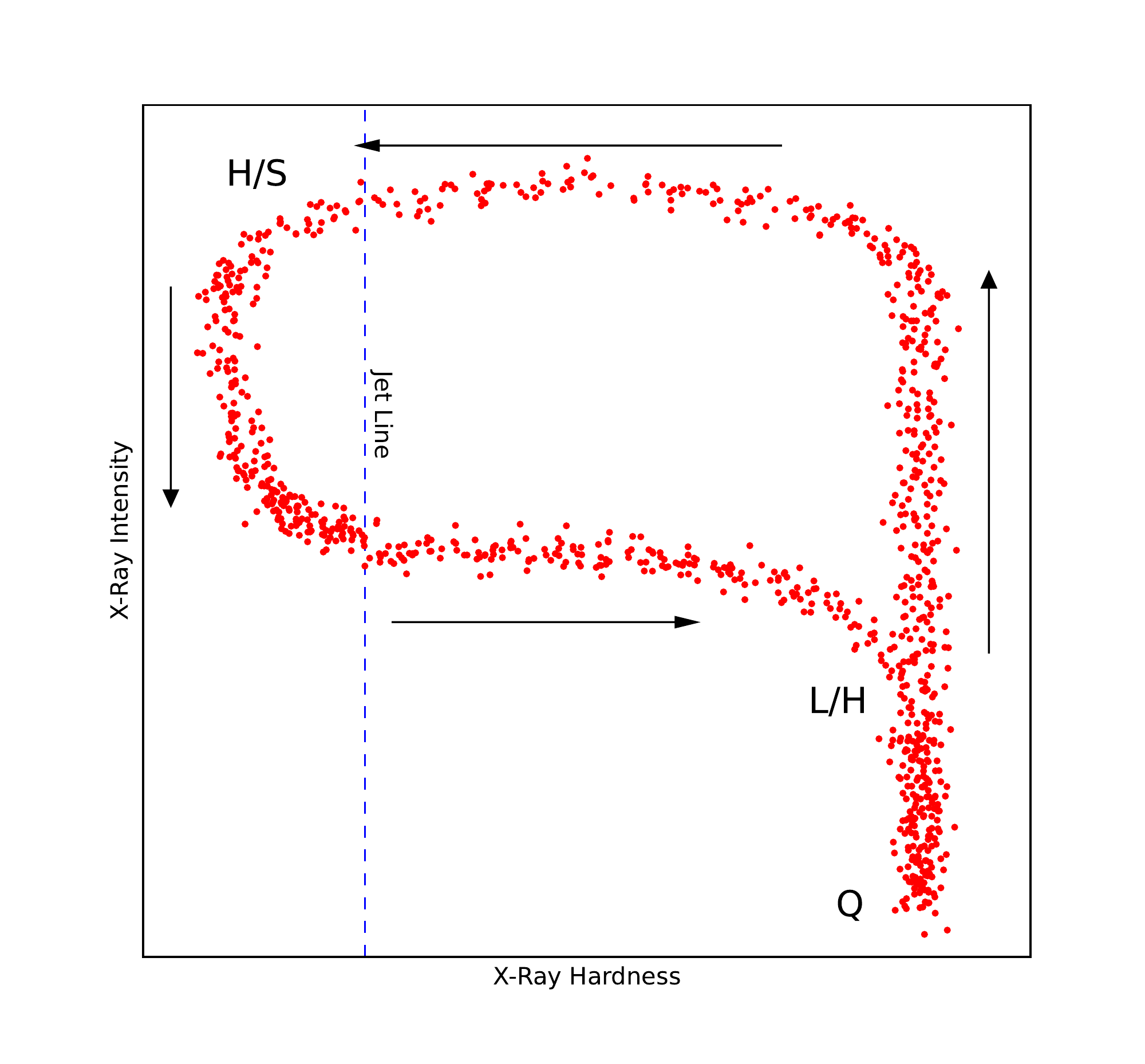}
    \captionsetup{singlelinecheck=off}
    \caption[A schematic hardness-intensity diagram adapted from \citet{Fender_UniJets}, showing the evolutionary path of a typical black hole LMXB outburst.]{A schematic hardness-intensity diagram\index{Hardness-intensity diagram} adapted from \citet{Fender_UniJets}, showing the evolutionary path of a typical black hole\index{Black hole} LMXB\index{X-ray binary!Low mass} outburst\index{Outburst} and roughly indicating the positions of quiescence (Q)\index{Quiescence} and the the Low/Hard (L/H)\index{Low/Hard state} and High/Soft (H/S)\index{High/Soft state} States.  The jet line roughly demarcates the portion of the outburst in which a jet\index{Jet} is observed (right of the line) from the portion in which it is not observed (left of the line).}
   \label{fig:Fender}
\end{figure}

\begin{figure}
   \centering
    \includegraphics[width=\columnwidth, trim = 0mm 0mm 0mm 0mm, clip]{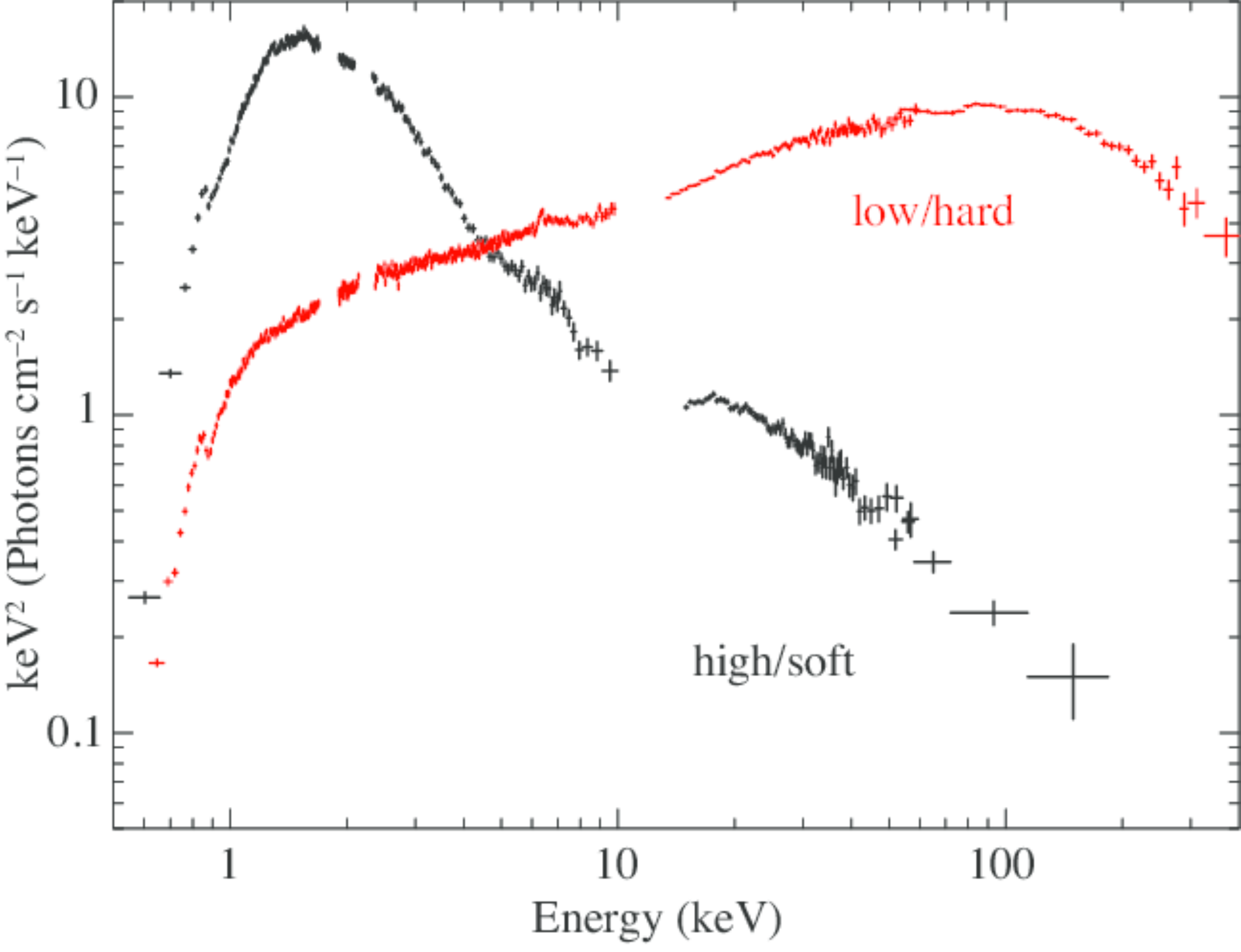}
    \captionsetup{singlelinecheck=off}
    \caption[Energy spectra of the black hole HMXB Cygnus X-1 in its low/hard and high/soft states, presented as typical spectra of a black hole XRB in these states.]{\textit{Suzaku}\indexsuzaku\ energy spectra of the black hole\index{Black hole} HMXB\index{X-ray binary!High mass} Cygnus X-1\index{Cyg X-1} in its low/hard\index{Low/Hard state} (black) and high/soft\index{High/Soft state} (red) states, presented as typical spectra of a black hole XRB in these states.  Figure taken from \citet{Yamada_Spec}.}
   \label{fig:Yamada_Spec}
\end{figure}

 XRB outbursts tend to follow predictable evolutionary paths, evolving through a number of different `states' as they progress.  I show some of the states associated with black hole\index{Black hole} LMXB\index{X-ray binary!Low mass} outbursts\index{Outburst} in Figure \ref{fig:Fender} on a so-called `hardness-intensity diagram'\index{Hardness-intensity diagram}\index{HID|see {Hardness-intensity diagram}}, which traces how the brightness and the spectral shape of a source evolve over time (see Section \ref{sec:hids} for more information on hardness-intensity diagrams).  At the start of a typical black hole LMXB outburst emission from the source is spectrally hard\index{Colour}\index{Hardness|see {Colour}}, i.e. dominated by higher-energy photons.  This part of the outburst is referred to as a Low/Hard State\index{Low/Hard state}\index{Hard state|see {Low/Hard state}} (bottom-right of Figure \ref{fig:Fender}), and a radio jet\index{Jet} is generally visible at this time.  The luminosity of the source gradually increases until it reaches some maximum, and then emission begins to become softer as the system heads towards the High/Soft State\index{High/Soft state}\index{Soft state|see {High/Soft state}} (top-right of Figure \ref{fig:Fender}).  During this transition, the system crosses the so-called `jet line', and the radio jet switches off.  Sources tend to spend a large portion of their outbursts in the high/soft state, appearing to meander in the hardness-intensity diagram.  This meandering may include additional crossings of the jet line, causing the radio jet to flicker on and off during this period.  The X-ray luminosity of the source then decreases, before the source returns to the hard state along a path of approximately constant luminosity.  The source then fades back into quiescence.  This typical outburst behaviour forms a distinctive `q' shape in the hardness-intensity diagram, as I show in Figure \ref{fig:Fender}, and can be thought of as the inner accretion disk\index{Accretion disk} filling with matter before draining onto the compact object\index{Compact object} or flowing out of the system in winds\index{Wind} or a jet\index{Jet} (e.g. \citealp{Fender_UniJets}).  I show typical spectra of an XRB in the low/hard\index{Low/Hard state} and high/soft\index{High/Soft state} states in Figure \ref{fig:Yamada_Spec}, taken from \citet{Yamada_Spec}.

\par
\begin{figure}
   \centering
    \includegraphics[width=\columnwidth, trim = 1mm 1mm 1mm 1mm, clip]{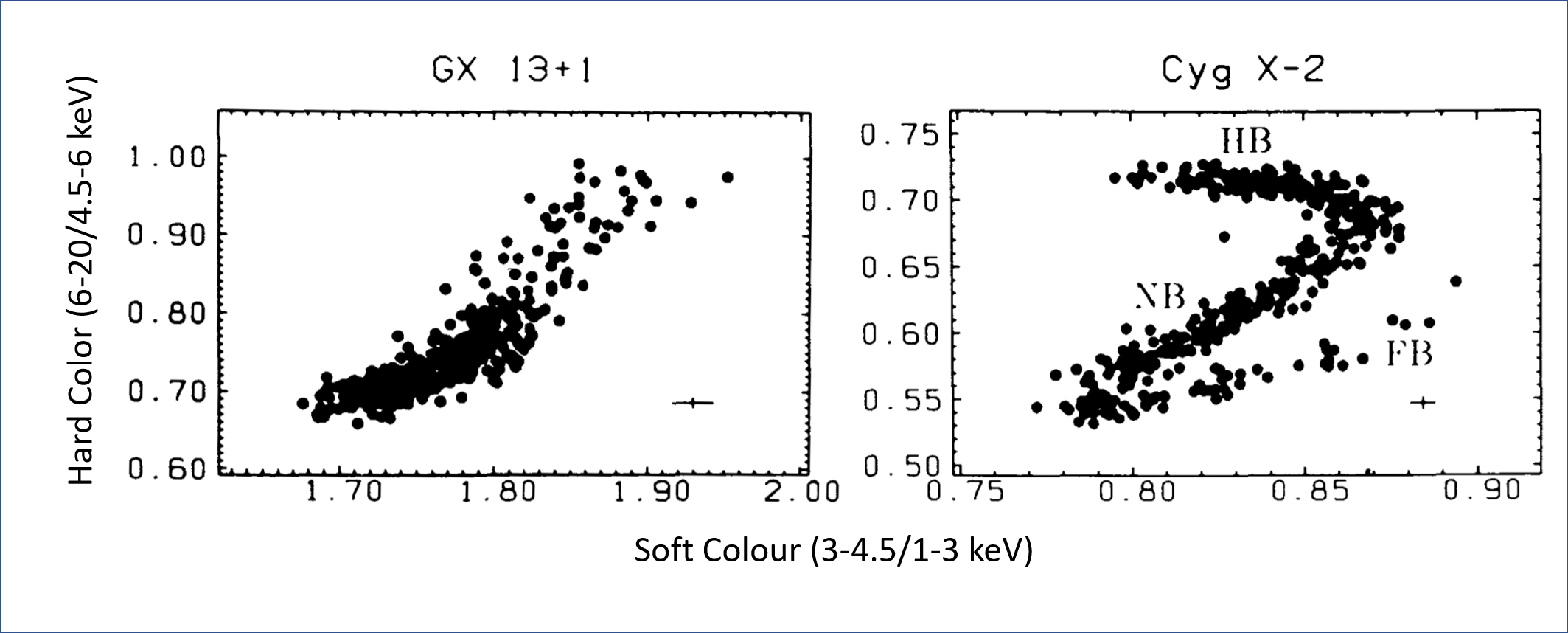}
    \captionsetup{singlelinecheck=off}
    \caption[Colour-Colour diagrams from \citet{vanderKlis_ZAtoll} showing typical evolutionary paths of Atoll-type and Z-type Neutron Star LMXBs.]{Colour-colour diagrams from \citet{vanderKlis_ZAtoll}, showing evolutionary paths of typical outbursts of Atoll-type\index{Atoll source} and Z-type\index{Z source} neutron star\index{Neutron star} LMXBs\index{X-ray binary!Low mass} (GX 13+1\index{GX 13+1} and Cyg X-2\index{Cyg X-2} respectively).  On the right-hand panel, the typical `branches' of a Z-type source are marked: the High Branch (HB), Normal Branch (NB) and Flaring Branch (FB).}
   \label{fig:Zatoll}
\end{figure}

 Neutron Star\index{Neutron star} LMXBs\index{X-ray binary!Low mass} on the other hand tend to follow one of two patterns during outburst\index{Outburst}, dividing them into so-called `Z sources'\index{Z source} and `atoll sources'\index{Atoll source} (e.g. \citealp{vanderKlis_ZAtoll}).  In Figure \ref{fig:Zatoll} I show examples of colour-colour diagrams\index{Colour-colour diagram}\index{CCD|see {Colour-colour diagram}} (which plots two different hardness ratios against each other, see Section \ref{sec:hids}) for typical Z-type and atoll-type sources.  Z sources trace out a number of `branches' during outburst, each corresponding to a period of different source behaviour.  Atoll sources on the other hand spend most of the time in the so-called `banana branch' on the colour-colour diagram, occasionally jumping over to the `island state' at larger values of hard and soft colour.  Unlike black hole LMXBs which trace out their characteristic evolutionary pattern once per outburst, Z and atoll sources trace out their evolutionary paths many times per outburst.  Z sources can complete the entire `z' over timescales of days.  Most Z sources are classified as persistent\index{Persistent source} objects, although some Z sources are transient\index{Transient source} \citep{Homan_TZ}.  On the other hand most atoll sources are transient, but some have been observed to be persistent (e.g. \citealp{Hasinger_PersAtoll}).  In addition to this, at least one source is known to change between Z- and atoll-like evolutionary patterns over time \citep{Barret_Flip}.  This complex evolution over the course of each outburst highlights the fact that accretion\index{Accretion} is not a simple process, and that understanding accretion gives us better understanding of a areas of the physics of matter in extreme environments.

\section{Relativistic Effects}

\par \index{General relativity}One of the most obvious exotic physical environments that accretion\index{Accretion} physics sheds light on is, of course, extreme gravitational fields\index{Gravitational field}.  General relativistic effects around compact objects\index{Compact object} are often expressed in relation to the gravitational radius\index{Gravitational radius} $r_g$, defined as:

\begin{equation}
r_g=\frac{GM}{c^2}
\end{equation}

Where $G$ is the gravitational constant, $c$ is the speed of light and $M$ is the mass of the compact object\index{Compact object}.  $2r_g$ is equal to the Schwarzchild radius\index{Schwarzchild radius}, or the radius of the event horizon\index{Event horizon} of a non-rotating black hole\index{Black hole} with mass $M$ \citep{Schwarzschild}.
\par One result of general relativity which is important when considering compact object\index{Compact object} accretion disks\index{Accretion disk} is the existence of an Innermost Stable Circular Orbit\index{Innermost stable circular orbit}\index{ISCO|see {Innermost stable circular orbit}}, or ISCO (e.g. \citealp{Misner_GravBook}).  This radius is at $6r_g$ from the centre of a non-rotating object, placing it well outside the event horizon\index{Event horizon} of a black hole\index{Black hole} and possibly above the surface of some neutron stars\index{Neutron star}.  It can be shown that any non-interacting point mass crossing this boundary from the outside will continue into the black hole, whereas any point mass crossing it from the inside will continue to infinity; as such, no stable orbit can exist with a periastron smaller than this radius. It can be shown that an accretion disk is also bounded by this radius \citep{Kozlowski_ISCO}, such that XRB accretion disks must all have an inner truncation radius at least this far from the compact object.  Within this radius, matter falls directly onto the compact object.
\par A black hole\index{Black hole} can be described with 3 parameters\footnote{This conjecture is often referred to as the `No-Hair' theorem.}: mass, angular momentum (or spin) and charge \citep{Israel_Bald}.  As the precursor stars to black holes are neutrally charged, it is expected that all astrophysical black holes are very close to being neutral as well.  However, these precursor stars also possess non-zero angular momentum.  As such, it is expected that most if not all astrophysical black holes are spinning.\index{Spin}  This spin is generally expressed as a number between 0 and 1, where 0 denotes a non-rotating black hole and 1 is the maximum permitted angular momentum the object can possess.
\par General relativity predicts that this spin\index{Spin} will also have a significant effect on accretion\index{Accretion} physics.  First of all, this spin changes the position of the ISCO\index{Innermost stable circular orbit}; moving it to a maximum of $9r_g$ for a retrograde black hole\index{Black hole} with spin of 1 \citep{Kerr_BH}.  A spinning black hole also distorts the space time around it, in a process known as frame-dragging \citep{Lense_Thirring}\index{Frame-dragging}.  This forces matter close to the black hole\index{Black hole} to orbit in the same plane as it.  As there is no reason to assume the outer disk\index{Accretion disk} orbits in the same plane as the black hole\index{Black hole}, this can lead to situations in which the accretion disk\index{Accretion disk} is warped, which in turn has implications for the flow of matter within it.
\par It is clear that general relativity should have observable implications on the flow of matter onto the accretion disk\index{Accretion disk}.  Studying the physics of accretion\index{Accretion} therefore allows us to measure parameters such as the spin\index{Spin} of black holes\index{Black hole} that would otherwise be inaccessible to us.  Additionally, a full understanding of the accretion onto the compact objects\index{Compact object} would allow us to look for discrepancies between what is observed and what is expected from relativity.  Therefore, a full understanding of accretion is one route to testing the theory of general relativity itself under some of the most extreme conditions in the universe.

\cleardoublepage

\chapter{The Physics of Accretion}

\label{sec:PhysAcc}

\epigraph{\textit{A black hole consumes matter, sucks it in, and crushes it beyond existence. When I first heard that, I thought that's evil in its most pure.}}{Alice Morgan -- \textit{Luther}}

\vspace{1cm}

\par\noindent The extreme environments in accreting\index{Accretion} systems lead to a variety of somewhat unintuitive physical effects and phenomena.  In this chapter I describe a number of these effects, and delve into the history of physical and mathematical models which have been proposed to explain the effects seen in X-ray binaries.

\section{The Shakura-Sunyaev Disk Model}

\par\index{Shakura-Sunyaev disk model} To try and understand the behaviour of accretion disks\index{Accretion disk}, a number of authors have constructed models.  Much of our understanding of the physics of astrophysical accretion disks stems from one of the earliest of these models, proposed by Nikolai Shakura and Rashid Sunyaev in 1973 \citep{Shakura_Disk}.  This model specifically considered the effects of accretion\index{Accretion} onto a black hole\index{Black hole}.  By showing that this would result in a system which would be bright in the X-ray, and describing how such a system would appear, this model proved pivotal in the scientific community's acceptance of the earliest XRB identifications (e.g. \citealp{Bolton_CygX1}).
\par \citeauthor{Shakura_Disk} model the accretion disk\index{Accretion disk} as a structure held up by centrifugal forces, generated by the large amount of angular momentum\index{Angular momentum} possessed by infalling matter due to the orbit of the binary system.  Frictional forces cause this angular momentum to be transferred outwards, heating up the disk and allowing matter to fall in towards the black hole\index{Black hole}.  The efficiency with which this angular momentum is transferred can be thought of as a measure of the viscosity\index{Viscosity} of the disk.
\par \citeauthor{Shakura_Disk} base their calculations on Newtonian mechanics; as such they ignore the region of the disk\index{Accretion disk} inwards of the ISCO\index{Innermost stable circular orbit} at $r=3r_g$, where relativistic effects become important.  They also assume that the disk in a steady state, that it is geometrically thin (such that height of the disk $H\ll r$ everywhere) and that it is cylindrically symmetric.  The last two assumptions allow us to write down formulae for the surface density $\Sigma$, mean radial bulk velocity $u_r$ and accretion rate\index{Accretion rate} $\dot{M}$ of the disk as a functions of radius $r$:
\begin{eqnarray}
\Sigma(r)&=&\int_{-H}^H\rho(r,z) dz\label{eq:base1}\\\nonumber \\
u_r(r)&=&\frac{1}{\Sigma(r)}\int_{-H}^H\rho(r,z)v_r(r,z)dz\label{eq:base2}\\\nonumber \\
\dot{M}(r)&=&-2\pi r\Sigma(r) u_r(r)\label{eq:base3}
\end{eqnarray}
Where $\rho(r,z)$ is the density at a radius $r$ and height $z$, and $v_r$ is the radial velocity of the gas at this point.
\par Now consider the Euler equations of hydrodynamics:
\begin{eqnarray}
\frac{\partial\rho}{dt}+\dv(\rho\bm{v})&=&0\label{eq:consmass}\\\nonumber \\
\rho\left(\frac{\partial\bm{v}}{dt}+(\bm{v}\cdot\dv) \bm{v}\right)&=&-\dv p\label{eq:fma}
\end{eqnarray}
Where Equation \ref{eq:consmass} is the conservation of mass and Equation \ref{eq:fma} is a differential form of Newton's second law of motion.  These equations can be cast in cylindrical co-ordinates to give 4 equations: the recast continuity equation and one motion equation for each of the radial ($r$), vertical ($z$) and azimuthal ($\theta$) directions:
\begin{eqnarray}
\frac{\partial\rho}{\partial t}+\frac{1}{r}\frac{\partial(r\rho v_r)}{\partial r}+\frac{1}{r}\frac{\partial v_\theta}{\partial\theta}+\frac{\partial v_z}{\partial z}&=&0\label{eq:ssc}\\\nonumber\\
\rho\left(\frac{\partial v_r}{\partial t}+v_r\frac{\partial v_r}{\partial r}+\frac{v_\theta}{r}\frac{\partial v_r}{\partial\theta}+v_z\frac{\partial v_r}{\partial z}-\frac{v_\theta^2}{r}\right)&=&\frac{-\partial p}{\partial r}\label{eq:ssr}\\\nonumber\\
\rho\left(\frac{\partial v_\theta}{\partial t}+v_r\frac{\partial v_\theta}{\partial r}+\frac{v_\theta}{r}\frac{\partial v_\theta}{\partial\theta}+v_z\frac{\partial v_\theta}{\partial z}+\frac{v_rv_\theta}{r}\right)&=&\frac{-\partial p}{\partial\theta}\\\nonumber\\
\rho\left(\frac{\partial v_z}{\partial t}+v_r\frac{\partial v_z}{\partial r}+\frac{v_\theta}{r}\frac{\partial v_z}{\partial\theta}+v_z\frac{\partial v_z}{\partial z}\right)&=&\frac{-\partial p}{\partial z}\label{eq:ssz}
\end{eqnarray}
By assuming that the disk\index{Accretion disk} is in a steady state and cylindrically symmetric, we can set all $\frac{\partial}{\partial\theta}$ and $\frac{\partial}{\partial t}$ terms to zero, simplifying equations \ref{eq:ssr} to \ref{eq:ssz}:
\begin{eqnarray}
\rho\left(v_r\frac{\partial v_r}{\partial r}+v_z\frac{\partial v_r}{\partial z}-\frac{v_\theta^2}{r}\right)&=&\frac{-\partial p}{\partial r}\label{eq:ssrs}\\\nonumber\\
\rho\left(v_r\frac{\partial v_\theta}{\partial r}+v_z\frac{\partial v_\theta}{\partial z}+\frac{v_rv_\theta}{r}\right)&=&0\label{eq:ssts}\\\nonumber\\
\rho\left(v_r\frac{\partial v_z}{\partial r}+v_z\frac{\partial v_z}{\partial z}\right)&=&\frac{-\partial p}{\partial z}\label{eq:sszs}
\end{eqnarray}
We can average the density term on left-hand side of Equation \ref{eq:ssc} in the $z$-direction, and substitute in the results from Equations \ref{eq:base1} to \ref{eq:base3} to find:
\begin{eqnarray}
\frac{1}{r}\frac{d}{dr}\left(r\int_{-H}^H\rho v_rdz\right)&=&0\\\nonumber\\
\frac{1}{r}\frac{d(r\Sigma u_r)}{dr}&=&0\\\nonumber\\
\frac{-1}{2\pi r}\frac{d\dot{M}}{dr}&=&0
\end{eqnarray}
Therefore the rate of inwards matter flow $\dot{M}$, or the accretion rate\index{Accretion rate}, is constant at all $r$.
\par Using the fact that the angular velocity $\omega$ of an element in the gas can be written as $\omega=v_\theta/r$, we can re-write Equation \ref{eq:ssrs} as:
\begin{equation}
\rho\left(v_r\frac{\partial v_r}{\partial r}-\omega^2r\right)=-\frac{\partial p}{\partial r}-\rho v_z\frac{GM}{r^2}
\end{equation}
Where the term $\frac{GM}{r^2}$ has been introduced to account for the fact that the gradient of the gravitational field\index{Gravitational field} in the $r$ direction is non-zero.  This leads to:
%We can rewrite $v_z\frac{\partial v_r}{\partial z}$ as $\frac{\partial v_r}{\partial z}\frac{\partial z}{\partial t}$, which is equal to the inwards acceleration of a gas element moving in the $z$-direction.  As the disk is assumed to be in a cylindrically symmetric gravitational potential, this acceleration can be given by $\dot{v}_r=\frac{GM}{r^2}$.
\begin{equation}
\rho\left(v_r\frac{\partial v_r}{\partial r}-\omega^2r\right)=-\frac{\partial p}{\partial r}-\rho\omega_k^2r
\end{equation}
Assuming that is thin and angular momentum is only transferred slowly, i.e. $v_r\frac{\partial v_r}{\partial r}\ll\omega$, this leads to:
\begin{equation}
\omega\approx\omega_k
\end{equation}
Showing that gas elements in the disk\index{Accretion disk} orbit at Keplerian\index{Keplerian motion} speeds.
\par Using similar logic, Equation \ref{eq:sszs} becomes:
\begin{equation}
\rho\omega_k^2 z=\frac{-\partial p}{\partial z}\label{eq:idgas}
\end{equation}
The ideal gas law $p=\rho RT$\footnote{$R$ is the specific gas constant, equal to the Boltzmann Constant $k_B$ divided by the mean molar mass of the gas.} can then be used to rewrite equation \ref{eq:idgas}:
\begin{equation}
\frac{p}{RT}\omega_k^2 z=\frac{-\partial p}{\partial z}\label{eq:idgas2}
\end{equation}
If we assume that the disk\index{Accretion disk} is chemically homogeneous and isothermal in the $z$-direction, then neither $R$ nor $T$ depend on $z$.  Equation \ref{eq:idgas2} then admits the solution:
\begin{equation}
\rho=\rho_0(r)e^{\left(\frac{-z^2\omega_k^2}{2RT}\right)}=\rho_0(r)e^{\frac{-z^2}{2H_s^2}}
\end{equation}
Where $\rho_0$ is the density at radius $r$ when $z=0$.  As such, the density of the disk\index{Accretion disk} has a Gaussian profile in the $z$-direction, with a scale-width $H_s$ given by:
\begin{equation}
H_s=\frac{\sqrt{RT}}{\omega_k}
\end{equation}
This shows that the scale height of the disk\index{Accretion disk} is finite for all $r$.  As the integral between $-\infty$ and $+\infty$ of a Gaussian with a finite scale-width is finite, the disk contains a finite amount of matter.
\par Finally, \citeauthor{Shakura_Disk} looked at the solutions to Equation \ref{eq:ssts}.  As every term in this equation depends on either $v_\theta$ or a derivative thereof, this equation admits the solutions $\rho=0$ or $v_\theta=0$.  Both of these solutions imply accretion rates\index{Accretion rate} of zero, as any matter in the disk\index{Accretion disk} must have a non-zero density and angular momentum\index{Angular momentum}.  In order to resolve this problem, \citet{Shakura_Disk} add the divergence of the viscous stress tensor \citep{Landau_Tensor} to the right-hand side of Equation \ref{eq:ssts} to represent the effects of viscosity\index{Viscosity} within the disk\index{Accretion disk}.  By doing this, they find the following two results:
\begin{eqnarray}
\dot{M}&=&\frac{4\pi H_s\eta_b r}{\omega}\frac{\partial\omega}{\partial r}\label{eq:diffrot}\\\nonumber\\
\dot{M}&=&6\pi\eta_b H_s\label{eq:viscos}
\end{eqnarray}
Equation \ref{eq:diffrot} confirms that the disk\index{Accretion disk} is a differential rotator, while Equation \ref{eq:viscos} confirms that accretion\index{Accretion} can only take place when $\eta_b$ (the bulk viscocity) is non-zero.
\par \citet{Shakura_Disk} found that molecular viscosity\index{Viscosity} alone cannot be high enough to result in the high values of $\dot{M}$ inferred for observed XRBs.  Instead, the authors assume that turbulence\index{Turbulence} is present in the disk\index{Accretion disk}.  Using formulae pertaining to turbulent hydrodynamics, and by ignoring supersonic perturbations, they find an upper bound on bulk viscosity $\eta$:
\begin{equation}
\eta_b\leq\frac{2}{3}\rho_0H\sqrt{RT}
\end{equation}
As such, they define a dimensionless viscosity parameter $\alpha$ as:
\begin{equation}
\alpha\equiv\frac{3\eta_b}{2\rho_0H\sqrt{RT}}\quad\quad\quad0<\alpha\leq1
\end{equation}

\subsection{The source of Turbulence}

\par \citet{Shakura_Disk} do not answer the question of what physical process causes the turbulence\index{Turbulence} required to stabilise accretion disks\index{Accretion disk}.  \citet{Balbus_MRI} were among the first to propose the Magnetorotational Instability \citep[MRI,][]{Velikhov_MRI,Chandrasekhar_MRI}\index{Magnetorotational instability} as the source of this turbulence.  MRI\index{MRI|see {Magnetorotational instability}} is a process which occurs in an ionised and differentially rotating disk.  Fluctuations in the material in the disk generate internal magnetic fields\index{Magnetic field}.  The field lines associated with these fields, in general, extend a finite distance in the radial direction, thus connecting gas elements at different radii.  As gas elements in a Shakura-Sunyaev accretion disk\index{Shakura-Sunyaev disk model}\index{Accretion disk} orbit the compact object\index{Compact object} at Keplerian speeds\index{Keplerian motion}, elements of gas at different radii move at different orbital speeds.  As such, these internal magnetic field lines become stretched as gas orbits the compact object.  This field line stretching imparts a torque on the gas elements, causing the outer, slower element to speed up and the inner, faster element to slow down.  As such, the net result of this process is an outwards transfer of angular momentum\index{Angular momentum}.
\par \citet{Balbus_MRI} found that the angular momentum transfer due to MRI\index{Magnetorotational instability} was more significant than that due to friction, hydrodynamic turbulence\index{Turbulence} or other sources in an accretion disk\index{Accretion disk}.  They suggest therefore that MRI is the main component of outwards angular momentum\index{Angular momentum} transfer, and thus of $\alpha$\index{Viscosity}, in astrophysical accretion disks.

\section{Accretion Phenomena}

\par The extreme physics involved in accretion\index{Accretion} onto compact objects\index{Compact object} leads to a number of non-intuitive physical phenomena.  In this section I describe a number of these theoretical effects, and explain how these phenomena manifest in physical LMXBs.  

\subsection{The Eddington Limit}

\label{sec:edd}

\par\index{Eddington limit}\index{Eddington luminosity|see {Eddington limit}}\index{Eddington rate|see {Eddington limit}} Consider an element of gas at distance $r$ from a compact object\index{Compact object}, with mass $m$.  This element of gas is acted on by a inwards-pointing gravitational force given by:
\begin{equation}
F_G=\frac{GMm}{r^2}
\end{equation}
Where $M$ is the mass of the compact object.
\par If we assume that a luminosity $L$ is emitted isotropically from the compact object\index{Compact object}, then the electromagnetic flux at distance $r$ is given by:
\begin{equation}
\phi(r)=\frac{L}{4\pi r^2}
\end{equation}
Electromagnetic radiation exerts a pressure on material corresponding to $\phi/c$.  As such, the radiation from the X-ray binary exerts an outwards force on our gas element corresponding to:
\begin{equation}
F_L=\frac{\kappa m\phi(r)}{c}=\frac{L\kappa m}{4\pi r^2c}
\end{equation}
Where $\kappa$ is the opacity\index{Opacity} of the cloud, or its surface area per unit mass.
\par If $F_G$ and $F_L$ are equal, then no net force is exerted on our cloud of matter and it will not accrete\index{Accretion} onto the compact object\index{Compact object}.  This happens when:
\begin{eqnarray}
F_G&=&F_L\\ \nonumber \\
\frac{GMm}{r^2}&=&\frac{L\kappa m}{4\pi r^2c} \\ \nonumber \\
L&=&\frac{GMm}{r^2}\frac{4\pi r^2c}{\kappa m} \\ \nonumber \\
L&=&\frac{4\pi GMc}{\kappa}
\end{eqnarray}
This luminosity, denoted as $L_E$, is the Eddington luminosity; the theoretical maximum isotropic luminosity an object can emit and still have spherically symmetric accretion\index{Accretion} take place.  It only depends on the mass of the compact object\index{Compact object} $M$ and the opacity\index{Opacity} of the accreting\index{Accretion} material $\kappa$, which in turn depends on the chemical composition of the accretion disk\index{Accretion disk}.  As accretion disks tend to be dominated by ionised hydrogen, $\kappa$ is usually assumed to be $\sigma_T/m_p$, where $\sigma_T$ is the Thomson scattering cross-section\index{Thomson scattering cross-section} of an electron and $m_p$ is the mass of a proton.  This assumption yields the final formula which only depends on the mass of the compact object:
\begin{equation}
L_E=\frac{4\pi GMm_pc}{\sigma_T}
\end{equation}
The luminosity due to matter falling into a compact object\index{Compact object} can be expressed as:
\begin{equation}
L=\eta\dot{M}c^2
\end{equation}
Where $\dot{M}$ is the accretion rate\index{Accretion rate} and $\eta$ is the efficiency at which the gravitational potential energy of infalling matter is converted to outgoing radiation.  As such, $L_E$ also corresponds to a limiting accretion rate $\dot{M}_E$.
\par However, a number of X-ray binaries have been seen to shine at luminosities far above this limit; in one of the most extreme cases, the confirmed neutron star XRB M82 X-1 has a luminosity of $\sim100L_E$ \citep{Bachetti_M82X1}.  This super-Eddington\index{Super-Eddington accretion} accretion is possible due to the fact that a number of assumptions made when calculating the Eddington limit do not apply to physical XRBs.  In particular, the calculation performed above assumes that both accretion\index{Accretion} on to the compact object\index{Compact object}, as well as electromagnetic emission from it, are isotropic.  An object may exceed the Eddington Limit\index{Eddington limit} if it is accreting anisotropically, as is the case for XRBs\index{X-ray binary} as these systems accrete\index{Accretion} from near-planar disks\index{Accretion disk}.  In this case the assumptions behind the calculation of the Eddington Limit break down, and more radiation can be emitted away from the plane of the disk, decreasing the radiation pressure on infalling material.  Anisotropically emitting systems may appear to further exceed the Eddington limit via beaming effects.  An XRB beaming its radiation in the direction of the Earth would lead us to infer an artificially high value of $L$, and thus overestimate its luminosity with respect to the Eddington Limit.
\par Despite these setbacks, the Eddington Luminosity\index{Eddington limit} is a useful tool to compare XRBs\index{X-ray binary} with different compact object\index{Compact object} masses.  By expressing the luminosity of an object as a fraction of its Eddington Limit, objects can be rescaled in such a way that we can compare how dominant radiation pressure\index{Radiation pressure} must be in each accretion disk\index{Accretion disk}.

\subsection{The Propeller Effect}

\label{sec:prop}

\par\index{Propeller effect} Another limit on accretion rate\index{Accretion rate} arises when one considers the effect of a strong neutron star\index{Neutron star} magnetic field\index{Magnetic field}.  To understand this effect, we must first define two characteristic radii of such a system.
\par First, assume that the magnetic field\index{Magnetic field} of the neutron star\index{Neutron star} can be approximated as a set of rigid field lines which are anchored to points on the neutron star surface.  The magnetic field can then be thought of as a `cage' which rotates with the neutron star at its centre.  The straight-line speed of a point on this rotating cage is given by:
\begin{equation}
v_\nu(r)=2\pi r\nu
\end{equation}
Where $r$ is the distance from the neutron star centre and $\nu$ is the rotation frequency of the neutron star.  This can be compared with the Keplerian speed, or the speed of a particle in a Keplerian orbit\index{Keplerian motion} around the compact object\index{Compact object}.  This is given by:
\begin{equation}
v_K(r)=\sqrt{\frac{GM}{r}}
\end{equation}
Where $M$ is the mass of the neutron star.  By setting these equal, we can find the radius at which the magnetic field is rotating at the same speed as a particle in a Keplerian\index{Keplerian motion} orbit:
\begin{eqnarray}
v_\nu(r)&=&v_K(r)\\ \nonumber \\
2\pi r\nu&=&\sqrt{\frac{GM}{r}}\\ \nonumber \\
r^3&=&\frac{GM}{4\pi^2\nu^2}\\ \nonumber \\
r&=&\sqrt[3]{\frac{GM}{4\pi^2\nu^2}}
\end{eqnarray}
This radius is denoted as $r_c$, the co-rotation radius\index{Co-rotation radius}.  Inside of this radius, a particle in an equatorial Keplerian orbit\index{Keplerian motion} has a greater velocity than the magnetic field lines; outside this radius, the magnetic field lines are moving faster.  To understand the significance of this radius, we must define another characteristic radius of the system.
\par In a neutron star\index{Neutron star} accretion disk\index{Accretion disk}, there are three significant sources of pressure: gas (or ram) pressure\index{Ram pressure} $P_g$, radiation pressure $P_\gamma$\index{Radiation pressure} and magnetic pressure $P_\mu$\index{Magnetic pressure}.  Whichever pressure is dominant in a given location will govern the physics of matter in that region.
\par Photon pressure\index{Radiation pressure} falls off sharply outwards from the inner disk\index{Accretion disk}, so it can be assumed to be negligible in the region of the disk considered here.  We can then calculate where in the disk each of the remaining two pressures dominates.
\par Assuming that the neutron star behaves as a magnetic dipole, the magnetic pressure at a point a distance $r$ above its equator can be given as:
\begin{eqnarray}
P_\mu&=&\frac{B^2}{2\mu_0}\\ \nonumber \\
B(r)&=&B_0\left(\frac{R_{NS}}{r}\right)^3\label{eq:NS}\\ \nonumber \\
\therefore\quad P_\mu&=&\frac{B_0^2}{2\mu_0}\left(\frac{R_{NS}}{r}\right)^6
\end{eqnarray}
Where $\mu_0$ is the vacuum permeability, $B_0$ is the equatorial magnetic field strength at the neutron star surface, $R_{NS}$ is the radius of the neutron star and Equation \ref{eq:NS} is the equation for the magnetic field strength above the equator of a dipole.
\par The functional form of the ram pressure depends on the assumed accretion\index{Accretion} geometry of the system.  As when calculating the Eddington Limit, one can assume the simplest possible case of spherically accreting\index{Accretion} free-falling matter.  The ram pressure\index{Ram pressure} is then given by:
\begin{equation}
P_g=\frac{\dot{M}}{4\pi r^2}\sqrt{\frac{2GM}{r}}
\end{equation}
When $P_\mu>P_g$, accreting\index{Accretion} material is dominated by magnetic pressure\index{Magnetic pressure} in such a way that material is `frozen' onto magnetic field\index{Magnetic field} lines \citep{Alfven_Waves}; this results in material flowing onto the neutron star\index{Neutron star} surface along magnetic field lines onto the poles, as described in section \ref{sec:NSintro}.  It is possible to express the region of the accretion disk\index{Accretion disk} within which matter is magnetically dominated:
\begin{eqnarray}
P_g&<&P_\mu\\ \nonumber \\
\frac{\dot{M}}{4\pi r^2}\sqrt{\frac{2GM}{r}}&<&\frac{B_0^2}{2\mu_0}\left(\frac{R_{NS}}{r}\right)^6\\ \nonumber \\
\frac{GM\dot{M}^2}{8\pi^2 r^5}&<&\frac{B_0^4}{4\mu_0^2}\left(\frac{R_{NS}}{r}\right)^{12}\\ \nonumber \\
r^7&<&\frac{2\pi^2}{G\mu_0^2}\frac{B_0^4R_{NS}^{12}}{M\dot{M}^2}\\ \nonumber \\
r&<&\sqrt[7]{\frac{2\pi^2}{G\mu_0^2}\frac{B_0^4R_{NS}^{12}}{M\dot{M}^2}}\label{eq:rmu}
\end{eqnarray}
The critical radius, the magnetospheric or Alfv\'en radius\index{Magnetospheric radius}\index{Alfv\'{e}n radius|see {Magnetospheric radius}}, is denoted as $r_\mu$.
\par
\begin{figure}
  \centering
  \includegraphics[width=0.7\linewidth, trim= 1mm 1mm 5mm 1mm, clip]{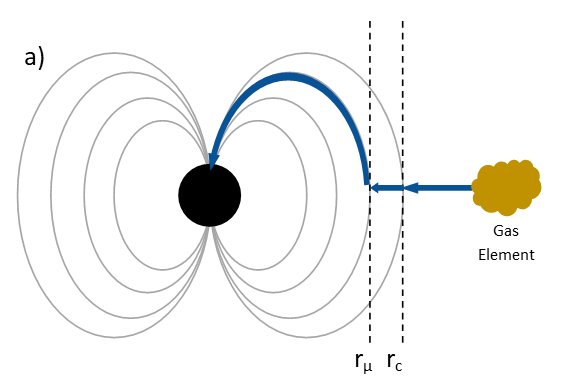}
  \includegraphics[width=0.7\linewidth, trim= 1mm 1mm 5mm 1mm, clip]{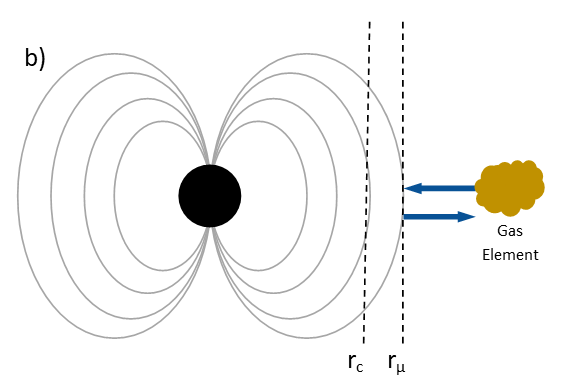}
  \caption[Diagrams showing the path of an element of gas in a neutron star accretion disk for different arrangements of the corotation and magnetospheric radii.]{Diagrams showing the path of an element of gas in a neutron star\index{Neutron star} accretion disk\index{Accretion disk} for different arrangements of the corotation\index{Co-rotation radius} ($r_c$) and magnetospheric\index{Magnetospheric radius} ($r_\mu$) radii.  In panel a), the gas falls freely inwards until it reaches $r_\mu$, at which point it freezes onto a magnetic field line.  As $r_\mu<r_c$, the magnetic field lines (grey lines) are rotating more slowly than the gas element at this point.  As such, the gas element experiences torque which slows it down, decreasing its altitude above the neutron star surface (black circle) and causing it to accrete\index{Accretion} along the magnetic field\index{Magnetic field} line.  In panel b), $r_\mu<r_c$.  As such, the magnetic field lines are rotating faster than the gas element at $r_\mu$.  The gas element experiences a torque which speeds it up at this point, increasing its altitude and preventing it from accreting\index{Accretion} onto the neutron star.}
  \label{fig:propdiag}
\end{figure}

 Now it is possible to consider what happens to matter approaching $r_\mu$ in two different physical regimes.  First of all, consider a system in which the corotation radius $r_c>r_\mu$.  In this case, which we show diagrammatically in panel A of Figure \ref{fig:propdiag}, magnetic field\index{Magnetic field} lines at $r_\mu$ are moving slower than the Keplerian speed\index{Keplerian motion}.  An element of matter approaching this radius from a Keplerian orbit will experience a torque slowing it down as it freezes onto the field lines.  This decrease in orbital speed causes the element's altitude above the neutron star surface to decrease.  This in turn pulls the element further into the magnetically-dominated regime and allows it to accrete\index{Accretion} freely along the field line onto the neutron star.

\par Now we can consider what happens when $r_c<r_\mu$.  In this case, which we show diagrammatically in panel B of Figure \ref{fig:propdiag}, field lines at $r_\mu$ are moving faster than the Keplerian speed\index{Keplerian motion}.  An element of matter approaching $r_\mu$ will therefore experience a torque speeding it up as it becomes frozen onto magnetic field lines.  This will increase its altitude, driving it back away from $r_\mu$.  In this case, the magnetospheric radius acts as a barrier to infalling matter, repelling any gas that approaches it and stopping accretion\index{Accretion} onto the neutron star\index{Neutron star} surface.  This set of circumstances is known as the `propeller regime'\index{Propeller effect}, due to the rapidly rotating field\index{Magnetic field} lines acting like a `propeller' which blows the inner part of the disk\index{Accretion disk} away.
\par As the propeller\index{Propeller effect} regime is expected to occur only for $r_c<r_\mu$, it is possible to work out what kind of system this should be observed in:
\begin{eqnarray}
r_c&<&r_\mu\\ \nonumber \\
\left(\frac{GM}{4\pi^2\nu^2}\right)^{1/3}&<&\left(\frac{2\pi^2}{G\mu_0^2}\frac{B_0^4R_{NS}^{12}}{M\dot{M}^2}\right)^{1/7}\\ \nonumber \\
M^{10/21}\dot{M}^{2/7}&<&k\nu^{2/3}B_0^{4/7}R_{NS}^{12/7}\label{eq:ineq}
\end{eqnarray}
Where $k$ is a constant.  Assuming that the radius and mass of neutron stars does not vary much, this inequality tells us that the propeller regime is more likely to be observed in neutron star XRBs with a high spin frequency and a high magnetic field.  The inequality shown in \ref{eq:ineq} also tells us that the propeller effect places a \textit{lower} limit on accretion\index{Accretion} in such systems: accretion is not possible unless infalling matter can apply enough ram\index{Ram pressure} pressure to push the magnetospheric radius\index{Magnetospheric radius} inside the corotation\index{Co-rotation radius} radius.
\par There are numerous problems with this relatively simplistic view of accretion\index{Accretion} in a highly magnetic\index{Magnetic field} regime.  Much like the formulation of the Eddington Limit I present in Section \ref{sec:edd}, the above formulation of the propeller effect\index{Propeller effect} depends on an unphysical spherical accretion\index{Accretion} geometry.  It also includes the assumption that the magnetic field lines can in no way be warped by the movement of ionised matter on them.
\par \citet{White_MRad} have shown that, in neutron stars\index{Neutron star}, the magnetospheric radius\index{Magnetospheric radius} may be close enough to the compact object\index{Compact object} that photon pressure\index{Radiation pressure} cannot be safely neglected.  \citeauthor{White_MRad} find two different possible behaviours of the magnetospheric radius in such a regime, depending on how $\alpha$\index{Viscosity} varies with $r$ and how the disk\index{Accretion disk} reacts to the magnetic field\index{Magnetic field}.  For a perfectly diamagnetic disk, they show that the magnetospheric radius should not depend on the accretion rate\index{Accretion rate}, preventing the formation of a propeller regime\index{Propeller effect} entirely.  For a case in which gas pressure\index{Ram pressure} is the dominant contributor to viscous stress, they find that the magnetospheric radius is up to $\sim30$ times smaller than that calculated by equation \ref{eq:rmu}.  Additionally, \citet{Ertan_Prop} has analytically shown that an optically thick accretion disk can only be in a stable propeller regime when the inner disk radius is $\gtrsim15$ times smaller that the $r_\mu$ na\"ively calculated in equation \ref{eq:rmu}.  This in turn results in a several orders of magnitude reduction in the critical accretion rate at the onset of the propeller regime in a given system, raising questions as to whether the effect would be observable at such low luminosities.
\par Despite these difficulties, an effect observationally similar to the propeller effect is observed in a number of astrophysical neutron star XRBs (e.g. the cessation of pulsed emission while the source is still in outburst and the neutron star is actively spinning up or down, \citealp{Fabian_Propex,Furst_Propex}) and other systems (e.g. the observation of a sudden steepening in lightcurves during the decays of outbursts, interpreted by e.g. \citealp{Campana_PropBorder} as the onset of the propeller phase).  Therefore it is likely that a propeller effect in some form is likely able to explain what we see in nature.

\subsection{Disk Instabilities}

\label{sec:diskinstab}

\par\index{Instability}\index{Disk instability|see {Instability}} A number of effects can cause an accretion disk\index{Accretion disk}, or portions of it, to become unstable.  Some of these instabilities can set up limit cycles of behaviour in the disk, resulting in quasi-periodic fluctuations\index{Quasi-periodic oscillation}\index{QPO|see {Quasi-periodic oscillation}} in the object's intensity or colour\index{Colour} as seen from Earth.  I describe a number of these instabilities here.
\par One of the first such instabilities to be described was discovered by \citet{Lightman_Instability}.  Using the assumptions present in the thin disk\index{Accretion disk}\index{Shakura-Sunyaev disk model} models of \citet{Shakura_Disk} and \citet{Novikov_Torque}, \citeauthor{Lightman_Instability} calculate the diffusion\index{Diffusion} of the gas in such a disk.  They show that the diffusion coefficient in the radial direction of radiatively dominated disk is negative.  As such, any initially smooth disk under these conditions tends to separate into thin, dense annuli.  As such any sufficiently thin disk, with $\alpha$ consistent with the prescription of \citet{Shakura_Disk} is unstable.
\par \citet{Shakura_Instab} described another instability which takes place in the radiation pressure-dominated region near the inner edge of accretion disks\index{Accretion disk}.  They find that steady state accretion\index{Accretion} in such a regime is only possible for a single value of $\alpha$\index{Viscosity}, and hence this region is unstable under small perturbations of viscosity.  They argue that an instability\index{Instability} due to this effect may take the form of propagating wavefronts in the inner disk, which in turn may cause some of the quasiperiodic\index{Quasi-periodic oscillation} fluctuations which are observed in these objects.

\label{sec:hic}

\par A further disk\index{Accretion disk} instability\index{Instability} arises by considering the propeller effect\index{Propeller effect} (see Section \ref{sec:prop}), and specifically considering neutron star\index{Neutron star} LMXBs\index{X-ray binary!Low mass} in which the magnetopsheric radius\index{Magnetospheric radius} and co-rotation radius\index{Co-rotation radius} are similar ($r_c\approx r_\mu$, e.g. \citealp{Spruit_Type2Mod}).  At this boundary, a small increase in global accretion rate\index{Accretion rate} from the donor star pushes $r_\mu$ inwards such that $r_\mu<r_c$.  In this regime, the neutron star accretes freely, and the system is relatively bright in X-rays.  However, a slight decrease in global accretion rate causes $r_\mu<r_c$: in this regime, accretion\index{Accretion} onto the compact object\index{Compact object}'s surface is halted and the system is relatively faint in X-rays.  This effect causes a small fluctuation in accretion rate to convert to a large fluctuation in luminosity between two quasi-stable values.  This effect is believed to be behind the so-called `hiccup accretion'\index{Hiccup accretion} seen in X-ray binaries such as IGR J18245-2452\index{IGR J18245-2452} \citep{Ferrigno_TMSPVar} and 1RXS J154439.4-112820\index{3FGL J1544.6-1125} \citep{Bogdanov_Proxy}.

\section{GRS 1915+105 and IGR J17091-3624}

\label{sec:1915}

\par One famous system in which disk\index{Accretion disk} instabilities\index{Instability} are extremely apparent is the black hole\index{Black hole} LMXB\index{X-ray binary!Low mass} GRS 1915+105\index{GRS 1915+105}.  GRS 1915+105 \citep{CastroTirado_GRS1915}, hereafter GRS 1915, is a black hole LMXB\index{X-ray binary!Low mass} which accretes\index{Accretion} at between a few tens and more than 100\% of its Eddington Limit\index{Eddington limit} (e.g. \citealp{Vilhu_SupEd,Done_GRS_HighAcc,Fender_DiskJet}).  The system lies at a distance of $8.6\pm2.0$\,kpc \citep{Reid_Parallax}, and consists of a 12.4$\pm$2.0\,M$_\odot$ black hole\index{Black hole} and a $<1$\,M$_\odot$ K-class giant companion star\index{Companion star} \citep{Reid_Parallax,Ziolkowski_GRSDonor}.
\par The components of GRS 1915 have the longest known orbital period\index{Orbital period} of any LMXB \citep{Greiner_BigDisk}, in turn implying that this system has the greatest orbital separation and the largest accretion disk\index{Accretion disk}.  GRS 1915 has been in outburst\index{Outburst} since its discovery in 1992 \citep{CastroTirado_GRS1915}, and the extreme length of this ongoing outburst is believed to be related to the large size of its accretion disk. 
\par
\begin{figure}
  \centering
  \includegraphics[width=\linewidth, trim= 0mm 0mm 0mm 0mm, clip]{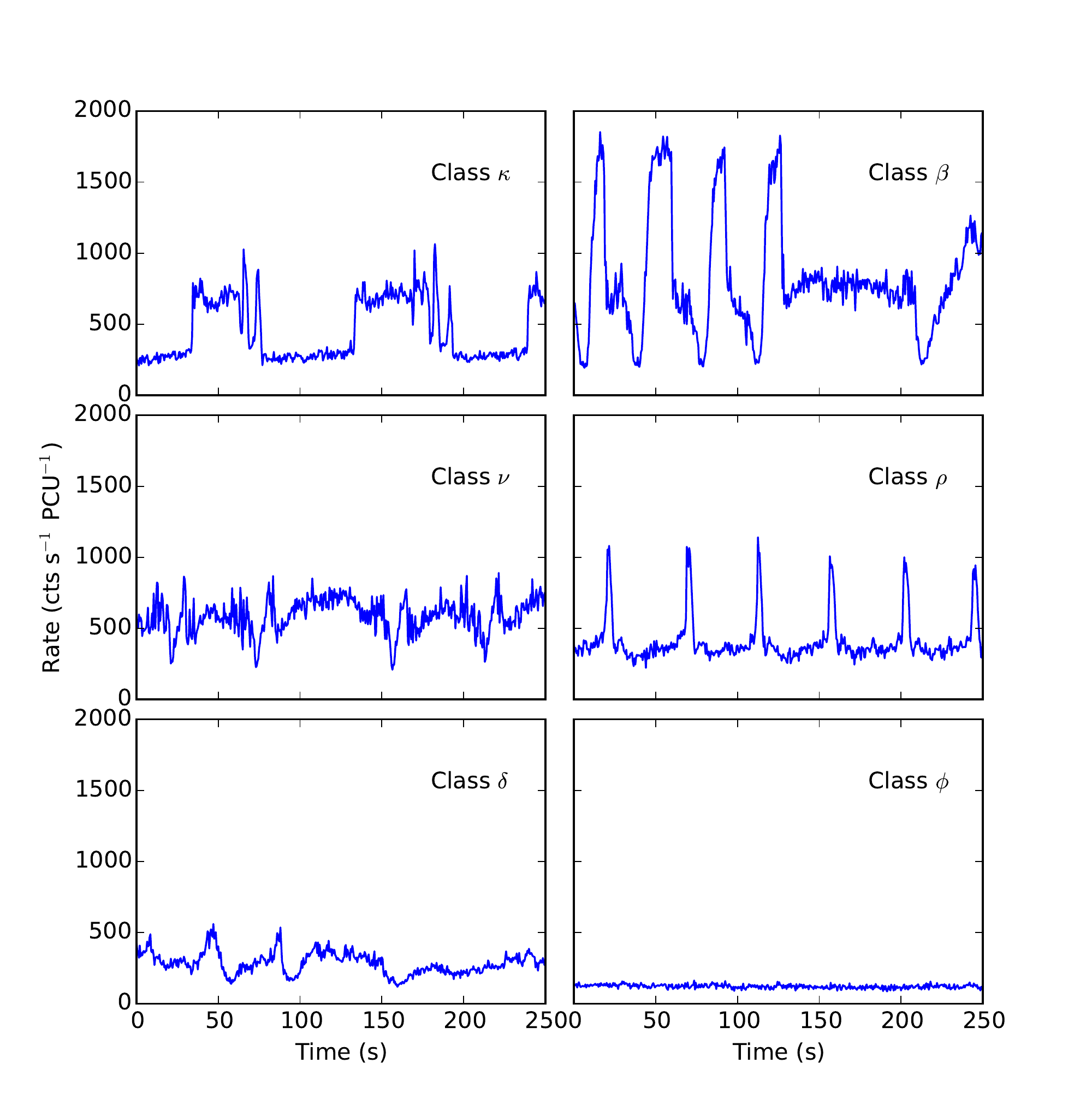}
  \caption[Typical lightcurves of a selection of variability classes seen in the LMXB GRS 1915+105.]{\indexkappa\indexnu\indexbeta\indexrho\indexdelta\indexphi Typical lightcurves\index{Lightcurve} of a selection of variability classes\index{Variability class} seen in GRS 1915, taken by the PCA instrument aboard \rxte\indexrxte\indexpca.  The classes are labelled according to the Greek letter names assigned to them in \citet{Belloni_GRS_MI}.}
  \label{fig:GRSsample}
\end{figure}

 GRS 1915 is also notable for the incredible variety and complexity of behaviours\index{Variability} it exhibits over timescales of seconds to minutes (e.g. \citealp{Yadav_GRSBursts,Belloni_GRS_MI}).  In total, at least 15 distinct `variability classes'\index{Variability class} have been described \citep{Belloni_GRS_MI,KleinWolt_OmegaClass,Hannikainen_NewClass, Pahari_NewClass}, a number of which I show lightcurves\index{Lightcurve}\footnote{A plot showing how the intensity of an object varies over time.} of in Figure \ref{fig:GRSsample}.  The system tends to stay in one variability class for no more than a few days but similar patterns are often repeated many months or years later, suggesting some capacity of the system to `remember' which variability classes it can show.

\par The variability classes of GRS 1915 consist of repeating patterns of flares\index{Flare}, dips\index{Dip} and periods of noisy fluctuation, with a range of amplitudes and timescales.  The  behaviour of the source during these classes, which are usually denoted by the Greek letter names assigned to them by \citet{Belloni_GRS_MI}, can range from highly quasi-periodic to apparently entirely unstructured.  The $\rho$\indexrho\ class, also referred to as the `heartbeat'\index{Heartbeat|see {Variability class, $\rho$}} class due to the similarity of its lightcurve to the output of an electrocardiagram, consists of sharp quasiperiodic flares\index{Quasi-periodic oscillation} with a recurrence time\index{Recurrence time} of a few tens of seconds (Middle-right panel of Figure \ref{fig:GRSsample}).  Other classes, such as class $\kappa$\indexkappa\ shown in the top-left panel of Figure \ref{fig:GRSsample}, consist of quasiperiodic fluctuations between two quasistable count rates: in the case of class $\kappa$, there is also a period of highly structured sub-second variability at each transition between these two classes.  Finally, two classes ($\chi$\indexchi\ and $\phi$\indexphi, an example of the latter is shown in the bottom-right panel of Figure \ref{fig:GRSsample}) show no significant variability other than red noise; these classes are separated from each other based on their spectral properties.  It has been suggested they they may be equivalent to the hard state\index{Low/Hard state} seen in other outbursting LMXBs \citep{VanOers_GRSHard}, providing a possible link between the behaviour of GRS 1915 and the behaviour of more typical LMXBs.
\par The dramatic variability seen in GRS 1915\index{GRS 1915+105} was long thought to be unique, driven by its unusually high accretion rate\index{Accretion rate} (e.g. \citealp{Belloni_Timescales}).  However in 2011, \citet{Altamirano_IGR_FH} unambiguously identified GRS 1915-like variability in a second object: the black hole\index{Black hole} LMXB\index{X-ray binary!Low mass} IGR J17091-3624\index{IGR J17091-3624} (hereafter IGR J17091).  This object is much fainter than GRS 1915: \citet{Altamirano_IGR_FH} showed that, assuming that this object accretes\index{Accretion} at its Eddington Limit\index{Eddington limit} by analogy with GRS 1915, the object may either be out in the halo of the Galaxy (at $\gtrsim20$\,kpc) or harbour the smallest mass black hole known to science ($\lesssim3$\,M$_\odot$).  The companion star\index{Companion star} to the black hole in this system has not been definitively identified \citep{Chaty_IGRCompanion}.
\par Much like GRS 1915\index{GRS 1915+105}, IGR J17091\index{IGR J17091-3624} displays a number of distinct classes of variability\index{Variability class} over time, and a number of these have been identified as being similar to the classes seen in GRS 1915 (e.g. \citealp{Altamirano_IGR_FH,Zhang_IGR}).  Unlike GRS 1915, IGR J17091 displays the pattern of outbursts\index{Outburst} and quiescence\index{Quiescence} more commonly seen in LMXBs; known outbursts of IGR J17091 occurred in 2011 and 2016, and GRS 1915-like\index{Variability} variability was observed in both \citep{Reynolds_2016HB}.
\par There are a number of notable differences between variability classes in GRS 1915 and IGR J17091.  In general, variability classes in IGR J17091 occur over shorter timescales than their counterparts in GRS 1915.  In addition to this, hard emission\index{Hard lag} tends to lag soft emission in the variability classes of GRS 1915 (e.g. \citealp{Janiuk_Lag}), while the opposite trend has been found in the `heartbeat'-like\indexrho\ class of IGR J17091 \citep{Altamirano_IGR_FH}.
\par In addition to GRS 1915 and IGR J17091\index{IGR J17091-3624}, there have been claims that a third LMXB\index{X-ray binary!Low mass} displays GRS 1915\index{GRS 1915+105}-like variability\index{Variability}.  \citet{Bagnoli_RB} report on two observations of MXB 1730-335\index{MXB 1730-335|see {Rapid Burster}}, also known as the `Rapid Burster'\index{Rapid Burster}, which show lightcurve patterns remarkably similar to those seen in the $\rho$\indexrho\ and $\theta$\indextheta\ classes of GRS 1915.  The presence of GRS 1915-like variability in the Rapid Burster is significant for a number of reasons: unlike GRS 1915\index{GRS 1915+105} or IGR J17091\index{IGR J17091-3624}, the Rapid Burster is known to contain a neutron star\index{Neutron star} accreting\index{Accretion} at no more than 20\% of its Eddington Limit\index{Eddington limit}, thereby ruling out any black hole\index{Black hole}-specific or near-Eddington-specific explanations for this behaviour.  In addition to this, the Rapid Burster is one of only 2 objects known to undergo so-called Type II X-Ray bursts\index{X-ray burst!Type II} (see Section \ref{sec:TypeII}), suggesting a possible link between these two phenomena.  However, as it has only been observed twice in the $\sim30$ years since the object was discovered, the true nature of the apparent GRS 1915-like variability in the Rapid Burster remains unclear.

\subsection{A History of Models of GRS 1915-like Variability}

\label{sec:models_GRS}

\par Over the years, a number of models and physical scenarios have been suggested to explain the complex variability seen in GRS 1915-like\index{GRS 1915+105} systems.  Successful models must also be able to explain why this type of variability is not seen in a wider array of sources.
\par One of the most best-studied classes of GRS 1915-like variability is Class $\rho$\indexrho, the `heartbeat' class.  This variability class is present in both GRS 1915 and IGR J17091\index{IGR J17091-3624} (e.g. \citealp{Altamirano_IGR_FH}), and has been the focus of many of the models proposed to explain GRS 1915-like variability.  It has been shown that hard X-ray photons lag soft X-ray photons\index{Hard lag} in this class (e.g. \citealp{Janiuk_Lag,Massaro_Lag}), suggesting that hard emission from this source is somehow caused by the softer emission.  Other classes in GRS 1915 which show quasi-periodic flaring behaviour also exhibit this phase lag.  Previous authors have established models to explain both the hard photon lag as well as the `heartbeat'-like flaring itself, generally based on the instability\index{Instability} in a radiation-dominated disk\index{Accretion disk} first formulated by \citet{Shakura_Instab} (see Section \ref{sec:diskinstab}).
\par \citet{Belloni_Model1} first proposed an empirical model for flaring\index{Flare} in GRS 1915.  They suggested that this behaviour is due to a rapid emptying of a portion of the inner accretion disk\index{Accretion disk}, followed by a slower refilling of this region over a viscous timescale.  \citeauthor{Belloni_Model1} divided data from a given observation into equal-sized 2-Dimensional bins in count rate-colour\index{Colour} space.  A spectral\index{Spectroscopy} model was then fit to each of these bins independently to perform `pseudo'-phase-resolved spectroscopy\index{Spectroscopy!Phase-resolved} (compare with the method outlined in Section \ref{sec:phasresspec}).  They showed that the time between flaring\index{Flare} events correlates with the maximum inner disk radius during the flare; i.e., a correlation between the amount of the disk which is emptied and the time needed to refill it.  They go on to suggest that their model is able to explain all flaring-type events seen in GRS 1915.
\par The scenario proposed by \citet{Belloni_Model1} was mathematically formalised by \citet{Nayakshin_GRSModel}, who found that it was not consistent with a `slim' accretion disk\index{Accretion disk} \citep{Abramowicz_Slim} or with a disk in which viscosity $\alpha$\index{Viscosity}\index{Alpha@$\alpha$|see {Viscosity}} is constant with respect to radius.  As such, their model consists of a cold accretion disk with a modified viscosity law, a non-thermal electron corona\index{Corona} and a transient jet\index{Jet} of discrete plasma emissions which are ejected when the bolometric luminosity approaches the Eddington Limit\index{Eddington limit}.  Using their model, \citet{Nayakshin_GRSModel} found that some formulations of $\alpha(r)$\index{Viscosity} result in the disk oscillating between two quasi-stable branches in viscosity-temperature space, over timescales consistent with those seen in the flaring of GRS 1915; they found that this occurs for accretion rates\index{Accretion rate} greater than 26\% of the Eddington limit\index{Eddington limit}.  They also found that by varying the functional form of $\alpha(r)$\index{Viscosity}, their model gives rise to a number of lightcurve morphologies which generally match what is seen in data from GRS 1915.  \citet{Janiuk_RadInstab} built on this model further by including the effect of the transient jet\index{Jet} in cooling the disk; an effect not considered in the model by \citet{Nayakshin_GRSModel}.  In this formulation, \citet{Janiuk_RadInstab} found that GRS 1915-like variability should occur at luminosities as low as 16\% of Eddington.  The criteria of a relatively low accretion rate\index{Accretion rate} in these models suggests that many more black hole\index{Black hole} LMXBs should show GRS 1915-like behaviour, which is at odds with observations.  Additionally, they are unable to explain GRS 1915-like behaviour in objects with even lower accretion rates, such as the $\rho$-like\indexrho\ behaviour reported by \citet{Bagnoli_RB} in the Rapid Burster\index{Rapid Burster}.
\par \citet{Belloni_GRS_MI} found that variability in GRS 1915 can be empirically described by transitions between three phenomenological states, which differ in luminosity and hardness ratio:
\begin{enumerate}
\item State B: high rate, high 5--13/2--5\,keV hardness ratio.
\item State C: low rate, low 5--13/2--5\,keV hardness ratio, variable 13--60/2--5\,keV hardness ratio.
\item State A: low rate, low 5--13/2--5\,keV hardness ratio, lowest 13--60/2--5\,keV hardness ratio.
\end{enumerate}
\citeauthor{Belloni_GRS_MI} find that no variability class shows a transition from state C to state B, and they suggest that this transition is forbidden.  The phenomenological scenario they establish is at odds with the model of \citet{Nayakshin_GRSModel}, which only results in two quasi-stable states; one high rate and one low rate state.
\par \citet{Nobili_Hotspot} tried to account for the hard X-Ray lag\index{Hard lag} by considering a scenario in which a significant proportion of the X-Ray disk\index{Accretion disk} variability comes from a single hotspot.  They suggest that the lag corresponds to a light travel time, after which a portion of this emission is Comptonised\index{Compton scattering} by the jet.  In this case, the geometric location of this hotspot determines the magnitude of this lag.  This scenario goes some way to explaining why GRS 1915 is special, as it requires the presence of a jet\index{Jet} during a soft-like\index{High/Soft state} state.
\par \citet{Tagger_MagneticFlood} propose a magnetic explanation for the ejection of the inner accretion disk\index{Accretion disk} required by \citet{Nayakshin_GRSModel} and \citet{Janiuk_RadInstab}.  In their scenario, they suggest the existence of a limit cycle in which a poloidal magnetic field\index{Magnetic field} is advected towards the inner disk during the refilling of this region.  Associated field lines are then destroyed in reconnection\index{Reconnection} events, releasing energy which results in the expulsion of matter from the inner disk.  They suggest that the three quasi-stable states proposed by \citealp{Belloni_GRS_MI} can be explained as states in the inner accretion disk with different values of plasma $\beta$\index{Plasma $\beta$}\footnote{The ratio of the plasma pressure to the magnetic pressure in an ionised medium.}.
\par \citealp{Janiuk_Lag} attempt to explain the hard lag in the heartbeats\indexrho\ of GRS 1915\index{GRS 1915+105} more simply, by proposing a model in which it is caused by the non-thermal corona\index{Corona} smoothly adjusting to changes in luminosity from the disk\index{Accretion disk}.  They base the variability\index{Variability} of the disk on the model of \citealp{Nayakshin_GRSModel}, and show that the presence of a non-thermal corona which reacts to this variability naturally reproduces the lag behaviour seen in Class $\rho$\indexrho\ in GRS 1915.
\par \citealp{Merloni_MagDom} also propose a magnetic explanation for the reformulation of $\alpha(r)$ required by the model of \citealp{Nayakshin_GRSModel}.  Assuming that the viscosity in the accretion disk\index{Accretion disk} is dominated by turbulence\index{Turbulence} due to the magnetorotational instability\index{Magnetorotational instability}, they find that allowing for a magnetically dominated corona naturally allows for the forms of $\alpha(r)$ required by \citealp{Nayakshin_GRSModel}.
\par \citealp{Zheng_Model} propose a model which suggests that, when the effects of a magnetic field are included, the accretion rate\index{Accretion rate} threshold for GRS 1915-like\index{GRS 1915+105} variability\index{Variability} should be $\sim50$\% of Eddington\index{Eddington limit}; significantly higher than the 16\% or 26\% reported by \citealp{Janiuk_RadInstab} or \citealp{Nayakshin_GRSModel}.  \citeauthor{Zheng_Model} go on to suggest that this type of variability is only seen in GRS 1915 due to this source having the highest accretion rate\index{Accretion rate} of all permanently soft-state sources.  As such, this scenario still relies on a high accretion rate to trigger GRS 1915-like variability, but it is more consistent with observations than the models of \citealp{Janiuk_RadInstab} or \citealp{Nayakshin_GRSModel}.%  However, magnetohydronamic simulations of a radiation dominated inner disk performed by \citealp{Hirose_Stable} suggest that the thermal instabilities required by models of heartbeat should not arise at any value of accretion rate.
%\par \citealp{Xue_Spin} derive a mathematical model of the evolution of a slim accretion disk around a Kerr black hole.  They hypothesize that the spin of the black hole, not the accretion rate, may be the driving factor behind GRS 1915-variability.  However, they find that that the morphology of X-ray lightcurves from such a disk only has a weak dependence on the spin of the black hole, ruling this out as a possible explanation.

\label{sec:Neilsen}

\par
\begin{figure}
  \centering
  \includegraphics[width=.9\linewidth, trim= 25mm 0mm 0mm 0mm]{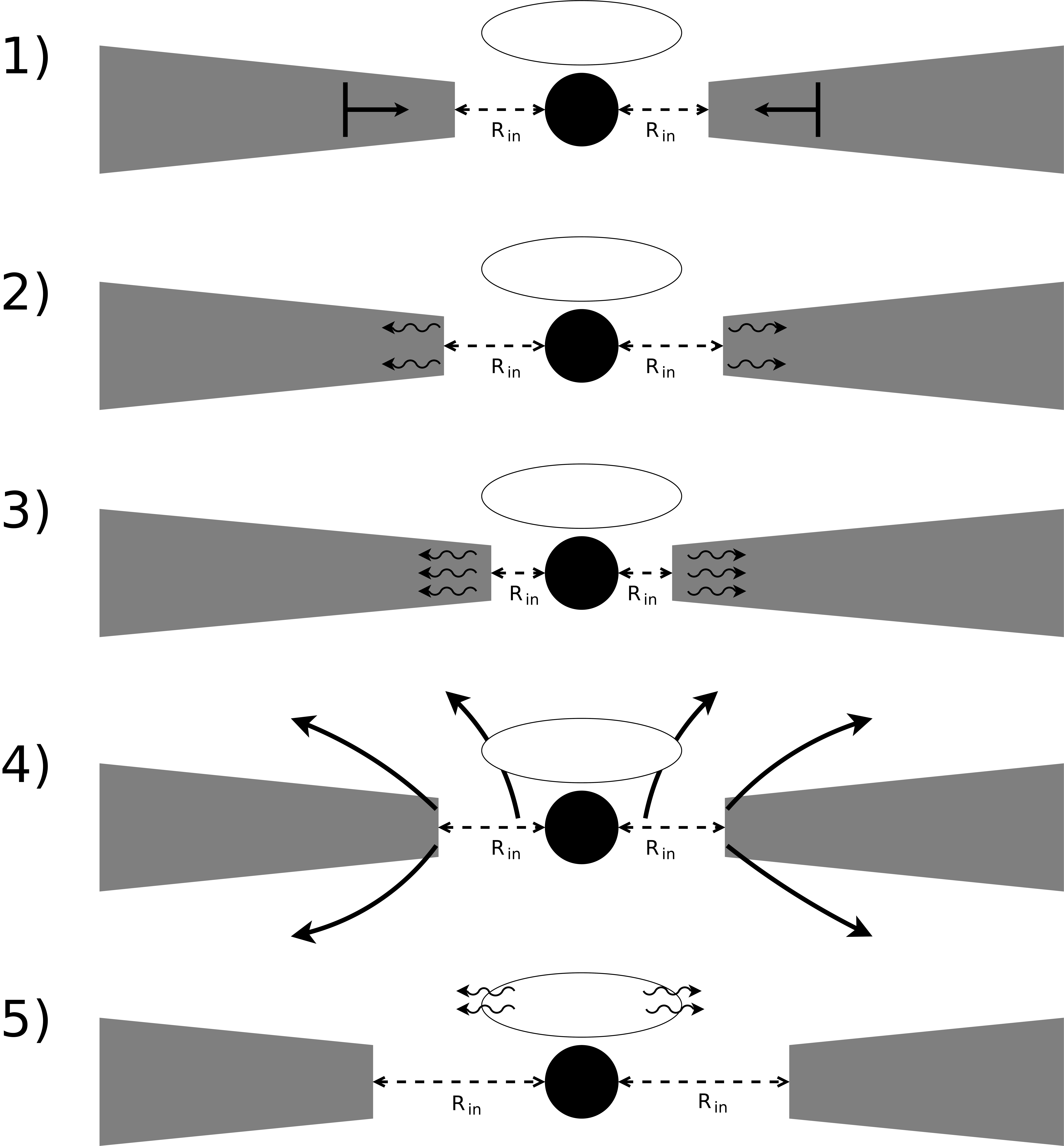}
  \caption[A schematic diagram illustrating the the process described by \citet{Neilsen_GRSModel} to describe the $\rho$ variability class in GRS 1915+105.]{A schematic diagram illustrating the the process described by \citealp{Neilsen_GRSModel} to describe the $\rho$\indexrho\ variability class in GRS 1915+105\index{GRS 1915+105}.  1) The X-ray emission from the system originates from both the accretion disk\index{Accretion disk} truncated at an inner radius $r_{in}$ (grey) and a corona of non-thermal electrons\index{Corona} (white ellipse).  At some time $t$, an overdensity in the accretion disk\index{Accretion disk} (formed by the instability\index{Instability} described by \citealp{Shakura_Instab}) propagates inwards towards $r_{in}$.  2) As the inner disc heats up, $r_{in}$ begins to slowly increase due to an increase in photon pressure\index{Radiation pressure}.  This destabilises\index{Instability} the disc.  3) At some critical density, the disc becomes too unstable and collapses inwards, greatly decreasing $r_{in}$ and raising the inner disc temperature.  4) The sudden increase in emission exceeds the local Eddington limit\index{Eddington limit} at $r_{in}$, ejecting matter from the inner accretion disc in the form of extreme winds\index{Wind}.  5) Having been excited by matter in the winds passing through it, the non-thermal electron cloud emits a hard Brehmsstrahlung `pulse'.}
  \label{fig:WindsModel}
\end{figure}

 \citealp{Neilsen_GRSModel} performed phase-resolved spectroscopy of the $\rho$\indexrho\ class in GRS 1915\index{GRS 1915+105}.  They find a hard `spike' after each flare, which they associate with the hard lag\index{Hard lag} in this class previously noted by e.g. \citealp{Janiuk_Lag}.  They propose a scenario in which high-velocity winds\index{Wind} formed by the ejection of matter from the inner disk\index{Accretion disk} interact directly with the corona\index{Corona} after a light travel time.  The corona then re-releases this energy as a hard bremsstrahlung pulse, causing the hard count rate spike seen in phase-resolved spectra.  This scenario is outlined in Figure \ref{fig:WindsModel}.  The authors expand on this scenario in \citealp{Neilsen_Rho} to suggest that this mechanism can explain all classes in GRS 1915 which display $\rho$-like flaring\index{Flare}.  However, this scenario still relies on the model of \citealp{Nayakshin_GRSModel} to generate the instability in the disk, and it implies that hard photons should always lag soft photons\index{Hard lag} in heartbeat-like variability classes.  Significantly, this scenario is therefore unable to explain the soft lags which have been observed in $\rho$-like variability in IGR J17091 \citep{Altamirano_IGR_FH}.

%\pagebreak
%\pagebreak

\par \citealp{Neilsen_GRSModel} also perform phase-resolved spectroscopy\index{Spectroscopy!Phase-resolved} (see Section \ref{sec:phasresspec}) of the flaring during $\rho$ variability.  In their fitting, \citealp{Neilsen_GRSModel} consider three spectral models:
\begin{enumerate}
\item An absorbed disk black body with a high energy cutoff, of which some fraction has been Compton upscattered
\item An absorbed disk black body with a high energy cutoff, plus a Compton component with a seed photon spectrum tied to the emission from the disk
\item An absorbed disk black body plus a Compton component with a seed photon spectrum tied to the emission from the disk and a bremsstrahlung component
\end{enumerate}
They find that the first of these models (Model 1) is the best fit to the data.
\par \citealp{Mineo_PhasRes} also performed psuedo-phase-resolved spectroscopy\index{Spectroscopy!Phase-resolved} of the $\rho$\indexrho\ class in GRS 1915, using a number of different spectral models to \citealp{Neilsen_GRSModel} but a significantly lower phase resolution.  In this work, the authors consider six models:
\begin{enumerate}
\item A multi-temperature disk black body plus a corona containing both thermal and non-thermal electrons (as formulated by \citealp{Poutanen_Hybrid}).
\item A multi-temperature disk black body plus a multi-temperature disk black body plus a power law.
\item A multi-temperature disk black body plus an independent Compton component.
\item A multi-temperature disk black body plus a power law plus reflection from the outer disk.
\item A model of Comptonization due to the bulk-motion of matter in the disk.
\item A multi-temperature disk black body plus a power law plus a standard black body.
\end{enumerate}
With the exception of Models 1 and 6, the authors find that none of these models are able to satisfactorily fit the data in each of their phase bins independently.  As there is no reasonable physical explanation behind Model 6, the authors only consider Model 1.  Their results suggest a large reduction of the corona luminosity during each heartbeat flare, which they interpret as the corona\index{Corona} condensing onto the disk\index{Accretion disk}.  They also find that their results are consistent with GRS 1915\index{GRS 1915+105} having a slim disk, but inconsistent with the hard lag\index{Hard lag} being caused by photon upscattering in the corona.
\par \citealp{Massa_MoveLag} found that the magnitude of the lag\index{Hard lag} between hard and soft photons in the $\rho$-class\indexrho\ of GRS 1915\index{GRS 1915+105} is not constant.  They found that the lag varies between $\sim3$--$10$\,s, and correlates strongly with count rate.  The magnitude of the lag, therefore, is too large to be simply due to a light travel time to the corona\index{Corona} from the disk\index{Accretion disk}.  The authors suggest that their results are instead consistent with the thermal adjustment of the inner disk itself as part of the instability limit cycle invoked to explain the flares\index{Flare}.
\par \citealp{Massaro_Numerical} constructed a set of differential equations to mathematically model the behaviour of the oscillator underlying $\rho$-like flaring in GRS 1915.  They find that a change between variability classes likely corresponds to a a change in global accretion rate\index{Accretion rate}, but that the global accretion rate within the $\rho$ class is constant.  This model reproduces the count rate-lag correlation reported by \citealp{Massa_MoveLag}, as well as a previously reported correlation between flare recurrence time\index{Recurrence time} and count rate \citep{Massaro_Lag}.
\par \citealp{Mir_LagModel} instead propose a model of variability\index{Variability} in the outer disk\index{Accretion disk} propagating inwards to the hotter inner disk.  They propose a model that explains both the hard lag of the fundamental frequency associated with the heartbeat flares, but also the hard lag\index{Hard lag} of the first harmonic\index{Harmonic}.  In contrast to the findings of \citealp{Massaro_Numerical}, their scenario requires a sinusoidal variation in the global accretion rate\index{Accretion rate} as a function of time.
\par More recently, \citealp{Zoghbi_Bulge} found that the reflection spectrum\index{Reflection spectrum} from GRS 1915\index{GRS 1915+105} does not match what would be expected from the inner disk\index{Accretion disk} behaviour assumed by e.g. \citealp{Nayakshin_GRSModel}.  They again perform phase-resolved spectroscopy\index{Spectroscopy!Phase-resolved} and fit a number of complex spectral models, finding that their data is best-described by a scenario involving the emergence of a bulge in the inner disk which propagates outwards during each flare.
\par The models and scenarios proposed for GRS 1915-like\index{GRS 1915+105} variability\index{Variability} all suffer from being based on observations of a single object: GRS 1915.  In Chapter \ref{ch:IGR} I perform a study of the variability in the GRS 1915-like object IGR J17091\index{IGR J17091-3624}.  Using my results from this second object, I discuss which of these scenarios are likely to best describe the physics of GRS 1915-like variability.

\section{Type II Burst Sources}

\label{sec:TypeII}

\par \index{X-ray burst!Type II}Type II Bursts are another dramatic form of second-to-minute scale X-Ray variability which are thought to be caused by disk\index{Accretion disk} instabilities\index{Instability} (e.g. \citealp{Lewin_TypeII}).  They are named by analogy to Type I X-Ray\index{X-ray burst!Type I} bursts; second-scale flashes of X-rays which are caused by thermonuclear explosions\index{Thermonuclear burning} on the surface of neutron stars\index{Neutron star} \citep{vanParadijs_TypeI,Lewin_Bursts}.
\par
\begin{figure}
  \centering
  \includegraphics[width=.9\linewidth, trim= 0mm 0mm 0mm 80mm,clip]{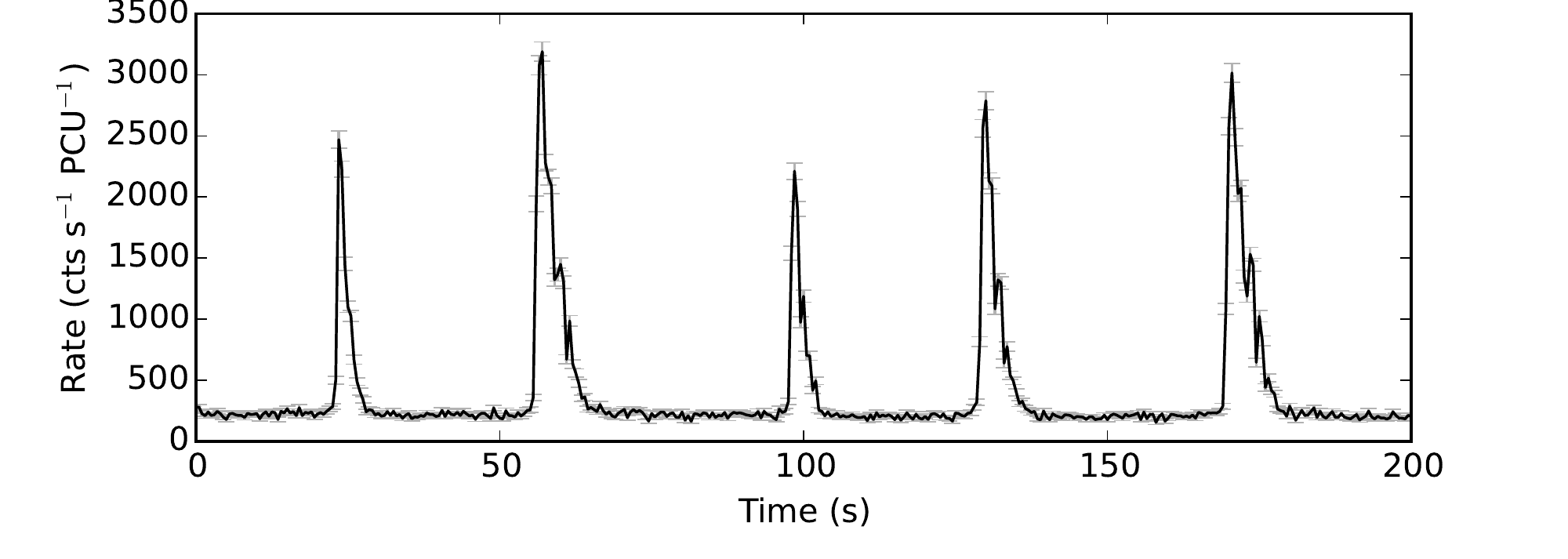}
  \caption[An \textit{RXTE}/PCA lightcurve\index{Lightcurve} of the Rapid Burster, showing a number of typical Type II X-ray bursts.]{An \textit{RXTE}/PCA\indexrxte\indexpca\ lightcurve of MXB 1730-335 (also known at the `Rapid Burster', showing a number of typical Type II X-ray bursts.}
  \label{fig:BgB}
\end{figure}

 In general, Type II bursts can be defined as second-to-minute scale X-ray bursts\index{X-ray burst} from neutron star LMXBs which are non-thermonuclear in origin; specifically, they lack the power-law-like decay profile \citep{intZand_Decay} and spectral cooling \citep{Hoffman_T1Cool} seen in Type I bursts.  In Figure \ref{fig:BgB}, I show lightcurves\index{Lightcurve} of a number of Type II bursts from the LMXB MXB 1730-335\index{Rapid Burster} \citep{Bagnoli_PopStudy}.  Type II bursts have a fast rise and a slow decay, and occur with separation times from tens of seconds to hours.

\par Type II bursts have definitively been observed in only two objects: the neutron star LMXBs MXB 1730-335\index{Rapid Burster} (also known as the `Rapid Burster', \citealp{Lewin_TypeII}) and GRO J1744-28\index{Bursting Pulsar}\index{GRO J1744-28|see {Bursting Pulsar}} (also known as the `Bursting Pulsar', \citealp{Paciesas_BPDiscovery}).  In both objects, Type II bursts have been observed during the soft state\index{High/Soft state} portion of multiple outbursts; this in turn suggests that the ability to produce Type II bursts is a property of the system, rather than the property of a specific outburst\index{Outburst}.  There have been claims of Type II-burst-like features during outbursts of a number of other LMXBs, such as SMC X-1\index{SMC X-1} \citep{Angelini_SMC}, but whether these features are the same phenomenon remains unclear.
\par The Rapid Burster\index{Rapid Burster} is an LMXB located in the globular cluster Liller 1\index{Liller 1} \citep{Lewin_TypeII}.  No pulsations\index{Pulsar} have been detected from the system, and as such the spin\index{Spin} of its compact object\index{Compact object} is not known.  However, the presence of Type I\index{X-ray burst!Type I} bursts from this object confirms that the compact object is a neutron star \citep{Hoffman_RB}.  Due to its location in a globular cluster, a number of infrared sources are consistent with the X-ray position of the Rapid Burster, and it is unclear which, if any, is the companion star\index{Companion star} in the system \citep{Homer_RBNoSec}.  However, also due to its association with Liller 1, the distance to the Rapid Burster is known to be 8.9--10\,kpc \citep{Ortolani_LillerD}.  Using this information, it has been shown that the persistent emission\index{Persistent emission} from the object during outburst peaks at no more than 20\% of its Eddington Limit\index{Eddington limit} \citep{Bagnoli_RB}.  The X-ray luminosity of the system at the peak of a Type II burst is around 100\% of its Eddington Limit \citep{Tan_RBBursts,Bagnoli_PopStudy}.  In addition to Type I and Type II bursts, variability\index{Variability} has been observed in the Rapid Burster which is remarkably similar to that associated with GRS 1915 and IGR J17091\index{GRS 1915+105}\index{IGR J17091-3624} (see Section \ref{sec:1915}), suggesting a possible link between these types of variability.
\par The Bursting Pulsar\index{Bursting Pulsar} is an LMXB located in a region of the sky very close to the Galactic centre.  Although Type I\index{X-ray burst!Type I} bursts have not been observed from this system, a coherent 2.14\,Hz X-ray pulsation seen from the object proves that the compact object\index{Compact object} is a pulsar\index{Pulsar} \citep{Kouveliotou_BPPulse} and hence a neutron star.  The distance to the object is $\sim3.4$--4.1\,kpc \citep{Sanna_BP}, and the nature of the companion star\index{Companion star} is unknown.  The persistent emission\index{Persistent emission} from the Bursting Pulsar is believed to peak at $\sim100$\% of its Eddington limit during outbursts, while its peak luminosity during Type II bursts greatly exceeds the Eddington limit\index{Eddington limit} \citep{Sturner_BPNature}.

\subsection{A History of Models of Type II Bursts}

\label{sec:TIImod}

\par \index{X-ray burst!Type II}No models have been proposed which can fully explain Type II bursting behaviour, but several models have been proposed in the context of Type II bursting from the Rapid Burster MXB 1730-33\index{Rapid Burster}.  A number of models invoke viscous instabilities\index{Instability} in the inner disk\index{Accretion disk} as the source of cyclical bursting: for a more detailed review of these models, see \citet{Lewin_Bursts}.
\par One such model was presented by \citet{Taam_Evo}.  They show that a disk\index{Accretion disk} that would be expected to be unstable due to the instability\index{Instability} described by \citealp{Shakura_Instab} can be stabilised by non-local energy transfer.  However they find that this effect is not sufficient to stabilise a disk in the case where viscous stress in the disk scales with local pressure.  In this case, they instead find that a limit cycles of behaviour can be set up, resulting in quasiperiodic flaring which the authors argue is similar to that seen in the Rapid Burster\index{Rapid Burster}.
\par \citet{Walker_Type2Mod} suggests that, for a neutron star\index{Neutron star} with a radius less than its ISCO\index{Innermost stable circular orbit}, a similar cycle of accretion\index{Accretion} can be set up when considering the effects of a high radiative torque.  In their scenario, \citeauthor{Walker_Type2Mod} find that pressure in the inner accretion disk\index{Accretion disk} of such an ultra-compact neutron star is entirely dominated by radiation stresses.  This leads to an unstable and highly non-linear region of the disk, leading to strong aperiodic variability.
\par \citet{Spruit_Type2Mod} (see also \citealp{Dangelo_Episodic1,Dangelo_Episodic2}) use a different approach.  Their model shows that, in some circumstances near the boundary of the propeller regime\index{Propeller effect}, the interaction between an accretion disk\index{Accretion disk} and a rapidly rotating magnetospheric boundary\index{Magnetospheric radius} can naturally set up a cycle of discrete accretion\index{Accretion} events rather than a continuous flow (for a description of this instability in a more general context, see Section \ref{sec:hic}\index{Hiccup accretion}).  The authors specifically discuss this flaring in the context of the Rapid Burster, noting a number of similarities between the output of their models and the properties of flares seen from the Rapid Burster.  However, they note a number of key ways in which their model differs from observations: the flares produced by their model are strictly periodic for a given accretion rate, and consequently the observed relationship between burst waiting time and burst fluence in the Rapid Burster cannot be reproduced.
\par In Chapter \ref{ch:BPbig} I perform a population study\index{Population study} of bursts from the Rapid Burster-like\index{Rapid Burster}\index{Bursting Pulsar} Bursting Pulsar, and use my results to better evaluate the models proposed to explain the Rapid Burster.  In Chapter \ref{ch:BPletter} I also consider an instability\index{Instability} similar to that proposed by \citet{Spruit_Type2Mod} to explain a previously undiscovered variability\index{Variability} during the late stages of outbursts from thr Bursting Pulsar.

\cleardoublepage

\chapter{Tools \& Methods}
\label{ch:methods}

\epigraph{\textit{The infinite is obvious and everywhere. To engage the finite takes courage.}}{Hunter Hunt-Hendrix -- \textit{Transcendental Black Metal}}

\vspace{1cm}

\par\noindent In this Chapter, I describe the tools and methods I employed as part of my studies.  In Section \ref{sec:sat} I describe the scientific instruments which were used to take the data I present in this thesis.  In Section \ref{sec:tec} I describe a number of methods and algorithms created by others which I make use of in my analysis.  I also present algorithms I have created as part of my studies.

\section{Instrumentation}

\label{sec:sat}

\par The atmosphere of the Earth is opaque to X-rays and gamma-rays, so we must use space-based observatories in order to study high-energy astrophysical phenomena.  A number of satellites dedicated to the study of X-rays have been launched over the years, starting with \textit{Uhuru}\index{Uhuru@\textit{Uhuru}} in 1970 \citep{Giacconi_Uhuru} and culminating, most recently, with NASA-operated \textit{NICER}\index{NICER@\textit{NICER}} \citep{Gendreau_Nicer} and the Chinese-operated \textit{Insight}\index{Insight@\textit{Insight}} \citep{Li_HXMT} in 2017.  I use data from a number of these missions in the research reported in this thesis; in particular I use data from the NASA satellites \textit{RXTE}, \textit{Swift}, \textit{Chandra} and \textit{NuSTAR}, the European satellites \textit{XMM-Newton} and \textit{INTEGRAL}, and the Japanese satellite \textit{Suzaku}\indexrxte\indexswift\indexchandra\indexnustar\indexxmm\indexintegral\indexsuzaku.  This section introduces the instruments used in my studies, as well as the tools used to extract their data for further analysis.
\index{Rossi X-Ray Timing Experiment@the \textit{Rossi X-Ray Timing Experiment}|see {\textit{RXTE}}}
\index{Neil Gehrels Swift Observatory@the \textit{Neil Gehrels Swift Observatory}|see {\textit{Swift}}}
\index{X-Ray Multi-Mirror Mission@the \textit{X-Ray Multi-Mirror Mission}|see {\textit{XMM-Newton}}}
\index{Nuclear Spectroscopic Telescope Array@the \textit{Nuclear Spectroscopic Telescope Array}|see {\textit{NuSTAR}}}
\index{International Gamma-Ray Astrophysics Laboratory@the \textit{International Gamma-Ray Astrophysics Laboratory}|see {\textit{INTEGRAL}}}
\index{Hubble Space Telescope@the \textit{Hubble Space Telescope}|see {\textit{HST}}}
\index{Neutron Star Interior Composition Explorer@the \textit{Neutron Star Interior Composition Explorer}|see {\textit{NICER}}}
\index{Advanced Telescope for High Energy Astrophysics@the \textit{Advanced Telescope for High Energy Astrophysics}|see {\textit{ATHENA}}}
\index{Compton Gamma Ray Observatory@the \textit{Compton Gamma Ray Observatory}|see {\textit{CGRO}}}
\index{Australian Telescope Compact Array@the Australian Telescope Compact Array|see {ATCA}}

\subsection{The \textit{Rossi X-Ray Timing Experiment}}

\par The \textit{Rossi X-Ray Timing Experiment}\indexrxte, more commonly known as \textit{RXTE}, was a NASA-operated satellite launched from Cape Canaveral in the United States on December 30, 1995 \citep{Bradt_RXTE}.  \textit{RXTE} was primarily an X-ray observatory, constructed specifically to study X-ray variability seen in X-ray Binaries \citep{Bradt_XTEaims}.  The observatory operated until January 5, 2012, when it was decommissioned.  \textit{RXTE} likely re-entered Earth's atmosphere over Venezuela on April 30, 2018.
\par \textit{RXTE} carried three scientific instruments.  The main instruments were a pair of X-ray telescopes: the Proportional Counter Array (PCA\indexpca, \citealp{Jahoda_PCA}) and the High Energy X-Ray Timing Experiment (HEXTE\indexhexte, \citealp{Gruber_HEXTE}).  The satellite also carried an X-ray All-Sky Monitor (ASM\indexasm, \citealp{Levine_ASM}).  PCA consisted of 5 Proportional Counting Units (PCUs) which were sensitive between $\sim2$--$60$\,keV.  The instrument had an excellent time resolution approaching 1\,$\mu$s, and an energy resolution of $\sim18\%$ at 6\,keV.  X-rays were guided onto the detectors by a collimator, resulting in an instrumental field of view with a full-width half-maximum of 1$^\circ$.  PCA had a 6500\,cm$^2$ collecting area, and no angular resolution \citep{Jahoda_PCA}.
\par The HEXTE\indexhexte\ instrument \citep{Gruber_HEXTE} provided complimentary coverage at higher energies, being sensitive in the $\sim15$--$250$\,keV range.  This instrument consisted of 8 detectors on two separate arms, with a total collecting area of 1600\,cm$^2$, and had a similar field of view to that of PCA.  The time resolution was 8\,$\mu$s, and the energy resolution was 15\% at 60\,keV \citep{Gruber_HEXTE}.
\par Finally, ASM\indexasm\ was a medium-energy X-ray all sky-monitor which covered 80\% of the sky every 90 minutes.  It was sensitive in the range 2--10\,keV, with a total collecting area of 90\,cm$^2$ and a spatial resolution of $3'\times15'$ \citep{Levine_ASM}.  Due to its near continual coverage of the sky, ASM was excellent for long-term monitoring of transients in the soft X-ray sky.

\subsubsection{Data Formatting}

\par Much of the work in this thesis is based largely on data from \textit{RXTE}/PCA, which is freely available through the HEASARC\index{HEASARC} archive maintained by NASA's Goddard Space Flight Centre\footnote{\url{https://heasarc.gsfc.nasa.gov/cgi-bin/W3Browse/w3browse.pl}}.  PCA, as well as many other X-ray instruments, records data in one of two forms:
\begin{itemize}
\item \textbf{Event-Mode Data\index{Event-mode data}:} A list of photon arrival times.  Depending on the instrument and observing mode, each of these times will have an associated channel\index{Channel}, information about where in the detector the photon hit and a flag indicating the pattern that the photon made on the detector.
\item \textbf{Binned Data\index{Binned data}:} A list of evenly spaced time bins with the number of photons which arrived during each.  Depending on the instrument and observing mode, this may be accompanied by some information on the channel distribution of photons arriving in each bin.
\end{itemize}
\par Both event-mode and binned-mode data are stored in a Flexible Image Transit System\index{Flexible Image Transit System|see {\texttt{FITS}}}\index{FITS@\texttt{FITS}} (\texttt{.fits}) format.  This is a hierarchical data format consisting of a number of `Header Data Units' (HDUs), each of which contains data in some format and a header with details of the format.  In addition to either an event list or a table of binned data, astronomical FITS files also contain a list of Good Time Intervals (GTIs) during which the satellite was functioning normally, as well as an amount of housekeeping information such as the start and end times of the observation.
\par The channel\index{Channel} a photon falls into is determined by its energy, although the channel-to-energy conversion for a particular instrument changes over time as the instrument degrades or settings are altered.  The approximate channel-to-energy conversions for PCA can be found at \url{https://heasarc.gsfc.nasa.gov/docs/xte/e-c_table.html}.
\par For PCA observations of faint objects, event mode data with full energy information (referred to as \texttt{goodxenon}-mode\indexgx\ data) is generally available.  However when brighter objects were observed, telemetry constraints sometimes prevented this full information from being transmitted to Earth.  In all observations, a number of alternative data products are available; \texttt{Standard1}\indexsto\ data (binned data with 0.125\,s time resolution but no energy information), \texttt{Standard2}\indexstt\ data (binned data with 16\,s time resolution, divided into 129 bins by channel) and a number of other data products with various time and energy resolutions.  \texttt{Standard2} data are useful for studying spectral variability over long timescales, while \texttt{Standard1} data are useful for studying fluctuations in X-ray luminosity over shorter timescales.  I use \texttt{goodxenon} data when available, as this allowed me to use the maximum possible time and energy resolutions.  When \texttt{goodxenon} was not I available I used various other datamodes, including \texttt{Standard1} and a number of different event-mode datamodes.

\subsubsection{Data Extraction \& Background Correction}

\par To perform science with PCA or other instruments, one must extract science products (such as lightcurves\index{Lightcurve}, power spectra\index{Fourier analysis}\index{Power spectrum|see {Fourier analysis}} and energy spectra\index{Spectroscopy}\index{Energy spectrum|see {Spectroscopy}}) from the raw data.  Tools to create lightcurves and power spectra from PCA\indexpca\ data are available as part of \texttt{FTOOLS}\index{FTOOLS@\texttt{FTOOLS}} \footnote{\url{https://heasarc.gsfc.nasa.gov/ftools/}}, a free NASA-maintained suite of software for manipulating \texttt{.fits}\index{FITS@\texttt{FITS}} formatted data.  These scripts make use of CALDBs: freely available databases of calibration files provided by NASA for a number of active and historical X-ray telescopes (e.g. \citealp{Graessle_ChaCALDB}).  I also wrote my own software \texttt{PANTHEON}\index{PANTHEON@\texttt{PANTHEON}} (Python ANalytical Tools for High-energy Event-data manipulatiON, presented in Appendix \ref{app:PAN}) to extract a number of additional products, such as power spectra and spectrograms.
\par In astronomy, the general way to subtract background\index{Background subtraction} from data is by selecting an empty piece of sky from the same observation as the source of interest, and then subtract the counts in one from the other.  However as PCA had no imaging capability, this is not possible with data from this instrument.  Instead, the \textit{RXTE} Guest Observatory Facility provides background models, which estimate the background of an observation based on the known X-ray background near the pointing direction and how the radiation environment of the spacecraft changes over its orbit.  Two background models are available, for faint\footnote{\url{http://heasarc.gsfc.nasa.gov/FTP/xte/calib_data/pca_bkgd/Faint/pca_bkgd_cmfaintl7_eMv20051128.mdl}} ($<40$\,cts\,s$^{-1}$\,PCU$^{-1}$) and bright\footnote{\url{http://heasarc.gsfc.nasa.gov/FTP/xte/calib_data/pca_bkgd/Sky_VLE/pca_bkgd_cmbrightvle_eMv20051128.mdl}} ($>40$\,cts\,s$^{-1}$\,PCU$^{-1}$) sources; these can be used in conjunction with the \texttt{pcabackest} tool in \texttt{FTOOLS}\index{FTOOLS@\texttt{FTOOLS}} to estimate the background as a function of time and energy.  This spectral model can then be subtracted from binned PCA data.
\par As the PCA background models do not subtract the contributions from other sources in the field of view, I also use a different technique to subtract background from observations of GRO J1744-28\index{Bursting Pulsar} (which is in a very crowded region of the sky near the Galactic centre).  To try and account for these other sources, I instead chose an observation of the region of GRO J1744-28 taken while this source was in quiescence\index{Quiescence}; I assume that all photons in this observation must be from the particle background, the cosmic background or another source in the field of view.  Although this method does subtract some of the background contributed from other sources in the field, it must be treated with caution as these other sources are likely also variable.
\par To compare photometry data from PCA with data from other instruments, I normalise the data by the flux from the Crab nebula\index{Crab nebula}.  The Crab is a commonly used reference source in astronomy due to its apparent brightness and low variability across a wide portion of the electromagnetic spectrum.  To Crab-normalise PCA data from a given observation, I take the PCA observation of the Crab which is closest in time to the observation of interest and in the same gain epoch.  This follows the method employed in \citet{Altamirano_CrabNorm}.
\par In addition to PCA, I also make use of data from \textit{RXTE}/ASM.  Long-term lightcurves from ASM are available on the ASM Light Curves Overview web page (\url{http://xte.mit.edu/asmlc/ASM.html}) maintained by MIT.

\subsection{The \textit{Neil Gehrels Swift Observatory}}

\par The \textit{Neil Gehrels Swift Observatory}\indexswift, formerly and more commonly known as \textit{Swift}, is a NASA-operated satellite launched from Cape Canaveral on November 20, 2004 \citep{Gehrels_Swift}.  \textit{Swift} was specifically designed to study Gamma Ray Bursts\index{Gamma ray burst} (GRBs), and is notable for its fast slew speed.
\par \textit{Swift} carries three instruments: the X-Ray Telescope (XRT\indexxrt, \citealp{Burrows_XRT}), the wide field-of-view hard X-ray Burst Alert Telescope (BAT\indexbat, \citealp{Krimm_BAT}) and an UltraViolet/Optical Telescope (UVOT\indexuvot, \citealp{Roming_UVOT}).  XRT is the primary instrument on \textit{Swift}: it is a focusing telescope with an effective energy range of 0.2--10\,keV.  Unlike PCA, XRT has imaging capabilities, with a field of view with a radius of 23.6' and an angular resolution of 18''.  The telescope has a minimum time resolution of 1.8\,ms and a minimum energy resolution of $\sim5$\% at 6\,keV.  XRT is operated in one of a number of `operating modes' during each observation, depending on the requirements of the observer.  The two main observing modes are:
\begin{enumerate}
\item Proportional Counting (PC) Mode: a full 2-dimensional image every 2.5\,s.
\item Windowed Timing (WT) Mode: a 1-dimensional image every 2.8\,ms.
\end{enumerate}
Both PC and WT modes also contain full energy information.
\par The main purpose of the wide area BAT\indexbat\ telescope is to identify gamma ray bursts\index{Gamma ray burst} as soon as possible after their onset, so that \textit{Swift} can then slew to them for follow-up observation with XRT.  Due to its large field of view (1.4\,sr) and effective energy range of 15--150\,keV, BAT also provides us with long-term hard X-ray lightcurves of many bright sources in the X-ray sky.  It has a detecting area of 5200\,cm$^2$ and, when operating in survey mode, a time resolution of 5 minutes.
\par The final instrument, UVOT\indexuvot, is intended to take simultaneous optical and ultraviolet observations of sources observed with XRT\indexxrt.  It observes in the wavelength range between 170--650\,nm.

\subsubsection{Data Extraction}

\par XRT\indexxrt\ and BAT\indexbat\ data on non-GRB\index{Gamma ray burst} transients\index{Transient source} are available via online portals maintained by the University of Leicester\footnote{\url{http://www.swift.ac.uk/user_objects/}} and the Goddard Space Flight Centre\footnote{\url{https://swift.gsfc.nasa.gov/results/transients/}} respectively.  The University of Leicester portal automatically extracts lightcurves\index{Lightcurve}, energy spectra\index{Spectroscopy}, images and source positions from raw XRT data of a given target, using the \texttt{xrtpipeline} provided in \texttt{FTOOLS}\index{FTOOLS@\texttt{FTOOLS}}.  The Goddard Space Flight Centre provides ready-made 15--50\,keV lightcurves of 1023\footnote{Count as of October 2018.} X-ray transients, with cadences of either 1 per day or 1 per \textit{Swift} orbit.

\subsection{The \textit{X-Ray Multi-Mirror Mission}}

\par The \textit{X-Ray-Multi Mirror Mission}\indexxmm\ (\textit{XMM-Newton}, \citealp{Jansen_XMM}) is an ESA-operated satellite which was launched from Kourou, French Guiana on December 10, 1999, and is still operating almost 20 years later.  Like \textit{RXTE} and \textit{Swift}, \textit{XMM-Newton} also carries a number of separate instruments: namely the European Photon Imaging Camera (EPIC\indexepic, \citealp{Bignami_EPIC}), the Reflection Grating Spectrometer (RGS\indexrgs, \citealp{denHerder_RGS}) and an Optical Monitor (OM\indexom, \citealp{Mason_OM}).  In the research presented in this thesis, I only make use of data from EPIC.
\par EPIC\indexepic\ consists of three CCD cameras which work independently: two metal-oxide semiconductor CCD cameras (EPIC-MOS1 and EPIC-MOS2) and a single pn CCD camera at the focus of the telescope (EPIC-pn).  All cameras observe in the energy range 0.15--15\,keV, with a Field of View of 30', an angular resolution of 6'' and a maximum energy resolution of $\sim5$\%.  The detectors can be operated in full frame, partial window or timing mode, each of which has a greater time resolution but narrower field of view than the last.  The maximum time resolution achievable by EPIC is 7\,$\mu$s which EPIC-pn is operated in burst mode; a special pn-only variant of timing mode.

\subsubsection{Data Extraction \& Processing}

\par \textit{XMM-Newton}\indexxmm\ data are extracted and processed using the \texttt{SAS}\index{SAS@\texttt{SAS}} software \citep{Ibarra_sas} provided by ESA\footnote{\url{https://www.cosmos.esa.int/web/xmm-newton/sas}}.  These make use of the continuously updated Current Calibration Files (CCF), also provided by ESA.
\par The process of extracting basic data products from the EPIC instruments can be reduced to a number of steps:
\begin{itemize}
\item Use the \texttt{SAS} command \texttt{cifbuild} to create a Calibration Index File (CIF), containing pointers to the information in the CCF needed to reduce the chosen dataset.
\item Use the \texttt{SAS} command \texttt{odfingest} to create a summary file, containing data corrected by the CCF and by the EPIC housekeeping files.
\item Construct a photon event list from EPIC-MOS1 and EPIC-MOS2 using the \texttt{SAS} command \texttt{emproc}, or from EPIC-pn using the command \texttt{epproc}.
\end{itemize}
The event lists\index{Event-mode data} that result from this process can then be filtered using \texttt{evselect}, which allows the user to sort photons by arrival time, spatial co-ordinate and energy channel, among other parameters.  These filtered event lists can then be used to create science data products, such as lightcurves\index{Lightcurve} and energy spectra\index{Spectroscopy}.

\subsection{\textit{Chandra}}

\par The \textit{Chandra X-Ray Observatory}\indexchandra\ (\textit{Chandra}, \citealp{Weisskopf_Chandra}) is a NASA-operated satellite which was launched from Cape Canaveral on July 23, 1999 aboard Space Shuttle \textit{Columbia}.  The mission is considered to be one of NASA's `Great Observatories', along with the \textit{Hubble Space Telescope} (\textit{HST}, e.g. \citealp{Holtzman_Hubble})\index{HST@\textit{HST}}, the \textit{Compton Gamma Ray Observatory} (\textit{CGRO}\index{CGRO@\textit{CGRO}}, \citealp{Gehrels_CGRO}) and the \textit{Spitzer Space Telescope} (\textit{Spitzer}\index{Spitzer@\textit{Spitzer}}, \citealp{Fanson_Spitzer}), which collectively observed the sky between infrared and gamma-ray wavelengths.  \textit{Chandra} was designed to study the X-ray sky between $\sim$0.1--10\,keV, and contains two instruments: the High Resolution Camera (HRC\indexhrc, \citealp{Kenter_HRCI}) and the Advanced CCD Imaging Spectrometer (ACIS\indexacis, \citealp{Nousek_ACIS}).  The spacecraft also carries High and Low Energy Transmission Gratings (HETG and LETG respectively, \citealp{Markert_HETG,Brinkman_LETG}), which can be used in conjunction with the aforementioned detectors to produce high-resolution energy spectra.
\par HRC\indexhrc\ contains two detectors: the HRC Imager (HRC-I) and the HRC Spectrogram (HRC-S).  HRC-I has the largest field of view of any instrument aboard Chandra (30$\times$30'), but no time resolution and only poor spectral resolution.  HRC-S is a long-thin detector strip which is intended to be used as the readout for the LETG.  This detector can also be used in Continuous Clocking mode, in which it has no energy resolution but a timing resolution of 16\,$\mu$s.
\par ACIS\indexacis\ is intended for use either as an imaging camera or as a detector for the output of the HETG.  It has a primary field of view of $16.9\times16.9$', and operates at a maximum time resolution of 2.85\,ms.

\subsubsection{Data Extraction \& Processing}
\par Like \textit{XMM-Newton}, \textit{Chandra}\indexchandra\ data are analysed using a purpose-built suite of tools.  The software for \textit{Chandra} analysis is named \texttt{CIAO}\index{CIAO@\texttt{CIAO}} \citep{Fruscione_Ciao}, and is freely provided by Harvard University\footnote{\url{http://cxc.harvard.edu/ciao/}}.  \texttt{Ciao} filters and bins data based on any of the four possible parameters stored for a photon event (time, energy and two spatial co-ordinates), and facilitates the production of lightcurves\index{Lightcurve}, images and energy spectra\index{Spectroscopy}.

\subsection{\textit{Suzaku}}

\par \textit{Suzaku}\indexsuzaku\ \citep{Mitsuda_Suzaku} was a JAXA-operated satellite which operated from its launch from the Uchinoura Space Center, Japan on July 10, 2005 until being decomissioned on September 2, 2015.  The mission was intended for X-ray spectroscopy; however the satellite's primary instrument, the X-Ray Spectrometer (XRS\indexxrs, \citealp{Kelley_XRS}), lost all of its liquid helium coolant within the first month of operation, rendering it effectively unusable.  The remaining instruments aboard \textit{Suzaku}, namely the X-Ray Imaging Spectrometers (XIS\indexxis, \citealp{Koyama_XIS}) and the Hard X-Ray Detector (HXD\indexhxd, \citealp{Takahashi_HXD}) were unaffected by the malfunction and continued to operate normally throughout the spacecraft's lifetime.
\par XIS\indexxis\ consists of four X-ray cameras, with a total field of view of $17.8\times17.8$' and a spatial resolution of $\geq1.6$''.  One of these cameras (XIS2) was damaged by a micrometeorite, and was switched off on November 9, 2006.  The instrument has a good spectral resolution over its operational energy range of 0.2--10\,keV, peaking at $\sim170$\,eV at the upper end of this range.  Standard XIS observation modes provide a time resolution of 8\,s, corresponding to the duration of a single CCD exposure.  This timing resolution can be improved by a factor of a few by sacrificing imaging information in other observation modes, such as the one-dimensional P-sum mode with a timing resolution of 7.8\,ms.
\par HXD\indexhxd\ complements XIS\indexxis\ at higher energies, with an effective energy range of 12--600\,keV.  The instrument has an energy resolution of $\sim3$\,keV below 60\,keV, and $\sim7$--$8\sqrt{E}$\% above 60\,keV, where $E$ is the energy in MeV.  The instrument has an optimum time resolution of 61\,$\mu$s.
\par As with most instruments, there exist standard procedures when reducing and analysing data from \textit{Suzaku}\indexsuzaku.  First of all, the data must be reprocessed using the \texttt{aepipeline} script available as a part of \texttt{FTOOLS}\index{FTOOLS@\texttt{FTOOLS}}.  Lightcurves, images and spectra can then be extracted using the standard multimission tools also available in FTOOLS\index{FTOOLS@\texttt{FTOOLS}}.  Note that, for XIS\indexxis, backgrounds for each of the four detectors should be extracted separately due to differential degradation over the lifetime of the mission.

\subsection{The \textit{Nuclear Spectroscopic Telescope Array}}

\par The \textit{Nuclear Spectroscopic Telescope Array}\indexnustar\ (\textit{NuSTAR}, \citealp{Harrison_NuStar}) is a NASA-operated satellite launched from the \textit{Stargazer} aircraft off the coast of the Marshall Islands on June 13, 2012.  The satellite carries two co-pointing X-ray telescopes, which are matched with Focal Plane Modules referred to as FPMA and FPMB.  These detectors are sensitive and calibrated in the range 3--78\,keV, and each has an effective area of $\sim450$\,cm$^2$ at $\sim10$\,keV.  The telescopes have a field of view of $12.2\times12.2$', and a full-width half-maximum angular resolution of $\gtrsim18$''.  Events are detected by \textit{NuSTAR} with a time resolution of 2\,$\mu$s, while the energy resolution at 50\,keV is around 0.4\,keV.
\par \textit{NuSTAR} is the first instrument able to focus hard X-rays ($\gtrsim10$\,keV) to produce relatively clear images.  Additionally \textit{NuSTAR} does not suffer from issues due to pile-up\index{Pile-up} (an instrumental effect in most instruments which overestimates the hard flux from bright objects, see Section \ref{sec:pude}), but it does suffer from significant dead-time\index{Dead-time} (which underestimates flux from bright objects, see also Section \ref{sec:pude}).  See also \citet{Bachetti_dt}.
\par For the research presented in this thesis, I reduce data from \textit{NuSTAR} using the \texttt{nupipeline} script from the freely available \textit{NuSTAR} Data Analysis Software (NuSTARDAS\index{NuSTARDAS@\texttt{NuSTARDAS}}\footnote{\url{https://heasarc.gsfc.nasa.gov/docs/nustar/analysis/nustar_swguide.pdf}}).  This script automatically runs all of the relevant tasks required to reduce \textit{NuSTAR} data, including flagging of events, flagging of bad pixels and correcting for detector gain.  It also calculates sky co-ordinates and energy for each event.

\subsection{The \textit{International Gamma-Ray Astrophysics Laboratory}}

\par The \textit{INTernational Gamma-Ray Astrophysics Laboratory} (\textit{INTEGRAL}\indexintegral, \citealp{Winkler_INTEGRAL}) is an ESA-operated satellite which was launched from Kazakhstan's Baikonur Cosmodrome on October 17, 2002.  The primary purpose of the mission is the spectroscopy of astrophysical sources in the hard X-ray and soft gamma-ray bands, between $\sim4$--10,000\,keV.
\par \textit{INTEGRAL} carries two main scientific instruments: the SPectrometer on \textit{INTEGRAL} (SPI\indexspi, \citealp{Vedrenne_SPI}) and the Imager on-Board \textit{INTEGRAL} (IBIS\indexibis, \citealp{Winkler_IBIS}).  The spacecraft also carries an X-ray monitor (JEM-X\indexjemx, \citealp{Schnopper_JEMX}) and an optical camera (OMC\indexomc, \citealp{Gimenez_OMC}).  SPI\indexspi\ is a high-resolution gamma-ray spectrometer, with an energy resolution of $\sim$2.2\,keV at 1.33\,MeV and an energy range of 18\,keV to 8\,MeV.  It has a field of view of $>14^\circ$, an angular resolution of 2.5$^\circ$, a time resolution of 0.129\,ms and a collecting area of $\sim500$\,cm$^2$.
\par IBIS\indexibis\ has a wider energy range than SPI\indexspi, from 15\,keV to 10\,MeV, and a larger collecting area at 2600\,cm$^2$ at 100\,keV.  The timing accuracy is 61\,$\mu$s, but the energy resolution peaks at only 8\% at $\sim100$\,keV.  As IBIS is an imager, it has a good angular resolution of $\sim12$' and a fully-coded field of view of $8\times8$''.  IBIS contains two detector planes stacked on top of each other; the top layer (ISGRI, \citealp{Lebrun_ISGRI}) is designed to detect low-energy gamma rays, while the lower layer (PICsIT, \citealp{Labanti_Picsit}) is designed to detect the higher-energy gamma rays which pass through ISGRI undetected.
\par Data products from all four instruments are available via the \textit{INTEGRAL} Heavens portal \citep{Lubinski_Heavens} maintained by the \textit{INTEGRAL} Science Data Centre \footnote{\url{https://www.isdc.unige.ch/heavens/}}.  This portal provides images, lightcurves\index{Lightcurve} and spectra\index{Spectroscopy} of data taken from archived \textit{INTEGRAL} observations.

\subsection{Dead-time and Pile-up}

\label{sec:pude}

\par All X-ray telescopes suffer from a number of instrumental biases, caused by a number of instrumental effects.  Two of the most significant of these are dead-time\index{Dead-time} and pile-up\index{Pile-up}, which are both caused by the limitations of CCD detectors.  When a photon is detected, a CCD takes a finite time to respond to it to form a digital signal.  During this response time, known as the `dead-time', the instrument is unable to respond to any additional photons.  This means that the instrument is `blind' to photons for a period of time after each registered event.  The dead-time of a given instrument is generally of the order of a few $\mu$s per event.  For high incident count rates, dead-time can lead to a significantly reduced reported count rate.  In addition to this there is an effect on the statistics of photon arrival times.  Photon arrival times from an astrophysical source are generally Poisson distributed\index{Poisson distribution}; however, the existence of dead-time means that two consecutive reported photon arrival times are no longer independent of each other.  This in turn can effect the level and the shape of the noise component seen when analysing Fourier spectra\index{Fourier analysis} of the data (see Section \ref{sec:fou}).
\par Similarly, pile-up\index{Pile-up} is an effect which is mostly seen in data from bright sources.  Pile-up occurs when two photons coincide both temporally and spatially in such a way that the detector interprets them as a single event.  This causes two photon events to be recorded as a single photon event with an energy equal to the sum of the two.  This effect causes the hard emission from a source to be over-reported, and the soft emission to be under-reported.  The exact magnitude of the effects from dead-time to pile-up varies from detector to detector; for most detectors the effects are well understood and can therefore be estimated and corrected for.

\section{Methods \& Techniques}

\label{sec:tec}

\par To extract meaningful physics from the data provided by the space-based observatories described in Section \ref{sec:sat}, I use a number of mathematical and analytical techniques.  In a nutshell, I analyse three main properties of the data:
\begin{enumerate}
\item \textbf{Lightcurve Morphology:}\index{Lightcurve} Describing how the intensity of an object in a given energy band varies as a function of time.
\item \textbf{Timing Analysis:} Using Fourier\index{Fourier analysis} and Lomb-Scargle\index{Lomb-Scargle periodogram} analyses to identify periodic and quasi-periodic\index{Quasi-periodic oscillation} variability\index{Variability} in the data, and how these change with energy and time.
\item \textbf{Energy Spectral Analysis\index{Spectroscopy}:} Measuring the shape of the energy spectrum of an object, particularly by using hardness ratios (see e.g. Section \ref{sec:hids}), and analysing how this changes over time.
\end{enumerate}
\par Some of the techniques I used to explore these properties are detailed in this section.

\subsection{Lightcurve Morphology}

\label{sec:LCMorph}

\par The morphology of a lightcurve\index{Lightcurve}, how the brightness of a source varies over time, can tell us about the physical processes at work in a system.  For example, the rise and fall-times of an X-ray burst\index{X-ray burst} can be matched with characteristic physical timescales of an accreting system to better understand which of them play roles in generating the bursts.  However, quantifying these shapes over short timescales, or in low-quality datasets, can be difficult.  As such, methods exist to help analyse the morphology of these difficult datasets, and I employ a number of them in the research presented in this thesis.

\subsubsection{Lightcurve Folding}

\label{sec:normfold}

\par In systems with periodic or quasi-periodic behaviour\index{Quasi-periodic oscillation}, it is important to understand the morphology of a single cycle of the behaviour.  In order to improve the statistics on such data, one can take the average of many cycles, resulting in a single averaged cycle with a greatly increased signal-to-noise ratio.  The process of obtaining this average cycle is known as `folding'\index{Folding} data.
\par To fold a periodic dataset with a known period $p$, the time $t$ associated with each datapoint must be converted into a phase $\phi$ (for $0\leq\phi<1$) such that datapoints at the same stage of different oscillations have the same $\phi$.  This can be done using the formula:
\begin{equation}
\phi(t)=\Phi(t)\mod1=(t-t_0)/p\mod1=(t-t_0)/p-N_t
\label{eq:simfold}
\end{equation}
Where $\Phi(t)$ is the fractional number of cycles which have elapsed between times $t_0$ and $t$ for an arbitrary start time $t_0$, and $N_t$ is the integer number of complete cycles which have occured between times $t_0$ and $t$.
\par
\begin{figure}
    \includegraphics[width=\columnwidth, trim = 0mm 220mm 0mm 0mm,clip]{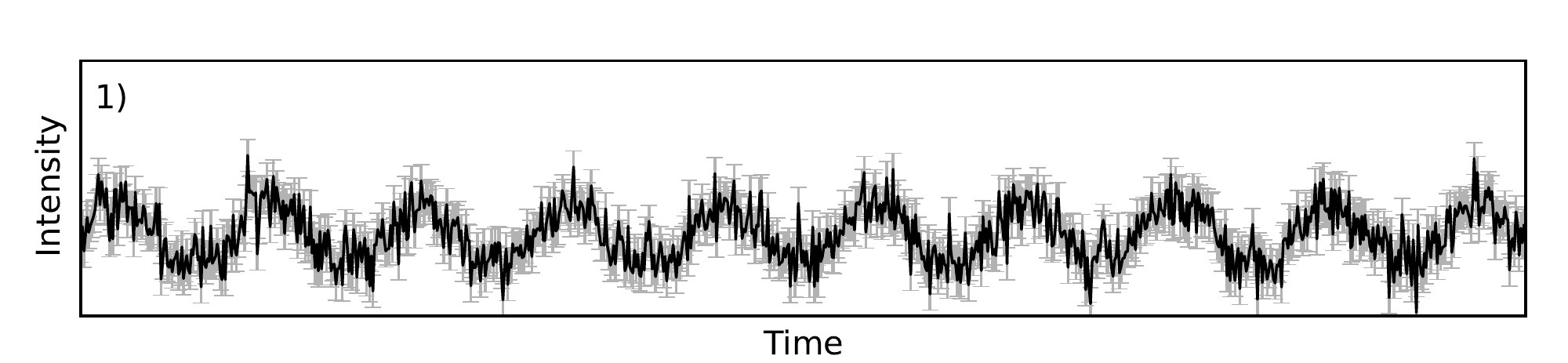}
    \includegraphics[width=\columnwidth, trim = 0mm 155mm 0mm 65mm,clip]{images/folding.eps}
    \includegraphics[width=\columnwidth, trim = 0mm 20mm 0mm 135mm,clip]{images/folding.eps}
    \captionsetup{singlelinecheck=off}
    \caption[A cartoon illustrating the process of folding a periodic lightcurve with a known period.]{A cartoon illustrating the process of folding\index{Folding} a periodic lightcurve with a known period, which I describe mathematically in Section \ref{sec:normfold}.  \textbf{1:} a simulated lightcurve with errors.  \textbf{2:} Divide the lightcurve into sections by cutting it at every time coordinate $Np$, where $p$ is the known period and $N$ is any integer.  Each data point may now be given a phase coordinate $\phi$ in addition to its time coordinate $t$, where $\phi=(t/p)-N$ for $N$ such that $0\leq\phi<1$.  \textbf{3:} The lightcurve segments can be realigned in phase-space, such that points with the same value of $\phi$ sit at the same $x$-coordinate.  \textbf{4:} All points within given bins in $\phi$-space are averaged to create a lightcurve corresponding to the averaged oscillations of the original lightcurve.  The folding has revealed a peak at $\phi=0.75$ which was not apparent in the unfolded data.}
   \label{fig:Folding}
\end{figure}

 This procedure can also be thought of as cutting a lightcurve into a number of segments each of length $p$.  Each datapoint in each segment can then be assigned a $\phi$ value, where $\phi$ corresponds to the datapoint's fractional position within its segment.  Once this information has been found for all datapoints, the data can be rebinned in $\phi$-space to `stack` every cycle on top of one another.  I illustrate this process visually in Figure \ref{fig:Folding}.  This method is useful for finding mean oscillation profiles when $p$ is very close to a constant, such as finding the mean pulse profile of a pulsar over a small number of rotations.  However in many cases, such as in quasi-periodic oscillations\index{Quasi-periodic oscillation} (QPOs), $p$ is not a constant.  More complex methods must then be used to find the mean pulse profile.

\subsubsection{Flare-Finding Algorithm}
\label{sec:Flares}

\par To fold a quasi-periodic oscillation\index{Quasi-periodic oscillation}, such as the `heartbeat'\indexrho\ flares\index{Flare} seen in GRS 1915+105\index{GRS 1915+105} and IGR J17091-3624\index{IGR J17091-3624}, it is first important to find the time-coordinates which characterise the beginning, end and peak of each flare.  To this end, I have created an algorithm to locate individual flares in a dataset containing non-periodic high-amplitude flares. The algorithm is performed as such:

\begin{enumerate}
  \item Choose some threshold values $T_L$ and $T_H$.  Set the y-value of all datapoints with $y<T_L$ to zero.
  \item Retrieve the time co-ordinate of the highest value remaining in the dataset.  Call this value $t_m$ and store it in a list.
  \item Set the $y$-value of the point at $t_m$ to zero.
  \item Scan forwards from $t_m$ by selecting the datapoint at $t_m+\Delta t$, where $\Delta t$ is the time resolution of the data.  If the selected point has a nonzero value, set it to zero and move to the next point.  If the selected point has a zero value, move to step 5.
  \item Scan backwards from $t_m$ by selecting the datapoint at $t_m-\Delta t$.  If the selected point has a nonzero value, set it to zero and move to the previous point.  If the selected point has a zero value, move to step 6.
  \item Retrieve the y-co-ordinate of the highest value remaining in the dataset.  Call this $y_m$.
  \item If $y_m>T_H$, repeat steps 2--7.  If $y_m<T_H$, proceed to step 8.
  \item Restore the original dataset.
  \item Retrieve the list of $t_m$ values found in step (ii).  Sort them in order of size.
  \item For each pair of adjacent $t_m$ values, find the $t$-coordinate of the datapoint between them with the lowest y-value.  Call these values $t_c$.
  \item This list of $t_c$ can now be used to demarcate the border between peaks.
\end{enumerate}

\begin{figure}
    \includegraphics[width=\columnwidth, trim = 0mm 30mm 0mm 28mm]{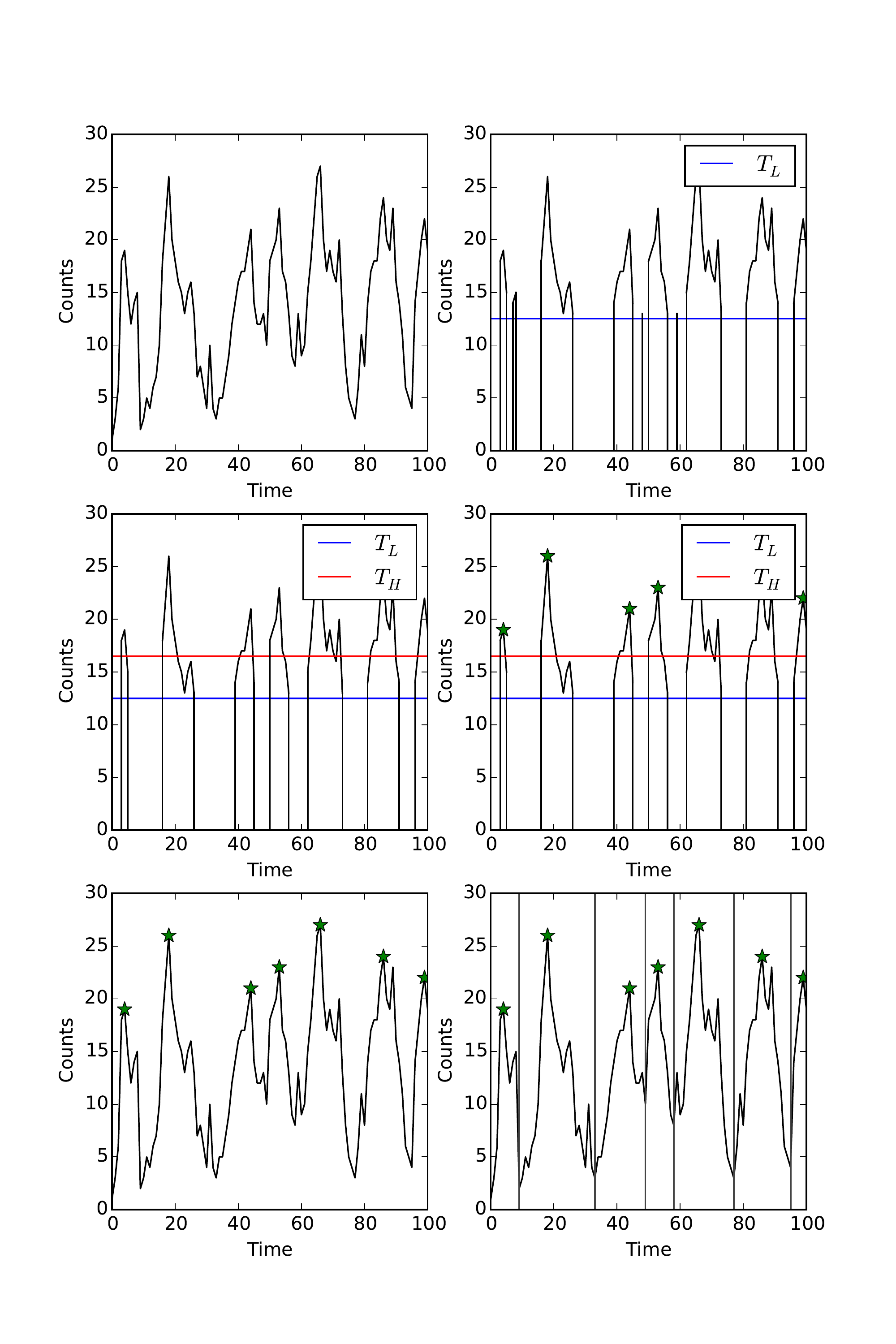}
    \captionsetup{singlelinecheck=off}
    \caption[A cartoon illustrating the procedure of the algorithm described in Section \ref{sec:Flares}.]{A cartoon illustrating the procedure of my algorithm which I describe in section \ref{sec:Flares}.  From top-left: (i) An untouched lightcurve\index{Lightcurve}.  (ii) The lightcurve with all $y<T_L$ removed.  (iii) The lightcurve with all contiguous nonzero regions with $\max(y)<T_H$ removed.  (iv) The $t$-coordinates of peak $y$-values $t_m$.  (v) The restored lightcurve with the $t_m$ highlighted.  (vi) The boundaries between adjacent peaks.}
   \label{fig:BurstAlg}
\end{figure}

The process can be thought of as using $T_L$ to divide the data into a number of discrete segments of non-zero data, and treating the peak of each segment as the peak of a flare.  I illustrate this process visually in Figure \ref{fig:BurstAlg}.

\par The threshold values $T_L$ and $T_H$ can also be procedurally generated for a given dataset:

\begin{enumerate}
  \item Select a small section of the dataset or a similar dataset (containing $\sim20$ peaks by eye) and note the time-coordinates $t_e$ of all peaks found by eye.
  \item Let $P_L$ and $P_H$ be two arbitrary values in the range $[0,100]$.
  \item Let $T_L$ ($T_H$) be the $P_L$th ($P_H$th) percentile of the y-values of the subsection of dataset.
  \item Run the flare-finding algorithm up to step 9.  Save the list of $t_m$.
  \item Split the dataset into bins on the x-axis such as the bin width $b\ll p$, where $p$ is the rough x-axis separation between peaks.
  \item For each bin, note if you found any value in $t_m$ falls in the bin and note if any value of $t_e$ falls in the bin.
  \item Using each bin as a trial, compute the Heidke Skill Score\index{Heidke skill score} \citep{Heidke_SKSC} of the algorithm with the method of finding peaks by eye:
  \begin{equation}HSS = \frac{2(AD-BC)}{(A+B)(B+D)+(A+C)(C+D)}
  \label{eq:HSS}
  \end{equation}
  Where $A$ is the number of bins that contain both $t_e$ and $t_m$, $B$ ($C$) is the number of bins that contain only $t_m$ ($t_e$) and $D$ is the number of bins which contain neither \citep{Kok_YesNo}.
  \item Repeat steps (iii)--(vii) for all values of $P_H>P_L$ for $P_L$ and $P_H$ in $[1,100]$.  Use a sensible value for the resolution of $P_L$ and $P_H$.  Save the HSS for each pair of values
  \item Locate the maximum value of HSS, and note the $P_L$ and $P_H$ values used to generate it.  Use these values to generate final $T_L$ and $T_H$ values.
\end{enumerate}

%I show an example of Heidke skill score grid for this algorithm, applied to a Class IV observation, in Figure \ref{fig:Heidke}.

%\begin{figure}
%    \includegraphics[width=\columnwidth, trim = 0mm 10mm 0mm 10mm]{images/HSS_J.eps}
%    \captionsetup{singlelinecheck=off}
%   \caption[The Heidke Skill score of a Class IV observation of IGR J17091-3624 for a selection of different values $P_L$ and $P_H$ (low and high threshold respectively).]{The Heidke Skill score of a Class IV observation of IGR J17091-3624 for a selection of different values $P_L$ and $P_H$ (low and high threshold respectively).}
%   \label{fig:Heidke}
%\end{figure}

\subsubsection{Variable Period Lightcurve Folding}

\label{sec:vfold}

\par With the values $t_m$ and $t_c$ found using the algorithm described above, it is possible to recast Equation \ref{eq:simfold} to fold\index{Folding} data over a high-amplitude but quasi-periodic oscillation\index{Quasi-periodic oscillation}.  I detail my method below:

\begin{enumerate}
  \item Take the ascending list of peak $t$-coordinates $t_m$.  Assign the first element a value $\Phi=0$.
  \item Assign each other point in $t_m$ an integer value $\Phi(t)$, such that the $\Phi$ value of the $i$th value of $t_m$ is defined as:
  \begin{equation}
  \Phi(t_m^{i})=\Phi(t_m^{i-1})+1, i\geq2
  \end{equation}
  \item If the troughs between peaks are well-defined, proceed to step 4.  Otherwise, skip to step 6.
  \item If the $t$-coordinate of the first datapoint in $t_c$ is less than the $t$-coordinate of the first datapoint in $t_m$, assign $\Phi(t_c^1)=-0.5$.  Otherwise, assign $\Phi(t_c^1)=0.5$.
  \item Assign each other point in $t_m$ a value $\Phi(x)$, such that the $\Phi$ value of the $i$th value of $t_c$ is defined as:
  \begin{equation}
  \Phi(t_c^{i})=\Phi(t_c^{i-1})+1, i\geq2
  \end{equation}
  \item Create a general function defining $\Phi$ for all $t$ by fitting the $t$ and $\Phi$ values of $t_m$ (and $t_c$, if used) with a monotonically increasing univariate cubic spline\index{Spline}\footnote{Computationally realised as \texttt{PchipInterpolator} in the \texttt{scipy} package for Python \citep{NumPy}.} $S(t)$.
  \item Define the phase $\phi(t)$ of an arbitrary time $t$ as $\phi(t)=S(t)\mod1$.
\end{enumerate}

\par With a phase defined for all points in time, the data can be manipulated as if it had been folded\index{Folding} in the usual way.  If the trough times in addition to the peak times are used to construct the spline, then the folded data are more accurate: however, by definition the rising part of each flare will occupy phases 0.5--1.0, while the falling part will occupy 0.0--0.5, so any asymmetry in the rise and fall times of the average flare is lost.
\par This method assumes that $d\phi/dt$ is continuous at all $t$, but this assumption is not necessarily true for cases in which each flare is a discrete event.  Consider for example the path of a juggling ball.  During each throw, the ball takes some time $\tau$ to complete its arc, moving from $\phi=0$ to $\phi=1$.  However, the value of $\tau$, and hence the value of $d\phi/dt$, depends on the impulse given to the ball at the moment of being thrown.  As such, $d\phi/dt$ is discontinuous at the point of the ball being thrown, and my method outlined here would not correctly fold a curve of its height as a function of time.

\subsection{Timing Analysis}

\label{sec:fou}

\par Another way of looking at the variability of an astrophysical source is by looking in the frequency domain.  Well-established mathematical techniques, in particular Fourier spectroscopy\index{Fourier analysis}, are able to deconvolve a time series into series of sine waves.  The amplitudes of these sine waves indicate how much variability in the system takes place at a given frequency.

\subsubsection{Fourier Analysis}

\par Fourier analysis\index{Fourier analysis} \citep{Fourier} is the most common way to perform frequency analysis on a time series.  The Fourier transform $\hat{f}(\nu)$ of a time series $f(t)$ is defined as:
\begin{equation}
\hat{f}(\nu)=\int_\infty^\infty f(t)e^{-2\pi it\nu} dt
\end{equation}
Where $\nu$ is the frequency to be probed and $i\equiv\sqrt{-1}$.  The magnitudes of the complex values $\hat{f}(\nu)$ describe the amplitude of the sine wave deconvolution at frequency $\nu$, while the arguments describe the relative phase of each of these sine waves.  As such, a plot of $|\hat{f}(\nu)|$ against $\nu$, known as a Fourier spectrum, can highlight the frequencies at which the time series shows oscillations.  A strictly periodic oscillation shows up in a Fourier spectrum as a delta spike at a single frequency $\nu_p$; if the oscillation is not strictly sinusoidal, then there may also be spikes present at the harmonic\index{Harmonic} frequencies $N\nu_p$ for any $N\in\mathbb{N}$.
\par An oscillation which is not strictly periodic is known as a quasi-periodic oscillation\index{Quasi-periodic oscillation}, or QPO.  The non-periodic component in a QPO can be related to its frequency (such as a spinning object which slows down over time), its amplitude (such as a damped harmonic oscillator) or some internal phase drift (such as the X-ray flux from an accreting X-ray pulsar\index{Pulsar} on which the hotspot is migrating, see e.g. \citealp{Patruno_Phase}).  A quasi-periodic oscillation shows up in a Fourier spectrum as a Lorentzian, defined by its amplitude and its quality factor $q$\indexq.  Quality factor is in turn defined as peak frequency divided by full-width half-maximum\footnote{The full-width half-maximum, or FWHM, of a QPO or spectral line is a measure of the width of the feature.  First calculate the amplitude $A$ of the feature above the local continuum level $k$.  The width of the feature in the $x$-direction at $y=k+\frac{A}{2}$ is its FWHM.}; for a QPO with a wandering frequency, this represents approximately the number of oscillations over which the QPO remains coherent.
\label{sec:convolver}
\par
\begin{figure}
    \includegraphics[width=\columnwidth, trim = 0mm 10mm 0mm 10mm]{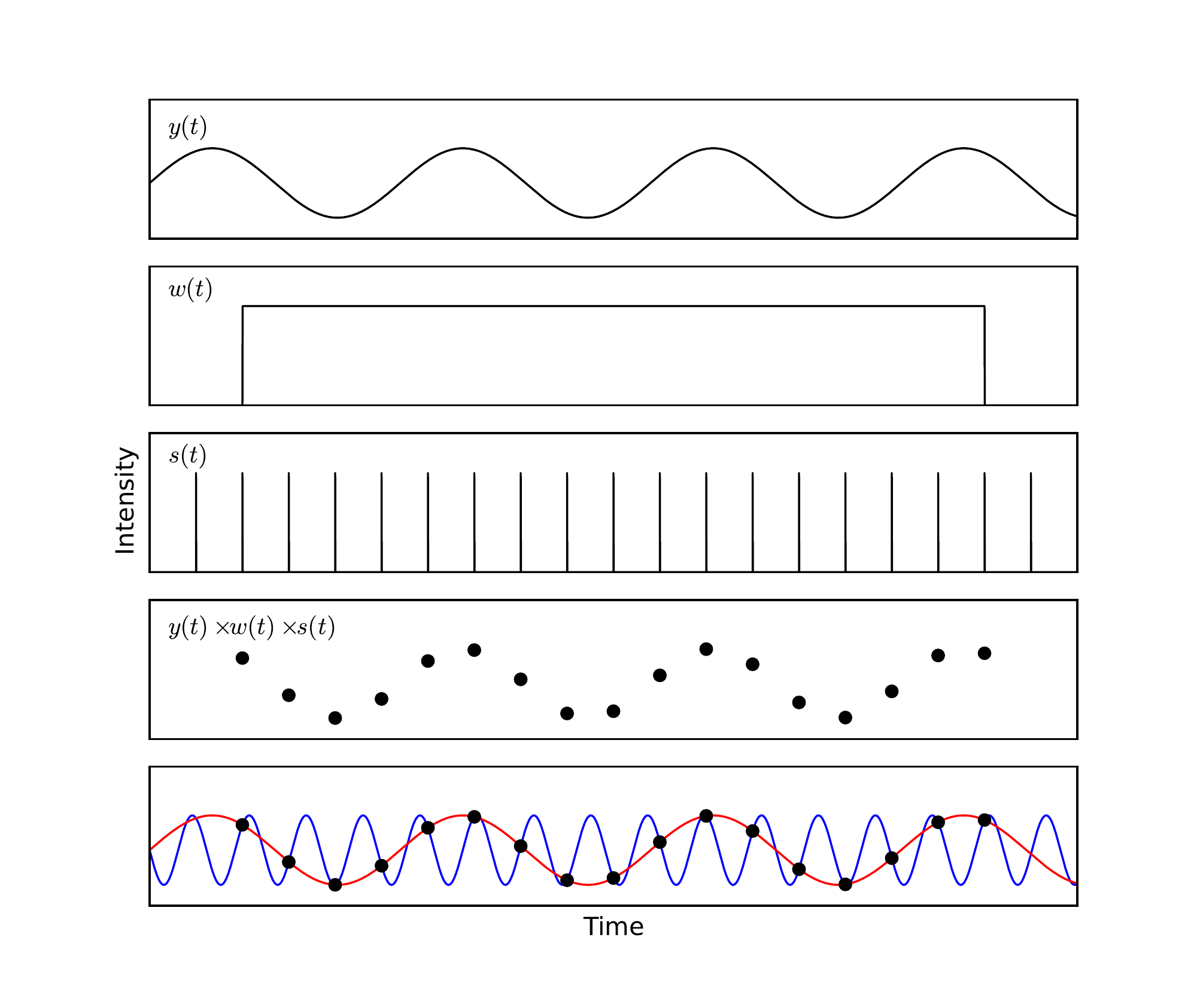}
    \captionsetup{singlelinecheck=off}
    \caption[A representation of how a continuous variable is convolved with a windowing function and a sampling function to yield physical data.]{A representation of how a continuous variable $y(t)$ is convolved with a windowing function $w(t)$ and a sampling function $s(t)$ to yield physical data.  I describe the effects of these convolutions in Section \ref{sec:convolver}.  The bottom panel shows how aliasing\index{Aliasing} arises, showing that sine waves of two different frequencies can be fit to the data: one with a frequency $\nu$ equal to that in the original dataset, and one of frequency $\sigma-\nu$ where $\sigma$ is the sampling frequency.  This explains the presence of aliased peaks in discrete data.}
   \label{fig:convolve}
\end{figure}

 Fourier analysis\index{Fourier analysis} was envisioned to analyse continuous, infinite data.  However, physical data differs from this ideal case in two important ways:
\begin{enumerate}
\item Physical data are discrete rather than continuous, consisting of samples taken at a finite rate $\sigma$.
\item Physical data are finite rather than infinite, being taken in some window of length $w$.
\end{enumerate}
As such, as I show in Figure \ref{fig:convolve}, physical data consists of a time series convolved with both a windowing function and a sampling function.  Each of these convolutions adds spurious features to the power spectrum produced from the data.

\par The convolution with a sampling function adds so-called `aliased'\index{Aliasing} peaks to the power spectrum of a given dataset.  For each peak in the power spectrum at frequency $\nu$, there will also be a peak present at a frequency of $\sigma-\nu$, where $\sigma$ is the sampling frequency.  This peak can be understood as the beat frequency between the oscillation in the data and the sampling frequency (see also the lower panel of Figure \ref{fig:convolve} for a visual explanation), and contains no additional information on the system.  To avoid these aliased peaks, values of $\hat{f}(\nu)$ outside of the range $0<\nu\leq\sigma/2$ are discarded.  The frequency $\sigma/2$, the maximum frequency at which one can extract useful information on a parameter sampled at constant frequency $\sigma$, is known as the Nyquist frequency\index{Nyquist frequency}.
\par In general, the convolution with a windowing function causes peaks in the power spectrum to be broadened; an effect known as `spectral leakage'\index{Spectral leakage}.  The form of this broadening depends on the windowing function which is being used.  Generally, physical data has been convolved with a so-called `boxcar' window; i.e., a function which takes a value of 1 during the period of measurement and 0 elsewhere.  A convolution with a boxcar window causes each peak in the power spectrum to be accompanied by a number of lower-amplitude sidelobes either side of it in frequency space; this serves to smear out a power spectrum and causes some information to be lost.  Other windows can be applied to data to attempt to lessen this effect; for example, convolving a dataset with a triangular or Gaussian window instead of a boxcar.  Many non-boxcar windows have been formulated to lessen the effect of spectral leakage, but it is impossible to remove the effect completely when working with a finite dataset.

\subsubsection{Fast Fourier Transform}

\par Taking the Fourier transform\index{Fourier analysis} of a time series is a computationally expensive procedure.  As such, it is common practice to instead use Fast Fourier Transform (FFT)\index{Fast Fourier transform} algorithms; computationally fast algorithms which specialise in finding the Fourier transform of evenly-spaced series.
\par One such FFT algorithm\index{Fast Fourier transform} is the Cooley-Tukey\index{Cooley-Tukey algorithm}\footnote{Computationaly realised as \texttt{fft} in the \texttt{scipy.fftpack} package for Python \citep{NumPy}.} algorithm \citep{Cooley_FFT}.  The Cooley-Tukey algorithm speeds up the Fourier transform process by recursively dividing a dataset in half to make many segments.  It uses the fact that the discrete Fourier transform of a single point is equal to itself, and then reconstructs the complete Fourier spectrum from these results. Unlike the basic Fourier transform, the Cooley-Tukey algorithm is only able to transform series which are evenly spaced in time and consisting of $2^N$ datapoints, for $N\in\mathbb{N}$.
\par The amplitude error of a Fast Fourier Transform\index{Fast Fourier transform} of a noise process is 100\%.  There are two ways to reduce this error to a level at which the data can be meaningfully analysed:
\begin{enumerate}
\item The original time series can be split into a number of equal-length windows.  The Fast Fourier-Transforms of these windows can be calculated independently of each other, and then averaged to create the mean FFT of the dataset.
\item The resultant power spectrum can be rebinned in frequency space.
\end{enumerate}
\par Propagating errors in the usual way, this results in a final error on Fourier power amplitude $\delta|\hat{f}(\nu)|^2$ of:
\begin{equation}
\delta|\hat{f}(\nu)|^2=\frac{|\hat{f}(\nu)|^2}{\sqrt{MW}}
\end{equation}
Where $W$ is the number of windows the original dataset was divided into, and $M$ is the number of frequency bins which were averaged to obtain the Fourier power at frequency $\nu$.  Increasing $W$ increases the minimum frequency at which the Fourier power of the dataset can be probed, while increasing $M$ decreases the resolution of the spectrum in frequency space.

\subsubsection{Normalising the Fourier Transform}

\par To understand the statistical significance of features in a power spectrum\index{Fourier analysis}, it is important to normalise the results in a standard and well-understood way.  One such method of normalisation is the `Leahy' normalisation\index{Leahy normalisation} \citep{Leahy_Norm}, defined as:
\begin{equation}
L(\nu)=\frac{2\times|\hat{f}(\nu)|^2}{n_{p}}
\end{equation}
Where $n_p$ is the total number of photon counts in the original dataset.  This normalisation has the property that pure Poisson\index{Poisson distribution} noise has a Leahy-normalised power of 2\footnote{In practise, due to instrumental dead-time effects meaning photon arrivals are not strictly independent, Poisson noise in astrophysical data tends to yield a Leahy-normalised power of slightly less than 2}.
\par I use one additional power spectrum normalisation in the work presented in this thesis: the RMS normalisation\index{RMS normalisation}.  This is defined as:
\begin{equation}
R(\nu)=\frac{(L(\nu)-2)r_s}{(r_s-r_b)^2}=\frac{2\left(|\hat{f}(\nu)|^2-Tr_s\right)}{T(r_s-r_b)^2}
\end{equation}
\par Where $T$ is the total time duration of all data used to produce the power spectrum, $r_s$ is the mean source count rate and $r_b$ is the mean background rate.  In this normalisation, uncorrelated Poisson noise corresponds to a power of zero.  Additionally, the power spectrum has the property that the integral of $R(\nu)$ between two frequencies is equal to the squared root-mean squared amplitude (RMS\index{RMS}\index{Root-mean squared variability|see {RMS}}$^2$) of the variability\index{Variability} of the original time series in that frequency band.

\subsubsection{Lomb-Scargle Periodograms}

\par Fast Fourier transforms\index{Fast Fourier transform} are unable to process unevenly spaced time series.  Additionally, while mathematical Fourier transforms\index{Fourier analysis} can in general process unevenly spaced datasets, the effects of aliasing\index{Aliasing} become increasingly complex and difficult to disentangle from real signal.  In these cases, a method known as the Lomb-Scargle periodogram\index{Lomb-Scargle periodogram}, based on proposals by \citet{Lomb_LombScargle} and \citet{Scargle_LombScargle}, can be used.
\par The Lomb-Scargle periodogram\index{Lomb-Scargle periodogram} can be thought of as the result of fitting sinusoids of frequency $\nu$ to a time series, and constructing a spectrum using the $\chi^2$ value of the fit of the sinusoid at each $\nu$.  Unlike a Fourier spectrum\index{Fourier analysis} of unevenly spaced data, the Lomb-Scargle periodogram of unevenly spaced data is statistically well-behaved as long as the noise component of the dataset is uncorrelated.
\par Unfortunately, due to dead-time\index{Dead-time} effects present in all X-ray telescopes, white noise in real datasets is not uncorrelated and so the statistical properties of the Lomb-Scargle spectrogram are generally not well-defined.  In this case, bootstrapping techniques can be used to estimate the significances of features in the power spectrum.

\subsection{Energy Spectral Analysis}

\par Energy spectral\index{Spectroscopy} analysis is another powerful tool available to understand the physical processes at work in astrophysical systems.  The distribution of arriving photons as a function of energy can be fit to physical models which, assuming a given system geometry, can provide estimates of various system parameters.
\par The disadvantage of spectral fitting is the aforementioned assumptions that one has to make.  A number of well-studied spectral models of LMXBs\index{X-ray binary!Low mass} exist, which are able to return estimates for values such as inner disk radius, black hole mass and spin when fit to data.  However, the values that different models return often contradict each other, and thus the values that a study infers for these parameters depend heavily on the system physics and geometry that the modeller assumes.

\subsubsection{Hardness-Intensity Diagrams}

\label{sec:hids}

\par A model-independent way to study the spectral properties of a source is by using colours\index{Colour}, also known as hardness ratios.  To obtain the colour of a source, I define two non-overlapping energy bands $A$ and $B$ with $B>A$.  The hardness ratio is then defined as $H(t)=r_B(t)/r_A(t)$, where $r_X(t)$ is the photon arrival rate in the band $X$.  The hardness ratio gives basic information on the shape of the energy spectrum without assuming a physical model.
\par Hardness ratios are often paired with intensity (the total flux of the object in some energy band which includes A and B) to create `hardness-intensity diagrams'\index{Hardness-intensity diagram} (HIDs) to explore how the source spectrally varies over time.  To explain what the shape of an HID can tell us about the spectral evolution of a source, consider the following examples of HIDs for a black body spectrum with temperature $T(t)$ and normalisation $n(t)$:
\begin{enumerate}
\item $T(t)=1, n(t)=\sin(t)$: in this example, the brightness of the source changes over time but the shape of its spectrum does not change.  As such the hardness is a constant, and the system traces a vertical line in hardness-intensity space (Figure \ref{fig:HIDexp}, Panel 1).
\begin{figure}
    \includegraphics[width=\columnwidth, trim = 0mm 25mm 0mm 25mm]{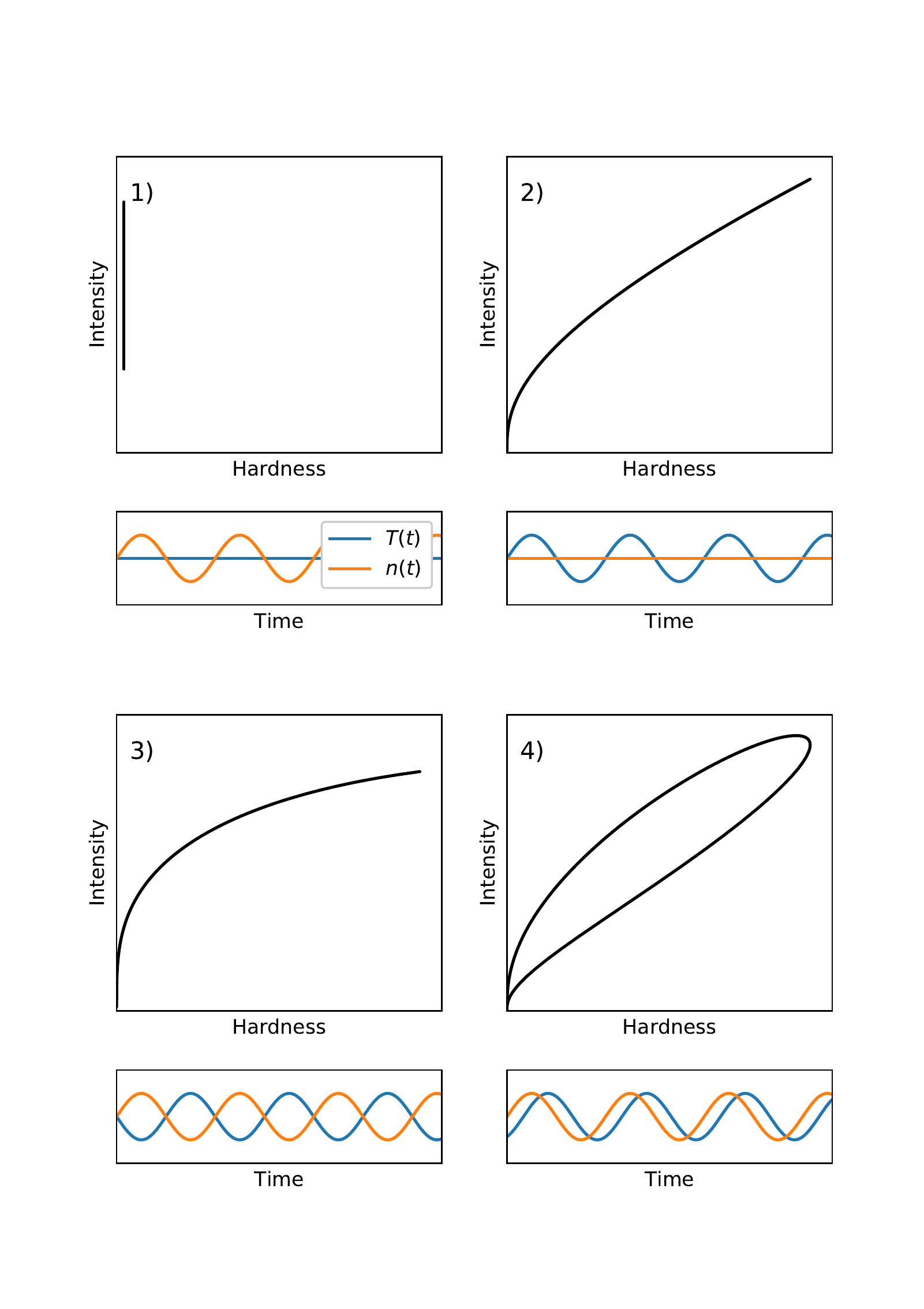}
    \captionsetup{singlelinecheck=off}
    \caption[Hardness-Intensity diagrams of black bodies with temperatures and normalisations described by various functional forms.]{Hardness-Intensity diagrams\index{Hardness-intensity diagram} of black bodies with temperatures and normalisations described by various functional forms $T(t)$ and $n(t)$ which I give in Section \ref{sec:hids}.  The plots show how HIDs differ between sources with \textbf{1)} changing brightness but no spectral change, \textbf{2)} changing temperature, \textbf{3)} changing temperature and normalisation in antiphase, and \textbf{4)} changing temperature and normalisation out of phase.  The shape of and direction of a loop in a HID can therefore give us information about the physical processes underlying spectral variability.}
   \label{fig:HIDexp}
\end{figure}
\item $T(t)=\sin(t), n(t)=1$: in this example, the spectrum of the source changes over time, resulting in a curved track in hardness-intensity space (Figure \ref{fig:HIDexp}, Panel 2).
\item $T(t)=\sin(t), n(t)=\sin(t-\pi)$: if two or more spectral parameters are varying at once, the track can become move complex.  If these parameters are varying in phase or antiphase, a single track is traced (Figure \ref{fig:HIDexp}, Panel 3).
\item $T(t)=\sin(t), n(t)=\sin\left(t+\frac{\pi}{3}\right)$: when parameters are varying out of phase with each other, the track of the object in a HID can take the form of a closed loop (Figure \ref{fig:HIDexp}, Panel 4).  I use$\frac{\pi}{3}$ here as an arbitrary phase shift.
\end{enumerate}
Case 4 is interesting, as it indicates the presence of a phase lag between two or more physical components of the system.  The direction in which the loop is executed over time can be used to infer the sign of this lag.  This in turn can give constraints on the causal links between components of a system, in turn giving constraints on physical models proposed to describe them.  The tracing of a loop in a hardness-intensity diagram\index{Hardness-intensity diagram} is known as hysteresis\index{Hysteresis}.

\subsubsection{Phase-Resolved Spectroscopy}

\label{sec:phasresspec}

\par Like lightcurves, HIDs\index{Hardness-intensity diagram} and time-resolved spectra can be difficult to analyse when constructed from data with poor statistics.  If the flux from a source is variable\index{Variability} in a periodic or quasi-periodic way\index{Quasi-periodic oscillation}, a modified version of the folding\index{Folding} algorithms detailed in Section \ref{sec:LCMorph} can be used to analyse the spectral evolution of an average cycle:
\begin{itemize}
\item Obtain the function $\phi(t)$ to describe how phase varies as a function of time.
\item Split the interval $[0,1)$ into a number of sub-intervals $i$.
\item For each sub-interval $i$, compile a list of good time intervals (GTIs) denoting periods of time during which $\phi(t)\in i$.
\item For each list of GTIs, filter the original dataset such that it only contains photons which arrived during one of the intervals.
\item From each new filtered dataset, a spectrum or hardness ratio can be calculated.  This can be compared with the spectra or hardness ratios taken from the other filtered datasets to analyse how the spectrum of the source varies as a function of phase.
\end{itemize}
This technique is known as phase-resolved spectroscopy\index{Spectroscopy!Phase-resolved}.  An example of the use of this method with the algoritm I describe in Section \ref{sec:vfold} is presented in \citet{Wang_Reflect}.

\cleardoublepage

\chapter{Variability in IGR J17091-3624: Classification}

\label{ch:IGR}

\epigraph{\textit{Song and call are useful aids to identification, and reference is made to vocalisation for each species.}}{Paul Sterry -- \textit{Collins Guide to British Birds}}

\vspace{1cm}

\par\noindent Accounting for the unusual X-ray variability\index{Variability} observed in LMXBs is required for a complete understanding of the physics of matter in their accretion disks\index{Accretion disk}.  The first step is to describe and categorise the types of variability in these objects, and to look for similarities and differences which may shed light on their physical origins.
\par In 2000, \citeauthor{Belloni_GRS_MI} performed a complete model-independent analysis of variability classes\index{Variability class} in GRS 1915\index{GRS 1915+105}.  This work highlighted the breadth and diversity of variability\index{Variability} in GRS 1915, and allowed these authors to search for features common to all variability classes.  For example, \citet{Belloni_GRS_MI} found that every variability class can be expressed as a pattern of transitions between three quasi-stable phenomenological states.
\par Previous works have noted that some of the variability classes\index{Variability} seen in IGR J17091\index{IGR J17091-3624} appear very similar to those seen in GRS 1915 (e.g. \citealp{Altamirano_IGR_FH, Zhang_IGR}).  However, although $\rho$-like\indexrho\ classes in the two objects both show lags between hard and soft X-rays photons\index{Hard lag}, these lags appear to possess different signs \citep{Altamirano_IGR_FH}.  Additionally, at least two variability classes have been reported in IGR J17091 which have not yet been reported in GRS 1915 \citep{Pahari_IGRClasses}.  Previous works have described some of the behaviour seen in IGR J17091 in the context of the variability classes described by \citealt{Belloni_GRS_MI} for GRS 1915 (e.g. \citealp{Altamirano_IGR_FH,Pahari_RhoDiff}).  To further explore the comparison between GRS 1915 and IGR J17091, here I perform the first comprehensive model-independent analysis of variability classes in IGR J17091 using the complete set of \indexrxte\rxte\ data taken of the 2011-2013 outburst\index{Outburst} of the object.  I also use data from all other X-ray missions that observed the source during this time to analyse the long-term evolution of the outburst.
\par \textbf{The results I present in this chapter have been published as \citet{IGR}.}

\section{Data and Data Analysis}

\label{sec:dex}

\par In this chapter, I report data from \indexrxte\rxte, \indexintegral\textit{INTEGRAL}, \indexswift\textit{Swift}, \indexchandra\textit{Chandra}, \indexxmm\textit{XMM-Newton} and \indexsuzaku\textit{Suzaku} covering the 2011-2013 outburst of IGR J17091.  Unless stated otherwise, all errors are quoted at the 1$\sigma$ level.
\par
\begin{figure}
    \includegraphics[width=\columnwidth, trim = {0.75cm 1.0cm 1.0cm 0.8cm},clip]{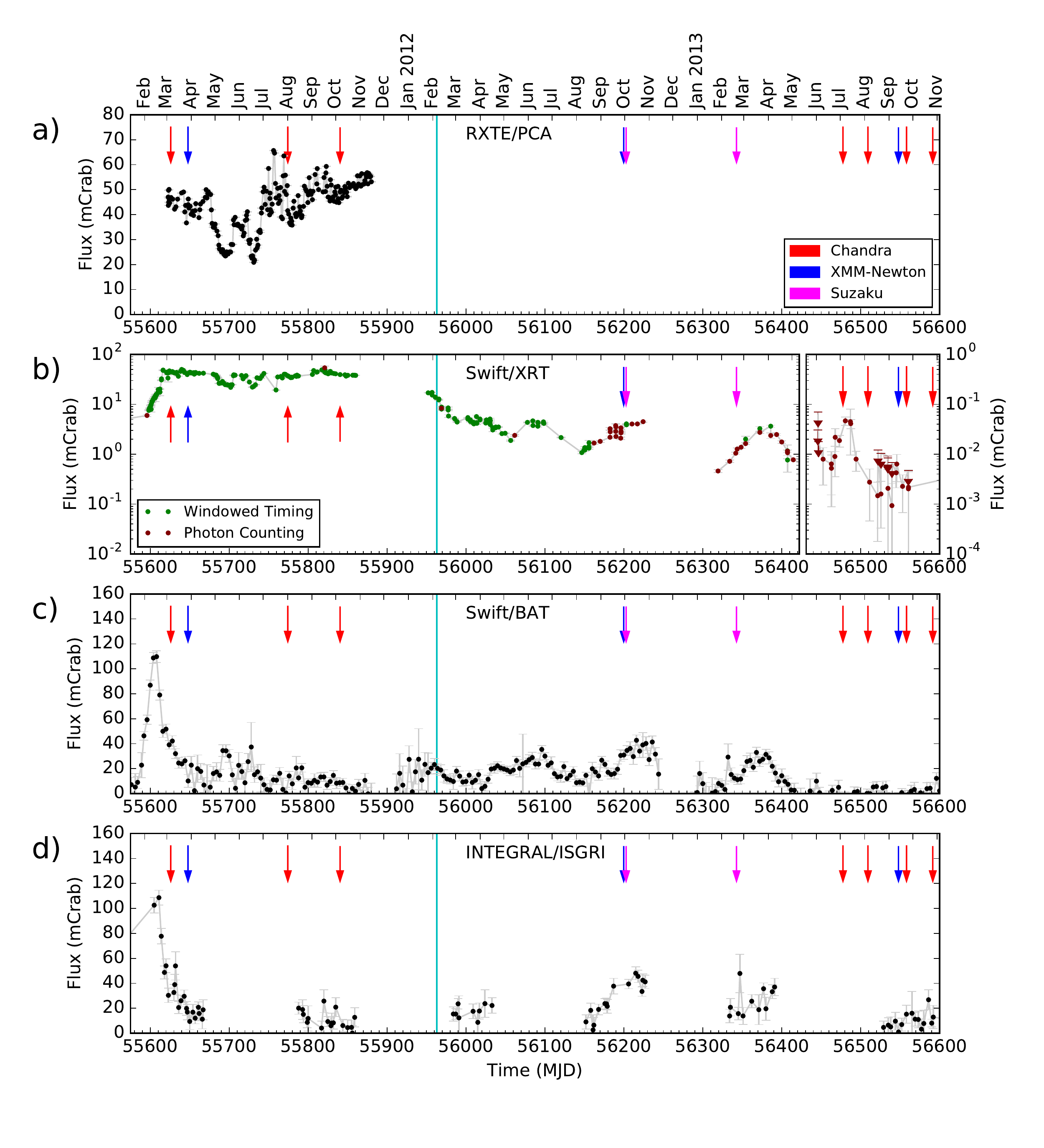}
    \captionsetup{singlelinecheck=off}
    \caption[Lightcurves of IGR J17091-3624, from a number of instruments, during its 2011-2013 outburst.]{\indexrxte\indexpca\rxte\ /PCA (Panel a), \indexswift\indexxrt\textit{Swift/XRT} (Panel b), \indexbat\textit{Swift/BAT} (Panel b) and \indexintegral\indexibis\textit{INTEGRAL}/IBIS (Panel d) lightcurves\index{Lightcurve} of IGR J17091-3624\index{IGR J17091-3624} during its 2011-2013 outburst.  Arrows mark times at which \indexxmm\textit{XMM-Newton} (blue), \indexchandra\textit{Chandra} (red) or \indexsuzaku\textit{Suzaku} (magenta) observed IGR J17091-3624.  The cyan line represents MJD 55963, the approximate time IGR J17091-3624 transitions from the soft\index{High/Soft state} to the hard\index{Low/Hard state} state \citep{Drave_Return}.  \rxte /PCA \citep{Jahoda_PCA} data are for the 2--16\,keV energy band and taken from \citep{Altamirano_IGR_FH}, \textit{Swift/BAT} \citep{Barthelmy_BAT} data are for 15--50\,keV, \textit{Swift/XRT} \citep{Burrows_XRT}  data are for 0.3--10\,keV and \textit{INTEGRAL/ISGRI} \citep{Ubertini_IBIS} data are for 20-40\,keV.  Note that the data from \textit{Swift/XRT} (Panel B) are shown with a logarithmic $y$-axis to better show the late time progression of the outburst.  Data points are coloured according to the observing mode used.  The \textit{Swift/XRT} data from times later than MJD 56422 are shown to a different scale to better represent the post-outburst evolution of the source.  All data are presented in 1 day bins, except for data from \textit{Swift/BAT} which is presented in 4 day bins.  See also Figure \ref{fig:WhereCls}, in which data from \rxte \textit{/PCA} is presented on a smaller scale.  The Crab\index{Crab nebula} count rates used to normalise these data were 2300 cts s$^{-1}$ PCU$^{-1}$, 747.5 cts s$^{-1}$, 0.214 cts s$^{-1}$ and 183.5 cts s$^{-1}$ for \rxte , \textit{Swift/XRT}, \textit{Swift/BAT} and \textit{INTEGRAL/ISGRI} respectively.  \rxte\ data have not been corrected for the 25' offset to avoid contamination from GX 349+2\index{GX 349+2}, and for all instruments \textsf{D.A.} and I implicitly assume that IGR J17091 presents a Crab-like spectrum.}
   \label{fig:allmissions}
\end{figure}

 In Figure \ref{fig:allmissions} I present long-term lightcurves from \rxte ,  \textit{INTEGRAL} and \textit{Swift} to show the behaviour of the source during this outburst.  I indicate when during the outburst \textit{Chandra}, \textit{XMM-Newton} and \textit{Suzaku} observations were made.

\subsection{\rxte}

\label{sec:XTEDA}

\par For this variability study, I focus on the data from \indexrxte\indexpca\textit{RXTE}/PCA.  I analysed all PCA observations of IGR J17091\index{IGR J17091-3624} during 2011, corresponding to ObsIDs\footnote{Observations IDs.} 96065-03, 96103-01 and 96420-01.  The observations taken for proposals 96065-03 and 96103-01 were contaminated by the nearby X-ray source GX 349+2\index{GX 349+2} \citep{Altamirano_IGR_FH,Rodriguez_Contamination}.  As such I only use observations performed for proposal 96420-01, corresponding to a total of 243 orbits from 215 separate observations.  This in turn corresponds to 470\,ks of data, which is $\sim2$\% of \rxte 's operational time over the duration of the observation period.  These were offset by 25' such that GX 349+2 was not in the $1^\circ$ \textit{PCA} field of view.  \rxte\ was decommissioned during a period of Sun constraint centred on MJD\footnote{Modified Julian Date: the number of days since 0h00, November 17, 1858.} 55907, and hence the last observation of IGR J17091 was taken on MJD 55879.
\par I extracted data from the native \texttt{FITS}\index{FITS@\texttt{FITS}} format using my own \texttt{PANTHEON}\index{PANTHEON@\texttt{PANTHEON}} software (presented in Appendix \ref{app:PAN}).  To perform medium- to high-frequency ($\gtrsim1$\,Hz) timing analysis, I merged files formatted in PCA's `Good Xenon'\indexgx\ data mode and extracted their data at the maximum time resolution ($\sim9.5\times10^{-7}$ s) without accounting for the background.  I divided these data into 128\,s segments as this allowed us to reach frequencies below $\sim0.015$\,Hz, partly sampling the high amplitude quasi-periodic flaring behaviour seen in many classes.  Using the Fast Fourier Transform (FFT)\index{Fast Fourier transform}, I produced the power spectrum\index{Fourier analysis} of each segment separately.  I then averaged these spectra to create a one co-added Power Density Spectrum (PDS) for each observation.
\par For low-frequency ($\leq1$\,Hz) timing and correlated spectral/timing analysis, I rebinned the data to 0.5\,s and normalised count rates by the number of proportional counters (PCUs) active in each observation.  My choice of 1\,Hz allows us to analyse high amplitude flaring\index{Flare} behaviour (seen at frequencies $\lesssim0.5$\,Hz) separately from the lower-amplitude behaviour seen at $\gtrsim5$\,Hz.
\par I split the data into three energy bands: A (\textit{PCA} channels 0--14, $\sim2$--$6$\,keV), B (\textit{PCA} channels 15--35, $\sim6$--$16$\,keV) and C (\textit{PCA} channels 36--255, $\sim16$--$60$\,keV).  I chose these energy bands to be consistent with the energy bands used by the model-independent classification of variability classes of GRS 1915\index{GRS 1915+105} in \citet{Belloni_GRS_MI}.  For each of the energy-filtered lightcurves produced I estimated background using \texttt{pcabackest} from the \texttt{FTOOLS} package \citep{Blackburn_FTools} with the \textit{PCA} faint source background model\footnote{\url{http://heasarc.gsfc.nasa.gov/FTP/xte/calib\_data/pca\_bkgd/Faint/pca\_bkgd\_cmfaintl7\_eMv20051128.mdl}}\index{Background subtraction}. In all observations, I found that counts in the C band were consistent with background.  I then created Lightcurves $L_A$ and $L_B$ from background-subtracted photons counted in the A and B bands respectively.  I used these lightcurves to define the full-band lightcurve ($L_T=L_A+L_B$) and the soft colour\index{Colour} ($C_1=L_B/L_A$) of each observation.  To complement the Fourier spectra, I also constructed Generalised Lomb-Scargle Periodograms\index{Lomb-Scargle periodogram} of $L_T$ from each dataset, a modified version of the standard Lomb-Scargle periodogram \citep{Lomb_LombScargle, Scargle_LombScargle} that takes into account errors in the dataset \citep{Irwin_LombScargle}.  Using the Lomb-Scargle periodogram instead of the Fourier periodogram here allows us to sample the low-frequency behaviour of lightcurves with data gaps.  This is important, for example, in lightcurves which show two populations of flares, as it allows each population to be studied independently by cropping the other from the lightcurve.
\par I also used data from \citealt{Altamirano_IGR_FH} to sample the long-term colour evolution of IGR J17091.  I use 2 hardness ratios defined by \citeauthor{Altamirano_IGR_FH}: $H_{A1}$ and $H_{A2}$, corresponding to the ratios of the 3.5--6\,keV band against the 2--3.5\,keV band and the 9.7--16\,keV band against the 6--9.7\,keV band respectively.
\par When possible, if low-frequency peaks were present in the Lomb-Scargle spectrum of an observation, I used the position of the highest-amplitude peak to define a value for a period.  This period was then used to fold\index{Folding} the data to search for reccurent hysteretic\index{Hysteresis} patterns in the hardness-Intensity diagram\index{Hardness-intensity diagram} (hereafter HID$_1$, a plot of $L_T$ against $C_1$).  I found that quasi-periodic oscillations\index{Quasi-periodic oscillation} in the observations I used tended to show significant frequency shifts on timescales shorter than the length of the observations.  As such, I employed the variable-period folding algorithm outlined in Section \ref{sec:Flares} where appropriate.  For cases in which this algorithm was not appropriate, I considered small sections of each lightcurve, with a length equivalent to small number of periods, before performing folding.
\par Additionally, in observations which showed a pattern of high-amplitude X-ray flaring in $L_T$, I used my own algorithm to find individual flares\index{Flare} (this algorithm is described in Section \ref{sec:Flares}) and collect statistics on the amplitude, duration and profile of these events.
\par A list of all observations used in this study can be found in Appendix \ref{app:Obsids}.

\subsection{\textit{Swift}}

\par IGR J17091\index{IGR J17091-3624} was observed with \indexswift\textit{Swift}/XRT for a total of 172 pointed XRT observations between MJDs 55575 and 56600, corresponding to Target IDs 31921, 34543, 30967, 30973, 31920, 35096, 67137, 81917, 522245, 677582 and 677981.  These observations were interrupted during sun constraints centred on MJDs 55907 and 56272.  I created a long-term 0.3--10\,keV \indexxrt\textit{Swift/XRT} light curve, with one bin per pointed observation, using the online light-curve generator provided by the UK Swift Science Data Centre (UKSSDC; \citealp{Evans_Swift1}).  I have also created a long-term 15--50\,keV lightcurve using the publicly available \indexbat\textit{Swift/BAT} daily-averaged lightcurve\footnote{\url{http://swift.gsfc.nasa.gov/results/transients/weak/IGRJ17091-3624/}}.  These are shown in Figure \ref{fig:allmissions} Panels (b) and (c) respectively.

\subsection{\textit{INTEGRAL}}

\par Dr. Chris Boone (\textsf{C.B.}) and I analyse all available observations of IGR J17091 with \indexintegral\indexibis\textit{INTEGRAL}/IBIS \citep{Ubertini_IBIS} between MJD 55575--55625 where the source is less than 12 degrees from the centre of the field of view and where there is more than 1\,ks of good ISGRI time per 2\,ks Science Window. This corresponds to the spectrally hardest period of the 2011-2013 outburst. The filtering of observations results in a total of 188 Science Windows which were processed using the Offline Science Analysis (OSA) software version 10.2 following standard data reduction procedures\footnote{http://www.isdc.unige.ch/integral/analysis} in four energy bands (20--40, 40--100, 100--150, 150--300\,keV). These bands were selected as they are standard energy bands used in the surveys of \citet{Bird_Survey} and \citet{Bazzano_Survey} and allow comparison to these previous works. Images were created at the Science Window level, as well as a single mosaic of all Science Windows in each energy band.

\subsection{\textit{XMM-Newton}}
\label{sec:xmmdata}

\par \indexxmm\textit{XMM/Newton} observed IGR J17091 thrice during the period from 2011--2013 (represented by the blue arrows in Figure \ref{fig:allmissions}).  One of these observations (ObsID 0721200101) was made on 12 September 2013; I do not consider this observation further as IGR J17091 had returned to quiescence by this time \citep{Altamirano_Quiescence}.  The remaining two observations, corresponding to ObsIDs 0677980201 and 0700381301 respectively, were taken on March 27 2011 (MJD 55647) and September 29 2012 (MJD 56199).
\par During observation 0677980201, \indexepic\textit{EPIC-pn} was operating in burst mode and \textit{EPIC-MOS} was operating in timing mode.  Given the low efficiency of burst mode, I only consider data from \textit{EPIC-MOS} for this observation.  During observation 0700381301, \textit{EPIC-pn} was operating in timing mode, and thus I use data from \textit{EPIC-pn} for this observation.
\par I used the \textit{XMM-Newton} Science Analysis Software version 15.0.0 (\index{SAS@\texttt{SAS}}\texttt{SAS}, see \citealp{Ibarra_sas}) to extract calibrated event lists from \textit{EPIC} in both observations.  I used these to construct lightcurves to study the X-ray variability, following standard analysis threads\footnote{\url{http://www.cosmos.esa.int/web/xmm-newton/sas-threads}}.

\subsection{\textit{Chandra}}

\par \indexchandra\textit{Chandra} made 7 observations of IGR J17091 during the period 2011--2013.  Four of these observations were taken after IGR J17091 returned to quiescence, and I do not consider these further in this chapter.  The Chandra observations log is reported in Table \ref{tab:Chandra}. 

\begin{table}
\centering
\begin{tabular}{lllllll}
\hline
\hline
\scriptsize ObsID &\scriptsize  Instrument &\scriptsize Grating &\scriptsize Exposure (ks) &\scriptsize Mode &\scriptsize MJD\\
\hline
12505  	& \textit{HRC-I}    &   NONE      &    1.13      & $I$ & 55626\\
12405  	& \textit{ACIS-S} &   HETG     &    31.21     & $C$ & 55774\\
12406  	& \textit{ACIS-S} &   HETG     &    27.29     & $T$ & 55840\\
\hline
\hline
\end{tabular}
\caption[\textit{Chandra} observations log covering the three observations considered in Chapter \ref{ch:IGR}.]{\textit{Chandra} observations log covering the three observations considered in this chapter.  $I$ refers to Imaging mode, $C$ refers to CC33\_Graded mode and $T$ refers to Timed Exposure Faint mode.  HETG refers to the High Energy Transmission Grating.}
\label{tab:Chandra}
\end{table}

\par Dr. Margarita Pereyra (\textsf{M.P.}) analysed these data using \index{CIAO@\texttt{CIAO}}\texttt{CIAO} version 4.8 \citep{Fruscione_Ciao}, following the standard analysis threads. In order to apply the most recent calibration files (CALDB 4.7.0, \citealp{Graessle_ChaCALDB}), \textsf{M.P.} reprocessed the data from the three observations using the \texttt{chandra\_repro} script\footnote{See e.g. \url{http://cxc.harvard.edu/ciao/ahelp/chandra_repro.html}}, and used this to produce data products following standard procedures.
\par The first Chandra observation (ObsID 12505) of this source was made shortly after it went into outburst in February 2011. It was a 1\,ks observation performed to refine the position of the X-Ray source, using the High-Resolution Camera in Imaging mode \index{HRC}(HRC-I). \textsf{M.P.} created the 0.06--10\,kev light curve accounting for the Dead-Time\index{Dead-time} Factor (DTF), to correct the exposure time and count rate using the \texttt{dmextract} tool in the \texttt{CIAO} software.
\par Two additional observations (ObsIDs 12405 and 12406) were performed within 214 days of this first observation, using the High Energy Transmition Grating Spectrometer (HETGS) on board \textit{Chandra}. The incident X-Ray flux was dispersed onto \indexacis\textit{ACIS} using a narrow strip array configuration (ACIS-S). Continuous Clocking and Time Exposure modes were use in each observation respectively (see \citealp{King_IGRWinds} for further details). \textsf{M.P.} excluded any events below 0.4\,keV, since the grating efficiency is essentially zero below this energy. In the case of the ObsID 12405 observations \textsf{M.P.} also excluded the Flight Grade 66 events in the event file, as they were not appropriately graded. \textsf{M.P.} extracted the 0.5-10\,kev HEGTS light curves, excluding the zeroth-order flux, adopting standard procedures.

\subsection{\textit{Suzaku}}

\par \indexsuzaku\textit{Suzaku} observed IGR J17091 twice during the period 2011--2013; a 42.1\,ks observation on October 2--3, 2012 (MJD 56202--56203, ObsID: 407037010) and an 81.9\,ks observation on February 19--21, 2013 (MJD 56342--56344, ObsID: 407037020). \indexxis\textit{XIS} consists of four X-ray CCDs (\textit{XIS} 0, 1, 2 and 3), and all them except for XIS 2 were operating in the 1/4 window mode which has a minimum time resolution of 2 seconds.
\par Professor Kazutaka Yamaoka (\textsf{K.Y.}) analysed the \textit{Suzaku} data using \index{FTOOLS@\texttt{FTOOLS}}\index{HEASOFT@\texttt{HEASOFT}|see {\texttt{FTOOLS}}}\texttt{HEASOFT} 6.19 in the following standard procedures after reprocessing the data with \texttt{aepipeline} and the latest calibration database (version 20160607).  \textsf{K.Y.} extracted \textit{XIS} light curves in the 0.7--10 keV range, and subtracted background\index{Background subtraction} individually for XIS 0, 1 and 3 and then summed these to obtain the total background.  \textsf{K.Y.} created power density spectra\index{Fourier analysis} (PDS) using {\tt powspec} in the {\tt XRONOS} package.

\section{Results}
\label{sec:results}

\subsection{Outburst Evolution}

\label{sec:igrobevo}

\par The onset of the 2011-2013 outburst\index{Outburst} of \index{IGR J17091-3624}IGR J17091 can be seen in the \indexbat\textit{Swift/BAT} lightcurve (Figure \ref{fig:allmissions} Panel c).  In a 22 day period between MJDs 55584 and 55608, the 15--50\,keV intensity from IGR J17091 rose from $\sim9$\,mCrab to a peak of $\sim110$\,mCrab.  This onset rise in intensity can also be seen in 0.3--10\,keV \indexxrt\textit{Swift/XRT} data and 20--40\,keV \indexibis\textit{INTEGRAL/ISGRI} data.
\par After peak intensity, the 15--50\,keV flux (\textit{Swift/BAT}) began to steadily decrease, until returning to a level of $\sim$20\,mCrab by MJD 55633.  A similar decrease in flux can be seen in the data obtained by \textit{INTEGRAL} at this time (Figure \ref{fig:allmissions} Panel (d).  However, there was no corresponding fall in the flux at lower energies; both the long-term 2--16\,keV \indexpca\rxte\ /PCA data and \textit{Swift/XRT} data (Panels a and b respectively) show relatively constant fluxes of 45\,mCrab between MJDs 55608 and 55633.
\par The significant decrease in high-energy flux during this time corresponds to IGR J17091 transitioning from a hard state\index{Low/Hard state} to a soft/intermediate state\index{High/Soft state} \citep{Pahari_RhoDiff}.  This transition coincides with a radio flare reported by \citet{Rodriguez_D} which was observed by the Australian Telescope Compact Array (ATCA\index{ATCA}).
\par \citealp{Altamirano_10Hz} first reported a 10\,mHz QPO\index{Quasi-periodic oscillation} in \rxte\ data on MJD 55634 , evolving into `Heartbeat-like'\indexrho\ flaring\index{Flare} by MJD 55639 \citep{Altamirano_Discovery}.  Between MJDs 55634 and 55879, the global \indexpca\rxte\ /PCA lightcurve shows large fluctuations in intensity on timescales of days to weeks, ranging from a minimum of $\sim20$\,mCrab on MJD 55731 to a maximum of $\sim66$\,mCrab on MJD 55756.  The \indexxrt\textit{Swift/XRT} lightcurve shows fluctuations that mirror those seen by \rxte\ during this period, but the amplitude of the fluctuations is significantly reduced.
\par \indexxrt\textit{Swift/XRT} was unable to observe again until MJD 55952.  Between this date and MJD 55989, \textit{Swift/XRT} observed a gradual decrease in intensity corresponding to a return to the low/hard state\index{Low/Hard state} \citep{Drave_Return}.
\par Between MJD 55989 and the end of the outburst on MJD 56445, there are secondary peaks in the \indexxrt\textit{Swift/XRT}, \indexbat\textit{Swift/BAT} and \indexibis\textit{INTEGRAL/ISGRI} lightcurves that evolve over timescales of $\lesssim100$ days.  Similar humps have been seen before in lightcurves from other objects, for example the black hole candidate XTE J1650-500 \citep{Tomsick_MiniOutbursts} and the neutron stars SAX J1808.4-3658 \citep{Wijnands_1808} and SAX J1750.8-2900 \citep{Allen_1750}.  These humps are referred to as `re-flares'\index{Re-flare} (also as `rebrightenings',  `echo-outbursts', `mini-outbursts' or a `flaring tail', e.g. \citealp{Patruno_Reflares2}).  I identify a total of 3 apparent re-flares\index{Re-flare} in the \indexbat\textit{Swift/BAT} data, centred approximately at MJDs 56100, 56220 and 56375.
\par The observation with \indexepic\textit{XMM-Newton/EPIC-pn} on MJD 56547 (12 September 2013) recorded a rate of 0.019 cts s$^{-1}$.  An observation with \textit{EPIC-pn} in 2007, while IGR J17091\index{IGR J17091-3624} was in quiescence\index{Quiescence} \citep{Wijnands_Quiescence}, detected a similar count rate of 0.020 cts s$^{-1}$.  Therefore I define MJD 56547 as the upper limit on the endpoint of the 2011-2013 outburst.  As such the outburst, as defined here, lasted for $\lesssim$952 days.
\par After the end of the 2011-2013 outburst, IGR J17091 remained in quiescence\index{Quiescence} until the start of a new outburst\index{Outburst} around MJD 57444 (26 February 2016, \citealp{Miller_2016Outburst}).

\subsection{\rxte}

\label{sec:IGRclassesintro}

\par Using the \rxte\indexrxte\
 data products described in Section \ref{sec:dex}, I assigned a model-independent variability class\index{Variability class} to each of the 243 \rxte\textit{/PCA}\indexpca\ orbits during which IGR J17091 was observed\index{IGR J17091-3624}.  To avoid bias, this was done without reference to the classes defined by \citet{Belloni_GRS_MI} to describe the behaviour of GRS 1915\index{GRS 1915+105}.
\par Classes were initially assigned based on by-eye analysis of lightcurve\index{Lightcurve} profiles, count rate, mean fractional RMS\index{RMS} \citep{Vaughan_RMS}, Fourier power spectra and Lomb-scargle periodograms,\index{Fourier analysis}{Lomg-Scargle periodogram} and hardness-intensity diagrams\index{Hardness-intensity diagram}.  For observations with significant quasi-periodic variability at a frequency lower than $\sim1$\,Hz, I also attempted to fold\index{Folding} lightcurves to analyse count rate and colour as a function of phase.  When flares\index{Flare} were present in the lightcurve, I used my algorithm (described in Section \ref{sec:Flares}) to sample the distribution of parameters such as peak flare count rate, flare rise time and flare fall time.  All parameters were normalised per active PCU, and fractional RMS\index{RMS} values were taken from 2--60\,keV lightcurves binned to 0.5\,s.  I identify nine distinct classes, labelled I to IX; I describe these in the following sections.
\par Although the criteria for assigning each class to an observation was different, a number of criteria were given the most weight.  In particular, the detection, $q$-value\indexq\ and peak frequency of a QPO\index{Quasi-periodic oscillation} in the range 2\,Hz--10\,Hz were used as criteria for all classes, as well as the presence or absence of high-amplitude quasi-periodic flaring with a frequency between 0.01--1\,Hz.  The folded\index{Folding} profile of these flares, as well as the presence of associated harmonics\index{Harmonic}, were also used as classification diagnostics in observations.  Additionally, the presence or absence of low count-rate 'dips'\index{Dip} in a lightcurve was used as a criterion for Classes VI\indexvi, VIII\indexviii\ and IX\indexix.  Detailed criteria for each individual class are given below in Sections \ref{sec:ClassI} to \ref{sec:ClassIX}.  As each observation lasted less than $\lesssim3$\,ks, significantly shorter than the timescale over which IGR J17091-3624 evolved between classes, a single class could be assigned to all observations\footnote{See however Figure \ref{fig:HybridClasses} for an example lightcurve of an observation which appeared to capture a transition between two classes.}.
\par For hardness-intensity diagrams\index{Hardness-intensity diagram}, I describe looping behaviour\index{Hysteresis}\index{Loops|see {Hysteresis}} with the terms `clockwise' and `anticlockwise'; in all cases, these terms refer to the direction of a loop plotted in a hardness-intensity diagram with colour on the $x$-axis and intensity on the $y$-axis.  I did not study these hysteretic loops until after I had established my set of variability classes, and hence the presence or direction of a loop was not used as a diagnostic feature to assign a class to an observation.
\par In Appendix \ref{app:Obsids}, I present a list of all orbits used in the study along with the variability classes I assigned to them.
\par
\begin{figure}
    \includegraphics[width=\columnwidth, trim = {1.3cm 2.0cm 1.8cm 1.8cm},clip]{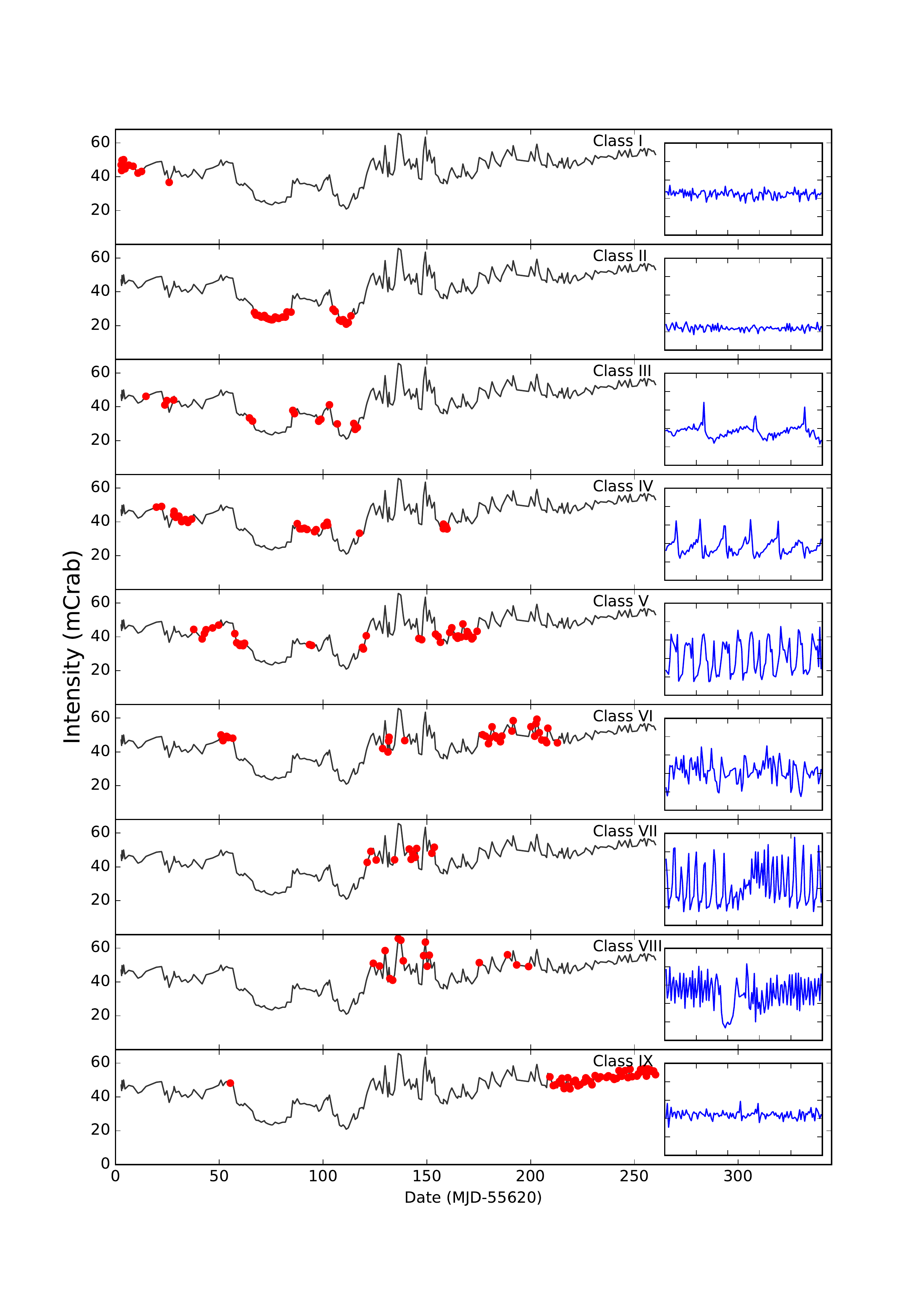}
    \captionsetup{singlelinecheck=off}
    \caption[Lightcurves of IGR J17091-3524 during the 2011-2013 outburst, showing when each of variability classes I-IX were observed.]{Global 2--3.5\,keV Lightcurves of IGR J17091-3524\index{IGR J17091-3624} during the 2011-2013 outburst, with each point corresponding to the mean Crab-normalised\index{Crab nebula} count rate of a single \indexrxte\rxte\ observation of the object (in turn corresponding to between 0.4 and 3.6 ks of data).  In each lightcurve, every observation identified as belonging to a particular class\index{Variability class} (indicated on the plot) is highlighted.  These are presented along with a characteristic lightcurve (inset) from an observation belonging to the relevant class.  Each lightcurve is 250\,s in length, and has a $y$-scale from 0 to 250\spcu .  Data taken from \citealt{Altamirano_IGR_FH}.}
   \label{fig:WhereCls}
\end{figure}

\begin{figure}
\centering
\subfloat[\textit{Colour-Colour Diagram}]{\includegraphics[width=0.8\columnwidth, trim = 0mm 0mm 0mm 8mm,clip]{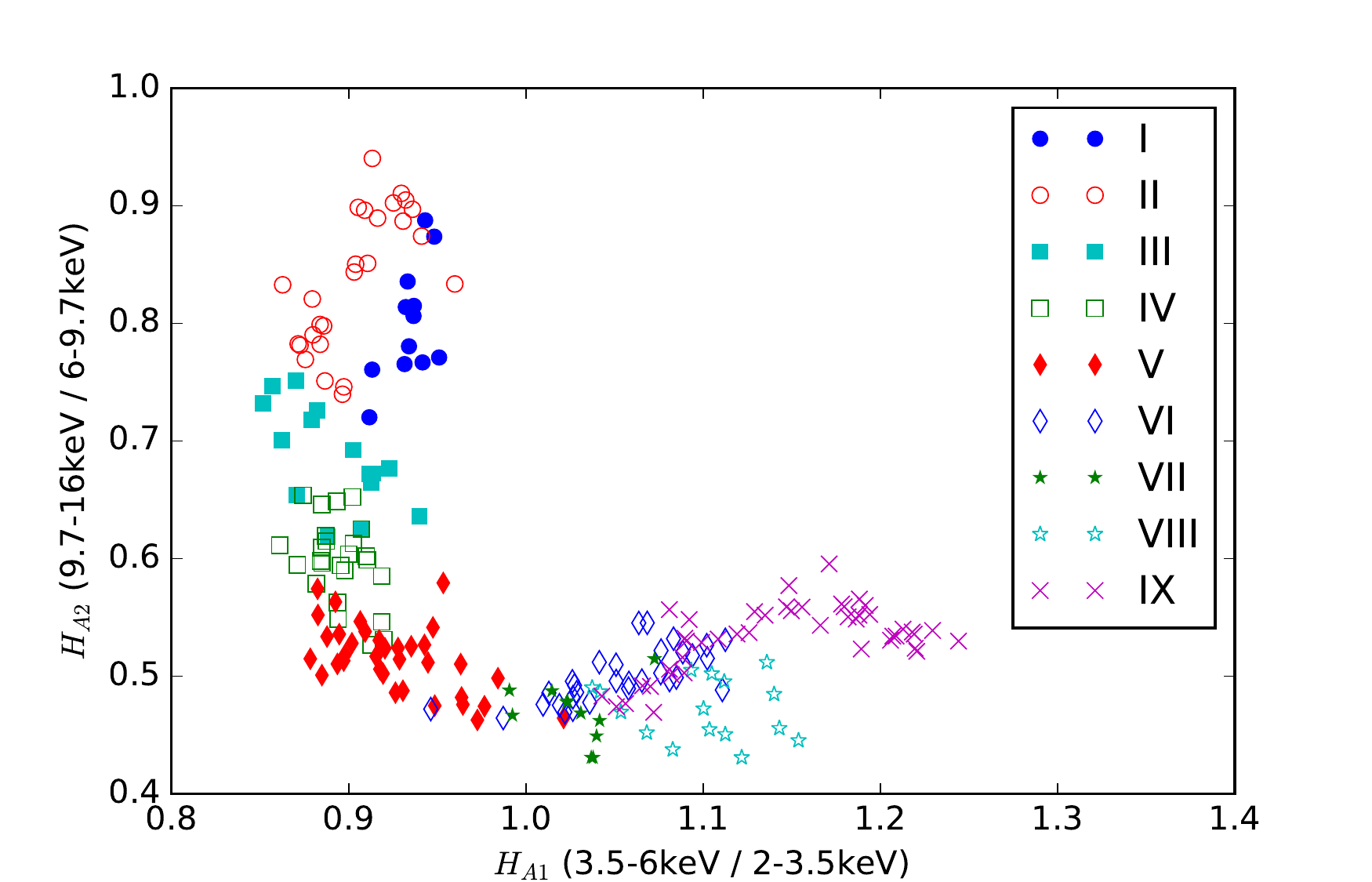}}\\
\subfloat[\textit{`Soft' ($H_{A1}$) Hardness-Intensity Diagram}]{\includegraphics[width=0.8\columnwidth, trim = 0mm 0mm 0mm 8mm,clip]{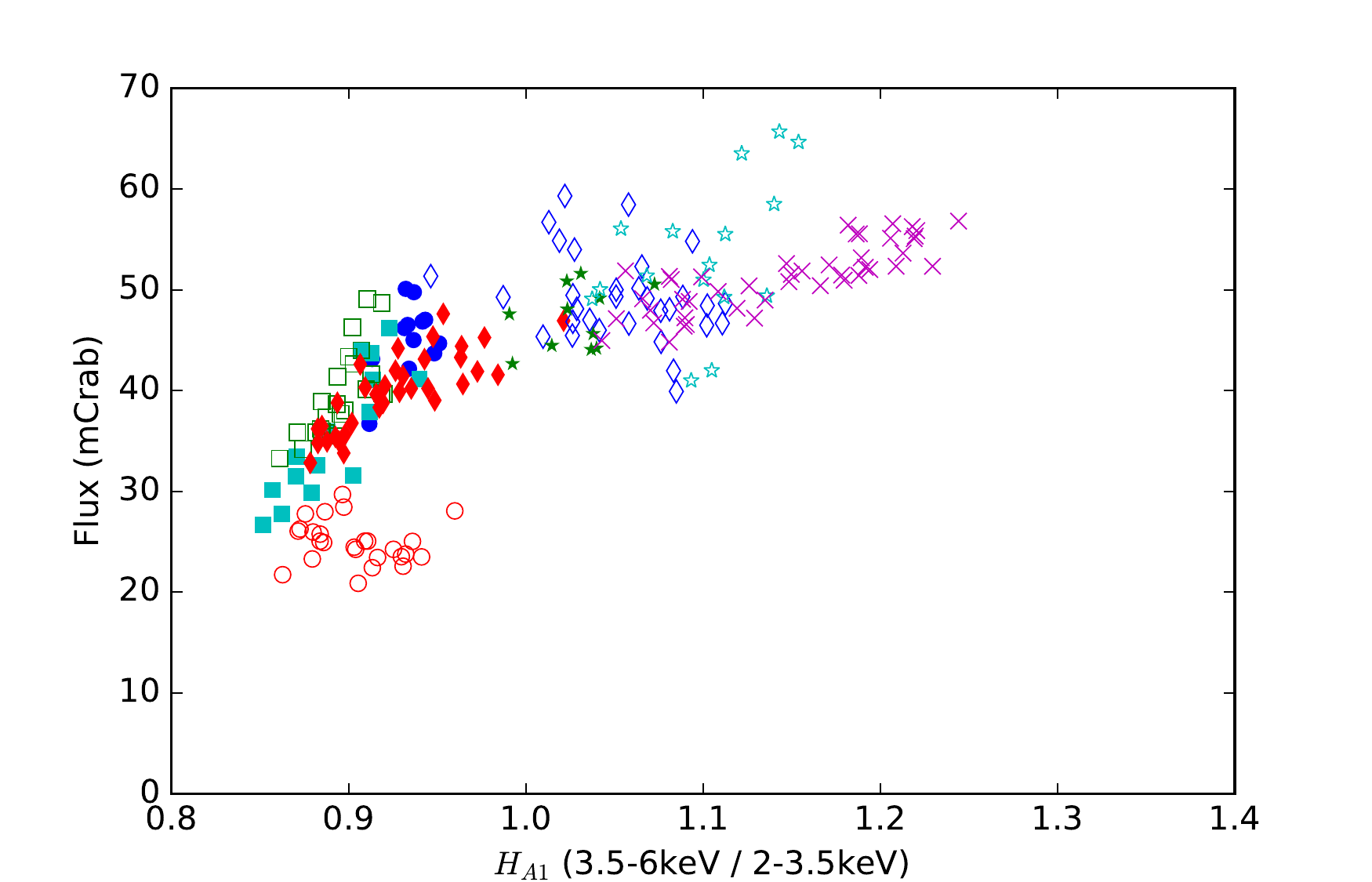}}\\
\subfloat[\textit{`Hard' ($H_{A2}$) Hardness-Intensity Diagram}]{\includegraphics[width=0.8\columnwidth, trim = 0mm 0mm 0mm 8mm,clip]{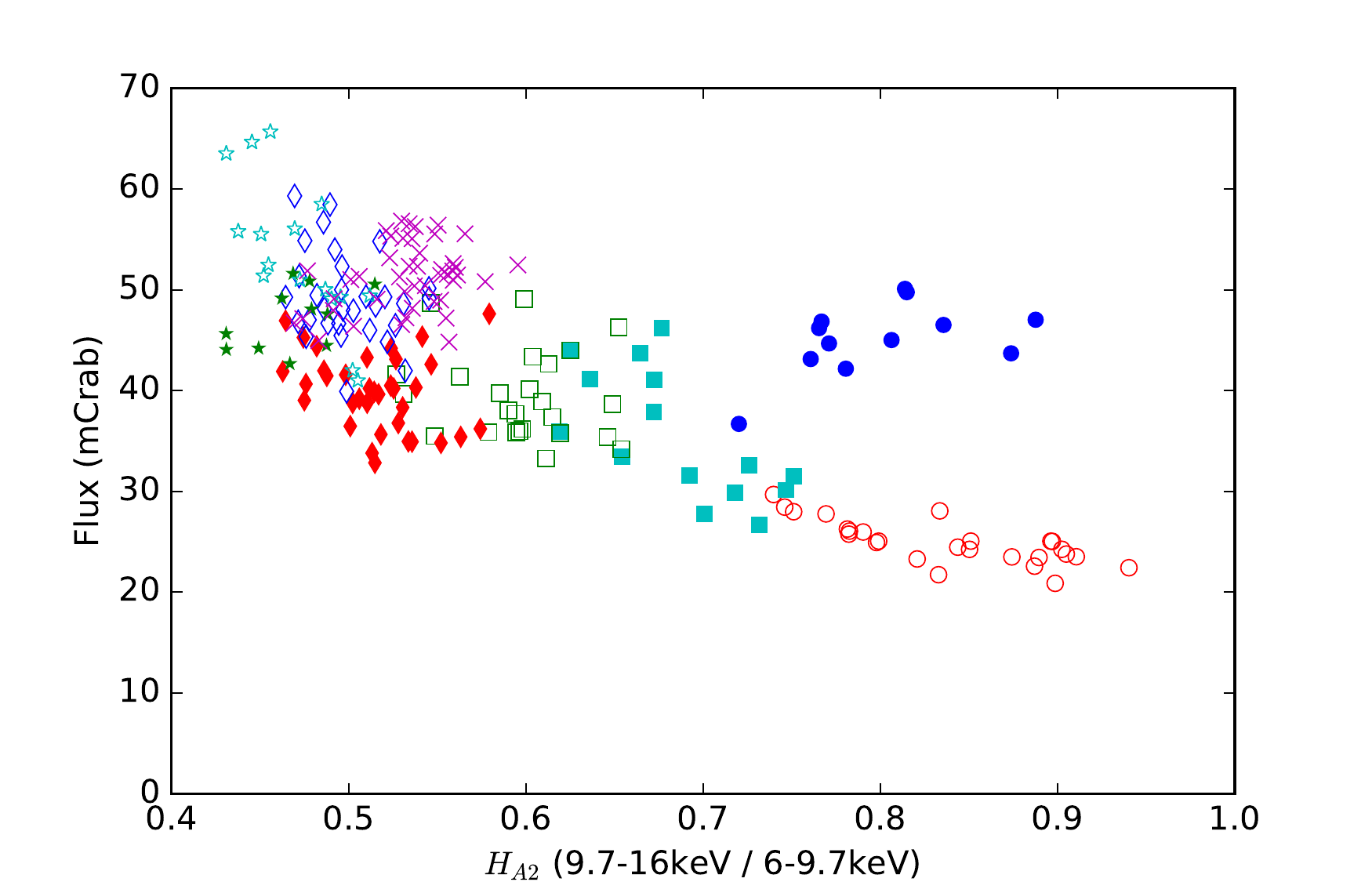}}\\
\captionsetup{singlelinecheck=off}
\caption[Hardness-Intensity diagrams of IGR J17091-3524 during the 2011-2013 outburst, showing when each of variability classes I-IX were observed.]{A global colour-colour diagram\index{Colour-colour diagram} (a), `soft' hardness-intensity diagram\index{Hardness-intensity diagram} (b) and `hard' hardness-intensity diagram (c) of the 2011-2013 outburst of IGR J17091, using the colours $H_{A1}$ and $H_{A2}$ defined previously.  Observations belonging to different classes have been highlighted in different colours.  Data taken from \citealt{Altamirano_IGR_FH}.}
\label{fig:IIIisHarder}
\end{figure}

 In Figure \ref{fig:WhereCls}, I show global 2--16\,keV\index{Lightcurve} lightcurves of IGR J17091 during the 2011-2013 outburst.  In each panel, all observations of a given class\index{Variability class} are highlighted in red.  A characteristic lightcurve is also presented for each class.  In Figure \ref{fig:IIIisHarder} panel (a), I show a plot of average hardness $H_{A2}$ against $H_{A1}$\index{Colour} for each observation, showing the long-term hysteresis\index{Hysteresis} of the object in colour-colour space\index{Colour-colour diagram}.  Again, observations belonging to each variability class are highlighted.  In Figure \ref{fig:IIIisHarder} panels (b) and (c), I show global hardness-intensity diagrams\index{Hardness-intensity diagram} for $H_{A1}$ and $H_{A2}$ respectively.
\par In Figure \ref{fig:IIIisHarder} Panel (a), we see that IGR J17091-3624\index{IGR J17091-3624} traces a two branched pattern in colour-colour\index{Colour-colour diagram} space corresponding to a branch which is soft ($\sim0.9$) in $H_{A1}$ and variable in $H_{A2}$ and a branch which is soft ($\sim0.5$) in $H_{A2}$ and variable in $H_{A1}$.  The `soft' HID shown in Figure \ref{fig:IIIisHarder} Panel (b) is dominated by a branch with a wide spread in $H_{A1}$ and intensities between $\sim40\mbox{--}60$\,mCrab.  A second branch exists at lower intensities, and shows an anticorrelation between intensity and $H_{A1}$.  Finally, the `hard' HID shown in Figure \ref{fig:IIIisHarder} Panel (c) shows an obvious anticorrelation between $H_{A2}$ and intensity, but there is also a secondary branch between $H_{A2}\approx 0.7\mbox{--}0.9$ at a constant intensity of $\sim40$\,mCrab.

\par For characteristic count rates and colours in each class, I quote the upper and lower quartile values \citep{Kenney_Quartile} instead of the mean.  This is due to the presence of high-amplitude but short-lived flares in many of the classes I describe.  Using the upper and lower quartiles as my measure of average and distribution means that my values will be less susceptible to outlier values of count rate and colour present in these flares.  All count rates have been background corrected\index{Background subtraction} (see Section \ref{sec:XTEDA}).
\par I have obtained mean values for these count rate quartiles, as well as values for colour\index{Colour} $C_1$ and fractional RMS\index{RMS}, by calculating these values individually for each orbit.  Histograms were then constructed from these datasets for each class, such that the mean and standard deviation of these values could be measured for each class.  These values are presented in Table \ref{tab:basicparams}.
\par I describe QPOs\index{Quasi-periodic oscillation} in terms of their $q$-value\indexq; a measure of coherence defined by the ratio of peak frequency and full-width half-maximum of each QPO.  I collected these values by fitting my power spectra with Lorentzians.

\begin{table}
\centering
\begin{tabular}{rllll} % four columns, alignment for each
\hline
\hline
\scriptsize Class &\scriptsize LQ Rate &\scriptsize  UQ Rate &\scriptsize Frac. RMS &\scriptsize Median C$_1$\\
\scriptsize &\scriptsize (cts s$^{-1}$) &\scriptsize (cts s$^{-1}$) & & \\
\hline
I\indexi&84--108&106--132&0.13--0.19&0.4--0.68\\
II\indexii&43--57&59--71&0.15--0.23&0.4--0.68\\
III\indexiii&64--84&80--110&0.17-0.23&0.35--0.45\\
IV\indexiv&63--81&92--122&0.27--0.37&0.32--0.4\\
V\indexv&49--67&88--134&0.44--0.54&0.28--0.46\\
VI\indexvi&64--98&111--155&0.29--0.47&0.33--0.61\\
VII\indexvii&65--79&128--140&0.45--0.57&0.32--0.42\\
VIII\indexviii&62--88&142--178&0.42--0.52&0.36--0.49\\
IX\indexix&87--111&114--144&0.16--0.24&0.42-0.6\\
\hline
\hline
\end{tabular}
\caption[A number of statistics averaged across all observations belonging to each IGR J17091 variability class.]{Lower and upper quartile count rates, fractional RMS\index{RMS} and median colour averaged across all observations belonging to each class\index{Variability class}.  Count rates and fractional RMS are taken from the full energy range of \indexpca\rxte\textit{/PCA}, and fractional RMS values are 2--60\,keV taken from lightcurves binned to 0.5\,s.  Count rates are normalised for the number of PCUs active during each observation.  All values are quoted as $1\sigma$ ranges.}
\label{tab:basicparams}
\end{table}

\par For each class\index{Variability class}, I present three standard data products; a 500\,s lightcurve\index{Lightcurve}, a variable-length lightcurve where the length has been selected to best display the variability associated with the class and a Fourier PDS\index{Fourier analysis}.  Unless otherwise stated in the figure caption, the 500\,s lightcurve and the Fourier PDS are presented at the same scale for all classes.  In Table \ref{tab:CPopD} I present a tally of the number of times I assigned each Variability Class to an \rxte\ orbit.

\begin{table}
\centering
\begin{tabular}{llll}
\hline
\hline
\scriptsize Class &\scriptsize  Orbits &\scriptsize Total Time (s) &\scriptsize Fraction \\
\hline
I\indexi\ & 31 &  69569 & 14.8\%\\
II\indexii\ & 26 &  50875 & 10.8\%\\
III\indexiii\ & 14 &  26228 & 5.6\%\\
IV\indexiv\ & 31 &  69926 & 14.9\%\\
V\indexv\ & 35 &  72044 & 15.3\%\\
VI\indexvi\ & 29 &  54171 & 11.5\%\\
VII\indexvii\ & 11 &  19241 & 4.1\%\\
VIII\indexviii\ & 16 &  26553 & 5.7\%\\
IX\indexix\ & 50 &  81037 & 17.3\%\\
\hline
\hline
\end{tabular}
\caption[A tally of the number of times I assigned each of my nine Variability Classes to an \rxte\ orbit observing IGR J17091.]{A tally of the number of times I assigned each of my nine Variability Classes\index{Variability class} to an \rxte\ orbit.  I have also calculated the amount of observation time corresponding to each class, and thus inferred the fraction of the time that IGR J17091\index{IGR J17091-3624} spent in each class.  Note: the values in the Total Time column assume that each orbit only corresponds to a single variability Class.}
\label{tab:CPopD}
\end{table}

\begin{figure}
    \includegraphics[width=0.8\columnwidth, trim = 0.6cm 0 3.9cm 0]{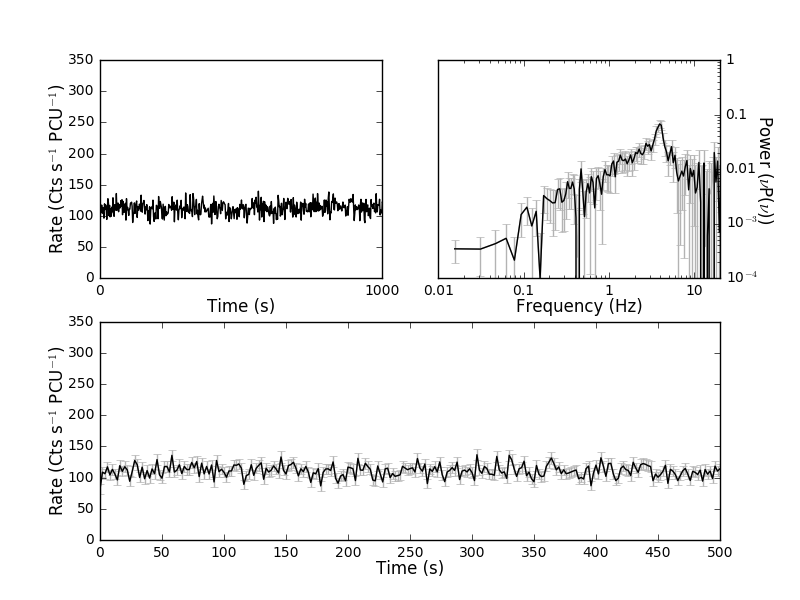}
    \captionsetup{singlelinecheck=off}
    \caption[Characteristic lightcurves and a power spectrum of Type I variability.]{Plots of the Class I\indexi\ observation 96420-01-01-00, orbit 0.  \textit{Top-left}: 1000\,s lightcurve\index{Lightcurve} binned on 2 seconds to show lightcurve evolution.  \textit{Top-right}: Fourier Power Density Spectrum\index{Fourier analysis}.  \textit{Bottom}: 500\,s lightcurve binned on 2 seconds.}
   \label{fig:Bmulti}
\end{figure}

\subsubsection{Class I --  Figure \ref{fig:Bmulti}}
\label{sec:ClassI}

In the 2\,s binned lightcurve\index{Lightcurve} of a Class I\indexi\ observation, there is no structured second-to-minute scale variability.  The Fourier PDS\index{Fourier analysis} of all observations in this class show broad band noise between $\sim1$--$10$\,Hz, as well as a weak QPO\index{Quasi-periodic oscillation} (with a $q$-value\indexq\ of $\sim5$) which peaks at around 5\,Hz.

\begin{figure}
    \includegraphics[width=0.8\columnwidth, trim = 0.6cm 0 3.9cm 0]{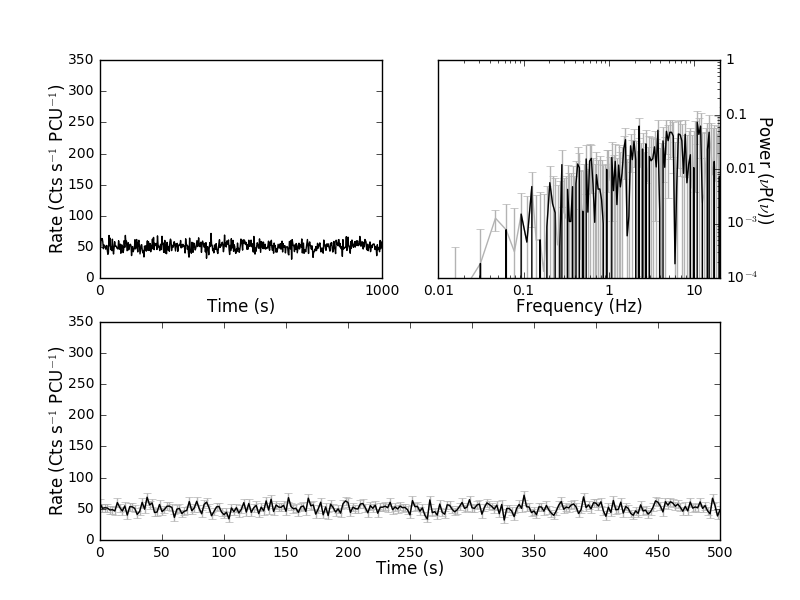}
    \captionsetup{singlelinecheck=off}
    \caption[Characteristic lightcurves and a power spectrum of Type II variability.]{Plots of the Class II\indexii\ observation 96420-01-11-00, orbit 0.  \textit{Top-left}:  1000\,s lightcurve\index{Lightcurve} binned on 2 seconds to show lightcurve evolution.  \textit{Top-right}: Fourier Power Density Spectrum\index{Fourier analysis}.  \textit{Bottom}: Lightcurve binned on 2 seconds.}
   \label{fig:Emulti}
\end{figure}

\subsubsection{Class II -- Figure \ref{fig:Emulti}}

\par Class II\indexii\ observations are a factor of $\sim2$ fainter in the $L_T$ band than Class I\indexi\ observations.  They also occupy a different branch in a plot of hardness $H_{A2}$\index{Hardness-intensity diagram} against intensity (see Figure \ref{fig:IIIisHarder}, panel c).  The PDS\index{Fourier analysis} shows no significant broad band noise above $\sim1$\,Hz unlike that which is seen in Class I.  The $\sim$5\,Hz QPO\index{Quasi-periodic oscillation} seen in Class I is absent in Class II.

\begin{figure}
    \includegraphics[width=0.8\columnwidth, trim = 0.6cm 0 3.9cm 0]{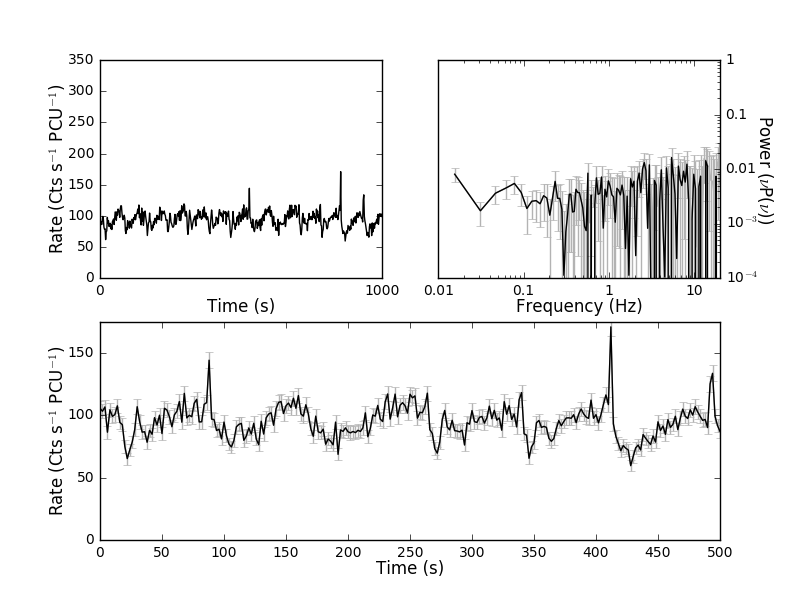}
    \captionsetup{singlelinecheck=off}
    \caption[Characteristic lightcurves and a power spectrum of Type III variability.]{Plots of the Class III\indexiii\ observation 96420-01-04-01, orbit 0.  \textit{Top-left}: 1000\,s lightcurve\index{Lightcurve} binned on 2 seconds to show lightcurve evolution.  \textit{Top-right}: Fourier Power Density Spectrum\index{Fourier analysis}.  \textit{Bottom}: Lightcurve binned on 2 seconds.  Note that, to emphasise the behaviour of the lightcurve in this class, I have magnified the 500\,s lightcurve y-scale by a factor of 2 compared with the lightcurves presented for other classes.}
   \label{fig:Gmulti}
\end{figure}

\subsubsection{Class III -- Figure \ref{fig:Gmulti}}
\label{sec:classIII}

\par Unlike Classes I \& II, Class III\indexiii\ lightcurves show structured flaring\index{Flare}, with a peak-to-peak recurrence time\index{Recurrence time} of $42$--$80$\,s.  Most flares consist of a steady $\sim60$\,s rise in count rate and then an additional and sudden rise to a peak count rate at $\gtrsim200$\spcu which lasts for $\lesssim$0.5\,s before returning to continuum level (I have magnified the y-scaling in the lightcurve of Figure \ref{fig:Gmulti} to emphasise this behaviour). This sudden rise is not present in every flare; in some observations it is absent from every flare feature.  No 5\,Hz QPO\index{Quasi-periodic oscillation} is present in the PDS\index{Fourier analysis} and there is no significant variability in the range between $\sim1\mbox{--}10$\,Hz.

\par
\begin{figure}
    \includegraphics[width=\columnwidth, trim = 0mm 0mm 0mm 0mm]{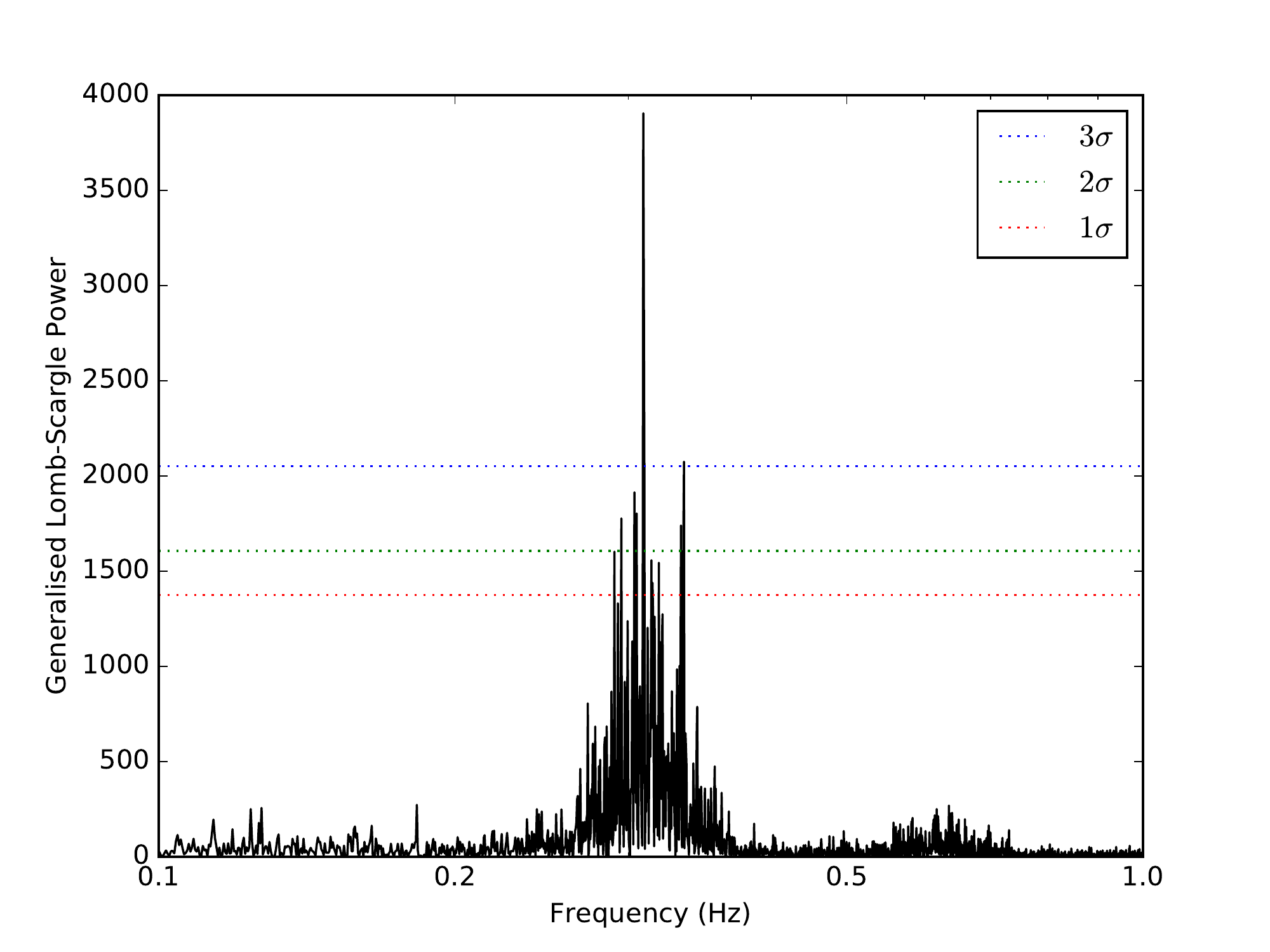}
    \captionsetup{singlelinecheck=off}
    \caption[The Lomb-Scargle periodogram of Class III observation 96420-01-19-01]{The Lomb-Scargle periodogram\index{Lomb-Scargle periodogram} of Class III\indexiii\ observation 96420-01-19-01, orbit 0, with significance levels of 1, 2 and 3$\sigma$ plotted.  The peak at 0.31\,Hz was used to define a QPO\index{Quasi-periodic oscillation} frequency when folding the data from this observation.}
   \label{fig:IIILS}
\end{figure}

\begin{figure}
    \includegraphics[width=\columnwidth, trim = 0mm 0mm 0mm 0mm]{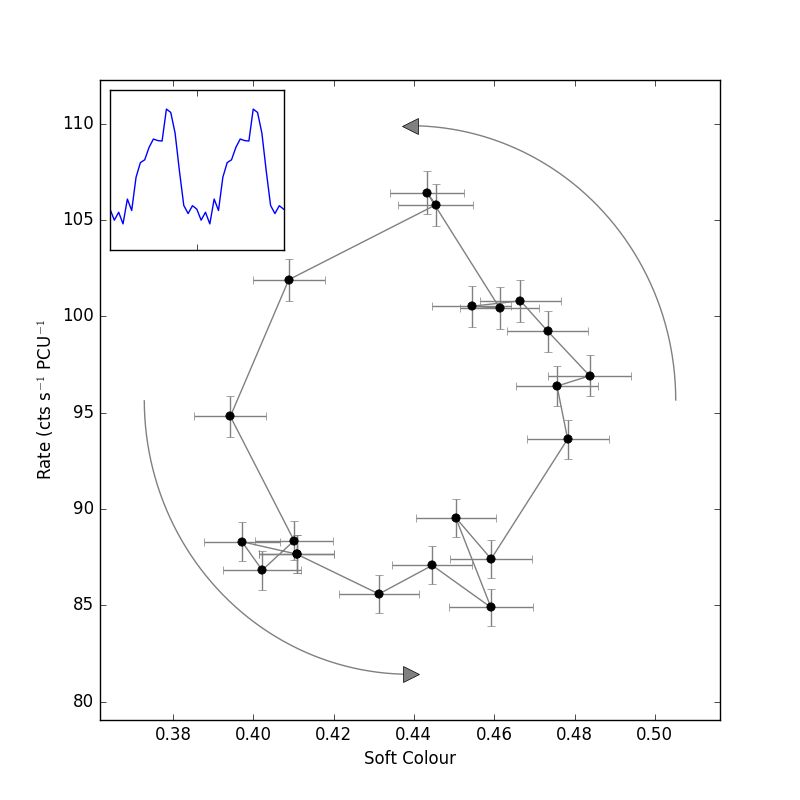}
    \captionsetup{singlelinecheck=off}
    \caption[A hardness-intensity diagram of the Class III observation 96420-01-04-01.]{The hardness-intensity diagram (HID$_1$)\index{Hardness-intensity diagram} of the Class III\indexiii\ observation 96420-01-04-01, orbit 0.  The data have been folded\index{Folding} over a period of 79.61 s, corresponding to the peak frequency in the Lomb-Scargle periodogram\index{Lomb-Scargle periodogram} of this observation.  Inset is the folded lightcurve\index{Lightcurve} of the same data.}
   \label{fig:LoopIII}
\end{figure}

 As this class has a well-defined periodicity, I folded\index{Folding} data in each observation to improve statistics using the best-fit period obtained from generalised Lomb-Scargle Periodogram Analysis; I show a representative Lomb-Scargle periodogram\index{Lomb-Scargle periodogram} in Figure \ref{fig:IIILS}.  I find an anticlockwise hysteretic\index{Hysteresis} loop in the folded HID$_1$\index{Hardness-intensity diagram} of all 15 Class III orbits.  In Figure \ref{fig:LoopIII} I show an example of one of these loops.

\begin{figure}
    \includegraphics[width=0.8\columnwidth, trim = 0.6cm 0 3.9cm 0]{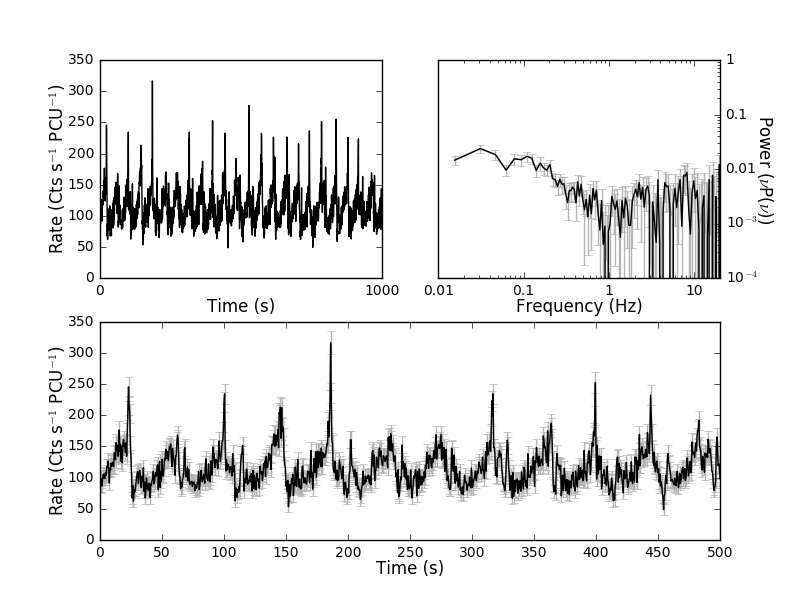}\\
    \captionsetup{singlelinecheck=off}
    \caption[Characteristic lightcurves and a power spectrum of Type IV variability.]{Plots of the Class IV\indexiv\ observation 96420-01-05-00, orbit 0.  \textit{Top-left}: 1000\,s lightcurve\index{Lightcurve} binned on 2 seconds to show lightcurve evolution.  \textit{Top-right}: Fourier Power Density Spectrum\index{Fourier analysis}.  \textit{Bottom}: Lightcurve binned on 0.5 seconds.}
   \label{fig:Jmulti}
\end{figure}

\subsubsection{Class IV -- Figure \ref{fig:Jmulti}}
\label{sec:classIV}

\par The lightcurves\index{Lightcurve} in this class\indexiv\ show regular variability\index{Variability} with a peak-to-peak recurrence time\index{Recurrence time} of $25$--$39$\,s.  I performed peak analysis (see Section \ref{sec:Flares}) on observations belonging to this class, finding that each flare\index{Flare} has a rise time with lower and upper quartile values of $19.5$ and $33.5$ s, a fall time with lower and upper quartile values of $4.6$ and $13.5$\,s and a peak count rate of $159$--$241$\spcu\ .  There are no significant QPOs\index{Quasi-periodic oscillation} in the Fourier PDS above $\sim1$\,Hz.
\par
\begin{figure}
    \includegraphics[width=\columnwidth, trim = 0mm 0mm 0mm 0mm]{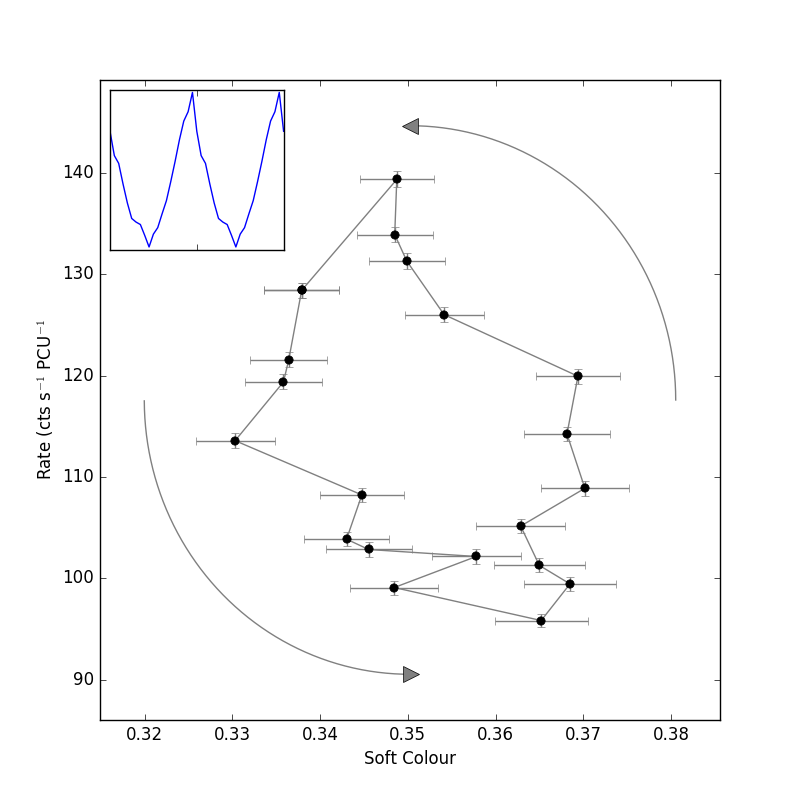}\\
    \includegraphics[width=0.5\columnwidth, trim = 0mm 0mm 0mm 0mm]{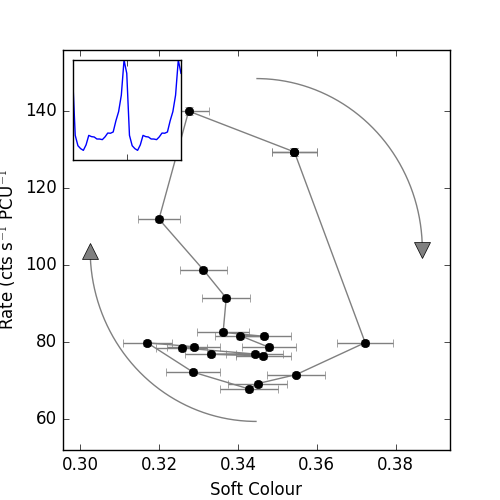}\includegraphics[width=0.5\columnwidth, trim = 0mm 0mm 0mm 0mm]{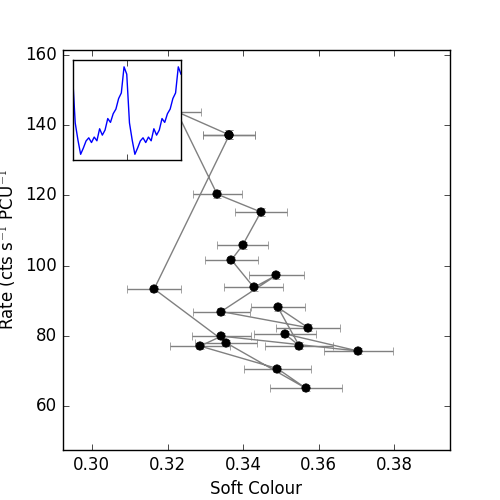}
    \captionsetup{singlelinecheck=off}
    \caption[The hardness-intensity diagram of the Class IV observation 96420-01-05-00, showing an anticlockwise loop.]{\textit{Top}: The hardness-intensity diagram\index{Hardness-intensity diagram} (HID$_1$) of the Class IV\indexiv\ observation 96420-01-05-00, orbit 0 showing an anticlockwise loop\index{Hysteresis}.  The data have been folded\index{Folding} over a variable period found with the algorithm described in Section \ref{sec:Flares}.  Inset is the folded lightcurve\index{Lightcurve} of the same data.  \textit{Bottom Left}: The hardness-intensity diagram of Class IV observations 96420-01-24-02 orbit 0, an example of a clockwise loop.  \textit{Bottom Right}: The hardness-intensity diagram of Class IV observation 96420-01-06-00 orbit 0, in which I was unable to ascertain the presence of a loop.}
   \label{fig:LoopIV}
\end{figure}

 I folded\index{Folding} individual Class IV\indexiv\ lightcurves and found anticlockwise hysteretic\index{Hysteresis} loops in the HID$_1$\index{Hardness-intensity diagram} of 14 out of 30 Class IV observations.  In the top panel of Figure \ref{fig:LoopIV} I show an example of one of these loops.  However, I also find clockwise hysteretic loops in 6 Class IV observations, and in 10 orbits the data did not allow us to ascertain the presence of a loop.  I provide an example of both of these in the lower panels of Figure \ref{fig:LoopIV}.  I note that the structure of clockwise loops are more complex than anticlockwise loops in Class IV, consisting of several lobes\footnote{In HIDs with multiple lobes, the loop direction I assign to the observation corresponds to the direction of the largest lobe.} rather than a single loop (Figure \ref{fig:LoopIV}, bottom-left).

\par Compared with Class III\indexiii, the oscillations in Class IV\indexiv\ occur with a significantly lower period, with a mean peak-to-peak recurrence time\index{Recurrence time} of $\sim30$\,s compared to $\sim60$\,s in Class III.
\par In Figure \ref{fig:IIIisHarder} I show that Classes III\indexiii\ and IV\indexiv\ can also be distinguished by average hardness\index{Colour}, as Class III tends to have a greater value of $H_{A2}$ than Class IV.

\begin{figure}
    \includegraphics[width=0.8\columnwidth, trim = 0.6cm 0 3.9cm 0]{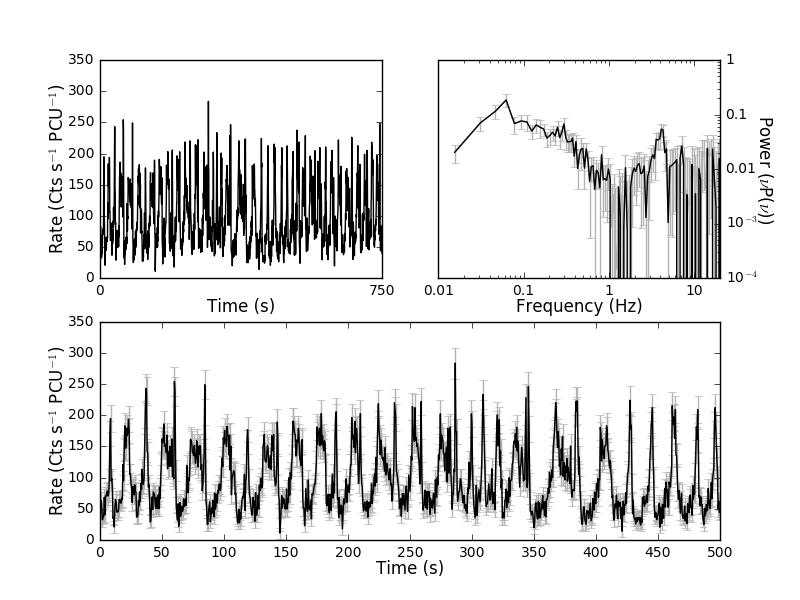}
    \captionsetup{singlelinecheck=off}
    \caption[Characteristic lightcurves and a power spectrum of Class V variability.]{Plots of the Class V\indexv\ observation 96420-01-06-03, orbit 0.  \textit{Top-left}: 750\,s lightcurve\index{Lightcurve} binned on 2 seconds to show lightcurve evolution.  \textit{Top-right}: Fourier Power Density Spectrum\index{Fourier analysis}. \textit{Bottom}: Lightcurve binned on 0.5 seconds.}
   \label{fig:Kmulti}
\end{figure}

\subsubsection{Class V -- Figure \ref{fig:Kmulti}}
\label{sec:classV}

\par The lightcurves\index{Lightcurve} in this class\indexv, like in Classes III\indexiii\ and IV\indexiv, show flaring\index{Flare} behaviour, with flares separated by a few tens of seconds.  At higher frequencies, the PDS\index{Fourier analysis} shows a prominent QPO\index{Quasi-periodic oscillation} centred at $\sim4$\,Hz with as $q$-value\indexq\ of $\sim3$.  There is also significant broad band noise between $\sim0.1$--$1$\,Hz
\par
\begin{figure}
    \includegraphics[width=\columnwidth, trim = 0mm 0mm 0mm 0mm]{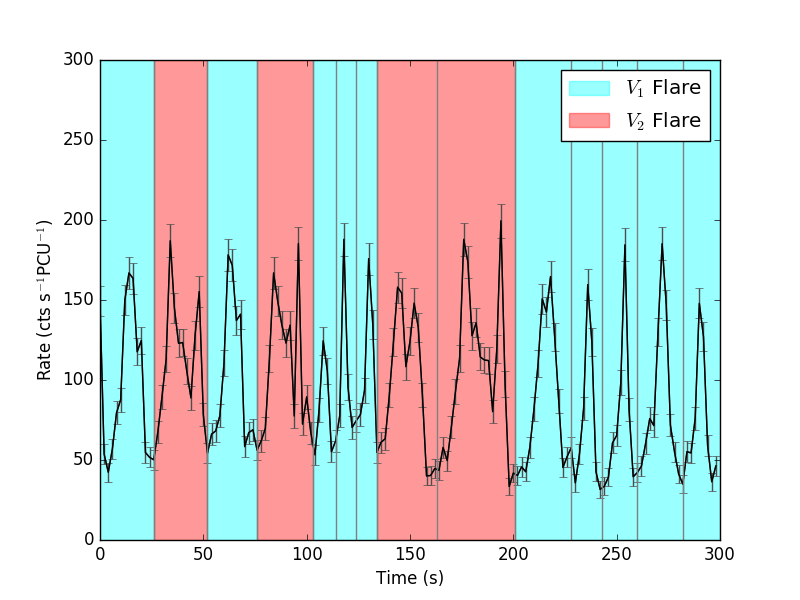}
    \captionsetup{singlelinecheck=off}
    \caption[A portion of the lightcurve of observation 96420-01-06-03 showing Type $V_1$ flares and Type $V_2$ flares.]{A portion of the lightcurve\index{Lightcurve} of observation 96420-01-06-03, orbit 0, showing Type $V_1$\indexv\index{Flare} flares (highlighted in cyan) and Type $V_2$ flares (highlighted in red).}
   \label{fig:id_flares_V}
\end{figure}

\begin{figure}
    \includegraphics[width=\columnwidth, trim = 0mm 0mm 0mm 0mm]{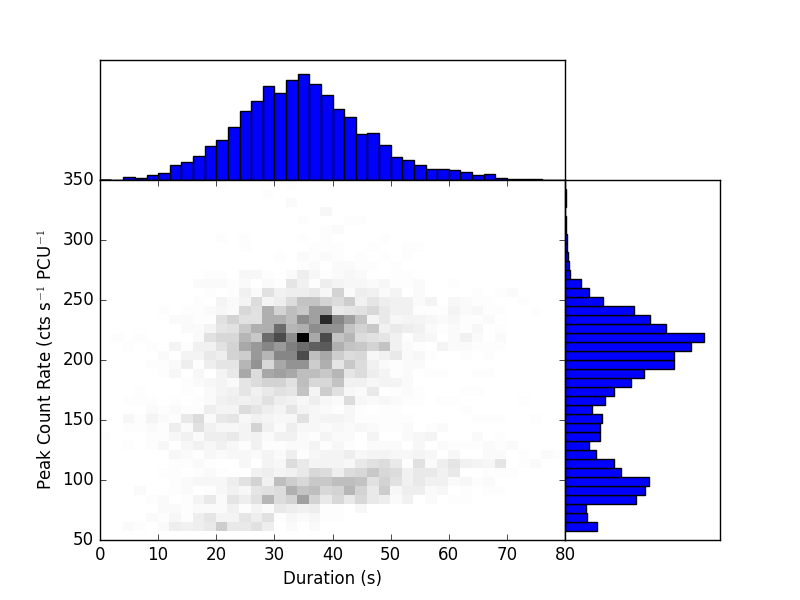}
    \captionsetup{singlelinecheck=off}
    \caption[Every flare in all observations identified as Class V, plotted in a two-dimensional histogram of flare peak count rate against flare duration to show the two-population nature of these events.]{Every flare\index{Flare} in all observations identified as Class V\indexv, plotted in a two-dimensional histogram of flare peak count rate against flare duration to show the two-population\index{Population study} nature of these events.}
   \label{fig:two_popV}
\end{figure}

 In Figure \ref{fig:id_flares_V} I show that the flaring in this class is more complex than that seen in Classes III and IV.  Class V lightcurves consist of short strongly peaked symmetrical flares\index{Flare} (hereafter Type $V_1$) and a longer more complex type of flare (hereafter Type $V_2$).  The Type $V_2$ flare consists of a fast rise to a local maximum in count rate, followed by a $\sim10$\,s period in which this count rate gradually reduces by $\sim50\%$ and then a much faster peak with a maximum count rate between 1 and 2 times that of the initial peak.  In both types of flare, I find that the increase in count rate corresponds with an increase in soft colour\index{Colour}.  The two-population nature\index{Population study} of flares in Class V can also clearly be seen in Figure \ref{fig:two_popV}, where I show a two-dimensional histogram of flare peak count rate against flare duration.
\par
\begin{figure}
    \includegraphics[width=\columnwidth, trim = 0mm 0mm 0mm 0mm]{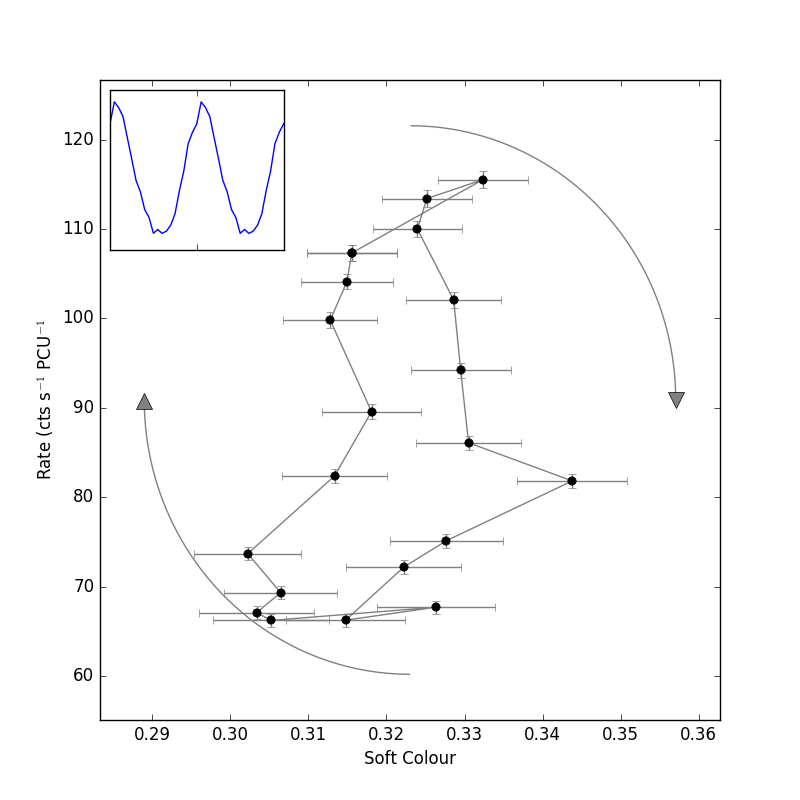}\\
    \includegraphics[width=0.5\columnwidth, trim = 0mm 0mm 0mm 0mm]{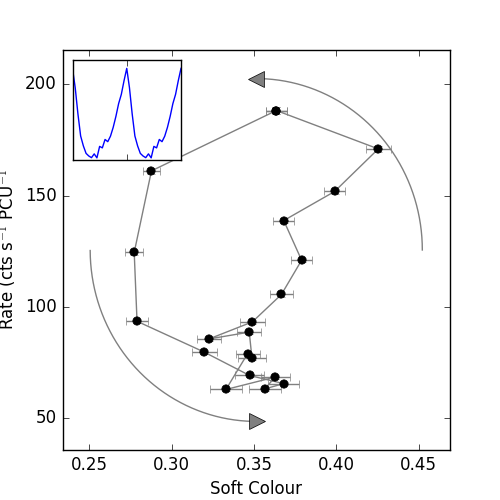}\includegraphics[width=0.5\columnwidth, trim = 0mm 0mm 0mm 0mm]{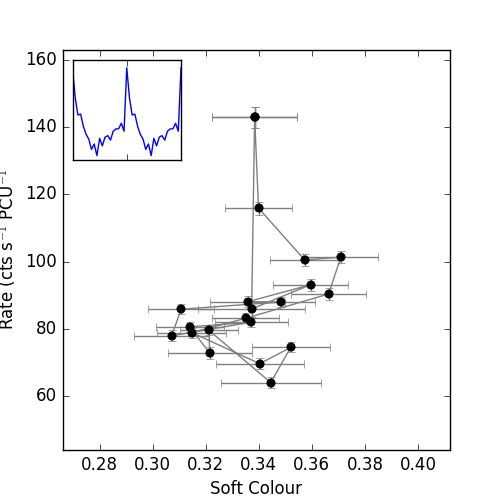}
    \captionsetup{singlelinecheck=off}
    \caption[The hardness-intensity diagram of a type $V_1$ flaring period in Class V observation 96420-01-07-00 showing a clockwise loop.]{\textit{Top}: The hardness-intensity diagram\index{Hardness-intensity diagram} (HID$_1$) of a type $V_1$\indexv\ flaring\index{Flare} period in Class V observation 96420-01-07-00, orbit 0 showing a clockwise loop\index{Hysteresis}.  The data have been folded\index{Folding} over a variable period found with the algorithm described in Section \ref{sec:Flares}.  Inset is the folded lightcurve\index{Lightcurve} of the same data. \textit{Bottom Left}: The hardness-intensity diagram of Class V observation 96420-01-25-05 orbit 0, an example of an anticlockwise loop.  \textit{Bottom Right}: The hardness-intensity diagram of Class V observation 96420-01-25-06 orbit 0, in which I was unable to ascertain the presence of a loop.}
   \label{fig:LoopV}
\end{figure}

 I folded\index{Folding} all individual Class V\indexv\ lightcurves, in each case cropping out periods of $V_2$ flaring.  I find clockwise hysteretic\index{Hysteresis} loops in the HID$_1$\index{Hardness-intensity diagram} of 30 out of 33 Class V observations, suggesting a lag\index{Hard lag} in the aforementioned relation between count rate and soft colour.  In the upper panel Figure \ref{fig:LoopV} I present an example of one of these loops.  In one observation however, I found an anticlockwise loop in the HID$_1$ (shown in Figure \ref{fig:LoopV} lower-left panel).  I was unable to ascertain the presence of loops in the remaining 2 orbits; for the sake of completeness, I show one of these in the lower-right panel of Figure \ref{fig:LoopV}.

\begin{figure}
    \includegraphics[width=0.8\columnwidth, trim = 0.6cm 0 3.9cm 0]{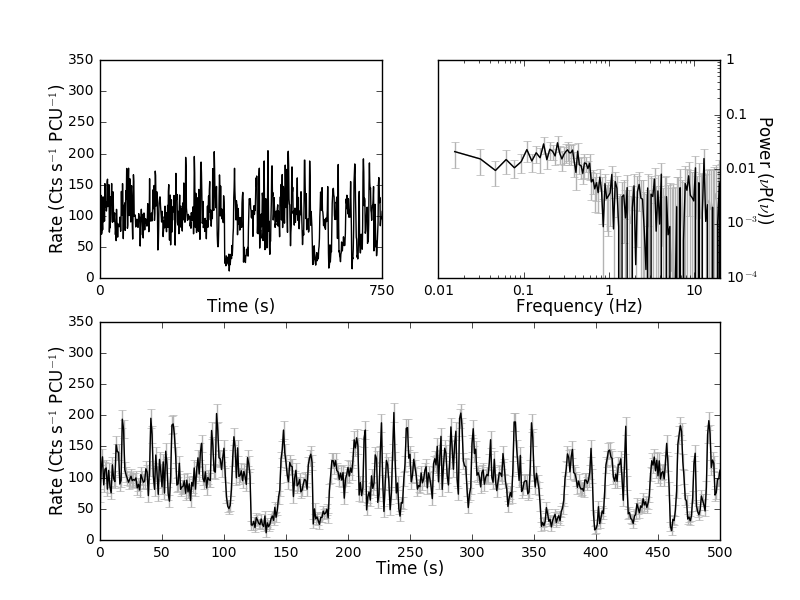}
    \captionsetup{singlelinecheck=off}
    \caption[Characteristic lightcurves and a power spectrum of Type VI variability.]{Plots of the Class VI observation 96420-01-09-00, orbit 0.  \textit{Top-left}: 750\,s lightcurve\index{Lightcurve} binned on 2 seconds to show lightcurve evolution.  \textit{Top-right}: Fourier Power Density Spectrum\index{Fourier analysis}.  \textit{Bottom}: Lightcurve binned on 1 second.}
   \label{fig:Lmulti}
\end{figure}

\subsubsection{Class VI -- Figure \ref{fig:Lmulti}}

\par The lightcurves\index{Lightcurve} of observations of this class\indexvi\ show large dips\index{Dip} in count rate; this can be seen in Figure \ref{fig:Lmulti} at, for example, $t\approx125$--$150$\,s .  These dips vary widely in duration, from $\sim5$ to $\sim50$ seconds, and the count rate in both $L_A$ and $L_B$ fall to a level consistent with background.  The dips' rise and fall times are fast, both lasting no longer than a second.  They do not appear to occur with any regular periodicity.
\par Aside from the dips, Class VI\indexvi\ observations show other structures in their lightcurves.  Large fluctuations in count rate, by factors of $\lesssim3$, occur on timescales of $\sim1\mbox{--}5$ s; no periodicity in these oscillations could be found.  This behaviour is reflected in the PDS\index{Fourier analysis}, which shows high-amplitude broad band noise below $\sim0.5$\,Hz with RMS-normalized power\index{RMS normalisation} \citep{Belloni_RMSNorm} of up to $\sim1.1 $\,Hz$^{-1}$.  As can be seen in Figure \ref{fig:Lmulti}, this feature takes the form of a broad shoulder of noise which shows either a weak peak or no clear peak at all.  The $\sim5$\,Hz QPO\index{Quasi-periodic oscillation} seen in the PDS of other classes is not present in Class VI observations.
\par
\begin{figure}
    \includegraphics[width=\columnwidth, trim = 0mm 0mm 0mm 0mm]{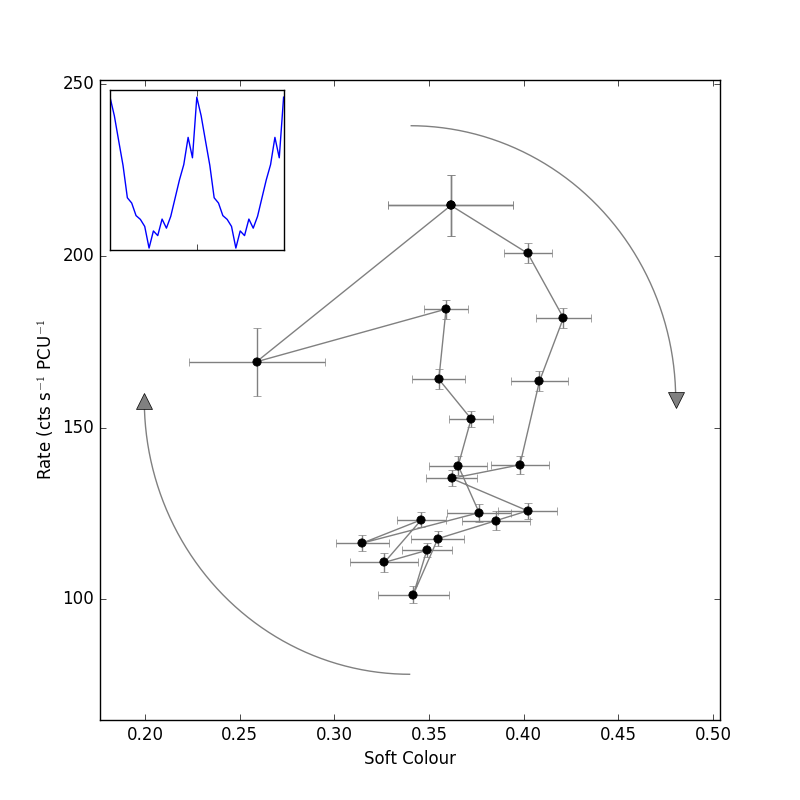}\\
    \includegraphics[width=0.5\columnwidth, trim = 0mm 0mm 0mm 0mm]{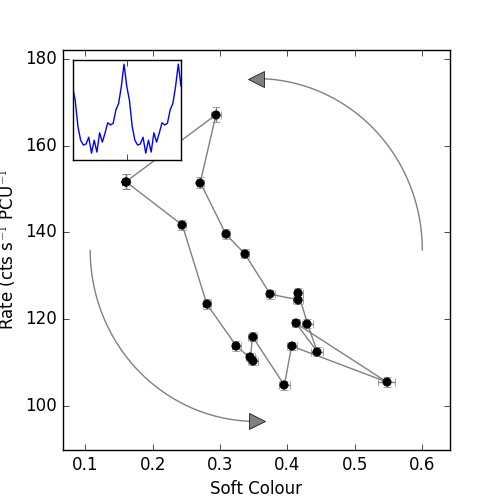}\includegraphics[width=0.5\columnwidth, trim = 0mm 0mm 0mm 0mm]{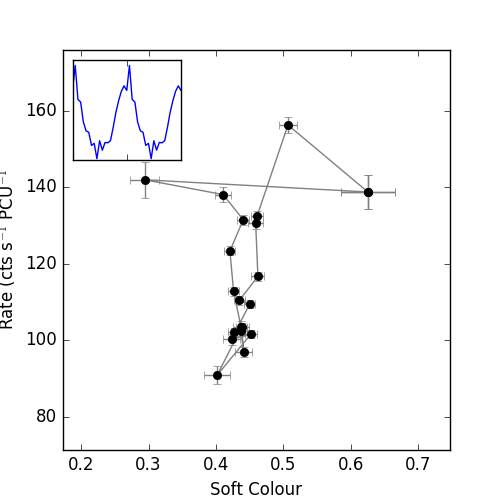}
    \captionsetup{singlelinecheck=off}
    \caption[The hardness-intensity diagram of the Class VI observation 96420-01-30-03, showing a clockwise loop.]{\textit{Top}: The hardness-intensity diagram\index{Hardness-intensity diagram} (HID$_1$) of the Class VI\indexvi\ observation 96420-01-30-03, orbit 0 showing a clockwise loop\index{Hysteresis}.  The data have been folded\index{Folding} over a variable period found with the algorithm described in Section \ref{sec:Flares}.  Inset is the folded lightcurve\index{Lightcurve} of the same data. \textit{Bottom Left}: The hardness-intensity diagram of Class VI observation 96420-01-30-04 orbit 0, an example of an anticlockwise loop.  \textit{Bottom Right}: The hardness-intensity diagram of Class VI observation 96420-01-09-03 orbit 0, in which I was unable to ascertain the presence of a loop.}
   \label{fig:LoopVI}
\end{figure}

 I attempted to fold\index{Folding} all individual Class VI\indexvi\ lightcurves, ignoring the sections of data corresponding to the large count rate dips described above.  In general, folding lightcurves belonging to this class is difficult; many orbits showed low-amplitude oscillations which were difficult to fold using my flare-finding algorithm (see Section \ref{sec:Flares}), while many others only showed oscillatory behaviour for a small number of periods between each pair of dips.  As such, I only succesfully folded 23 of the 40 Class VI orbits.  Of these, 19 showed clockwise loops\index{Hysteresis} in the HID$_1$\index{Hardness-intensity diagram} (top panel, Figure \ref{fig:LoopVI}), 3 showed anticlockwise loops (bottom-left panel, Figure \ref{fig:LoopVI}).  In the remaining 1 observation, the data did not allow us to ascertain the presence of loops (bottom-right panel, Figure \ref{fig:LoopVI}).
\par Like in Class VI, I note that the clockwise loops in Class VI appear more complex than clockwise loops.  Again, the clockwise loop shown in Figure \ref{fig:LoopVI} appears to have a 2-lobe structure; this is repeated in all clockwise loops found in this class.

\begin{figure}
    \includegraphics[width=0.8\columnwidth, trim = 0.6cm 0 3.9cm 0]{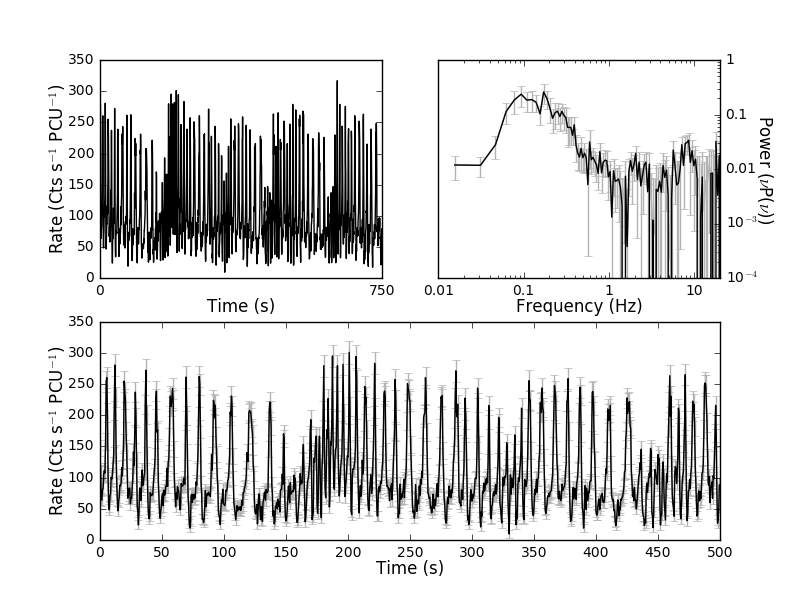}
    \captionsetup{singlelinecheck=off}
    \caption[Characteristic lightcurves and a power spectrum of Type VII variability.]{Plots of the Class VII\indexvii\ observation 96420-01-18-05, orbit 0.  \textit{Top-left}: 750\,s lightcurve\index{Lightcurve} binned on 2 seconds to show lightcurve evolution.  \textit{Top-right}: Fourier Power Density Spectrum\index{Fourier analysis}.  \textit{Bottom}: Lightcurve binned on 0.5 seconds.}
   \label{fig:Nmulti}
\end{figure}

\subsubsection{Class VII -- Figure \ref{fig:Nmulti}}

\par
\begin{figure}
    \includegraphics[width=0.8\columnwidth, trim = 0.6cm 0 3.9cm 0]{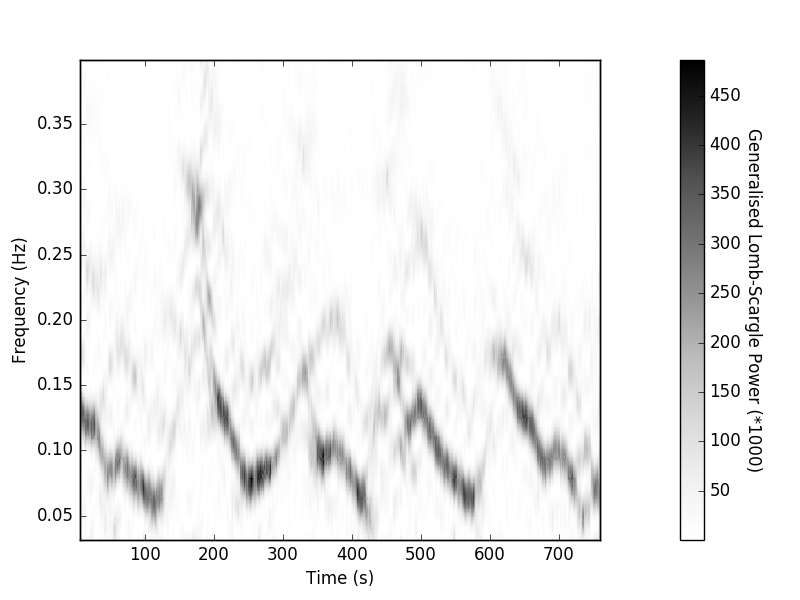}
    \captionsetup{singlelinecheck=off}
    \caption[A sliding window Lomb-Scargle spectrogram of Class VII observation 96420-01-18-05.]{A sliding window Lomb-Scargle periodogram\index{Lomb-Scargle periodogram} of Class VII\indexvii\ observation 96420-01-18-05, showing power density spectra from an overlapping 32\,s window moved 1\,s at a time.  The peak frequency of this low frequency QPO\index{Quasi-periodic oscillation} itself appears to oscillate with a frequency of $\sim5$mHz.}
   \label{fig:spect}
\end{figure}

 Class VII\indexvii\ shows high-amplitude flaring\index{Flare} behaviour with a peak-to-peak recurrence time\index{Recurrence time} of $6$--$12$\,s.  In Figure \ref{fig:spect} I show a dynamical Lomb-Scargle spectrogram\index{Lomb-Scargle periodogram} of a Class VII observation, showing that the fast flaring behaviour has a frequency which moves substantially over time.  This in turn accounts for the large spread in the value of the flare peak-to-peak recurrence time.
\par In Figure \ref{fig:spect} I show that the peak frequency of the QPO\index{Quasi-periodic oscillation} also varies in a structured way.  I also suggest that the variabilitity of the frequency is itself a QPO with a period of $\sim150$\,s.
\par At higher frequencies, the PDS\index{Fourier analysis} shows a weak QPO\index{Quasi-periodic oscillation} centred at $\sim8$\,Hz, with a $q$-value\indexq\ of $\sim2$.
\par I used my flare-finding algorithm (see Section \ref{sec:Flares}) to perform variable-frequency folding\index{Folding} of Class VII\indexvii\ orbits.  I find clockwise loops\index{Hysteresis} in 9 out of 11 Class VII orbits.  In the remaining two observations, the oscillations were extremely fast.  As a result, the errors in the HID$_1$\index{Hardness-intensity diagram} of these two observations were too large to succesfully select peaks, and I am unable to confirm or reject the presence of loops.

\begin{figure}
    \includegraphics[width=0.8\columnwidth, trim = 0.6cm 0 3.9cm 0]{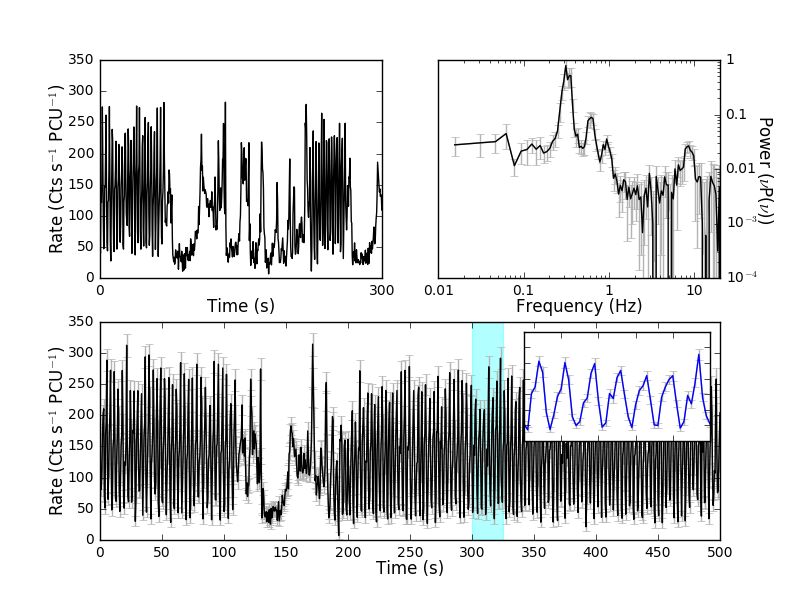}
    \captionsetup{singlelinecheck=off}
    \caption[Characteristic lightcurves and a power spectrum of Type VIII variability.]{Plots of the Class VIII\indexviii\ observation 96420-01-19-03, orbit 0.  \textit{Top-left}: 300\,s lightcurve\index{Lightcurve} binned on 2 seconds to show lightcurve evolution.  \textit{Top-right}: Fourier Power Density Spectrum\index{Fourier analysis}.  \textit{Bottom}: Lightcurve binned on 0.5 seconds.  Inset is a zoom of the 25\,s portion of the lightcurve highlighted in cyan, to show the second-scale structure in the lightcurve.}
   \label{fig:Omulti}
\end{figure}

\subsubsection{Class VIII -- Figure \ref{fig:Omulti}}

\par The lightcurve\index{Lightcurve} of this variability class\indexviii\ shows the dipping\index{Dip} behaviour seen in Class VI\indexvi, as can be seen in Figure \ref{fig:Omulti} at $t\approx125$--$150$\,s.  The dips are less frequent than in Class VI.  The behaviour outside of the dips is dominated by highly structured high-amplitude oscillations consisting of flares\index{Flare} with a peak to peak separation of $3.4\pm1.0$\,s.  The PDS\index{Fourier analysis} shows this behaviour as a very significant ($q$-value > 20\indexq) QPO\index{Quasi-periodic oscillation}; two harmonics\index{Harmonic} of this QPO are also visible.  The PDS also shows a strong ($q$-value$\sim5$\indexq) QPO at $\sim9$\,Hz.
\par I attempted to fold\index{Folding} Class VIII\indexviii\ lightcurves, ignoring the portions of data corresponding to dips\index{Dip}, using my flare-finding algorithm.  The high frequency of the dominant oscillation in Class VIII resulted in large errors in the peak times of individual flares, which translated to large errors in all HID$_1$s\index{Hardness-intensity diagram}; however, I was able to ascertain the presence in loops\index{Hysteresis} in 8 out of 16 orbits.  All 8 of these loops are clockwise.

\begin{figure}
    \includegraphics[width=0.8\columnwidth, trim = 0.6cm 0 3.9cm 0]{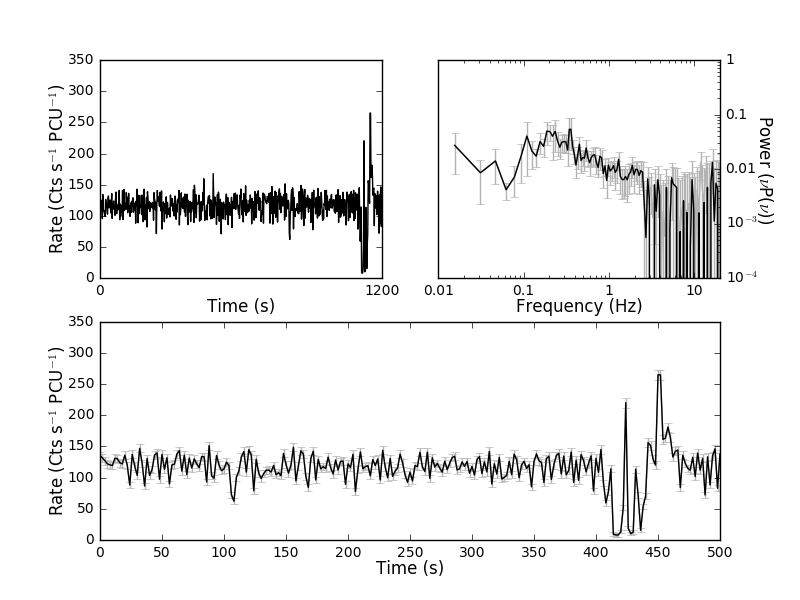}
    \captionsetup{singlelinecheck=off}
    \caption[Characteristic lightcurves and a power spectrum of Type IX variability.]{Plots of the Class IX\indexix\ observation 96420-01-35-02, orbit 1.  \textit{Top-left}: 1200\,s lightcurve\index{Lightcurve} binned on 2 seconds to show lightcurve evolution.  \textit{Top-right}: Fourier Power Density Spectrum\index{Fourier analysis}.  \textit{Bottom}: Lightcurve binned on 2 seconds.}
   \label{fig:Qmulti}
\end{figure}

\subsubsection{Class IX -- Figure \ref{fig:Qmulti}}
\label{sec:ClassIX}

\par The 1\,s lightcurve\index{Lightcurve} of a Class IX\indexix\ observation is superficially similar to the lightcurve of a Class I\indexi\ observation, with little obvious structured variability\index{Variability} at timescales larger than 2\,s; however, large count rate dips\index{Dip} like those seen in Classes VI\indexvi\ and VIII\indexviii\ (e.g. the feature at $t\approx410$\,s in the lightcurve of Figure \ref{fig:Qmulti}) are very occasionally observed.  These dips may in turn be coupled to short second-scale flares\index{Flare} in which count rate briefly increases by a factor of 2--3.
\par Outside of these dips and flares, the lightcurve of a Class IX\indexix\ observation is indistinguishable from the lightcurve\index{Lightcurve} of a Class I\indexi\ or Class II\indexii\ observation.  However, in Figure \ref{fig:IIIisHarder}, I show that Class IX occupies a very different part of the global $H_{A2}$/$H_{A1}$ colour-colour diagram\index{Colour-colour diagram}.  Class IX observations show a significantly larger $H_{A2}$\index{Colour} than Class I and II observations, but a significantly lower $H_{A1}$.
\par The PDS\index{Fourier analysis}\index{PDS|see {Fourier analysis}} reveals significant broad band noise peaked at $\sim$0.3 Hz, and the $\sim5$\,Hz QPO seen in other classes is absent.  \citet{Altamirano_HFQPO} discovered high frequency ($\sim66$\,Hz) QPOs\index{Quasi-periodic oscillation} in observations corresponding to this variability class.

\subsection{Swift}

\par Observations with \indexswift\textit{Swift} took place throughout the 2011-2013 outburst of IGR J17091-3624\index{IGR J17091-3624}.  Between MJDs 55622 and 55880, 17 \indexxrt\textit{Swift/XRT} were at least partly simultaneous with an \rxte\indexrxte\ observation, corresponding to at least one observation of all 9 classes.  In each case, the \textit{Swift} and \rxte\ lightcurves were similar.  The remainder of the \textit{Swift/XRT} observations during this time were also consistent with belonging to one of my nine classes.  Given that the \rxte\ data have higher count rate and time resolution, I do not further discuss the \textit{Swift} observations taken before MJD 55880.
\par Between MJD 55952 and 56445, \indexswift\textit{Swift} observations showed IGR J17091-3624\index{IGR J17091-3624} decreasing in flux.  For all observations longer than 500\,s, I rebinned the lightcurves to 10\,s and calculated the fractional RMS\index{RMS}.  I find the lower and upper quartiles of the fractional RMS in these measurements to be 18.3\% and 21.7\% respectively.  \textit{INTEGRAL} observations taken as part of a scan programme of the Galactic Plane \citep{Fiocchi_PlaneScan} and reported by \citet{Drave_Return} suggest that IGR J17091-3624 returned to the hard state\index{Low/Hard state} between MJDs 55952 and 55989.  Therefore these observations sample IGR J17091-3624 in the hard state.

\subsection{INTEGRAL}

\par The results of the \indexintegral\indexibis\textit{INTEGRAL}/IBIS analysis are presented in Table \ref{tab:IBIS_results}. \textsf{C.B.} finds clear detections of IGR J17091-3624\index{IGR J17091-3624} in all energy bands during the hardest period (MJD 55575--55625) of the 2011--2013 outburst\index{Outburst}. Conversion from detected counts to flux was achieved using an \textit{INTEGRAL}/IBIS observation of the Crab taken between MJD 57305.334 and 57305.894. Conversion from Crab units to standard flux units was obtained by conversion factors listed in \citet{Bird_Survey} and \citet{Bazzano_Survey}.

\begin{table*}
\begin{tabular}{cccccc}
\hline
\hline
Energy 		& Intensity 		& Significance 	& Exposure 	& Flux 				& Flux					\\
(keV)		& (cts/s)			& $\sigma$		& (ks)		& (mCrab) 			& (10$^{-10}$ergs~s$^{-1}$~cm$^{-2}$) 	\\
\hline
20--40		& 12.39$\pm$0.05	& 247			& 115		& 93.5$\pm$0.38		& 7.08$\pm$0.03			\\
40--100		& 7.06$\pm$0.05		& 157			& 163		& 83.5$\pm$0.60		& 7.87$\pm$0.06			\\
100--150	& 1.05$\pm$0.03		& 40			& 173		& 66.9$\pm$1.91		& 2.14$\pm$0.06			\\
150--300	& 0.23$\pm$0.03		& 7.6			& 179		& 46.6$\pm$5.96		& 2.24$\pm$0.29			\\	
\hline
\hline
\end{tabular}
\caption[Results from the IBIS/ISGRI analysis of the 2011--2013 Outburst of IGR J17091.]{Results from the \indexibis IBIS/ISGRI analysis of the 2011--2013 Outburst\index{Outburst} of IGR J17091\index{IGR J17091-3624}. The 20--40\,keV flux is given in units of mCrab and (10$^{-11}$\ergf ). Conversion between counts and mCrab was obtained using an observation of the Crab taken during Revolution 1597 between MJD 57305.334 and 57305.894 and the conversion factors of \citet{Bird_Survey} and \citet{Bazzano_Survey}.}
\label{tab:IBIS_results}
\end{table*}

\par
\begin{figure}
    \includegraphics[width=0.7\columnwidth, trim = 0.6cm 0 3.9cm 0]{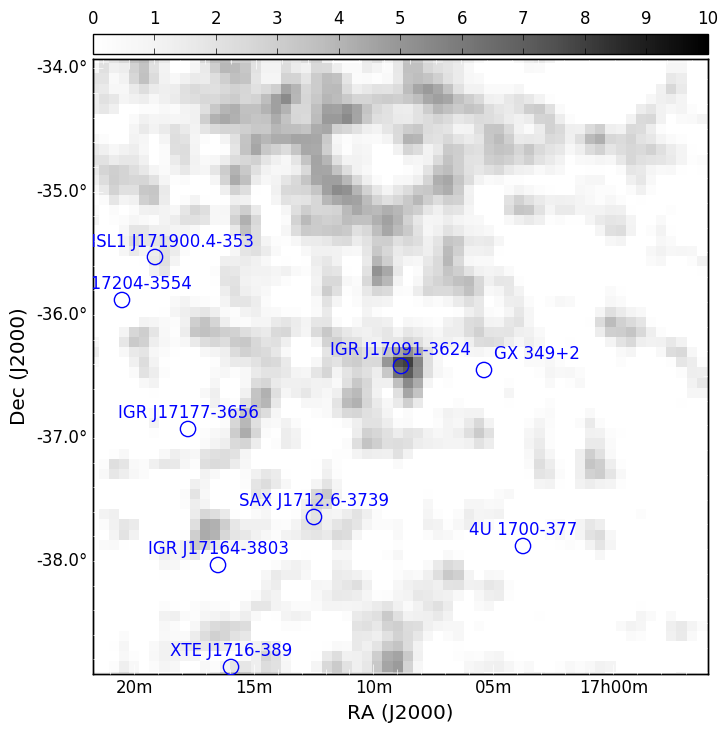}
    \captionsetup{singlelinecheck=off}
    \caption[\textit{INTEGRAL}/ISGRI 150--300\,keV significance map of a $2^\circ$ region centred on the position of IGR J17091-3624.]{\indexibis\textit{INTEGRAL}/ISGRI 150--300\,keV significance map of a $2^\circ$ region centred on the position of IGR J17091-3624\index{IGR J17091-3624}, showing the first significant detection of this source above 150\,keV.  The detection significance is 7.6 $\sigma$.}
   \label{fig:sigmap}
\end{figure}

 Comparing these results with those of \citet{Bazzano_Survey}, we see that IGR J17091\index{IGR J17091-3624} is detected for the first time above 150\,keV with a detection significance of 7.6\,$\sigma$, corresponding to a flux of $2.24\pm0.29\times10^{-10}$\ergf\ (Figure \ref{fig:sigmap}).

\subsection{Chandra}

\par
\begin{figure}
    \includegraphics[width=0.8\columnwidth, trim = 0.6cm 0 3.9cm 0]{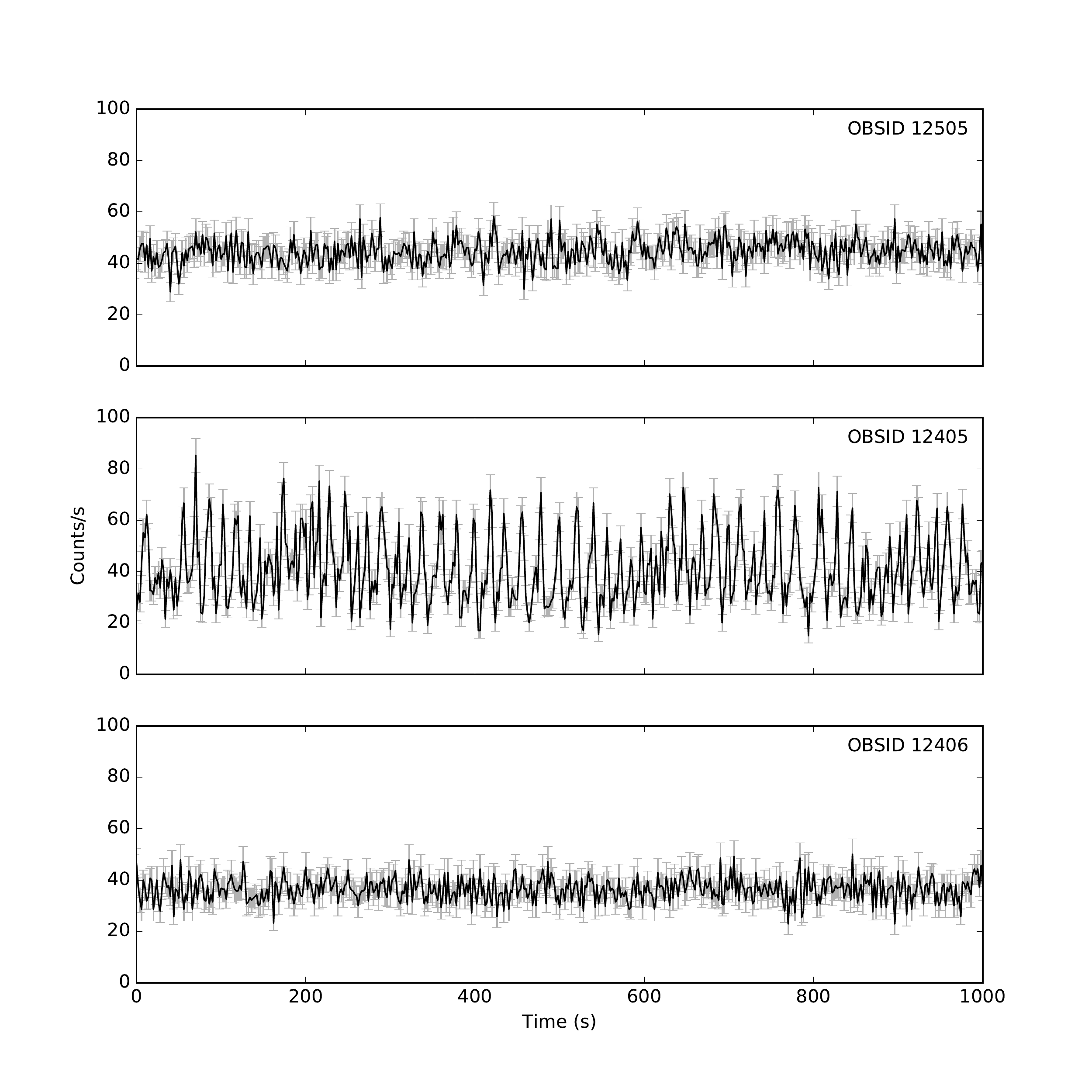}
    \captionsetup{singlelinecheck=off}
    \caption[\textit{Chandra} lightcurves showing examples of Class I, VII and IX variability.]{1 ks segments of lightcurves taken from \indexchandra\textit{Chandra} observations 12505, 12405 and 12406, showing Class I\indexi, Class VII\indexvii\ and Class IX\indexix\ variability respectively.  The lightcurve presented for observation 12505 is for the energy range 0.06-10\,keV, while the other two lightcurves are for the energy range 0.5-10\,keV.  All three lightcurves are binned to 0.5\,s.}
   \label{fig:Cha_lc}
\end{figure}

 In Figure \ref{fig:Cha_lc}, I present lightcurves from the three \indexchandra\textit{Chandra} observations considered in this chapter (see also Table \ref{tab:Chandra} for details of these observations).

\par Observation 12505 was performed within 24 hours of \indexrxte\rxte\ observation 96420-01-02-01, which showed Class I\indexi\ variability.  No structured variability is seen in the lightcurve of ObsID 12505 (Figure \ref{fig:Cha_lc}, upper panel), which is consistent with Class I.  Note that I consider the energy range 0.06-10\,keV for this observation but 0.5-10\,keV for observations 12405 and 12406.
\par Observation 12405 was performed within 24 hours of \indexrxte\rxte\ observation 96420-01-23-03, which showed Class V\indexv\ variability.  The two observations were not simultaneous; ObsID 12405 began $\sim8.4$ ks after ObsID 96420-01-2303 finished.  The lightcurve of \indexchandra\textit{Chandra} ObsID 12405 (shown in Figure \ref{fig:Cha_lc}, middle panel) shows a mean count rate of 41\,cts\,s$^{-1}$.  The lightcurve shows fast flaring\index{Flare} behaviour (with a recurrence time\index{Recurrence time} on the order of 10s of seconds) in which the frequency changes widely on timescales of $\sim1000$\,s.  This observation strongly resembles a Class VII\indexvii\ lightcurve, but with its characteristic timescales increased by a factor of $\sim4$.  This leads to the possibility that the low number of Class VII\indexvii\ \rxte\ observations I identify is due to a selection effect; we would not have been able to see this observation's long-term Class VII-like behaviour if the observation had been shorter than $\sim2$ ks.
\par Observation 12406 was performed within 24 hours of \indexrxte\rxte\ observation 96420-01-32-06, which showed Class IX\indexix\ variability.  The lightcurve presented for \indexchandra\textit{Chandra} ObsID 12406 shows a mean count rate (36 cts s$^{-1}$), which is consistent with IGR J17091\index{IGR J17091-3624} being harder in this observation than in Observation 12505.  This, combined with the lack of variability seen in its lightcurve, suggests that Observation 12505 is consistent with Class IX.

\subsection{XMM-Newton}

\par
\begin{figure}
    \includegraphics[width=0.8\columnwidth, trim = 0.6cm 0 3.9cm 0]{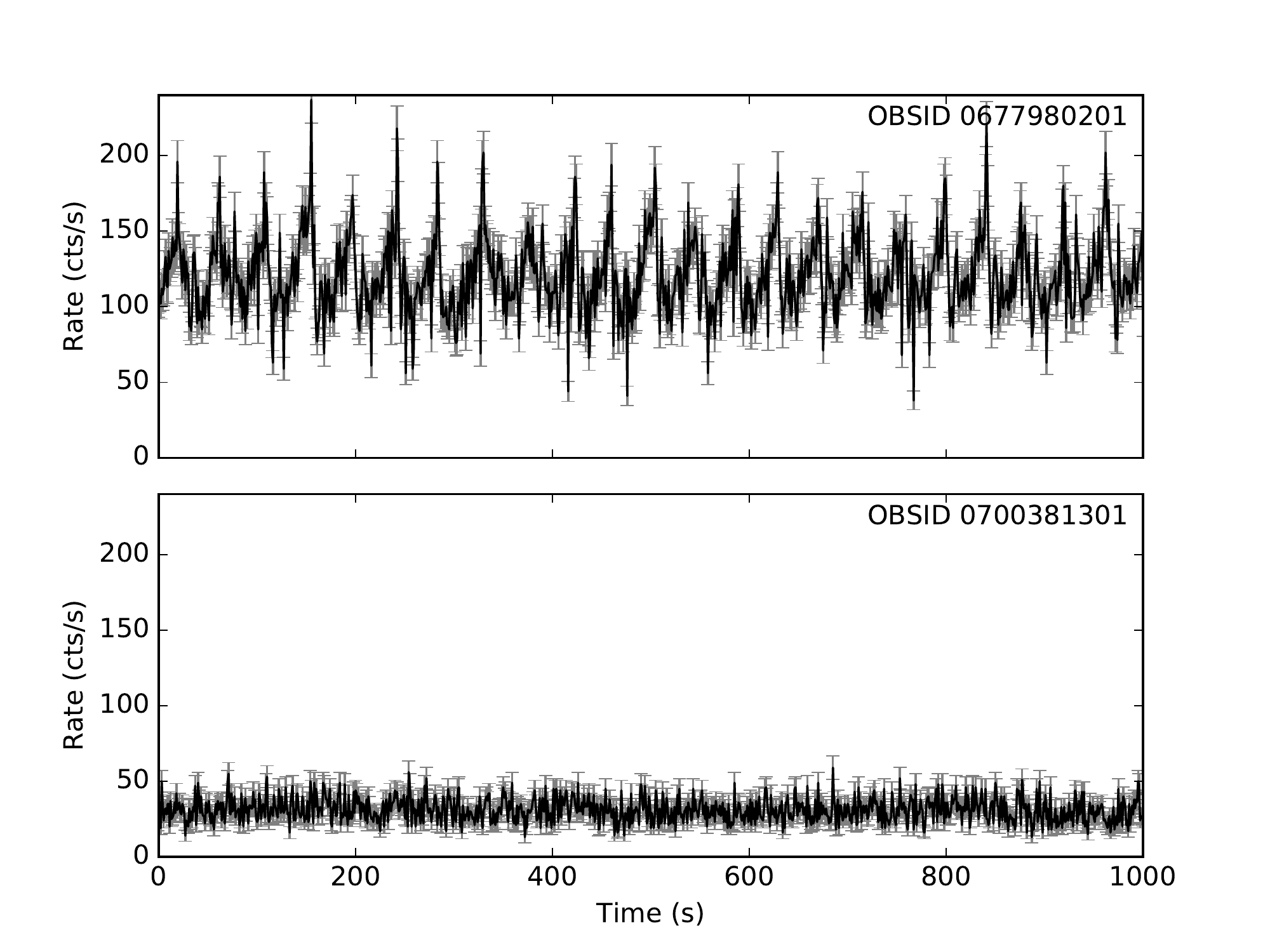}
    \captionsetup{singlelinecheck=off}
    \caption[\textit{XMM-Newton} lightcurves showing an example of Class IV variability and the hard state.]{Lightcurves of \indexxmm\textit{XMM-Newton} observations 0677980201 and 0700381301, showing Class IV\indexiv\ variability and the hard\index{Low/Hard state} state respectively.  Both lightcurves binned to 2\,s.  Data for observation 0677980201 is taken from \indexepic\textit{EPIC-MOS2} and data for observation 0700381301 is taken from \textit{EPIC-pn}.}
   \label{fig:XMM}
\end{figure}

 In Figure \ref{fig:XMM} I show lightcurves from two \indexxmm\textit{XMM-Newton} observations.  The lightcurve of \textit{XMM-Newton} observation 0677980201, shown in the upper panel of Figure \ref{fig:XMM}, shows the regular flares characteristic of Class IV variability.  A simultaneous \indexrxte\rxte\ observation (ObsID 96420-01-05-000) also showed Class IV\indexiv\ variability.

\par
\begin{figure}
    \includegraphics[width=0.9\columnwidth, trim = 0cm 0cm 0.5cm 1.0cm, clip]{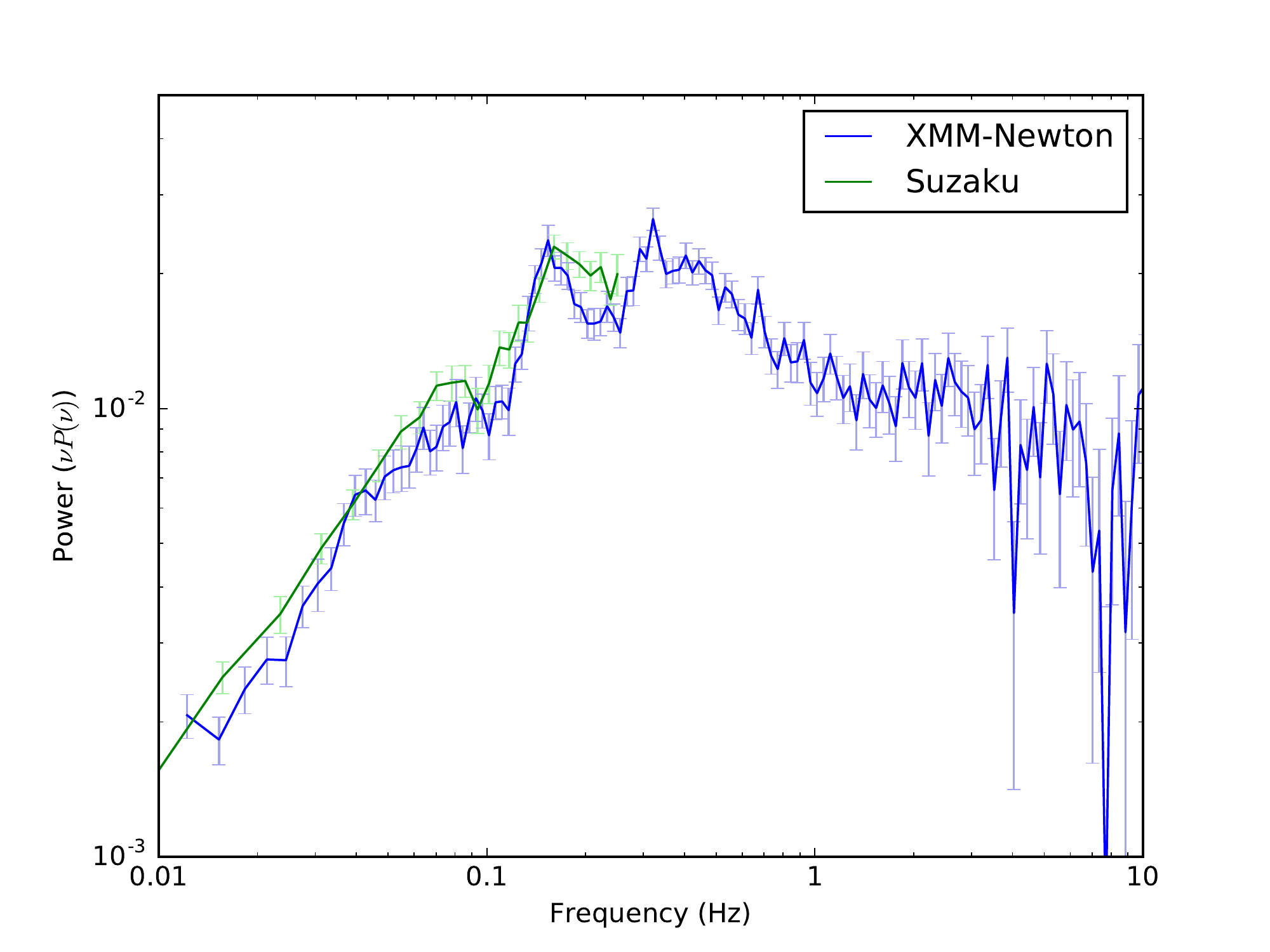}
    \captionsetup{singlelinecheck=off}
    \caption[$\nu P(\nu)$-normalised co-added power density spectra of \textit{XMM-Newton} observation 0700381301 and \textit{Suzaku} observation 407037010.]{RMS-normalised\index{RMS normalisation} co-added power density spectra\index{Fourier analysis} of \indexxmm\textit{XMM-Newton} observation 0700381301 and \indexsuzaku\textit{Suzaku} observation 407037010.  Both observations were taken simultaneously on September 29 2012 (MJD 56199).  I sample observation 0700381301 up to a frequency of 10\,Hz, while the 2\,s time resolution of observation 407037010 results in a Nyquist frequency\index{Nyquist frequency} of 0.25\,Hz.}
   \label{fig:xmmqpo}
\end{figure}

 \indexxmm\textit{XMM-Newton} observation 070038130, shown in the lower panel of Figure \ref{fig:XMM}, was made after the end of \indexrxte\rxte\ observations IGR J17091-3624\index{IGR J17091-3624}.  As such it cannot be compared with contemporaneous \rxte\ data.  The 5\,s binned lightcurve shows no apparent variability, but a Fourier PDS of the observation (shown in Figure \ref{fig:xmmqpo}) reveals a QPO\index{Quasi-periodic oscillation} centred at around $\sim0.15$\,Hz and a broad band noise component at lower frequencies.  \citet{Drave_Return} reported that IGR J17091 transited to the hard state\index{Low/Hard state} in February 2012, seven months before this observation was taken.  As such, I find that observation 0677980201 samples the hard state in IGR J17091 and is thus beyond the scope of my set of variability classes.

\subsection{\textit{Suzaku}}

\par The two \indexsuzaku\textit{Suzaku} observations of IGR J17091-3624\index{IGR J17091-3624} considered here, ObsIDs 407037010 and 407037020, were performed during the 2nd and 3rd re-flares\index{Re-flare} of the hard state\index{Low/Hard state} phase of the 2011--2013 outburst.  ObsID 407037010 was taken simultaneously with \indexxmm\textit{XMM-Newton} observation 0700381301.  The\indexxis\ XIS 0 count rates are 7.8 cts\,s$^{-1}$ and 2.5\,cts\,s$^{-1}$ respectively.
\par Neither lightcurve shows `heartbeats'\indexrho\ or any other type of GRS 1915-like\index{GRS 1915+105} variability\index{Variability}.  However, \textsf{K.Y.} and I find evidence of a low frequency QPO\index{Quasi-periodic oscillation} feature at $\sim$0.15 Hz in the ObsID 407037010; this QPO is also seen in \indexxmm\textit{XMM-Newton} observation 0700381301 (Figure \ref{fig:xmmqpo}).  The presence of a QPO below 1\,Hz and flat-topped power density spectrum\index{Fourier analysis} confirm that IGR J17091\index{IGR J17091-3624} was in the hard state\index{Low/Hard state} at this time.

\section{Discussion}

\par Using observations from \indexxmm\textit{XMM-Newton}, \indexrxte\rxte\ and \indexchandra\textit{Chandra}, I describe the complex variability\index{Variability} seen in IGR J17091\index{IGR J17091-3624} as a set of nine variability `classes'\index{Variability class}, labelled I to IX.  These classes are distinguished from each other by values of upper and lower quartile (i.e. 25\textsuperscript{th} and 75\textsuperscript{th} percentile) count rates, mean RMS\index{RMS}, the presence of QPOs\index{Quasi-periodic oscillation} in Fourier PDS\index{Fourier analysis}, the shape of flare\index{Flare} and dip\index{Dip} features in the lightcurve\index{Lightcurve} and the presence of loops\index{Hysteresis} in the 6--16/2--6 keV hardness-intensity diagram HID$_1$\index{Hardness-intensity diagram}.  See Section \ref{sec:results} for a full description of these classes.
\par
\begin{figure}
    \includegraphics[width=\columnwidth, trim =0cm 0 0cm 0]{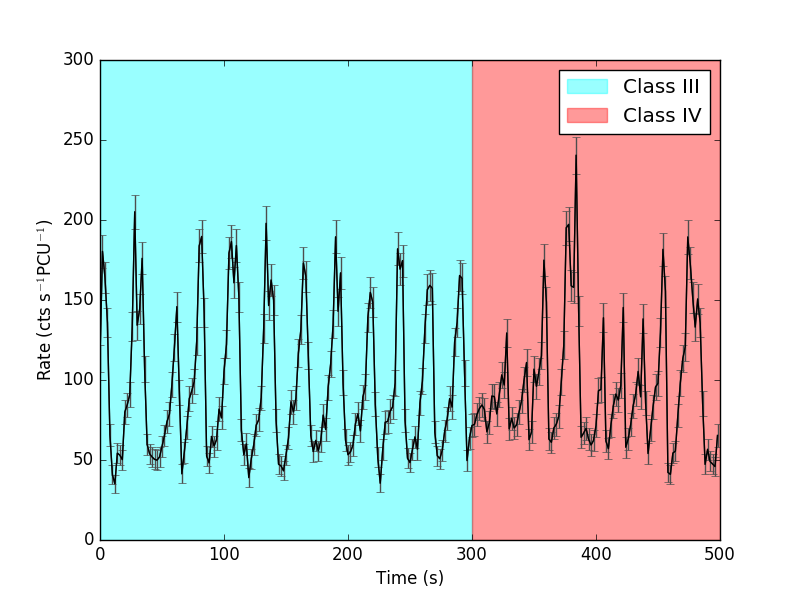}
    \captionsetup{singlelinecheck=off}
    \caption[A lightcurve of observation 96420-01-06-02, showing a transition in behaviour between Classes IV and V.]{A lightcurve\index{Lightcurve} of observation 96420-01-06-02, orbit 0, showing a transition in behaviour between Class III\indexiii\ (in cyan, see Section \ref{sec:classIII}) and Class IV\indexiv\ (in red, see Section \ref{sec:classIV}).}
   \label{fig:HybridClasses}
\end{figure}

 The classification of some observations is clearer than others.  Some orbits were too short to definitively quantify the behaviour of the source, whereas some other orbits contain a transition between two classes\index{Variability class}.  An example lightcurve\index{Lightcurve} showing a transition from Class III\indexiii\ to Class IV\indexiv\ is presented in Figure \ref{fig:HybridClasses}.

\par My set of classes\index{Variability class} is analogous to, but not based upon, the set of variability classes defined by \citealt{Belloni_GRS_MI} to describe the behaviour of the similarly complex LMXB\index{X-ray binary!Low mass} GRS 1915\index{GRS 1915+105}.  This ensures that my set of classes is not biased by an \textit{a priori} assumption that the two objects are similar.  However if we do assume that wide range of variability\index{Variability} seen in these two objects are driven by the same physical processes, a direct comparison between the variability classes in the two systems can further our understanding of the physics that drive these exotic objects.
%\par I also use all 2011-2013 IGR J17091-3624 data from \rxte , \textit{XMM-Newton}, \textit{Chandra}, \textit{Swift}, \textit{INTEGRAL} and \textit{Suzaku} to analyse the long-term evolution of the 2011--2013 outburst.  This in turn corresponds to all available X-ray data taken during this outburst.

\subsection{Variability Classes: IGR J17091 vs. GRS 1915}

\label{sec:IGRcomp}

\par As observations of IGR J17091\index{IGR J17091-3624} and GRS 1915\index{GRS 1915+105} suffer from different values of interstellar absorption\index{NH@$N_H$}\index{Absorption|see {$N_H$}} $N_H$\footnote{$N_H$, or the interstellar absorption, is a measure of the surface density of hydrogen atoms along a column between the object in question and the Earth.  A high value of $N_H$ causes low-energy X-rays to be supressed more than high-energy X-rays, increasing the apparent colour of a source.  $N_H$ can be estimate by fitting models to the energy spectrum of a source.}, I cannot directly compare the absolute colours\index{Colour} of these two objects.  However, I can compare the evolution of colour both over time and as a function of count rate.  I therefore use these parameters, along with power spectra\index{Fourier analysis} and lightcurve\index{Lightcurve} morphology, when comparing GRS 1915 with IGR J17091.
\par For seven of my classes\index{Variability class}, I was able to assign the closest matching class described by \citealt{Belloni_GRS_MI} for GRS 1915\index{GRS 1915+105} (see Table \ref{tab:class_assign}).  I am unable to find analogues to my classes VII\indexvii\ and VIII\indexviii\ in observations of GRS 1915, and I suggest that these classes are unique to IGR J17091\index{IGR J17091-3624}.

\begin{table}
\centering
\caption[The nine variability classes of IGR J17091-3624, showing the name of the closest corresponding variability class in GRS 1915+105.]{The nine variability classes\index{Variability class} of IGR J17091-3624\index{IGR J17091-3624}, showing the name of the closest corresponding variability class in GRS 1915+105\index{GRS 1915+105}.  The names of GRS 1915+105 classes are taken from \citet{Belloni_GRS_MI}, where more detailed descriptions can be found.  Eight additional classes of GRS 1915+105 have been described; I do not find analogies to these classes in IGR J17091-3624.}
\label{tab:class_assign}
\begin{tabular}{cc} % four columns, alignment for each
\hline
\hline
IGR J17091-3624 Class & GRS 1915+105 Class\\
\hline
I\indexi&$\chi\indexchi$\\
II\indexii&$\phi\indexphi$\\
III\indexiii&$\nu\indexnu$\\
IV\indexiv&$\rho\indexrho$\\
V\indexv&$\mu$\indexmu\\
VI\indexvi&$\lambda\indexlambda$\\
VII\indexvii&\textit{None}\\
VIII\indexviii&\textit{None}\\
IX\indexix&$\gamma\indexgamma$\\
\hline
\hline
\end{tabular}
\end{table}

\par Below, I evaluate my mapping between GRS 1915\index{GRS 1915+105} and IGR J17091\index{IGR J17091-3624} classes\index{Variability class}, and interpret the differences between each matched pair.

\subsubsection{Classes I and II -- Figures \ref{fig:Bmulti}, \ref{fig:Emulti}}

\label{sec:DisI}

\par Classes I\indexi\ and II\indexii\ both show low count rates and little structure in their lightcurves\index{Lightcurve}.  The two classes in GRS 1915\index{GRS 1915+105} that also show this lightcurve behaviour are Class $\chi$\indexchi\footnote{Note that, in GRS 1915+105, Class $\chi$ is further subdivided into four classes based on hard colour \citep{Belloni_GRS_MI,Pahari_Chi}.  As I cannot obtain hard colour\index{Colour} for IGR J17091\index{IGR J17091-3624}, I treat $\chi$ as a single variability class here.} and Class $\phi\indexphi$.  \citealt{Belloni_GRS_MI} differentiate between Classes $\phi$ and $\chi$ based on the hard colour (corresponding to $C_2$), as Class $\chi$ has a significantly higher value for this colour than Class $\phi$.

\par Data from \indexrxte\rxte\  indicates that the transition from the hard\index{Low/Hard state} state to the soft intermediate\index{High/Soft state} state between MJDs 55612 and 55615 \citep{Drave_Return}.  This was confirmed by a radio spectrum taken on MJD 55623 which was consistent with an observation of discrete ejecta \citep{Rodriguez_D}.  This observation of discrete ejecta at the transition between the hard state and the intermediate state has been reported in other LMXBS\index{X-ray binary!Low mass} (e.g. XTE J1550-564\index{XTE J1550-564}, \citealp{Rodriguez_XTE}), and has also been associated with transitions to the $\chi$\indexchi\ Class in GRS 1915\index{GRS 1915+105} (\citealp{Rodriguez_Ejection}, see also review by \citealp{Fender_Jets}).

\par Using Fourier PDS\index{Fourier analysis}, I conclude that Class I\indexi\ is analogous to Class $\chi$\indexchi\ in GRS 1915, while Class II\indexii\ is analogous to Class $\phi$\indexphi.  In Class $\chi$ observations of GRS 1915, broad band noise between $\sim1-10$\,Hz and a QPO\index{Quasi-periodic oscillation} at around 5\,Hz are seen in the PDS.  I find that both of these are present in Class I observations of IGR J17091\index{IGR J17091-3624}.  On the other hand, I find that Class $\phi$ observations of GRS 1915\index{GRS 1915+105} do not show this broad band noise, and show either a weak ($q$-value\indexq\ $\lesssim 3$) QPO at $\sim5$\,Hz or no QPO at all.  I find that the weak QPO and lack of broad band noise are also seen in the PDS of Class II observations.

\subsubsection{Classes III and IV -- Figures \ref{fig:Gmulti}, \ref{fig:Jmulti}}

\par
\begin{figure}
    \includegraphics[width=\columnwidth, trim = 0mm 0mm 0mm 0mm]{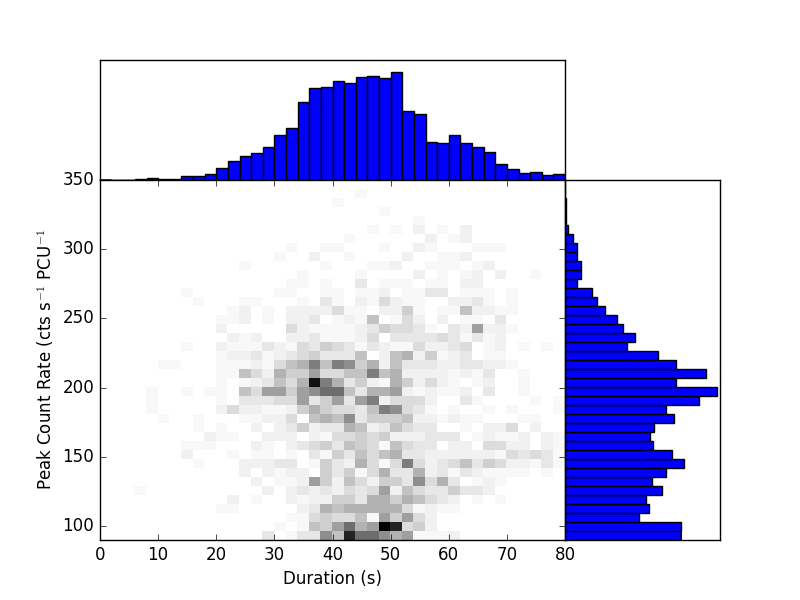}
    \captionsetup{singlelinecheck=off}
    \caption[Every flare in all observations identified as Class III or Class IV, plotted in a two-dimensional histogram of flare peak count rate against flare duration to show the two-population nature of these events.]{Every flare\index{Flare} in all observations identified as Class III\indexiii\ or Class IV\indexiv, plotted in a two-dimensional histogram of flare peak count rate against flare duration to show the two-population\index{Population study} nature of these events.  Flares belonging to Class IV occupy the distribution at higher peak rate and lower duration, whereas flares belonging to Class III occupy the distribution at lower peak rate and higher duration.}
   \label{fig:III_IV_burst}
\end{figure}

 Classes III\indexiii\ and IV\indexiv\ both show highly regular flaring\index{Flare} activity in their lightcurves\index{Lightcurve}, but they differ in terms of timescale and pulse profile.  As can be seen in lightcurves in Figure \ref{fig:Jmulti}, flares in Class IV occur every $\sim32$\,s and are nearly identical to each other in shape.  On the other hand, as can be seen in Figure \ref{fig:Gmulti}, flares in Class III occur every $\sim61$\,s and may or may not end in a much faster sharp peak which is never seen in Class IV.  In Figure \ref{fig:III_IV_burst} I show a two-dimensional histogram of flare peak count rate against flare duration, showing all flares in all observations classified as Class III or Class IV.  In this figure, I can see that flares tend to group in one of two regions in count rate-duration space; a region between $\sim90\mbox{--}110$ \spcu and $\sim35\mbox{--}55$\,s, corresponding to flares seen in Class III, and a region between $\sim150\mbox{--}250$ \spcu and $\sim20\mbox{--}55$\,s, corresponding to flares seen in Class IV.  From this plot, I conclude that the flares seen in Class III exist in a different population\index{Population study} to the flares seen in Class IV.
\par The GRS 1915\index{GRS 1915+105} classes that show behaviour most similar to these are $\rho$\indexrho\ and $\nu$\indexnu; both produce similar structures in their lightcurve, but Class $\nu$ is differentiated from Class $\rho$ by the presence of a secondary count rate peak which occurs $\sim5$\,s after the primary \citep{Belloni_GRS_MI}.

\par
\begin{figure}
    \includegraphics[width=\columnwidth, trim = 0mm 0mm 0mm 0mm]{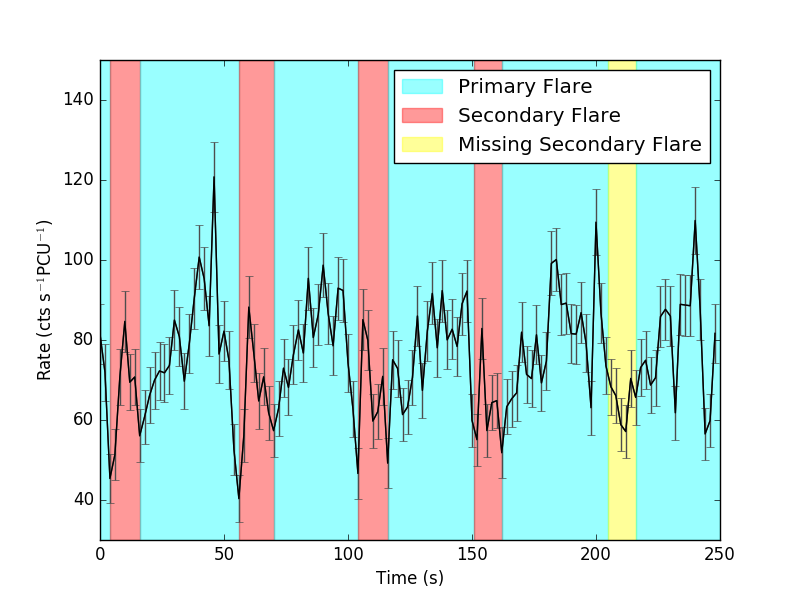}
    \captionsetup{singlelinecheck=off}
    \caption[Lightcurve from Class III observation 96420-01-10-01 of IGR J17091-3624, with pairs of primary and secondary count rate spikes highlighted.]{Lightcurve\index{Lightcurve} from Class III\indexiii\ observation 96420-01-10-01 of IGR J17091-3624\index{IGR J17091-3624}, with pairs of primary and secondary count rate spikes highlighted in cyan and red respectively.  The yellow region highlights a primary count rate spike that did not produce a secondary.}
   \label{fig:III_IV_spike}
\end{figure}

 The secondary peak is present in most Class III observations and some Class IV observations (Figure \ref{fig:III_IV_spike}), suggesting that both classes consist of a mix of $\rho$-like and $\nu$-like observations.  However, the poor statistics sometimes make the presence of this secondary peak difficult to detect.  As such, I do not use the presence or absence of this peak as a criterion when assigning classes.  Instead I choose to separate Classes III and IV based on the larger-scale structure in their lightcurves (see Section \ref{sec:classIV}).  Due to the aforementioned difference in burst populations\index{Population study} between the two classes, I suggest that classes III and IV do represent two distinct classes rather than a single class with a period that drifts over time.  I suggest that Classes $\rho$ and $\nu$ in GRS 1915 could also be re-partitioned in this way.

\par However, HID$_1$\index{Hardness-intensity diagram} loops\index{Hysteresis} are found to generally execute in an anticlockwise direction in Classes III\indexiii\ and IV\indexiv\ (previously noted by e.g. \citealp{Altamirano_IGR_FH}); the opposite direction to the clockwise loops in Classes $\rho$\indexrho\ and $\nu$\indexnu\ reported by e.g. \citealp{Belloni_GRS_MI} and repeated by us using the same method I apply to data from IGR J17091-3624\index{IGR J17091-3624} (see Section \ref{sec:dex}).  This suggests that Classes III and IV could be generated by a different physical mechanism to Classes $\rho$ and $\nu$.  Alternatively, Classes III and IV could be generated by the same mechanism as $\rho$ and $\nu$ if some other unknown process was able to alter the spectral evolution of flares\index{Flare} in these classes.

\subsubsection{Class V -- Figure \ref{fig:Kmulti}}

\par The lightcurve\index{Lightcurve} of a Class V\indexv\ observation appears similar to that of a Class $\mu$\indexmu\ observation of GRS 1915\index{GRS 1915+105}, as both are characterised by rapid $\rho$\indexrho-like flares which occur less regularly than in Class $\rho$.  In addition to this, flares in Class $\mu$ fall into two clear populations\index{Population study}, as do the flares in Class V.  However, significant differences exist between Class V and Class $\mu$.  Class $\mu$ observations are characterised by long ($\sim100$ s) excursions to plateaus\index{Plateau} of high count rate, a behaviour which is not seen in any Class V observation thus far.
\par I note that the HID$_1$\index{Hardness-intensity diagram} in Class V\indexv\ observations displays a loop\index{Hysteresis} in the clockwise direction; the opposite direction to the looping seen in Classes III\indexiii\ and IV\indexiv\ but the same direction seen in Class $\mu$\indexmu.
\par Regarding the two-population\index{Population study} nature of flares\index{Flare} seen in this class (see Section \ref{sec:classV}), I suggest that V$_2$ flares may simply be two V$_1$ flares that occur close together in time, such that the second flare starts during the decay of the first flare.  This would result in an apparent two-peaked flare structure, as we see in type V$_2$ flares.  This interpretation also accounts for the bimodal distribution of flare duarations shown in the 2D histogram of Figure \ref{fig:two_popV}, as this could be caused by the misinterpretation of two-flare V$_2$ events as a single event.  This also accounts for the Gaussian distribution of peak flare intensities seen in Figure \ref{fig:two_popV}), as the constituents of each V$_2$ event would be from the same population as V$_1$ flares.

\subsubsection{Class VI -- Figure \ref{fig:Lmulti}}

\par Class VI\indexvi\ is dominated by long flaring\index{Flare} periods which separate periods of low count rate, as can be seen in the lightcurve\index{Lightcurve} presented in Figure \ref{fig:Lmulti}.  Similar behaviour is seen in the lightcurves of observations of GRS 1915 belonging to Classes $\lambda$\indexlambda\ and $\omega$\indexomega\ \citep{KleinWolt_OmegaClass}.  However, the long count rate `dips'\index{Dip} are far less regular in Class VI than in Classes $\lambda$ and $\omega$, and I also note long periods of medium count rate during which neither flares nor dips occur.  This variability class\index{Variability class} is noted by \citet{Pahari_IGRClasses} who suggest that this class is unique to IGR J17091\index{IGR J17091-3624}\footnote{\citet{Pahari_IGRClasses} refers to Class VI as Class C2.}.  However, \citet{Pahari_ClassVI} show that, in a plot of burst decay time against burst rise time, Classes VI and $\lambda$ fall in a straight line, suggesting a similar physical origin for both.
\par While it is cetainly true that Class VI\indexvi\ is not a perfect analogue of either Class $\lambda$\indexlambda\ or Class $\omega$\indexomega\, Class VI only differs noticeably from Class $\lambda$ during the extended low-variability portions of its lightcurves.  As such, I associate Class VI with Class $\lambda$.

\subsubsection{Class VII -- Figure \ref{fig:Nmulti}}

\par I am unable to find an analogue of Class VII\indexvii\ in observations of GRS 1915\index{GRS 1915+105}.  This class, and its apparent uniqueness, have previously been noted by \citealp{Pahari_IGRClasses}\footnote{\citet{Pahari_IGRClasses} refers to Class VII as Class C1.}.  \citeauthor{Pahari_IGRClasses} found  that the $C_2$ hard colour\index{Colour} in this class increases during count rate dips and decreases during count rate peaks.  Here I reproduced the results of \citeauthor{Pahari_IGRClasses} and found that the anti-correlation between hard-colour and intensity is not physical, but due to the definition of $C_2$: the count rate in band $L_C$ is approximately constant and consistent with background, and therefore $C_2=L_C/L_A \propto L_A^{-1}$, which will naturally anticorrelate with intensity.
%\par Although a correlation between QPO\index{Quasi-periodic oscillation} frequency and count rate has been noted in the $\sim5$\,Hz QPO seen in GRS 1915 (e.g. \citealp{Markwardt_FluxFreqGRS,Vignarca_FluxFreqGRS}), this QPO is also seen in Class VII observations at the same time as the $\sim0.1$\,Hz QPO.  As such, the flux-frequency relationship in the very low frequency ($\sim0.1$\,Hz) QPO in Class VII is apparently unique amongst the classes of both IGR J17091 and GRS 1915.

\subsubsection{Class VIII -- Figure \ref{fig:Omulti}}

\par I am unable to find an analogue of Class VIII\indexviii\ in observations of GRS 1915\index{GRS 1915+105}.  When it is flaring\index{Flare}, the lightcurve\index{Lightcurve} waveform is similar to that seen in Class $\rho$\indexrho, with rapid regular spikes in count rate.  The lightcurve also shows irregular dips\index{Dip} in count rate similar to those seen in Class VI\indexvi\ and in Class $\lambda$\indexlambda\ in GRS 1915.
\par However, the amplitude of the flares in Class VIII\indexviii\ is much larger, and the frequency much higher, than in Classes VI\indexvi\ or $\lambda$\indexlambda.  The amplitude of the flares\index{Flare} in Class VIII can approach $\sim350$\,cts s$^{-1}$\,PCU$^{-1}$, while the flare separation time of 4--5\,s makes Class VIII the fastest flaring activity seen in any class of IGR J17091\index{IGR J17091-3624} or GRS 1915\index{GRS 1915+105}.  As such, I consider this variability class distinct from both Class VI and Class $\lambda$. 

\subsubsection{Class IX - Figure \ref{fig:Qmulti}}

\label{sec:DisIX}

\par Class IX\indexix\ is defined by long periods of high amplitude but unstructured variability\index{Variability} (with a broad peaked noise component in the Fourier spectrum\index{Fourier analysis} peaked at $\sim$0.3 Hz) punctuated with infrequent irregular short-duration `spikes' in which the count rate increases by a factor of $\sim2$--$3$.  A similarity between this Class and Class $\gamma$\indexgamma\ in GRS 1915\index{GRS 1915+105} has been previously noted by \citet{Altamirano_HFQPO}.  However, the irregular spikes seen in some Class IX lightcurves are not reproduced in Class $\gamma$ lightcurves of GRS 1915.

\subsection{General Comparison with GRS 1915+105}

\par Overall, variability\index{Variability} in IGR J17091\index{IGR J17091-3624} tends to be faster than structurally similar variability in GRS 1915\index{GRS 1915+105}, as can be noted in Classes III\indexiii\ and IV\indexiv\ compared to Classes $\rho$\indexrho\ and $\nu$\indexnu\ (see also \citealp{Altamirano_IGR_FH}).  Additionally, IGR J17091 also displays highly structured variability unlike anything yet seen in GRS 1915, with classes VII\indexvii\ and VIII\indexviii\ in particular showing very fine detail in their lightcurves.
\par In total I find 2 variability classes\index{Variability class} which are seen in IGR J17091\index{IGR J17091-3624} but not in GRS 1915\index{GRS 1915+105}, compared with 8 that are seen in GRS 1915 but not in IGR J17091.  As relatively little data exists on GRS 1915-like variability in IGR J17091, the presence of classes in GRS 1915 that are not seen in IGR J17091 could simply be an observational effect.  It is unknown how long each variability class lasts for and, as such, additional variability classes could have occurred entirely while IGR J17091 was not being observed.  However, GRS 1915 has displayed variability classes consistently since its discovery in 1992 (see e.g. see \citealp{Huppenkothen_ML}), implying that the two classes seen only in IGR J17091 are either completely absent in GRS 1915 or that they occur with a much lower probability.  In either case, this implies physical differences between methods of generating GRS 1915-like variability\index{Variability} in the two objects.  
\par As noted in section \ref{sec:DisI}, variability classes\index{Variability class} seen in both IGR J17091\index{IGR J17091-3624} and GRS 1915\index{GRS 1915+105} show differences between the different objects.  In particular, I note the presence of irregular flares\index{Flare} in Class IX\indexix\ which are not seen in the analogous Class $\gamma$\indexgamma.  If these classes are indeed generated by the same processes in both objects, the differences between them must represent physical differences between the objects themselves.
\par It has previously been noted that, while the hardness ratios\index{Colour} in IGR J17091\index{IGR J17091-3624} and GRS 1915\index{GRS 1915+105} during $\rho$\indexrho-like classes are different, the fractional hardening between the dip\index{Dip} and peak of each flare\index{Flare} is consistent with being the same in both objects \citep{Capitanio_peculiar}.  This suggests that the same physical process is behind the `heartbeats' seen in both objects.
\par I note the presence of hysteretic HID$_1$\index{Hardness-intensity diagram} loops\index{Hysteresis} in some classes\index{Variability class} of both objects.  Although these loops are always clockwise in GRS 1915\index{GRS 1915+105}, they can be executed in either direction in IGR J17091\index{IGR J17091-3624}.  Classes in IGR J17091 that show loops all have a preferred loop direction: anticlockwise in Classes III\indexiii\ and IV\indexiv\ and clockwise in classes V\indexv, VI\indexvi, VII\indexvii\ and VIII\indexviii.  In cases where the loop direction was opposite to that expected for a given class, loop detections were generally only marginally significant.  In particular, I note that Classes IV and V tend to show loops in opposite directions, despite the similarities between their lightcurves and the $\rho$, $\nu$ and $\mu$ classes in GRS 1915.   The fact that IGR J17091 can show HID$_1$ loops in both directions suggests that an increase in soft emission can either precede or lag\index{Hard lag} a correlated increase in hard emission from IGR J17091.  Whether soft emission precedes or lags hard emission is in turn is dependent on the variability class.
\par There are also non-trivial similarities between variability\index{Variability} in the two objects.  I note the presence of a $\sim5$\,Hz QPO\index{Quasi-periodic oscillation} in many of the classes seen in IGR J17091\index{IGR J17091-3624}, and this same 5\,Hz QPO is seen in data from GRS 1915\index{GRS 1915+105}.  Similarly \citet{Altamirano_HFQPO} reported the discovery of a 66\,Hz QPO in IGR J17091; a very similar frequency to the 67\,Hz QPO observed in GRS 1915 \citep{Morgan_QPO}.  It is not clear why these QPOs would exist at roughly the same frequencies in both objects when other variability in IGR J17091 tends to be faster.

\subsection{Comparison with the Rapid Burster}

\par In 2015, \citet{Bagnoli_RB} reported the discovery of two GRS 1915-like\index{GRS 1915+105} variability classes\index{Variability class} in the neutron star\index{Neutron star} binary\index{X-ray binary!Low mass} MXB 1730-335, also known as the `Rapid Burster'\index{Rapid Burster}.  Specifically, \citet{Bagnoli_RB} note the presence of variability similar to Classes $\rho$\indexrho\ and $\theta$\indextheta\ in GRS 1915.
\par Class $\theta$\indextheta-like variability\index{Variability}, seen in \indexrxte\rxte\ observation 92026-01-20-02 of the Rapid Burster\index{Rapid Burster}, is not closely matched by any of the classes I identify for IGR J17091\index{IGR J17091-3624}.  However, the lightcurves of a Class $\theta$ observation feature large dips\index{Dip} in count rate similar to those seen in Classes VI\indexvi\ and VIII\indexviii\ in IGR J17091.
\par Conversely, Class $\rho$-like\indexrho\ variability\index{Variability} is seen in all three objects.  \citet{Bagnoli_RB} note that the variability of the $\rho$-like flaring\index{Flare} is slower in the Rapid Burster\index{Rapid Burster} than in either GRS 1915\index{GRS 1915+105} or IGR J17091\index{IGR J17091-3624}. It has previously been suggested that the maximum rate of flaring in LMXBs\index{X-ray binary!Low mass} should be inversely proportional to the mass of the compact object\index{Compact object} (e.g. \citealp{Belloni_Timescales,Frank_Timescales}).  In this case, the fact that variability is faster in IGR J17091 than in GRS 1915 could simply be due to a lower black hole\index{Black hole} mass in the former object \citep{Altamirano_IGR_FH}.  However if variability in the Rapid Burster is assumed to be physically analogous to variability in these two black hole objects, then a correlation between central object mass and variability timescale no longer holds.

\subsection{Comparison with \citealp{Altamirano_IGR_FH}}

\label{sec:Alta}
\par \citet{Altamirano_IGR_FH} identify 5 GRS 1915\index{GRS 1915+105} variability classes\index{Variability class} in a subset of observations from the 2011-2013 outburst\index{Outburst} of IGR J17091\index{IGR J17091-3624}: six of these observations are presented in Table \ref{tab:me_Diego} along with the best-fit GRS 1915 class that I assign it in this chapter (see also Table \ref{tab:class_assign}).

\begin{table}
\centering
\caption[The six IGR J17091-3624 ObsIDs explicitly classified in \citet{Altamirano_IGR_FH}.]{The six IGR J17091-3624\index{IGR J17091-3624} ObsIDs explicitly classified in \citet{Altamirano_IGR_FH}.  I also present the GRS 1915\index{GRS 1915+105} class\index{Variability class} with which I implicitly label each ObsID in this chapter.}
\label{tab:me_Diego}
\begin{tabular}{ccc} % four columns, alignment for each
\hline
\hline
ObsID & Altamirano \textit{et al.}& My Class\\
&Class&(implied)\\
\hline
96420-01-04-03&$\alpha\indexalpha$&$\rho\indexrho/\nu\indexnu$\\
96420-01-05-00&$\nu$&$\rho/\nu$\\
96420-01-06-00&$\rho$&$\rho/\nu$\\
96420-01-07-01&$\rho$&$\mu\indexmu$\\
96420-01-08-03&$\beta\indexbeta/\lambda\indexlambda$&$\lambda$\\
96420-01-09-06&$\mu$&$\lambda$\\
\hline
\hline

\end{tabular}
\end{table}

\par I acknowledge differences between the classifications assigned by me and by \citet{Altamirano_IGR_FH}.  I ascribe these differences to the different approaches we have used to construct our classes\index{Variability class}.  In particular while I have constructed an independent set of variability classes for IGR J17091\index{IGR J17091-3624} which I have then compared to the \citeauthor{Belloni_GRS_MI} classes for GRS 1915\index{GRS 1915+105}, \citeauthor{Altamirano_IGR_FH} applied the \citeauthor{Belloni_GRS_MI} classes for GRS 1915 directly to IGR J17091.
\par In general, the variability classes\index{Variability class} I find to be present in IGR J17091\index{IGR J17091-3624} are broadly the same as those noted by \citet{Altamirano_IGR_FH}.  I do not associate any class with Class $\alpha$\indexalpha in GRS 1915\index{GRS 1915+105}, but I find examples of all of the other variability classes posited by \citeauthor{Altamirano_IGR_FH} to exist in IGR J17091.
\par \citealp{Altamirano_IGR_FH} noted the presence of an anticlockwise loop\index{Hysteresis} in the HID\index{Hardness-intensity diagram} of `heartbeat'\indexrho-like observations of IGR J17091\index{IGR J17091-3624}, opposed to the clockwise loop seen in HIDs of $\rho$-class observations of GRS 1915\index{GRS 1915+105}.  This is consistent with my finding that hysteretic loops in classes III\indexiii\ and IV\indexiv\ also tend to execute in an anticlockwise direction.  However, I additionally find that hysteretic loops in classes V\indexv, VI\indexvi, VII\indexvii\ and VIII\indexviii\ tend to execute in a clockwise direction.  This is also different from GRS 1915, in which the loop is executed in the same direction in all classes.  I also additionally report that clockwise loops tend to be more complex than anticlockwise loops in IGR J17091, with many showing a multi-lobed structure not seen in GRS 1915.  This apparent inconsistency between the objects strengthens the suggestion in \citealp{Altamirano_IGR_FH} that the heartbeat-like classes in GRS 1915 and IGR J17091 may be generated by physically different mechanisms.

\subsection{New Constraints on Accretion Rate, Mass \& Distance}
\label{sec:newmass}

\par The constraints that \citealp{Altamirano_IGR_FH} placed on the mass and distance of IGR J17091\index{IGR J17091-3624} assumed that the object emitted at its Eddington luminosity\index{Eddington limit} at the peak of the 2011--2013 outburst\index{Outburst}.  They report a peak 2--50\,keV flux of $4\times10^{-9}$\ergf\ during flares\index{Flare} in `heartbeat'\indexrho-like lightcurves\index{Lightcurve} during this time.  The correction factor\index{Bolometric correction factor} $C_{Bol,Peak}$ to convert 2--50\,keV flux to bolometric flux is not well constrained, but \citealp{Altamirano_IGR_FH} suggest an order-of-magnitude estimate of $\lesssim3$, corresponding to a peak bolometric flux of $\lesssim1.2\times10^{-8}$\ergf .
\par \citealp{Maccarone_2pct} performed a study of the soft\index{High/Soft state} to hard\index{Low/Hard state} transitions in 10 LMXBs\index{X-ray binary!Low mass} with well-constrained distances and compact object masses.  They found that all but one perform this transition at a luminosity consistent with between 1\% and 4\% of their Eddington limit\index{Eddington limit}.  By assuming that all LMXBs complete their soft-to-hard transitions at Eddington fractions of $\sim1--4$\%, it is then possible to estimate the Eddington fraction of an object at any point during its outburst, even if its distance and compact object mass are not known.
\par I use \indexswift\textit{Swift} observation 00031921058 taken on MJD 55965 to create a spectrum of IGR J17091\index{IGR J17091-3624} during the approximate time of its transition from a soft to a hard state \citep{Drave_Return}.  I fit this spectrum\index{Spectroscopy} above 2\,keV with a power-law, and extrapolated to find a 2--50\,keV flux of $8.56\times10^{-10}$\ergf .  Assuming that the transition bolometric correction factor $C_{Bol,Tran}$ is also $\lesssim3$, this corresponds to a bolometric flux of $\lesssim2.5\times10^{-9}$\ergf .
\par
\begin{figure}
    \includegraphics[width=\columnwidth, trim = 0mm 0mm 0mm 0mm]{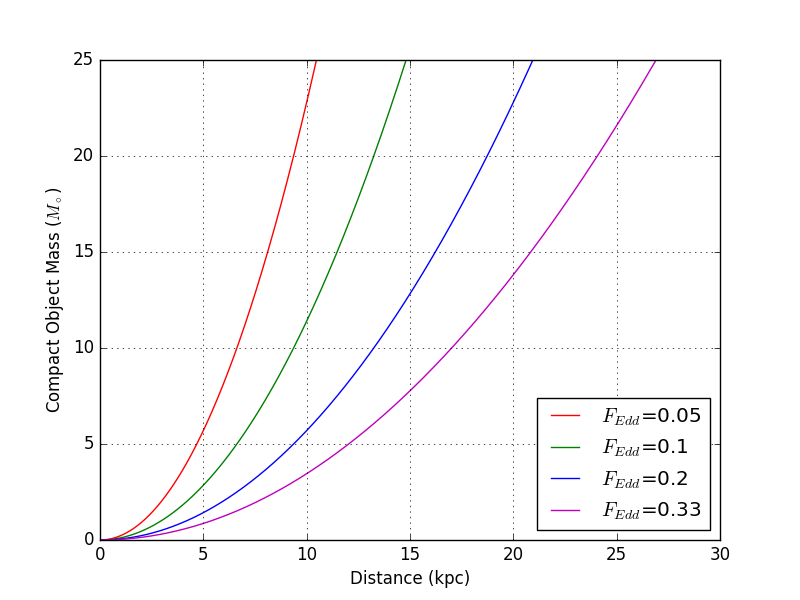}
    \captionsetup{singlelinecheck=off}
    \caption[Mass of the compact object in IGR J17091-3624 plotted against its distance, for values of peak Eddington fractions of $F_{Edd}=$0.05, 0.1, 0.2 and 0.33.]{Mass of the compact object\index{Compact object} in IGR J17091-3624\index{IGR J17091-3624} plotted against its distance, for values of peak Eddington\index{Eddington limit} fractions of $F_{Edd}=$0.05, 0.1, 0.2 and 0.33.}
   \label{fig:IGRMass}
\end{figure}

 By comparing this with the results of \citealp{Maccarone_2pct} and \citealp{Altamirano_IGR_FH}, I find that IGR J17091\index{IGR J17091-3624} was likely emitting at no more than $\sim5$--20\% of its Eddington Limit\index{Eddington limit} at its peak.  This number becomes $\sim6\mbox{--}25$\% if I instead use $C_{Bol,Tran}=2.4$, or $\sim8\mbox{--}33$\% if $C_{Bol,Tran}=1.8$.  With this new range of values, I am able to re-derive the compact object\index{Compact object} mass as the function of the distance (Figure \ref{fig:IGRMass}).  I find that for a black hole\index{Black hole} mass of $\sim10$\ms , as suggested by \citealp{Iyer_Bayes}, IGR J17091 is within the Galaxy at a distance of 6--17\,kpc.  This is consistent with the estimated distance of $\sim11\mbox{--}17$\,kpc estimated by \citealp{Rodriguez_D} for a compact object mass of 10\ms .

\subsection{Implications for Models of `Heartbeat' Variability}

\label{sec:IGRimp}

\par I have found that hysteretic\index{Hysteresis} HID\index{Hardness-intensity diagram} loops can execute in both directions in IGR J17091\index{IGR J17091-3624} (e.g. Section \ref{sec:Alta}), as well as found a revised estimate that IGR J17091 accretes\index{Accretion rate} at $\lesssim20$\% Eddington (Section \ref{sec:newmass}).  Both of these findings have implications for physical models of GRS 1915\index{GRS 1915+105}-like variability\index{Variability} in this source.
\par Firstly, I find that Eddington-limited\index{Eddington limit} accretion\index{Accretion} is neither necessary nor sufficient for GRS 1915-like\index{GRS 1915+105} variability\index{Variability}.  The discovery of GRS 1915-like variability in the sub-Eddington Rapid Burster\index{Rapid Burster} \citep{Bagnoli_RB,Bagnoli_PopStudy} provided the first evidence that Eddington-limited accretion may not be a driving factor in this type of variability.  I strengthen this case by finding that IGR J17091-3624 is also likely sub-Eddington.  As such, I further rule out any scenario in which Eddington-limited accretion is required for GRS 1915-like variability in black hole\index{Black hole} LMXBs\index{X-ray binary!Low mass} specifically.
\par Secondly, by using the direction of hysteretic\index{Hysteresis} HID\index{Hardness-intensity diagram} loops, I find that hard photon lag\index{Hard lag} in `heartbeat'\indexrho-like classes of IGR J17091\index{IGR J17091-3624} can be either positive or negative.  This could mean that we must rule out the causal connection between soft and hard emission being common to all classes.
\par In either case, I find that scenarios that require high global accretion rates\index{Accretion rate} or predict a consistent hard photon lag\index{Hard lag} (e.g. \citealp{Neilsen_GRSModel,Janiuk_Lag}), are not able to explain GRS 1915\index{GRS 1915+105}-like variability\index{Variability} in IGR J17091\index{IGR J17091-3624} unless they also feature geometric obscuration in a subset of variability classes.  I note that simulations by \citealp{Nayakshin_GRSModel} require an Eddington fraction of $\gtrsim0.26$ before GRS 1915-like variability occurs, a value which falls in the range $\sim0.05\mbox{--}0.33$ that I find for the peak Eddington fraction of IGR J17091.
\par An alternative way to explain the reversal of the direction of HID\index{Hardness-intensity diagram} hysteresis\index{Hysteresis} is by considering the information propagation timescales in GRS 1915\index{GRS 1915+105} and IGR J17091\index{IGR J17091-3624}.  A number of proposed models and scenarios to explain GRS 1915-like variability\index{Variability}, such as the scenario of \citet{Neilsen_GRSModel} which we describe in Section \ref{sec:Neilsen}, rely on information being propagated from one component of the LMXB\index{X-ray binary!Low mass} system to another; in the scenario of \citeauthor{Neilsen_GRSModel}, this propagation takes the form of a disk wind\index{Wind} which interacts with a geometrically displaced corona\index{Corona}.  Such a propagation takes a finite time.  If the timescale of the propagation of information is similar to or greater than the characteristic timescale of heartbeat\indexrho\ flares\index{Flare}, then each hard pulse from the corona could take place immediately before the flare subsequent to the flare which triggered it.  As heartbeats are a relatively coherent quasiperiodic\index{Quasi-periodic oscillation} phenomenon, it would appear to an observer that each hard pulse \textit{precedes} a soft flare, even though in reality the causality is reversed.  If this scenario was behind the hysteretic reversal seen in IGR J17091, then we would expect to see that only the fastest variability classes exhibited loops in the `wrong' direction, indicating soft lags.  However, I find that loops in Classes V\indexv, VI\indexvi, VII\indexvii\ and VIII\indexviii\ in IGR J17091 show hard lags\index{Hard lag}, whereas the slower classes III\indexiii\ and IV\indexiv\ show soft lags.  Therefore I rule out an information propagation timescale-based explanation for the difference in HID hysteresis between IGR J17091 and GRS 1915.
\par In addition to being near its Eddington limit\index{Eddington limit} GRS 1915\index{GRS 1915+105} also has the largest orbital period\index{Orbital period} of any known LMXB\index{X-ray binary!Low mass} (e.g. \citealp{McClintock_BHBs}).  \citealp{Sadowski_MagField} have also shown that thin, radiation dominated regions of disks\index{Accretion disk} in LMXBs require a large-scale threaded magnetic field\index{Magnetic field} to be stable, and the field strength required to stabilise such a disk in GRS 1915 is higher than for any other LMXB they studied.  I suggest that one of these parameters is more likely to be the criterion for GRS 1915-like variability\index{Variability}.  If better constraints can be placed on the disk size and minimum stabilising field strength in IGR J17091, it will become clear whether either of these parameters can be the unifying factor behind LMXBs that display GRS 1915-like variability\index{Variability}.

\section{Conclusions}

\par I have constructed the first model-independent set of variability classes\index{Variability class} for the entire portion of the 2011--2013 outburst\index{Outburst} of IGR J17091\index{IGR J17091-3624} that was observed with \indexrxte\rxte .  I find that the data are well-described by a set of 9 classes;  7 of these appear to have direct counterparts in GRS 1915\index{GRS 1915+105}, while two are, so far, unique to IGR J17091.  \textsf{D.A.} and I find that variability\index{Variability} in IGR J17091 is generally faster than in the corresponding classes of GRS 1915, and that patterns of quasi-periodic\index{Quasi-periodic oscillation} flares\index{Flare} and dips\index{Dip} form the basis of most variability in both objects.  Despite this, I find evidence that `heartbeat'\indexrho-like variability in both objects may be generated by different physical processes.  In particular, while hard photons always lag\index{Hard lag} soft in GRS 1915, I find evidence that hard photons can lag or precede soft photons in IGR J17091 depending on the variability class.
\par I also report on the long-term evolution of the 2011--2013 outburst\index{Outburst} of IGR J17091\index{IGR J17091-3624}, in particular noting the presence of 3 re-flares\index{Re-flare} during the later part of the outburst.  Using an empirical relation between hard\index{Low/Hard state}-soft\index{High/Soft state} transition luminosity and Eddington\index{Eddington limit} luminosity \citep{Maccarone_2pct}, I estimate that IGR J17091 was likely accreting\index{Accretion rate} at no greater than $\sim33$\% of its Eddington limit at peak luminosity.
\par I use these results to conclude that any model of GRS 1915\index{GRS 1915+105}-like variability\index{Variability} which requires a near-Eddington\index{Eddington limit} global accretion rate\index{Accretion rate} is insufficient to explain the variability we see in IGR J17091\index{IGR J17091-3624}.  As such I suggest that an extreme value of some different parameter, such as disk size or minimum stabilising large-scale magnetic field\index{Magnetic field}, may be the unifying factor behind all objects which display GRS 1915-like variability.  This would explain why sub-Eddington sources such as IGR J17091 and the Rapid Burster\index{Rapid Burster} do display GRS 1915-like variability, while other Eddington-limited sources such as GX 17+2\index{GX 17+2} and V404 Cyg\index{V404 Cyg} do not.

\cleardoublepage

\chapter{The Evolution of X-ray Bursts in the `Bursting Pulsar' GRO J1744--28}

\label{ch:BPbig}

\epigraph{\textit{The fountains of the great deep came bursting through, and the windows of heaven were open.}}{Genesis 7:11}
\vspace{1cm}

\par\noindent In Chapter \ref{ch:IGR}, I present a new way to classify variability\index{Variability} in the LMXB\index{X-ray binary!Low mass} IGR J17091-3624\index{IGR J17091-3624}.  I compare this object with GRS 1915\index{GRS 1915+105}; although I find a number of differences between variability in the two systems, I conclude that the same broad phenomenon is likely behind variability in both.  I also find that IGR J17091 is likely significantly sub-Eddington\index{Eddington limit} during periods in which it displays GRS 1915-like variability.  This result can be seen as yet another piece of evidence that near-Eddington accretion is neither sufficient or necessary for GRS 1915-like behaviour.
\par To try and better constrain what does unite GRS 1915\index{GRS 1915+105}-like objects, the next step is to look for analogous behaviour in other systems.  As previously mentioned, \citet{Bagnoli_RB} reported variability\index{Variability} similar to GRS 1915 in \indexrxte\textit{RXTE} lightcurves from the Rapid Burster\index{Rapid Burster}.  As such the Rapid Burster, and its sister system the Bursting Pulsar\index{Bursting Pulsar}, are natural places to look for evidence of GRS 1915-like variability\index{Variability}.  Type II\index{X-ray burst!Type II} bursts seen in the Rapid Burster and the Bursting Pulsar are believed to be caused by viscous instabilities\index{Instability} in the accretion disk\index{Accretion disk} \citep{Lewin_TypeII}, as is the X-ray variability seen GRS 1915 and IGR J17091\index{IGR J17091-3624}.  However, as I discuss in Section \ref{sec:TIImod}, the exact details of the mechanism responsible for Type II bursts remain unclear.
\par The Type II\index{X-ray burst!Type II} bursting behaviour in the Rapid Burster\index{Rapid Burster} has been extensively studied (see e.g. \citealp{Lewin_TypeII,Hoffman_RB}).  \citet{Bagnoli_PopStudy} performed a full population study of all Type II bursts observed in this object by \textit{RXTE}.  Their results suggest that gating of the accretion\index{Accretion} by a strong magnetic field\index{Magnetic field} plays some role in the creation of Type II bursts: as this scenario requires a highly magnetised compact object\index{Compact object}, it cannot be employed to explain the variability\index{Variability} seen in the black hole\index{Black hole}-primary GRS 1915\index{GRS 1915+105} or IGR J17091\index{IGR J17091-3624}.  To further probe the physics behind Type II X-ray bursts, in this chapter I perform a similar population study\index{Population study} on bursts from the Bursting Pulsar\index{Bursting Pulsar}.
\par Previous work by \citet{Giles_BP} indicated that Type II\index{X-ray burst!Type II} bursts in the 1995--1996 outburst\index{Outburst} of the Bursting Pulsar\index{Bursting Pulsar} could be separated into a number of distinct populations\index{Population study} based on peak flux.  This is a notable difference from the Rapid Burster\index{Rapid Burster}, in which all Type II bursts have peak fluxes approximately equal to or less than object's Eddington Luminosity\index{Eddington limit} \citep{Tan_RBBursts}.  In this chapter I expand on the work of \citet{Giles_BP} and analyze \indexrxte\textit{RXTE}, \indexnustar\textit{NuSTAR}, \indexchandra\textit{Chandra}, \indexxmm\textit{XMM-Newton}, \indexswift\textit{Swift} and \indexintegral\textit{INTEGRAL} data to fully quantify the population of Type II bursts in the Bursting Pulsar during all 3 outbursts\index{Outburst} in which they have been observed.  I study how the bursting in this object evolves over time throughout each outburst, and I link this behaviour to the long-term evolution of the source.  I also perform basic timing\index{Fourier analysis}, morphology\index{Lightcurve} and spectral\index{Spectroscopy} analysis on bursts, to try and understand the physical processes behind these phenomena.
\par \textbf{The results I present in this chapter have been published as \citet{BPpaper}.}

\section{Data and Data Analysis}

\par Since discovery, the Bursting Pulsar\index{Bursting Pulsar} has undergone three bright outbursts\index{Outburst}, which began in 1995, 1997 and 2014.  I refer to these outbursts as Outbursts 1, 2 and 3.  I do not consider the faint outburst in 2017 in this chapter \citep{Sanna_BPOutburst}, as no Type II bursts\index{X-ray burst!Type II} were observed during this time, nor do I analyse data taken while the source was in quiescence\index{Quiescence}.  See \citet{Daigne_BPQ}, \citet{Wijnands_BPQ} and \citet{Degenaar_BPQuiescence} for studies of the Bursting Pulsar during quiescence.
\par I analysed data from all X-ray instruments which observed the Bursting Pulsar\index{Bursting Pulsar} during these outbursts\index{Outburst}.  Specifically, I analysed lightcurves\index{Lightcurve}, the evolution of hardness ratios\index{Colour} as a function of time and of count rate, and performed statistical analysis of properties associated with each individual burst\index{X-ray burst!Type II}.

\subsection{\textit{RXTE}}

\par I analysed data from \indexrxte\indexpca\textit{RXTE}/PCA corresponding to the Outbursts\index{Outburst} 1 \& 2 of the Bursting Pulsar\index{Bursting Pulsar}.  This in turn corresponded to observation IDs starting with 10401-01, 20077-01, 20078-01, 20401-01 and 30075-01, between MJDs 50117 and 51225.  This resulted in a total of 743\,ks of data over 300 observations, which I have listed in Appendix \ref{app:obs}. Lightcurve\index{Lightcurve} data were extracted from \texttt{fits}\index{FITS@\texttt{FITS}} files using \texttt{FTOOLS}\index{FTOOLS@\texttt{FTOOLS}}\footnote{\url{https://heasarc.gsfc.nasa.gov/ftools/ftools_menu.html}}.  Errors were calculated and quoted at the 1$\,\sigma$ level.
\par I also use data from the \indexasm\textit{RXTE}/ASM to monitor the long-term evolution of the source.  ASM data were taken from MIT's ASM Light Curves Overview website\footnote{\url{http://xte.mit.edu/ASM_lc.html}}.

\subsubsection{Long-Term Evolution}

\par To analyse the long-term evolution of the source during its outbursts, I extracted 2--16\,keV count rates from the \texttt{Standard2}\indexstt\ PCA\indexpca\ data in each observation.  Following \citet{Altamirano_CrabNorm}, I normalised the intensity estimated in each observation by the intensity of the Crab nebula\index{Crab nebula}, using the Crab observation that is the closest in time but within the same PCA gain epoch as the observation in question (see \citealp{Jahoda_Calibrate}).

\subsubsection{Burst Identification and Analysis}

\label{sec:burst_diff}

\par To perform population studies\index{Population study} on the Type II\index{X-ray burst!Type II} bursts in the Bursting Pulsar\index{Bursting Pulsar}, I first extracted lightcurves\index{Lightcurve} from the \texttt{Standard1}\indexsto\ data in each observation, as this data is available for all \indexpca\textit{RXTE}/PCA observations.  I used my own \texttt{PANTHEON}\index{PANTHEON@\texttt{PANTHEON}} software to search these lightcurves and return a list of individual bursts, using the algorithm described in Section \ref{sec:Flares}.  I manually cleaned spurious detections from my sample.  I defined a `burst'\index{X-ray burst}\index{Burst|see {X-ray burst}} as an event that lasted at least 3 seconds during which the 1\,s binned count rate exceeded 3 standard deviations above the persistent emission\index{Persistent emission} level and reached a maximum of at least five standard devations above the persistent emission level.  I did not subtract background\index{Background subtraction}, as all count rate-related parameters I analyse are persistent emission subtracted, automatically removing background contribution.
\par During the analysis, Arianna Albayati (\textsf{A.A.}) and I discovered a number of different burst\index{X-ray burst} `classes', similar to the multiple classes of burst described by \citet{Giles_BP}.  Our classes varied significantly in terms of overall structure, and as such needed to be treated separately; I show representative lightcurves\index{Lightcurve} from each of our classes in Figure \ref{fig:classes}.  These classes were separated from one another by a number of criteria including peak count rate and recurrence time\index{Recurrence time} (the time between peaks of consecutive bursts).
\par The vast majority of detected bursts\index{X-ray burst} resembled the Type II bursts\index{X-ray burst!Type II} seen in the Rapid Burster (referred to as `Normal Bursts'\index{Normal burst} in Section \ref{sec:Results}) in terms of shape, duration and amplitude.  I rebinned the data corresponding to these Normal Bursts to 0.5\,s.  I sampled the persistent emission\index{Persistent emission} before the burst, and defined the start of the burst as the first point at which count rate exceeded 5 standard deviations above the persistent emission before the burst.  The end of the burst was defined similarly, but instead sampling the persistent emission after the burst; by doing this, I avoid making the implicit assumption that the persistent emission is equal before and after the burst.  I fitted phenomenologically-motivated lightcurve\index{Lightcurve} models to each of these bursts (described in detail in Section \ref{sec:struc}), and used these fits to extract a number of parameters which characterise the shape and energetics of a burst (such as burst duration, total photon counts associated with a burst and persistent emission count rate).
\par Due to the high peak count rates of Normal Bursts\index{Normal burst}, data were affected by dead-time\index{Dead-time} (compare e.g. \textit{GRANAT}\index{GRANAT@\textit{GRANAT}} data presented in \citealp{Sazonov_BPGranat}).  I calculate the approximate Dead-Time Factors (DTFs) for a number of the brightest Normal Bursts in my sample, using 1\,s binned data, using the following formula in the \indexrxte\textit{RXTE} Cookbook\footnote{\url{https://heasarc.gsfc.nasa.gov/docs/xte/recipes/pca_deadtime.html}}:
\begin{equation}
\Delta=\frac{C_{Xe}+C_{Vp}+C_{Rc}+15C_{VL}}{N_{PCU}}\times10^{-5}
\end{equation}
Where $\Delta$ is the fractional detector deadtime, $C_{Xe}$ is the Good Xenon\indexgx\ count rate, $C_{Vp}$ is the coincident event count rate, $C_{Rc}$ is the propane layer count rate, $C_{VL}$ is the very large event count rate and $N_{PCU}$ is the number of PCUs active at the time.
\par I estimate that dead-time\index{Dead-time} effects reduce the peak count rates of Normal Bursts by no more than $\sim12$\%; however, due to the sharply-peaked nature of bursts\index{X-ray burst} from the Bursting Pulsar\index{Bursting Pulsar}, the deadtime effect depends on the binning used.  Due to this ambiguity I do not correct for dead-time in Normal Bursts.  The dead-time corrections required for the count rates seen in other classes of burst are minimal, as they are orders of magnitude fainter \citep{Giles_BP}.
\par To test for correlations between parameters in a model-independent way, I used the Spearman's Rank\index{Spearman's rank correlation coefficient} correlation coefficient (as available in \texttt{Scipy}, \citealp{NumPy}).  This metric only tests the hypothesis that an increase in the value of one parameter is likely to correspond to an increase in the value of another parameter, and it is not affected by the shape of the monotonic correlation to be measured.  Although dead-time\index{Dead-time} effects lead to artificially low count rates being reported, a higher intensity still corresponds to a higher reported count rate.   As such, using this correlation coefficient removed the effects of dead-time on my detection of any correlations.
\par To calculate the distribution of recurrence times\index{Recurrence time} between consecutive bursts\index{X-ray burst}, I considered observations containing multiple bursts.  If fewer than 25\,s of data gap exists between a pair of bursts, I considered them to be consecutive and added their recurrence time to the distribution.  I choose this maximum gap size as this is approximately the timescale over which a Normal Burst\index{Normal burst} occurs.
\par When \texttt{SB\_62us\_0\_23\_500ms} and \texttt{SB\_62us\_24\_249\_500ms} data were available, I divided my data into two energy bands: A (PCA channels 0--23\index{Channel}, corresponding to $\sim2$--$7$\,keV\footnote{In \textit{RXTE} gain epoch 1, corresponding to dates before MJD 50163.  This corresponds to $\sim2$--$9$\,keV in epoch 2 (MJDs 50163--50188) and $\sim2$--$10$\,keV in epoch 3 (MJDs 50188--51259).}) and B (channels 24--249\index{Channel}, corresponding to $\sim8$--$60$\,keV\footnote{In \textit{RXTE} gain epoch 1.  This corresponds to $\sim9$--$60$\,keV in epoch 2 and $\sim10$--$60$\,keV in epoch 3.}).  The evolution of colour\index{Colour} (defined as the ratio of the count rates in B and A) throughout a burst\index{X-ray burst} could then be studied.  Due to the very high count rates during Normal Bursts\index{Normal burst}, I did not correct for background\index{Background subtraction}.  During fainter types of burst I estimate the background in different energy bands by subtracting count rates from \indexrxte\textit{RXTE} observation 30075-01-26-00 of this region, when the source was in quiescence\index{Quiescence}.  Unlike using the \textit{RXTE} background model, this method subtracts the contributions from other sources in the field.  However, as it is unclear whether any of the rest of these sources are variable, the absolute values of colours I quote should be treated with caution.  I created hardness-intensity diagrams\index{Hardness-intensity diagram} to search for evidence of hysteretic\index{Hysteresis} loops in hardness-intensity space.
\par Following \citet{Bagnoli_PopStudy}, I used the total number of persistent emission-subtracted counts as a proxy for fluence for all bursts\index{X-ray burst} other than Normal Bursts\index{Normal burst}.  As the contribution of the background does not change much during a single observation, this method also automatically subtracts background\index{Background subtraction} counts from my results.

\subsubsection{Detecting Pulsations}

\par The Bursting Pulsar\index{Bursting Pulsar} is situated in a very dense region of the sky close to the Galactic centre, and so several additional objects also fall within the 1$^\circ$ \indexpca\textit{RXTE}/PCA field of view.  Therefore it is important to confirm that the variability\index{Variability} I observe in my data does in fact originate from the Bursting Pulsar.
\par To ascertain that all bursts\index{X-ray burst} considered in this study are from the Bursting Pulsar\index{Bursting Pulsar}, Dr. Andrea Sanna (\textsf{A.S.}) analysed the coherent X-ray pulse at the pulsar\index{Pulsar} spin frequency to confirm that the source was active.  \textsf{A.S.} first corrected the photon time of arrivals of the \indexpca\textit{RXTE}/PCA dataset, and barycentred this data using the \texttt{faxbary} tool available in \texttt{FTOOLS}\index{FTOOLS@\texttt{FTOOLS}} (DE-405 Solar System ephemeris).  \textsf{A.S.} corrected for the binary motion by using the orbital parameters reported by \citet{Finger_Pulse}.
\par For each \indexpca\ PCA observation \textsf{A.S.} investigated the presence of the $\sim 2.14$\,Hz coherent pulsation by performing an epoch-folding\index{Folding} search of the data using 16 phase bins and starting with the spin frequency value $\nu=2.141004$ Hz, corresponding to the spin frequency measured from the 1996 outburst\index{Outburst} of the source\index{Bursting Pulsar} \citep{Finger_Pulse}, with a frequency step of $10^{-5}$\,Hz for 10001 total steps. \textsf{A.S.} detected X-ray coherent pulsations in all PCA observations performed during Outbursts 1 \& 2.

\subsection{\textit{Swift}}
\par In this study, I made use of data from \indexxrt\ XRT and \indexbat\ BAT aboard \indexswift\textit{Swift}.  I extracted a long-term 0.3--10\,keV \textit{Swift}/XRT lightcurve\index{Lightcurve} of Outburst\index{Lightcurve} 3 using the lightcurve generator provided by the UK Swift Science Data Centre (UKSSDC, \citealp{Evans_Swift1}).  I also make use of \textit{Swift}/BAT lightcurves from the Swift/BAT Hard X-ray Transient website\footnote{\url{https://swift.gsfc.nasa.gov/results/transients/}} (see \citealp{Krimm_BAT}).

\subsection{\textit{INTEGRAL}}

\par I also made use of data from IBIS\indexibis\ aboard \textit{INTEGRAL}\indexintegral.  I extracted 17.3--80\,keV IBIS/ISGRI lightcurves of the Bursting Pulsar during Outburst 3\index{Outburst} using the \textit{INTEGRAL} Heavens portal.  This is provided by the \textit{INTEGRAL} Science Data Centre \citep{Lubinski_Heavens}.

\subsection{\textit{Chandra}}

\par The Bursting Pulsar\index{Bursting Pulsar} was targeted with \indexchandra\textit{Chandra} three times during Outburst\index{Outburst} 3 (Table \ref{tab:ChandraBP}).  One of these observations (OBSID 16596) was taken simultaneously with a \indexnustar\textit{NuSTAR} observation (80002017004).  In all three observations data were obtained with the HETG, where the incoming light was dispersed onto the \indexacis\ ACIS-S array. The ACIS-S was operated in continued clocking (CC) mode to minimize the effects of pile-up. The Chandra/HETG observations were analysed using standard tools available within \texttt{ciao}\index{CIAO@\texttt{CIAO}} v. 4.5 \citep{Fruscione_Ciao}. Dr. Nathalie Degenaar (\textsf{N.D.}) extracted 1\,s binned lightcurves\index{Lightcurve} from the \texttt{evt2} data using \texttt{dmextract}, where the first order positive and negative grating data from both the Medium Energy Grating (MEG; 0.4-5 keV) and the High Energy Grating (HEG; 0.8--8\,keV) were combined.

\begin{table}
\centering
\begin{tabular}{lllllll}
\hline
\hline
\scriptsize  OBSID &\scriptsize Exposure (ks) &\scriptsize MJD &\scriptsize Reference \\
\hline
16596  	& 10 &  56719      &   \citet{Younes_Expo} \\
16605  	& 35 &   56745    &    \citet{Degenaar_BPSpec}\\
16606  	& 35 &   56747    &    \citet{Degenaar_BPSpec}\\
\hline
\hline
\end{tabular}
\caption[Information on the three \textit{Chandra} observations of the Bursting Pulsar during its 2014 outburst.]{Information on the three \indexchandra\textit{Chandra} observations of the Bursting Pulsar during Outburst\index{Outburst} 3.  All other observations of the Bursting Pulsar in the Chandra archive were obtained at times that the source was in quiescence\index{Quiescence}.}
\label{tab:ChandraBP}
\end{table}

\subsection{\textit{XMM-Newton}}

\par A single pointed \indexxmm\textit{XMM-Newton} observation of the Bursting Pulsar\index{Bursting Pulsar} was taken during Outburst\index{Outburst} 3 on MJD 56722 (OBSID 0729560401) for 85\,ks.  I extracted a 0.5--10\,keV lightcurve\index{Lightcurve} from EPIC\indexepic-PN at 1\,s resolution using \texttt{SAS}\index{SAS@\texttt{SAS}} version 15.0.0.  During this observation, EPIC-PN was operating in Fast Timing mode.  I use EPIC-PN as the statistics are better than in MOS1 or MOS2.

\subsection{\textit{Suzaku}}
\par \indexsuzaku\textit{Suzaku} observed the Bursting Pulsar\index{Bursting Pulsar} once during Outburst\index{Outburst} 3 on MJD 56740 (OBSID 908004010).  To create a lightcurve\index{Lightcurve}, \textsf{K.Y.} reprocessed and screened data from the X-ray Imaging Spectrometer\indexxis\ (XIS, \citealp{Koyama_XIS}) using the \texttt{aepipeline} script and the latest calibration database released on June 7, 2016.  The attitude correction for the thermal wobbling was made by \texttt{aeattcor2} and \texttt{xiscoord} \citep{Uchiyama_SuzPSF}. The source was extracted within a radius of 250 pixels corresponding to 260'' from the image center.  The background was extracted from two regions near either end of the XIS chip, and subtracted from the source.

\subsection{\textit{NuSTAR}}

\par \indexnustar\textit{NuSTAR} observed the Bursting Pulsar\index{Bursting Pulsar} three times during its outbursts\index{Outburst}, all times in Outburst 3.  One of these observations was taken while the Bursting Pulsar was not showing X-ray bursts\index{X-ray burst}, and the other two are shown in Table \ref{tab:NuS}.  I extracted lightcurves\index{Lightcurve} from both of these observations using \texttt{nupipeline} and \texttt{nuproducts}, following standard procedures\footnote{See \url{https://www.cosmos.esa.int/web/xmm-newton/sas-threads}.}.

\begin{table}
\centering
\begin{tabular}{lllllll}
\hline
\hline
\scriptsize  OBSID &\scriptsize Exposure (ks) &\scriptsize MJD &\scriptsize Reference \\
\hline
80002017002 	& 29 & 56703 &  \citet{Dai_Hlags}  \\
80002017004 	& 9 & 56719 & \citet{Younes_Expo}\\
\hline
\hline
\end{tabular}
\caption[Information on two \textit{NuSTAR} observations of the Bursting Pulsar during its 2014 outburst.]{Information on the two \indexnustar\textit{NuSTAR} observations of the Bursting Pulsar\index{Bursting Pulsar} during the main part of Outburst\index{Outburst} 3.}
\label{tab:NuS}
\end{table}

\section{Results}
\label{sec:Results}

\subsection{Outburst Evolution}

\par
\begin{figure}
  \centering
  \includegraphics[width=.9\linewidth, trim={0cm 0 0cm 0},clip]{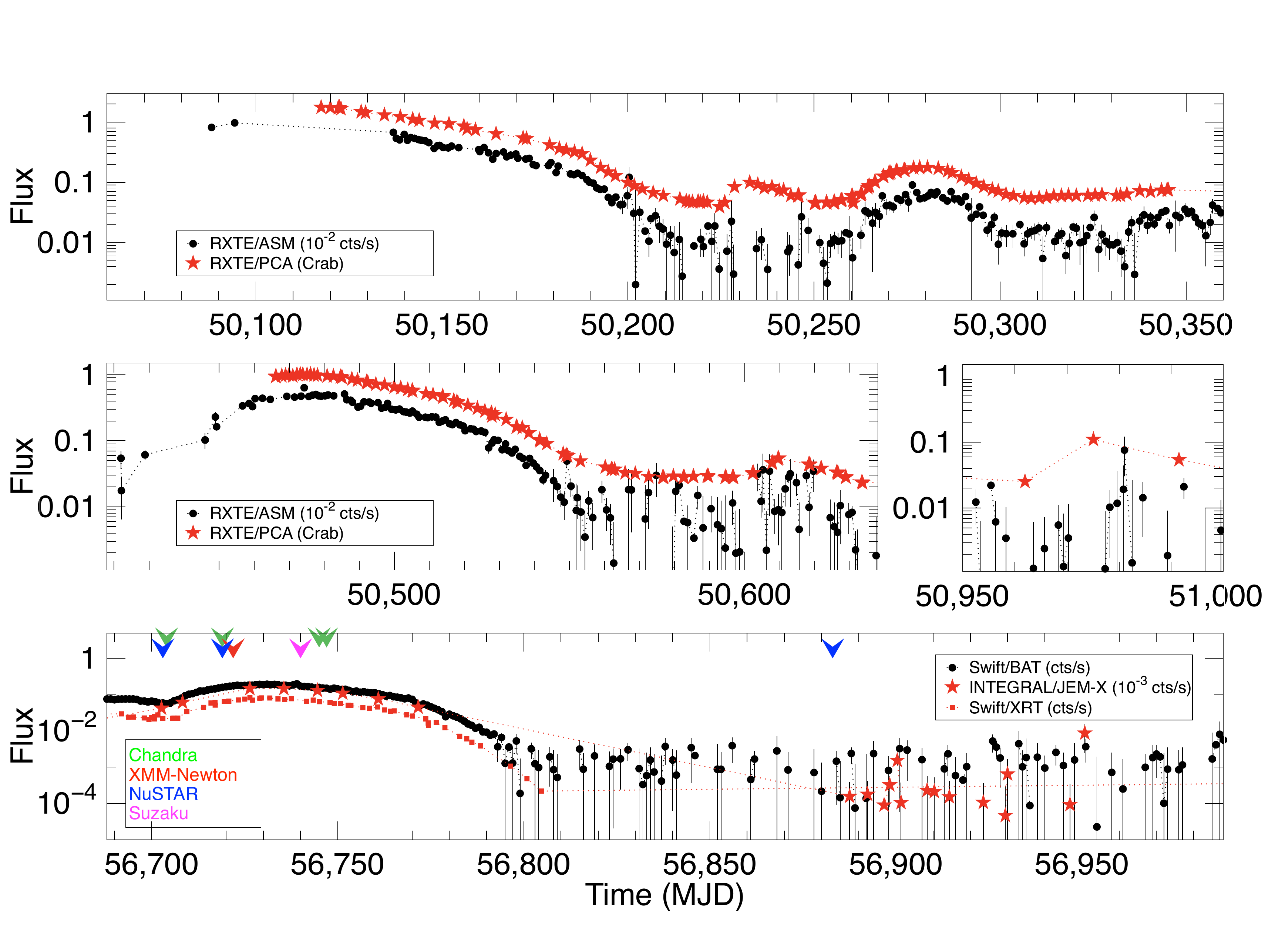}
  \caption[Comparisons of three outbursts of the Bursting Pulsar.]{\small \indexbat\indexxrt\indexrxte\indexasm\indexpca\indexintegral\indexjemx Comparisons of the three outbursts\index{Outburst} of the Bursting Pulsar\index{Bursting Pulsar} reported on in this chapter.  Times corresponding to pointed observations with \indexchandra\textit{Chandra}, \indexnustar\textit{NuSTAR}, \indexsuzaku\textit{Suzaku}, \indexswift\textit{Swift} and \indexxmm\textit{XMM-Newton} are marked.}
  \label{fig:global_ob}
\end{figure}

 I show the long-term monitoring lightcurves\index{Lightcurve} of Outbursts\index{Outburst} 1, 2 and 3 in Figure \ref{fig:global_ob}, as well as mark the dates of pointed observations with various instruments.

\par The Bursting Pulsar\index{Bursting Pulsar} was discovered already in outburst\index{Outburst} on December 12 1995 \citep{Fishman_Discovery}; \index{CGRO@\textit{CGRO}!BATSE}\index{CGRO@\textit{CGRO}}\textit{CGRO}/BATSE data suggest that this outburst began several days earlier on December 3 \citep{Paciesas_BPDiscovery,Bildsten_Rev}.  The main outburst ended around May 10 1996 \citep{Woods_PulseBursts}.  I show the global lightcurve\index{Lightcurve} of this outburst in Figure \ref{fig:global_ob}, Panel 1.  As \indexrxte\textit{RXTE} did not observe the object before or during the peak of Outburst, I can only obtain a lower limit of $\sim1.75$\,Crab for the peak 2--16\,keV flux.
\par There are at least two major rebrightening\index{Re-flare}\index{Rebrightening event|see {Re-flare}} events in the tail of Outburst\index{Outburst} 1, which can be seen clearly  in Figure \ref{fig:global_ob} centred at MJDs of $\sim50235$ and $\sim50280$.  During these rebrightening events, the 2--16\,keV flux peaked at $\sim0.10$ and $\sim0.18$\,Crab respectively.

\par Outburst\index{Outburst} 2 began on December 1 1996 and ended around April 7 1997 \citep{Woods_OB2}.  The 2--16\,keV flux peaked at 1.02\,Crab on MJD 50473; I show the global lightcurve\index{Lightcurve} of this outburst in Figure \ref{fig:global_ob}, Panel 2.  Type II-like\index{X-ray burst!Type II} bursts are seen in \indexpca\textit{RXTE}/PCA lightcurves from Outburst 2 between MJDs 50466 and 50544.  One rebrightening\index{Re-flare} event occurred during the tail of Outburst 2, centred at an MJD of $\sim50615$ with a peak 2--16\,keV flux of $\sim54$\,mCrab.  A second possible rebrightening event occurs at MJD 50975, with a peak 2--16\,keV flux of 11\,mCrab, but the cadence of \textit{RXTE}/PCA observations was too low to unambiguously confirm the existence of a re-flare at this time.

\par Outburst 3\index{Outburst} began on January 31, 2014 \citep{Negoro_OB3,Kennea_BPOutburst} and ended around April 23 (e.g. \citealp{Dai_OB3}).  The daily 0.3--10\,keV Swift/XRT rate peaked at 81\,cts\,s$^{-1}$ on MJD 56729, corresponding to 0.4\,Crab.  I show the global lightcurve\index{Lightcurve} of this outburst in Figure \ref{fig:global_ob}, Panel 3.
\par During the main part of Outburst 3, \indexswift\textit{Swift}, \indexxmm\textit{XMM-Newton} and \indexsuzaku\textit{Suzaku} made one pointed observation each, \indexchandra\textit{Chandra} made four observations, and \indexnustar\textit{NuSTAR} made three observations.  The \textit{Chandra} observation on March 3 2014 was made simultaneously with one of the \textit{NuSTAR} observations (see \citealp{Younes_Expo}).  After the main part of the outburst, the source was not well-monitored, although it remained detectable by \indexbat\textit{Swift}/BAT, and it is unclear whether any rebrightening events occured.  A single \textit{NuSTAR} observation was made during the outburst tail on August 14 2014.

\par As can be seen in Figure \ref{fig:global_ob}, the main section of all three outbursts\index{Outburst} follow a common profile, over a timescale of $\sim150$ days.  A notable difference between outbursts 1 \& 2 is the number of rebrightening\index{Re-flare} events; while I find two re-flares associated with Outburst 1, I only find one associated with Outburst 2 unless I assume the event at MJD 50975 is associated with the outburst.  Additionally, Outburst 2 was at least a factor $\sim1.7$ fainter at its peak than Outburst 1 (see also \citealp{Woods_OB2}), while Outburst 3 was a factor of $\gtrsim4$ fainter at peak than Outburst 1.

\subsubsection{Pulsations}

\par
\begin{figure}
  \centering
  \includegraphics[width=.9\linewidth, trim={0.4cm 1cm 0cm 1cm},clip]{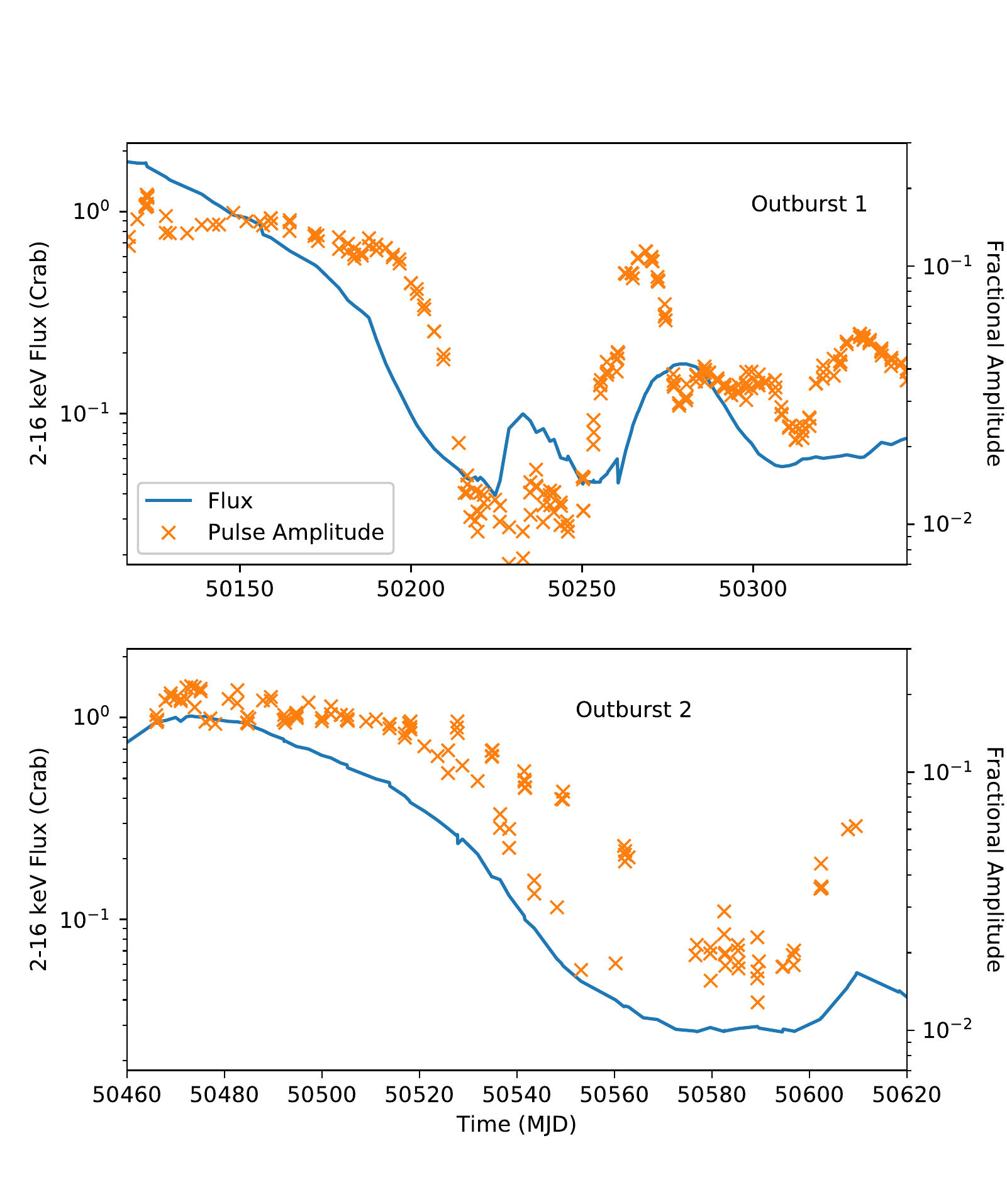}
  \caption[\textit{RXTE}/PCA lightcurves of Outbursts 1 \& 2 of the Bursting Pulsar, overlaid with plots showing how the fractional RMS of the 2.4\,Hz pulsation changes as a function of time.]{\small 2--16\,keV \indexpca\textit{RXTE}/PCA lightcurves\index{Lightcurve} of Outbursts\index{Outburst} 1 \& 2 of the Bursting Pulsar\index{Bursting Pulsar} (solid blue), overlaid with plots showing how the fractional RMS\index{RMS} of the 2.4\,Hz pulsation associated with the pulsar\index{Pulsar} changes as a function of time during these outbursts (orange crosses).}
  \label{fig:pulsovertime}
\end{figure}

 \textsf{A.S.} found pulsations in PCA\indexpca\ data throughout the entirety of Outbursts\index{Outburst} 1 \& 2.  This confirms that the Bursting Pulsar\index{Bursting Pulsar} was active as an X-ray pulsar\index{Pulsar} in all of my observations, leading us to conclude that all the types of X-ray burst\index{X-ray burst!Type II} that we see are from the Bursting Pulsar.  In Figure \ref{fig:pulsovertime}, I show that the amplitude of these pulsations approximately followed the intensity of the source in both outbursts, but there were significant deviations from this trend.  These deviations will require further investigation, and a comparison with other accreting\index{Accretion} pulsar systems.  Previous studies have shown that pulsations were also present during Outburst 3 (e.g. \citealp{Sanna_BP}).

\subsubsection{Bursting Behaviour}

\label{sec:bburstevo}

\par Bursts\index{X-ray burst} are seen in \indexpca\textit{RXTE}/PCA lightcurves\index{Lightcurve} from the start of the Outburst\index{Outburst} 1 (e.g. \citealp{Kouveliotou_BP}).  These bursts occur until around MJD 50200, as the source flux falls below $\sim0.1$\,Crab in the 2--16\,keV band.  
\par During the latter part of the first rebrightening\index{Re-flare} after Outburst 1\index{Outburst}, between MJDs 50238 and 50246, \textsf{A.A.} found Type II\index{X-ray burst!Type II}-like bursts\index{X-ray burst} with amplitudes $\sim2$ orders of magnitude smaller than those found during the main outburst event.  These gradually increased in frequency throughout this period of time until evolving into a period of highly structured variability\index{Variability} which persisted until MJD 50261.
\par In Outburst\index{Outburst} 2, Type II\index{X-ray burst!Type II} bursts occured between MJDs $\sim50466$ and $50542$.  Low-amplitude Type II-like bursts\index{X-ray burst} were seen during the latter stages of the main outburst, between MJDs 50562 and 50577.  These again evolved into a period of highly structured variability\index{Variability}; this persisted until MJD 50618, just after the peak of the rebrightening\index{Re-flare} event.
\par High-amplitude Type II\index{X-ray burst!Type II} bursts were also seen in Outburst\index{Outburst} 3 (e.g. \citealp{Linares_NewBurst}).  As no soft ($\lesssim10\,$keV) X-ray instrument was monitoring the Bursting Pulsar\index{Bursting Pulsar} during the latter part of Outburst 3, it is unknown whether this Outburst showed the lower-amplitude bursting behaviour seen at the end of Outbursts 1 \& 2.  Low amplitude bursting behaviour is not seen in the pointed \textit{NuSTAR} observation which was made during this time.

\subsection{Categorizing Bursts}

\label{sec:classes}

\par \textsf{A.A.} and I found that bursts\index{X-ray burst} in the Bursting Pulsar\index{Bursting Pulsar} fall into a number of discrete classes, lightcurves\index{Lightcurve} from which I show in Figure \ref{fig:classes}.  These classes are as follows:

\begin{figure}
  \centering
  \includegraphics[width=.9\linewidth, trim={0.7cm 2.1cm 1.5cm 3.4cm},clip]{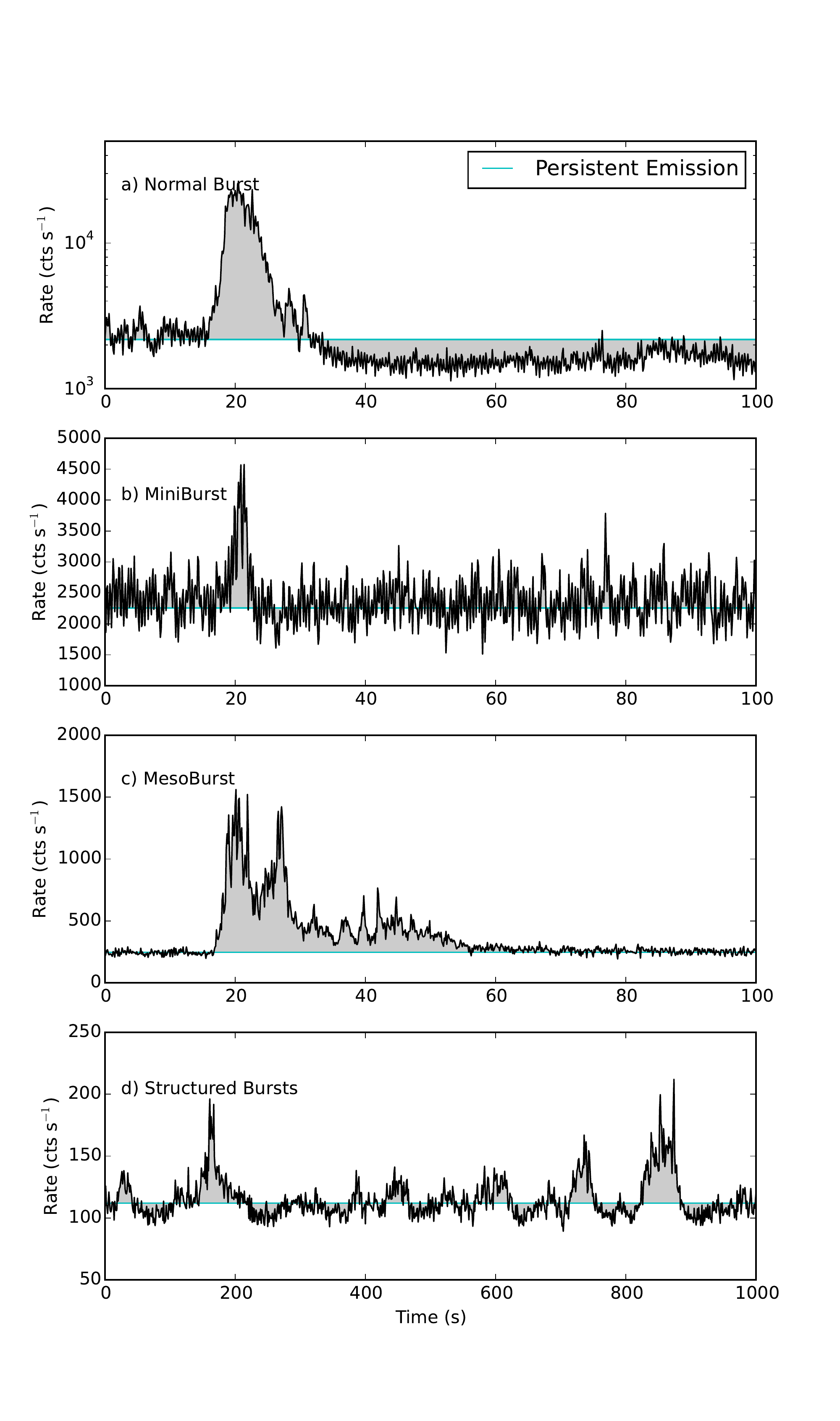}
  \caption[Lightcurves for the four classes of bursting behaviour identified in the Bursting Pulsar.]{\small 2--49\,keV lightcurves\index{Lightcurve} for the four classes of bursting\index{X-ray burst} behaviour identified in this chapter: \textbf{a)} Normal Burst\index{Normal burst}, \textbf{b)} Miniburst\index{Miniburst}, \textbf{c)} Mesoburst\index{Mesoburst}, \textbf{d)} Structured Bursts\index{Structured bursting}.  Note that Panel \textbf{d} is plotted with a different time scaling to the other panels so as to better show the behaviour of Structured Bursting.  On all figures the median count rate, which I use as a proxy for the persistent emission\index{Persistent emission}, is plotted in cyan.  Lightcurves \textbf{a}-\textbf{c} are binned to 0.125\,s, while lightcurve \textbf{d} is binned to 1\,s.}
  \label{fig:classes}
\end{figure}

\begin{itemize}
\item Normal Bursts\index{Normal burst} (Figure \ref{fig:classes}, Panel a): the brightest bursts\index{X-ray burst} seen from this source, with peak count 1\,s binned rates of $\sim10000$\,cts\,s$^{-1}$\,PCU$^{-1}$, and recurrence\index{Recurrence time} timescales of order $\sim1000$\,s.  These bursts are roughly Gaussian\index{Gaussian profile} in shape with durations of $\sim10$\,s, and are followed by a `dip'\index{Dip} in the persistent emission count rate with a duration of order 100\,s (see also e.g. \citealp{Giles_BP}).
\item Minibursts\index{Miniburst} (Figure \ref{fig:classes}, Panel b): faint bursts\index{X-ray burst} with 1\,s-binned peak count rates of $\sim2$ times the persistent emission\index{Persistent emission} count rate.  Minibursts are variable, with duration timescales between $\sim5$--50\,s.  These bursts are also sometimes followed by dips\index{Dip} similar to those seen after Normal Bursts.
\item Mesobursts\index{Mesoburst} (Figure \ref{fig:classes}, Panel c): Type II-like\index{X-ray burst}\index{X-ray burst!Type II} bursts.  These bursts differ from Normal\index{Normal burst} Bursts in that they do not show well-defined subsequent `dips'\index{Dip}.  They are also fainter than Normal Bursts, with peak count 1\,s binned count rates of $\sim1000$\,cts\,s$^{-1}$\,PCU$^{-1}$.  Their burst profiles show fast rises on timescales of seconds, with slower decays and overall durations of $\sim50$\,s.  The structure of the bursts is very non-Gaussian\index{Gaussian profile}, appearing as a small forest of peaks in lightcurves\index{Lightcurve}.
\item Structured Bursts\index{Structured bursting} (Figure \ref{fig:classes}, Panel d): the most complex class of bursting\index{X-ray burst} behaviour we observe from the Bursting Pulsar\index{Bursting Pulsar}, consisting of patterns of flares\index{Flare} and dips\index{Dip} in the X-ray lightcurve\index{Lightcurve}.  The amplitudes of individual flares are similar to those of the faintest Mesobursts\index{Mesoburst}.  The recurrence timescale\index{Recurrence time} is of the order of the timescale of an individual flare, meaning that is it difficult to fully separate individual flares of this class.
\end{itemize}

\par
\begin{figure}
  \centering
  \includegraphics[width=.8\linewidth, trim={1.3cm 0.4cm 2.0cm 0.8cm},clip]{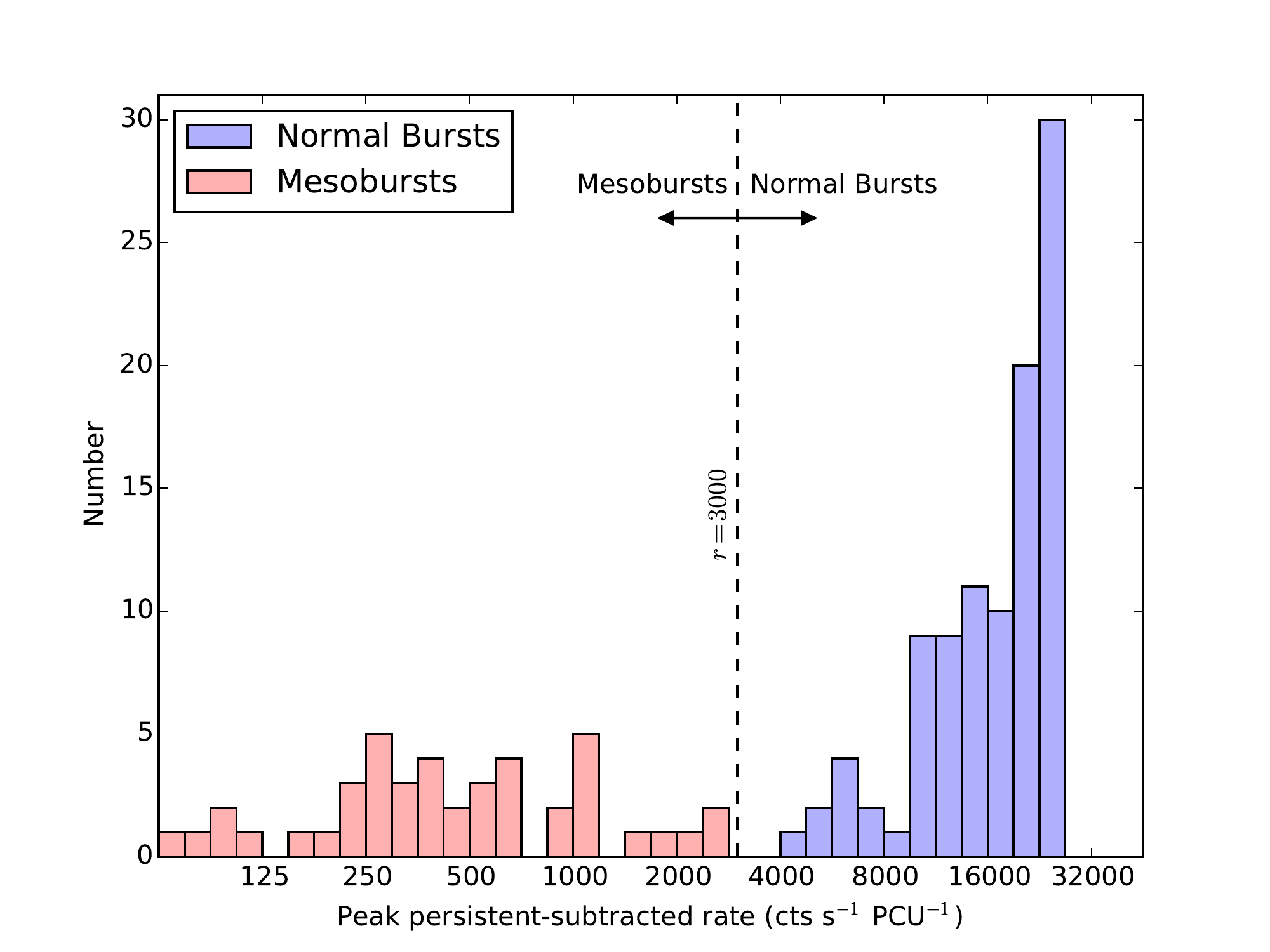}
  \includegraphics[width=.8\linewidth, trim={1.3cm 0.4cm 2.0cm 0.8cm},clip]{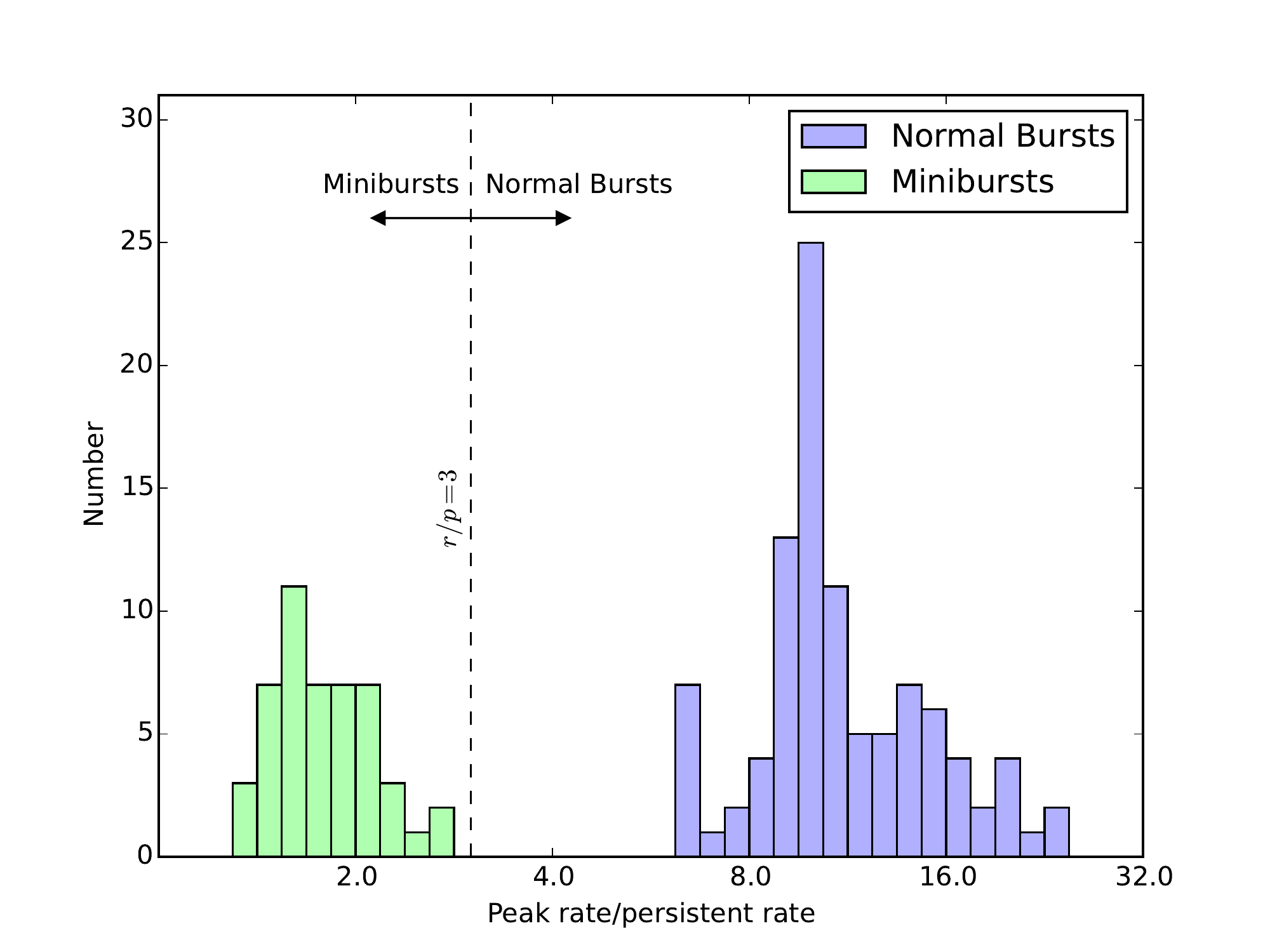}
  \caption[Histogram of the peak 1\,s binned peak count rates during Normal Bursts, Mesobursts and Minibursts as seen by \textit{RXTE}]{\small \textbf{Upper Panel:} A histogram of the peak 1\,s binned peak count rates of the joint population\index{Population study} of all Normal\index{Normal burst} and Mesobursts\index{Mesoburst} seen by \indexrxte\textit{RXTE}.  The dashed line indicates the position of the threshold above which I consider a Type II\index{X-ray burst!Type II}-like burst\index{X-ray burst} to be a Normal Burst.  The resultant split of the population into Normal and Mesobursts is indicated by blue and red shading respectively.  The skewed shape of the distribution of Normal Bursts is due to the effects of dead-time\index{Dead-time} putting an effective cap on their maximum observed intensity.
 \textbf{Lower Panel:} A histogram of the peak 1\,s binned peak count rates of the joint population of all Normal and Minibursts\index{Miniburst} seen by \textit{RXTE}, divided by the persistent emission count rate at that time.  The dashed line indicates the position of the threshold below which I consider a burst to be a Miniburst.  The resultant split of the population into Normal and Minibursts is indicated by blue and green shading respectively.
  Note that the $x$-axis of both plots is logarithmic, and so number density is not preserved.}
  \label{fig:jointhist}
\end{figure}

 In the upper panel of Figure \ref{fig:jointhist} I show a histogram of persistent-emission\index{Persistent emission}-subtracted peak count rates for all Normal\index{Normal burst} and Mesobursts\index{Mesoburst} observed by \indexrxte\textit{RXTE}.  I split these two classes based on the bimodal distribution in peak count rate as well as the lack of dips\index{Dip} in Mesobursts.  In the lower panel of Figure \ref{fig:jointhist}, I show the histogram of peak count rates for all Normal\index{Normal burst} and Minibursts\index{Miniburst} observed by \indexrxte\textit{RXTE} as a fraction of the persistent emission at that time.  I split these two classes based on the strongly bimodal distribution in fractional amplitude.

\par I also find 6 bursts\index{X-ray burst} with fast ($\sim1$\,s) rises and exponential decays that occur during the lowest flux regions of the outburst\index{Outburst} ($\lesssim50$\,mCrab).  \citet{Strohmayer_BPFieldTypeI} and \citet{Galloway_TypeI} have previously identified these bursts as being Type I X-ray\index{X-ray burst!Type I} bursts from another source in the \indexrxte\textit{RXTE} field of view.  To show that these unrelated Type I bursts would not be confused with Minibursts\index{Miniburst}, I add examples of the Type I bursts to lightcurves from observations containing Minibursts.  I find that the peak count rates in Type I bursts are roughly equal to the amplitude of the noise in the persistent\index{Persistent emission} flux in these observations, hence they would not be detected by my algorithms.
\par
\begin{figure}
  \centering
  \includegraphics[width=.9\linewidth, trim={9.5cm 0cm 10cm 0cm},clip]{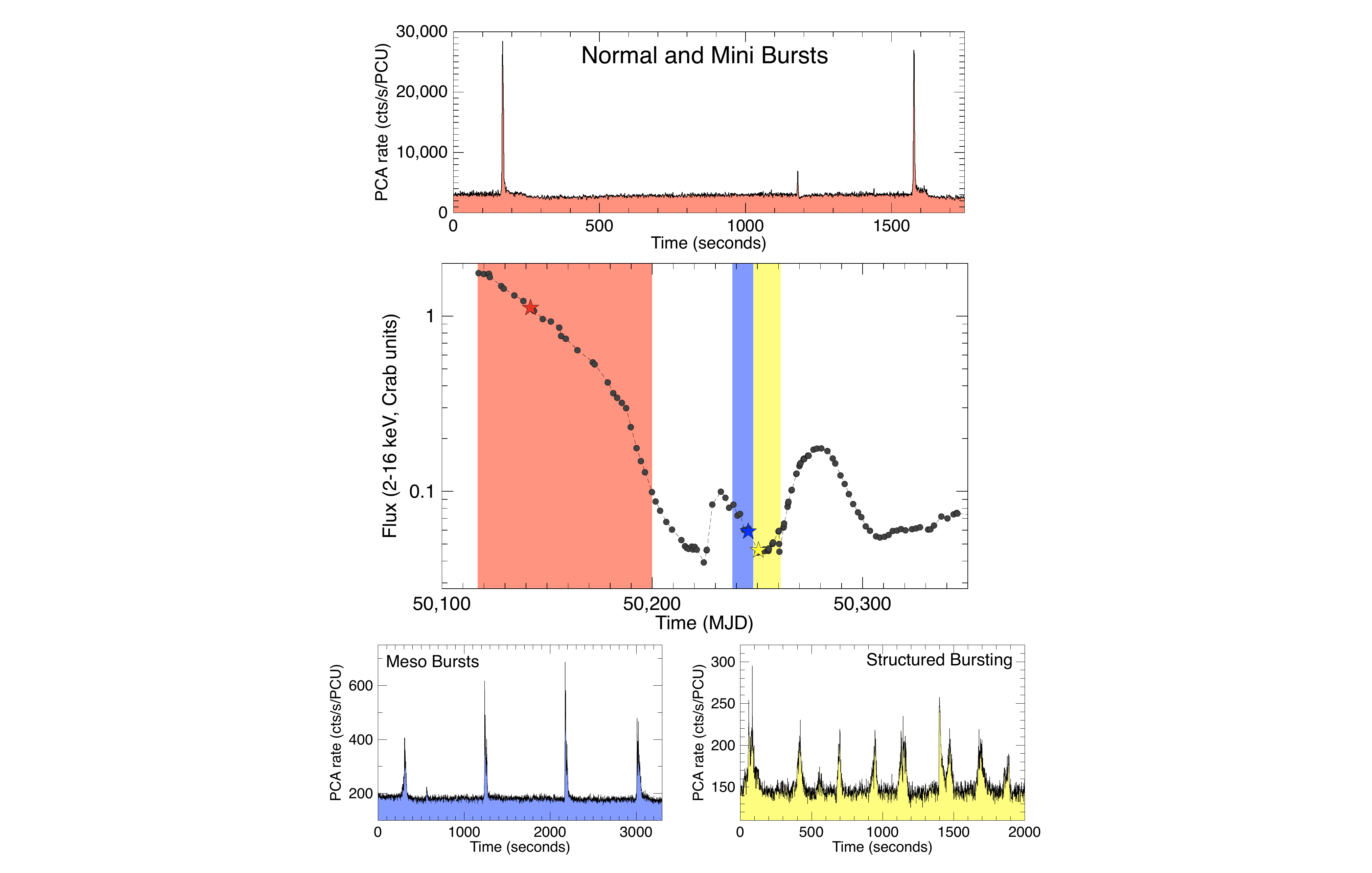}
  \caption[Lightcurve of the 1995--1996 outburst of the Bursting Pulsar, highlighting periods of time during which Mesobursts, Structured Bursts or Normal and Mini bursts are observed.]{\small Central panel shows the global 2--16\,keV \indexpca\textit{RXTE}/PCA lightcurve\index{Lightcurve} of the 1995--1996 outburst\index{Outburst} of the Bursting Pulsar\index{Bursting Pulsar}, highlighting periods of time during which Mesobursts\index{Mesoburst} (blue) Structured Bursts\index{Structured bursting} (yellow) or Normal\index{Normal burst} and Mini\index{Miniburst} bursts (red) are observed.  A single Mesoburst was also observed on MJD 50253, during the period of the outburst highlighted in yellow (see Figure \ref{fig:meso_in_struc}).  Other panels show example lightcurves which contain the aforementioned types of bursting\index{X-ray burst} behaviour.  See section \ref{sec:classes} for a detailed treatment of burst classification.  Fluxes reported in units of Crab.}
  \label{fig:ob_evo1}
\end{figure}

\begin{figure}
  \centering
  \includegraphics[width=.9\linewidth, trim={9.5cm 0cm 10cm 0cm},clip]{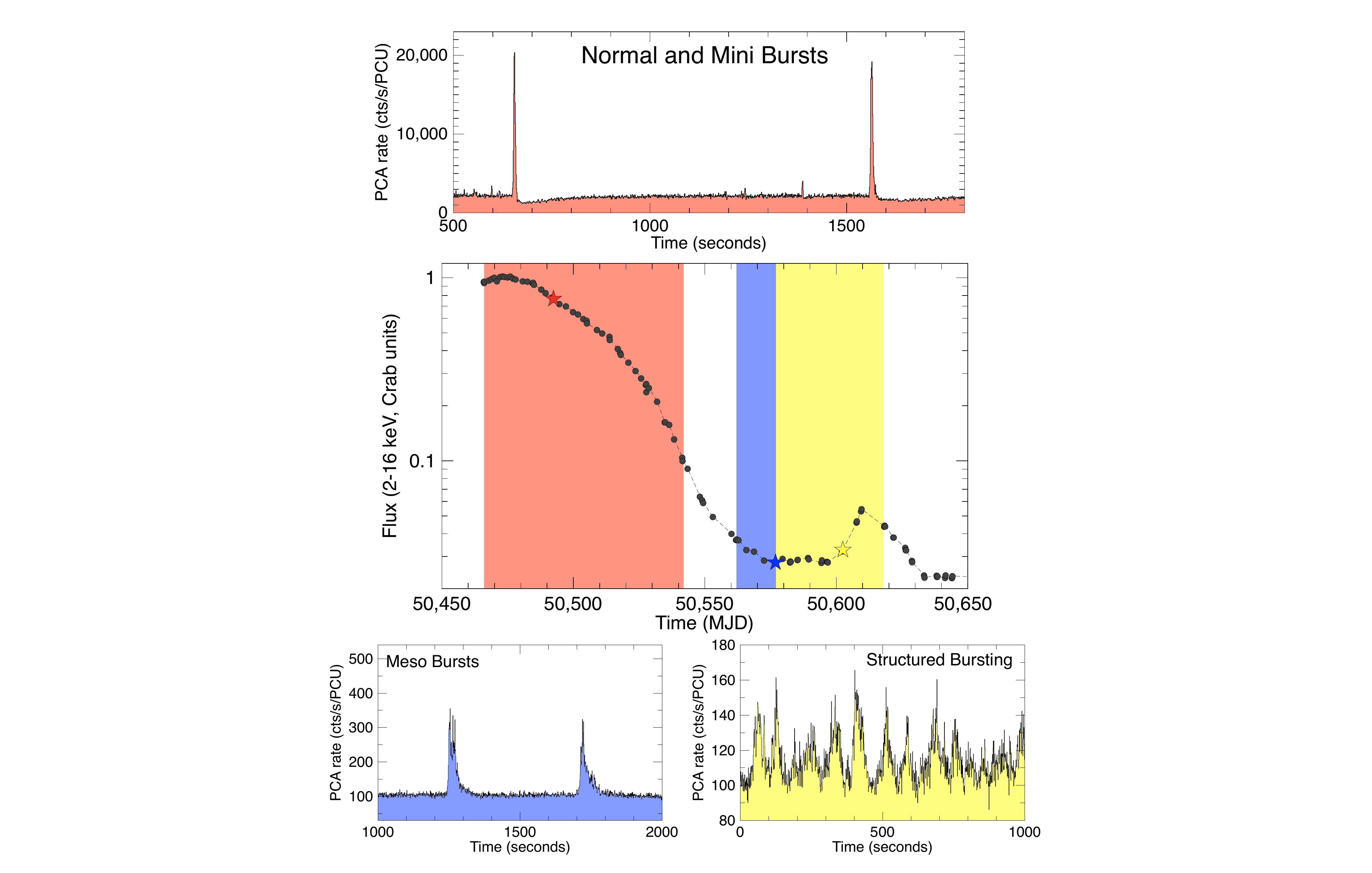}
  \caption[Lightcurve of the 1997--1999 outburst of the Bursting Pulsar, highlighting periods of time during which Mesobursts, Structured Bursts or Normal and Mini bursts are observed.]{\small Central panel shows the global 2--16\,keV \indexpca\textit{RXTE}/PCA lightcurve\index{Lightcurve} of the 1997--1999 outburst\index{Outburst} of the Bursting Pulsar\index{Bursting Pulsar}, highlighting periods of time during which Mesobursts\index{Mesoburst} (blue) Structured Bursts\index{Structured bursting} (yellow) or Normal\index{Normal burst} and Mini\index{Miniburst} bursts (red) are observed.  Other panels show example lightcurves which contain the aforementioned types of bursting behaviour\index{X-ray burst}.}
  \label{fig:ob_evo2}
\end{figure}

 I show when in Outbursts\index{Outburst} 1 \& 2 each type of burst was observed in Figures \ref{fig:ob_evo1} and \ref{fig:ob_evo2} respectively.  Normal Bursts\index{Normal burst} and Minibursts\index{Miniburst} (red) occur during the same periods of time from around the peak of an outburst until the persistent emission\index{Persistent emission} falls beneath $\sim0.1$\,Crab; assuming an Eddington Limit\index{Eddington limit} of $\sim1$\,Crab (e.g \citealp{Sazonov_BPGranat}), this corresponds to an Eddington ratio of $\sim0.1$.  After this point, bursting\index{X-ray burst} is not observed for a few tens of days.  Mesobursts\index{Mesoburst} (blue) begin at the end of a rebrightening\index{Re-flare} event in Outburst 1 and during the final days of the main part of the outburst in Outburst 2.  Structured Bursts\index{Structured bursting} (yellow) occur during the first part of a rebrightening event in both outbursts.  Although there was a second rebrightening event after Outburst 1, neither Mesobursts nor Structured Bursts were observed at this time.  Based on this separation, as well as differences in structure, I treat each class of burst separately below.

\subsection{Normal Bursts}

\label{sec:Normal_Bursts}

\par I define Normal Bursts\index{Normal burst} as the set of all bursts\index{X-ray burst} with a persistent-emission\index{Persistent emission}-subtracted peak 1\,s binned \indexpca\textit{RXTE}/PCA-equivalent count rate above 3000\,cts\,s$^{-1}$\,PCU$^{-1}$.  Normal Bursts account for 99 out of the 190\footnote{This number does not include Structured Bursts\index{Structured bursting} as their complex structure makes them difficult to separate.} bursts identified for this study.  They are observed during all three outbursts\index{Outburst} covered in this study.  They occurred between MJDs 50117 and 50200 in Outburst 1, and between 50466 and 50542 in Outburst 2; during these intervals, \indexrxte\textit{RXTE} observed the source for a total of 192\,ks.  See Table \ref{tab:staretimes} to compare these with numbers for the other classes of burst identified in this study.  Normal Bursts occur during the same time intervals in which Minibursts\index{Miniburst} are present.  In both of these outbursts, the period of Normal and Minibursts correspond to the time between the peak of the outburst and and the time that the persistent intensity falls below $\sim0.1$\,Crab.

\begin{table}
\centering
\begin{tabular}{llll}
\hline
\hline
\scriptsize  Bursting Mode &\scriptsize Bursts &\scriptsize Total Exposure (ks) &\scriptsize Duration (d) \\
\hline
Normal Bursts\index{Normal burst} & 99  & 192 & 76\\
Minibursts\index{Miniburst} & 48 & 192  & 76\\
Mesobursts\index{Mesoburst} & 43 &44 &25\\
Structured Bursts\index{Structured bursting} & - &80 &54 \\
\hline
\hline
\end{tabular}
\caption[Statistics on the population of bursts in the 1996 and 1997 outbursts of the Bursting Pulsar.]{Statistics on the population\index{Population study} of bursts\index{X-ray burst} I use for this study, as well as the duration and integrated \indexpca\textit{RXTE}/PCA exposure time of each mode of bursting.  All numbers are the sum of values for Outbursts\index{Outburst} 1 and 2.  As Normal\index{Normal burst} and Minibursts\index{Miniburst} happen during the same period of time in each outburst, the exposure time and mode duration for these classes of bursting are equal.}
\label{tab:staretimes}
\end{table}

\subsubsection{Recurrence Time}

\par
\begin{figure}
  \centering
  \includegraphics[width=.9\linewidth, trim={0.4cm 0 1.1cm 0},clip]{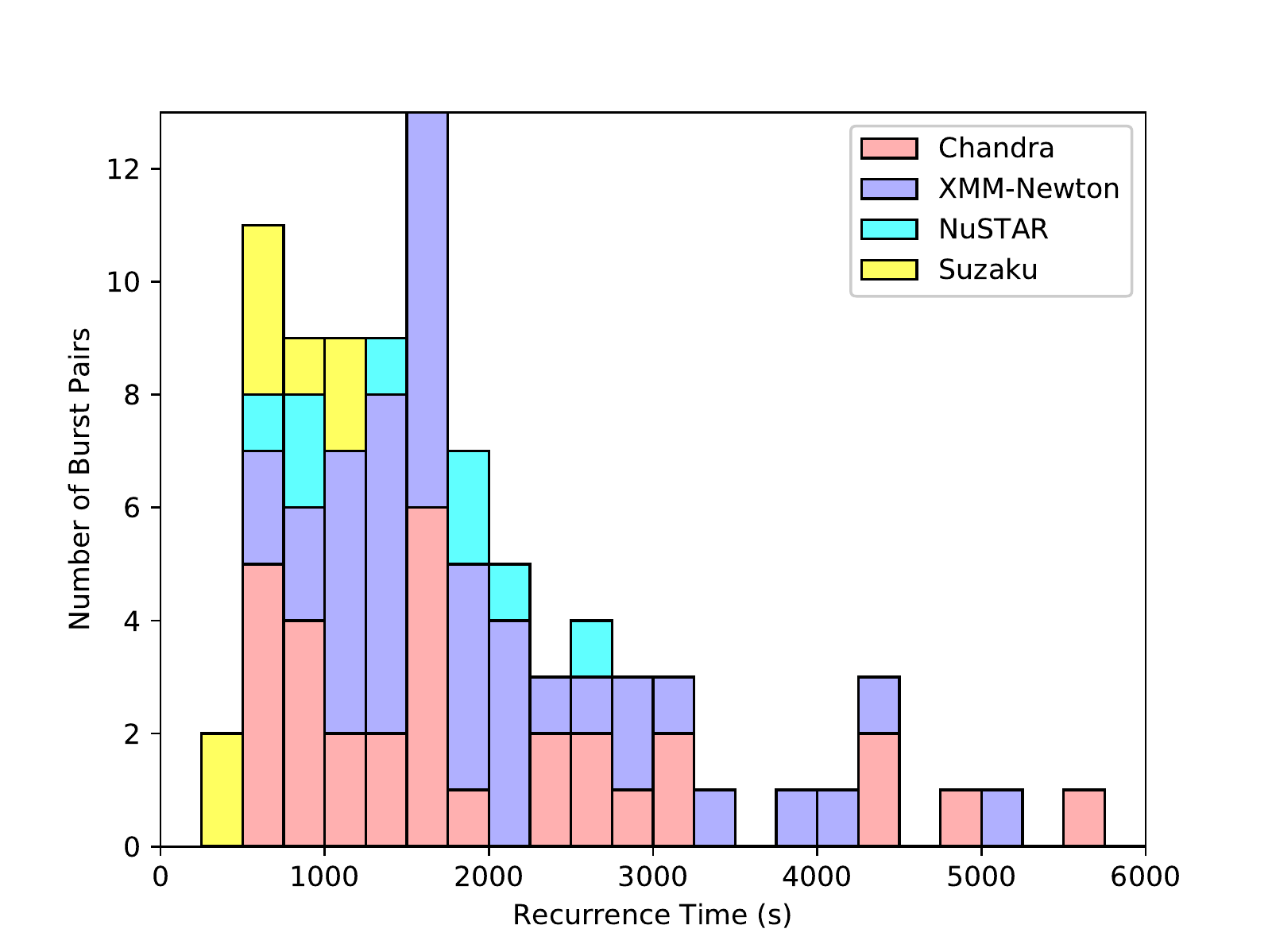}
  \caption[The distribution of recurrence times between consecutive Normal Bursts in Outburst 3.]{\small The distribution of recurrence times\index{Recurrence time} between consecutive Normal Bursts\index{Normal burst} seen in pointed \indexchandra\textit{Chandra}, \indexxmm\textit{XMM-Newton}, \indexnustar\textit{NuSTAR} and \indexsuzaku\textit{Suzaku} observations of Outburst\index{Outburst} 3 of the Bursting Pulsar\index{Bursting Pulsar}.  Distributions of bursts observed by different instruments are stacked on top of each other and colour coded.}
  \label{fig:sep}
\end{figure}

 Using Outburst\index{Outburst} 3 data from \indexchandra\textit{Chandra}, \indexxmm\textit{XMM-Newton}, \indexnustar\textit{NuSTAR} and \indexsuzaku\textit{Suzaku}, I find minimum and maximum Normal Burst\index{Normal burst} recurrence times\index{Recurrence time} of $\sim345$ and $\sim5660$\,s respectively\footnote{To avoid double-counting peak pairs, I do not use \textit{NuSTAR} observation 80002017004, which was taken simultaneously with \textit{Chandra} observation 16596.}.  I show the histogram of recurrence times from Outburst 3 in Figure \ref{fig:sep}, showing which parts of the distribution were observed with which observatory.  Compared to data from \textit{Chandra} and \textit{XMM-Newton}, data from \textit{Suzaku} generally suggests shorter recurrence times.  This is likely due to \textit{Suzaku} observations consisting of a number of $\sim2$\,ks windows; as this number is of the same order of magnitude as the recurrence time between bursts, there is a strong selection effect against high recurrence times in the \textit{Suzaku} dataset.
\par From the \indexrxte\textit{RXTE} data I find minimum and maximum Normal Burst\index{Normal burst} recurrence times\index{Recurrence time} of $\sim250$ and $\sim2510$\,s during Outburst\index{Outburst} 1, and minimum and maximum recurrence times of $\sim250$ and $\sim2340$\,s during Outburst 2.  As the length of an \textit{RXTE} pointing ($\lesssim3$\,ks) is also of the same order of magnitude as the recurrence time between bursts, selection effects bias us against sampling pairs of bursts with longer recurrence times, and hence this upper value is likely an underestimate.

\par To test whether consecutive Normal Bursts\index{Normal burst} are independent events, I tested the hypothesis that bursts are randomly distributed in time in a Poisson distribution\index{Poisson distribution} \citep{Poisson_Distribution}.  Assuming my hypothesis, as well as assuming that the frequency of Normal Bursts does not change during an outburst\index{Outburst} (e.g. \citealp{Aptekar_Recur}), I could concatenate different observations and the resultant distribution of burst times should still be Poissonian.  For each of Outbursts 1 \& 2, I concatenated all \indexrxte\textit{RXTE} data during the Normal Bursting part of the outburst into a single lightcurve\index{Lightcurve}.  I split this concatenated lightcurve into windows of length $w$ and counted how many bursts were in each, forming a histogram of number of bursts per window for the combined set of all bursts.  I fit this histogram with a Poisson probability density function, obtaining the value $\lambda$ which is the mean number of bursts in a time $w$.  $\lambda/w$ is therefore an expression of the true burst frequency per unit time, and should be independent of my choice of $w$.  I tried values of $w$ between 100 and 10000\,s for both outbursts, and found that in all cases $\lambda/w$ depends strongly on $w$.  Therefore my assumptions cannot both be valid, and I rejected the hypothesis that these bursts are from a Poisson distribution with constant $\lambda$.  This in turn suggests at least one of the following must be correct:
\begin{enumerate}
\item The average recurrence time\index{Recurrence time} of bursts was not constant throughout the outburst.  Or:
\item The arrival time of a given burst depends on the arrival time of the preceding burst, and therefore bursts are not independent events.
\end{enumerate}

\subsubsection{Burst Structure}

\label{sec:struc}

\par
\begin{figure}
  \centering
  \includegraphics[width=.9\linewidth, trim={0.4cm 0 1.1cm 0},clip]{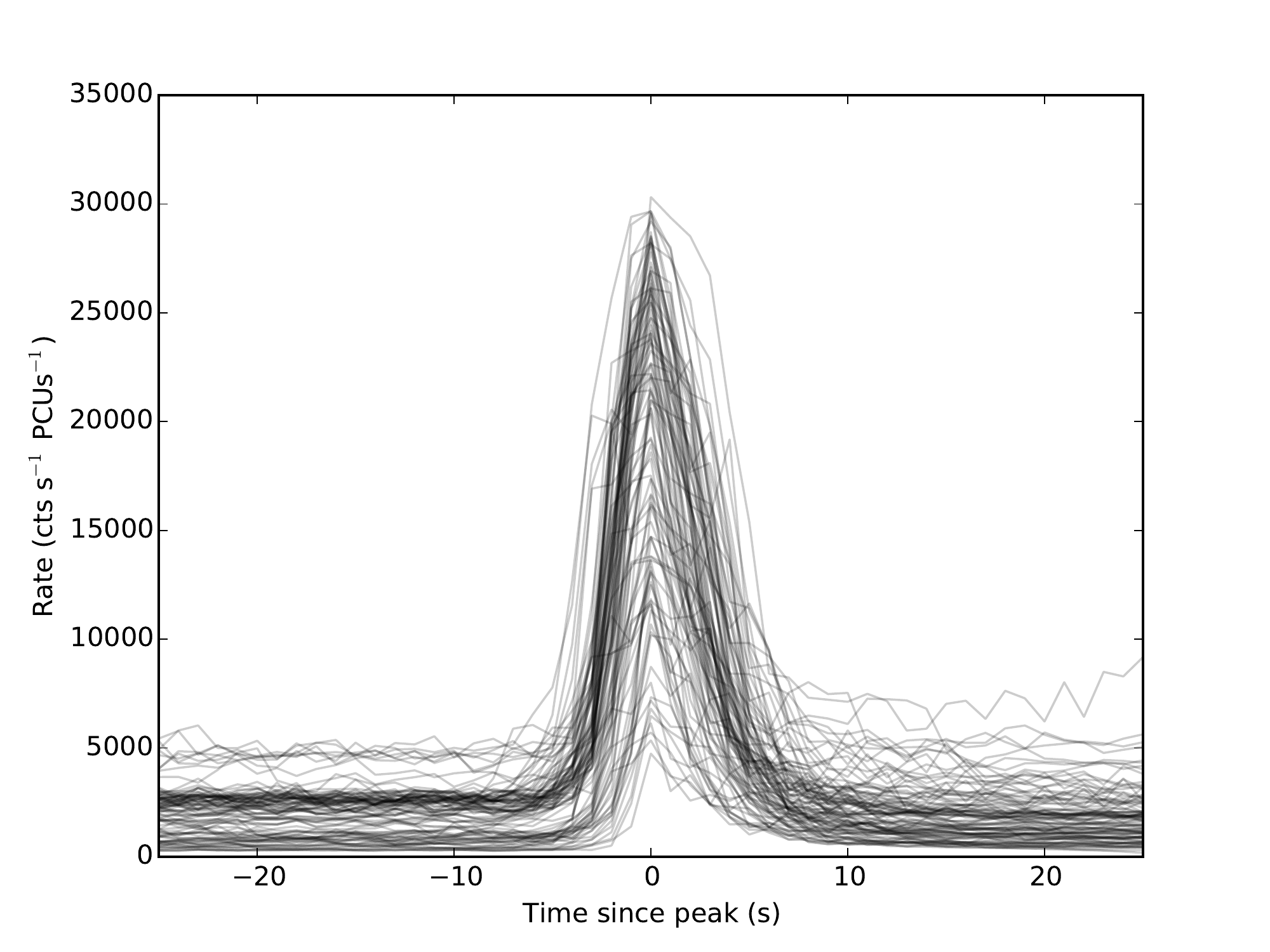}
  \includegraphics[width=.9\linewidth, trim={0.4cm 0 1.1cm 0},clip]{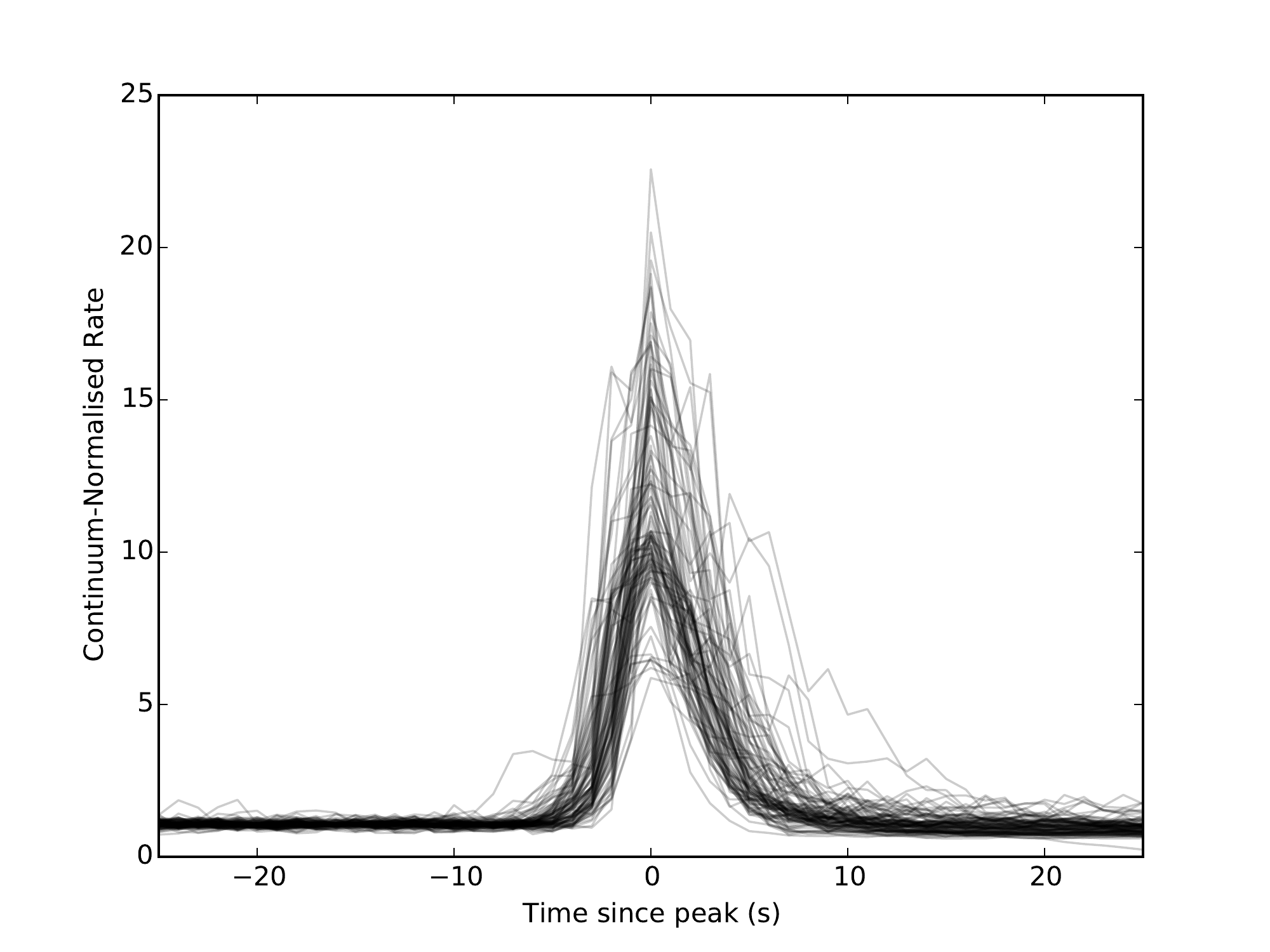}
  \caption[A plot of every Normal Burst, centred by the time of its peak, overlaid on top of each other to show the existence of a common pulse profile.]{\small \textbf{Top:} a lightcurve\index{Lightcurve} of every Normal Burst\index{Normal burst}, centred by the time of its peak, overlaid on top of each other to show the existence of a common pulse profile.  \textbf{Bottom:} a lightcurve of every Normal Burst in which count rates have been normalised by the persistent emission\index{Persistent emission} count rate during the observation from which each burst was observed.  As the bursts are on average closer to the average pulse profile in this metric, this suggests that the intensity of a burst is roughly dependent on the persistent emission rate.  Some persistent emission-normalised count rates may be artificially low due to dead-time\index{Dead-time} effects.}
  \label{fig:norm_overlay}
\end{figure}

 In the top panel of Figure \ref{fig:norm_overlay} I show a plot of all Normal Bursts\index{Normal burst} observed with \indexrxte\textit{RXTE} overlayed on top of one another.  I find that all Normal Bursts follow a similar burst profile with similar rise and decay timescales but varying peak intensities.  In the lower panel of Figure \ref{fig:norm_overlay} I show a plot of Normal Bursts overlaid on top of each other after being normalised by the persistent emission\index{Persistent emission} count rate in their respective observation.  The bursts are even closer to following a single profile in this figure, suggesting a correlation between persistent emission level in an outburst and the individual fluence of its bursts.

\par The structure of the lightcurve\index{Lightcurve} of a Normal Burst\index{Normal burst} can be described in three well-defined parts:

\begin{enumerate}
\item The main burst: roughly approximated by a skewed Gaussian\index{Gaussian profile} (see e.g. \citealp{Azzalini_Dist}).
\item A `plateau'\index{Plateau}: a period of time after the main burst during which count rate remains relatively stable at a level above the pre-burst rate.
\item A `dip'\index{Dip}: a period during which the count rate falls below the persistent level, before exponentially decaying back up towards the pre-burst level (e.g. \citealp{Younes_Expo}).
\end{enumerate}

\par
\begin{figure}
  \centering
  \includegraphics[width=.9\linewidth, trim={0.8cm 0 1.6cm 0},clip]{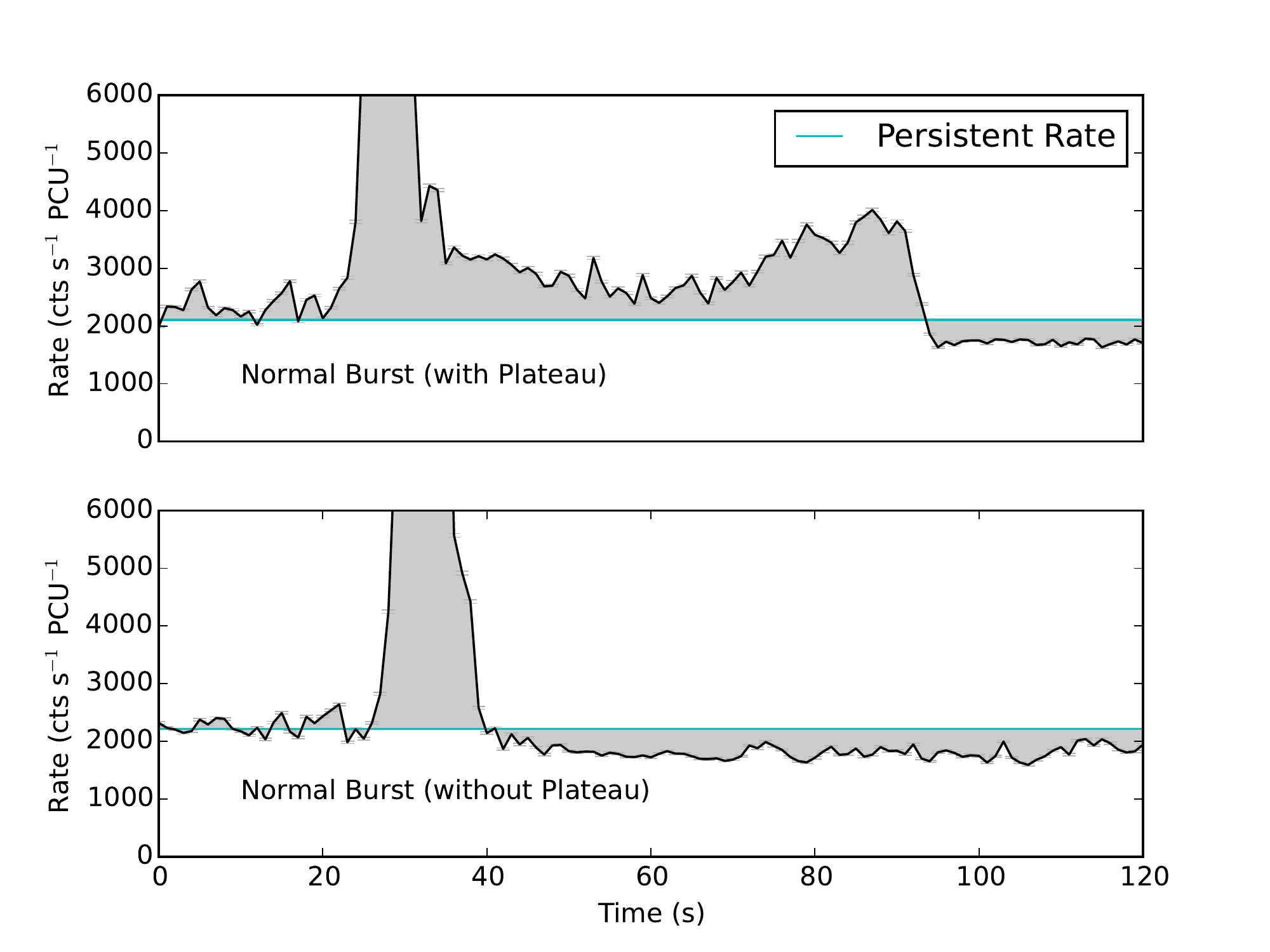}
  \caption[\textit{RXTE} lightcurves of Normal Bursts with (top) and without (bottom) `plateau' features, showing the burst structure in each case.]{\small \indexrxte\textit{RXTE} lightcurves\index{Lightcurve} of Normal Bursts\index{Normal burst} with (top) and without (bottom) `plateau'\index{Plateau} features, showing the burst structure in each case.  The median count rate, which I use as a proxy for the persistent emission\index{Persistent emission}, is plotted in cyan to highlight the presence of the count rate `dip'\index{Dip} after each burst.}
  \label{fig:w_wo}
\end{figure}

 The dip\index{Dip} is present after every Normal Burst\index{Normal burst} in my \indexrxte\textit{RXTE} sample from Outbursts\index{Outburst} 1 \& 2, whereas the plateau\index{Plateau} is only seen in 39 out of 99.  I show example lightcurves\index{Lightcurve} of bursts with and without plateaus in Figure \ref{fig:w_wo}, which also show that the dip is present in both cases.

\par
\begin{figure}
  \centering
  \includegraphics[width=.9\linewidth, trim={1.9cm 0 2.0cm 0},clip]{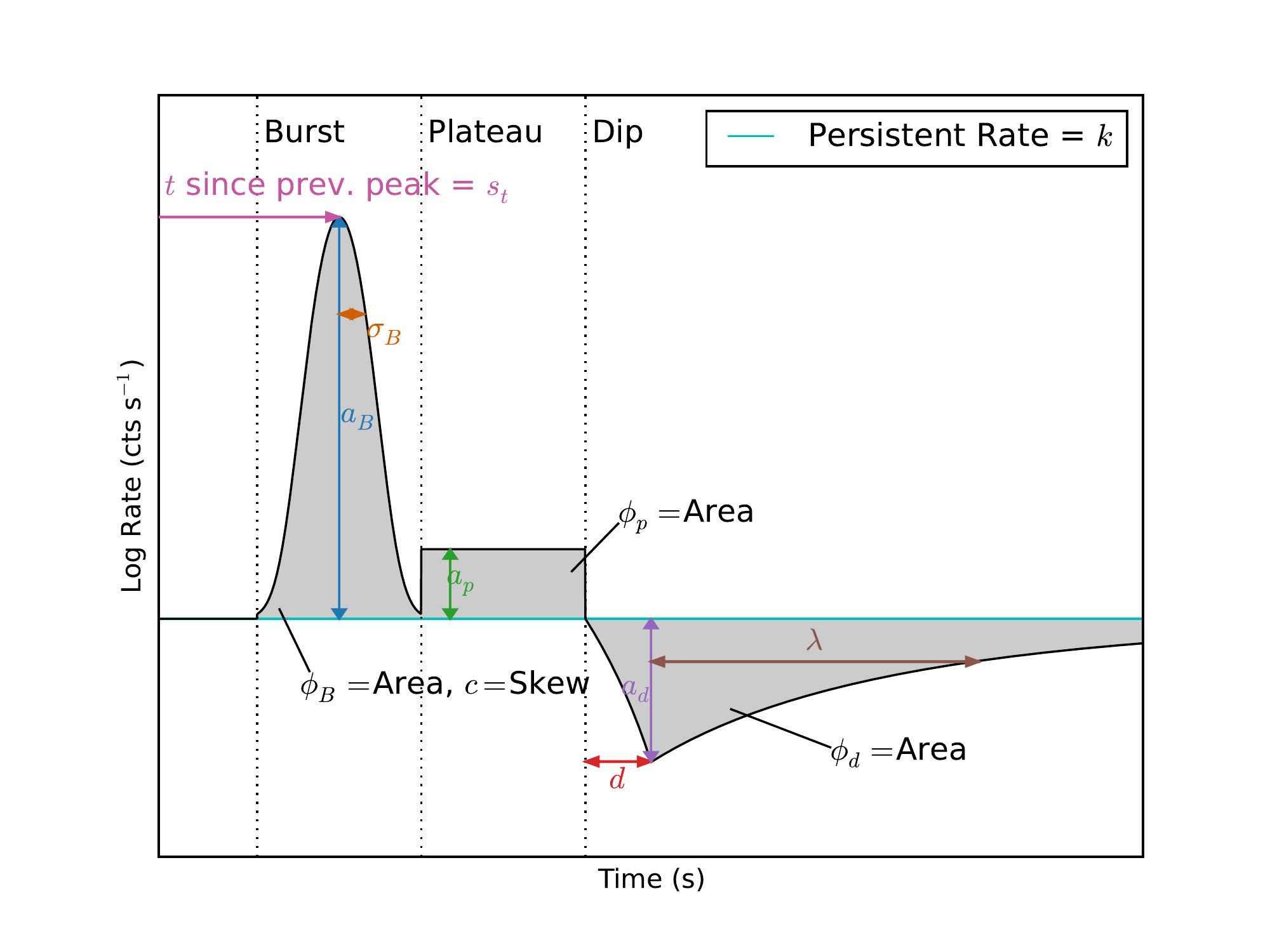}
  \caption[A schematic explaining the origin of the 12 Normal Burst parameters used in this study.]{\small A schematic explaining the origin of the 12 Normal Burst\index{Normal burst} parameters used in this study, as well as showing the functional forms of both the skewed Gaussian\index{Gaussian profile} fit to a burst and the `dipper function'\index{Dipper function} (Equation \ref{eq:dipper}) fit to a dip\index{Dip}.  Note that I do not fit a function to the plateau\index{Plateau}, and I calculate its fluence by summing the persistent rate\index{Persistent emission}-subtracted counts.  Diagram is for explanation only and the burst pictured is neither based on real data nor to scale.}
  \label{fig:explain}
\end{figure}

 In order to study Normal Bursts\index{Normal burst}, I fit the burst profiles with phenomenologically-motivated mathematical functions.  In Figure \ref{fig:explain} I show a schematic plot of my model, as well as annotations explaining the identities of the various parameters I use.  I fit the main burst with a skewed Gaussian\index{Gaussian profile}, centred at $t=x_0$ with amplitude $a_b$, standard deviation $\sigma_B$ and skewness\footnote{A measure of how far the peak of the Gaussian is displaced from its centre.} $c$, added to the persistent emission\index{Persistent emission} rate $k$.  I fit the `dip'\index{Dip} with the continuous piecewise function `Dipper function'\index{Dipper function} $f(t)$:

\begin{equation}
f(t)=
\begin{dcases}
k-\frac{a_d(t-t_0)}{d-t_0}, & \text{if } t\leq d\\
k-a_d\exp\left(\frac{d-t}{\lambda}\right), & \text{otherwise}
\end{dcases}
\label{eq:dipper}
\end{equation}

Where $t$ is time, $t_0$ is the start time of the dip\index{Dip}, $a_d$ is the amplitude of the dip, $d$ is the time at the local dip minimum and $\lambda$ is the dip recovery timescale.  This function is based on the finding by \citet{Younes_Expo} that dip count rates recover exponentially, but has the added advantage that the start of the recovery phase can also be fit as an independent parameter.  Using this fit, I can estimate values for burst fluence $\phi_B$, burst scale-length $\sigma_B$, `missing' dip fluence $\phi_D$ and dip scale-length $\lambda$ and compare these with other burst parameters.  When present, I also calculate the fluence of the plateau\index{Plateau} $\phi_p$ by summing the persistent emission\index{Persistent emission}-subtracted counts during the region between the end of the burst (as defined in Section \ref{sec:burst_diff}) and the start of the dip.  For each pair of parameters, I do not consider datapoints when the magnitude of the error on a parameter is greater than the value of the parameter.

\par I only extract these parameters from Normal Bursts\index{Normal burst} observed by \indexrxte\textit{RXTE} during Outbursts\index{Outburst} 1 \& 2.  This ensures that the resultant parameter distributions I extracted are not affected by differences between instruments.

\subsubsection{Parameter Distributions}

\label{sec:hists}

\par I extracted a total of ten parameters from my fit to each burst\index{Normal burst}: the parameters $a_d$, $d$ and $\lambda$ of the fit to the dip\index{Dip}, the missing fluence $\phi_D$ of the dip, the parameters $a_b$, $\sigma_B$ and $c$ of the skewed Gaussian\index{Gaussian profile} fit to the main burst, the main burst fluence $\phi_B$, the maximum persistent emission-subtracted\index{Persistent emission} rate in the plateau\index{Plateau} $a_p$ and the plateau fluence $\phi_P$.
\par Using my \indexrxte\textit{RXTE} sample of Normal Bursts\index{Normal burst}, I can construct distributions for all of the burst parameters described in Section \ref{sec:struc} for bursts in Outbursts\index{Outburst} 1 \& 2.  I give the mean and standard deviation for each parameter in each outburst in Table \ref{tab:params_perob}, and histograms for each can be found in Appendix \ref{app:hists}.

\begin{table}
\centering
\begin{tabular}{r c c c c c c}
\hline
\hline
 & \multicolumn{2}{c}{\scriptsize Outburst 1} & \multicolumn{2}{c}{\scriptsize Outburst 2} & \multicolumn{2}{c}{\scriptsize Outbursts 1\&2}  \\
 &Mean&S.D.&Mean&S.D.&Mean&S.D.\\
\hline
$\phi_B$&2.74e6&7.8e5&2.25e6&7.6e5&$2.43\mathrm{e}6$&$8.0\mathrm{e}5$\\
$a_B$&3.18e5&8.4e4&2.72e5&9.9e4&$2.90\mathrm{e}5$&$9.6\mathrm{e}4$\\
$\sigma_B$&3.39&0.35&3.42&0.59&3.41&0.52\\
$c$&2.68&1.9&2.79&2.0&2.75&2.0\\
$\phi_d$&1.74e6&1.3e6&1.17e6&3.6e5&$1.38\mathrm{e}6$&$8.7\mathrm{e}5$\\
$a_d$&550&335&536&307&541&318\\
$d$&49&46&20&22&31&36\\
$\lambda$&294&176&229&124&254&150\\
$\phi_p$&1.89e5&2.3e5&7577&5707&1.4e5&1.8e5\\
$a_p$&1289&1113&767&463&1063&928\\
\hline
\hline
\end{tabular}
\caption[A table showing the mean and standard deviation of 10 Normal Burst parameters of \textit{RXTE}-sampled bursts.]{A table showing the mean and standard deviation of 10 Normal Burst\index{Normal burst} parameters of \indexrxte\textit{RXTE}-sampled bursts.  In each case, I give the values for populations from only Outburst\index{Outburst} 1, from only Outburst 2 and from the combined population from both outbursts.  Histograms for each parameter can be found in Appendix \ref{app:hists}.}
\label{tab:params_perob}
\end{table}

\par The mean value of most parameters differs by no more than $\sim50$\% between outbursts\index{Outburst}.  Notable exceptions are $d$, $\phi_p$, $\phi_d$ and $a_p$, which are $\sim2.5$, $\sim2.5$ $\sim1.5$ and $\sim1.7$ times greater in Outburst 1 than in Outburst 2 respectively.  The less significant differences between values of $\phi_B$ and $a_B$ in Outbursts 1 \& 2 are expected, as the amplitude of a burst correlates with persistent rate\index{Persistent emission} $k$ which was generally higher in Outburst 1 than in Outburst 2.

\subsubsection{Correlations}

\label{sec:NormCorr}

\par In total, I extracted 12 parameters for each Normal Burst\index{Normal burst} in my \textit{RXTE} sample: the 10 burst parameters listed in Section \ref{sec:hists}, the recurrence time\index{Recurrence time} $s_t$ until the next burst and the persistent emission\index{Persistent emission} rate $k$ at the time of the burst.
\par As the amplitude of all 3 components in a burst scale with the persistent emission\index{Persistent emission} level, I rescaled my values of $a_b$, $a_d$, $\phi_B$, $\phi_D$ and $\phi_P$ by a factor $\frac{1}{k}$.  I show the covariance matrix with all 66 possible pairings of these normalised parameters in Figure \ref{fig:corr_n} (we present the covariance matrix of these parameters before being rescaled in Appendix \ref{app:corr}).  Using the Spearman's Rank Correlation Coefficient\index{Spearman's rank correlation coefficient}, I find the following $\geq5\,\sigma$ correlations which are highlighted in Figure \ref{fig:corr_n}:

\begin{itemize}
\item Persistent emission\index{Persistent emission} $k$ anticorrelates with normalised burst fluence $\phi_B/k$ ($>10\,\sigma$) and normalised burst amplitude $a_b/k$ ($>10\,\sigma$).
\item Normalised burst fluence $\phi_B/k$ correlates with normalised burst amplitude $a_B/k$ ($8.0\,\sigma$).
\item Normalised dip fluence\index{Dip} $\phi_d/k$ correlates with dip recovery timescale $\lambda$ ($6.3\,\sigma$).
\item Normalised dip amplitude $a_d/k$ anticorrelates with dip falltime $d$ ($5.7\,\sigma$) and dip recovery timescale $\lambda$ ($7.1\,\sigma$).
\item Normalised plateau\index{Plateau} fluence $\phi_p/k$ correlates with normalised plateau amplitude $a_p$ ($6.4\,\sigma$).
\end{itemize}

As $\phi_B$ can be approximated to first order as a product of $a_B$ and $\sigma$, the correlation between $\phi_B$ and $a_B$ is expected as they are not independent parameters.  Similarly, the correlations between $\phi_d$ \& $\lambda$ and $\phi_p$ and $a_p$ are likely due to these pairs of parameters not being independent.

\begin{figure}
  \centering
  \includegraphics[width=\linewidth, trim={2.1cm 2cm 3.5cm 3cm},clip]{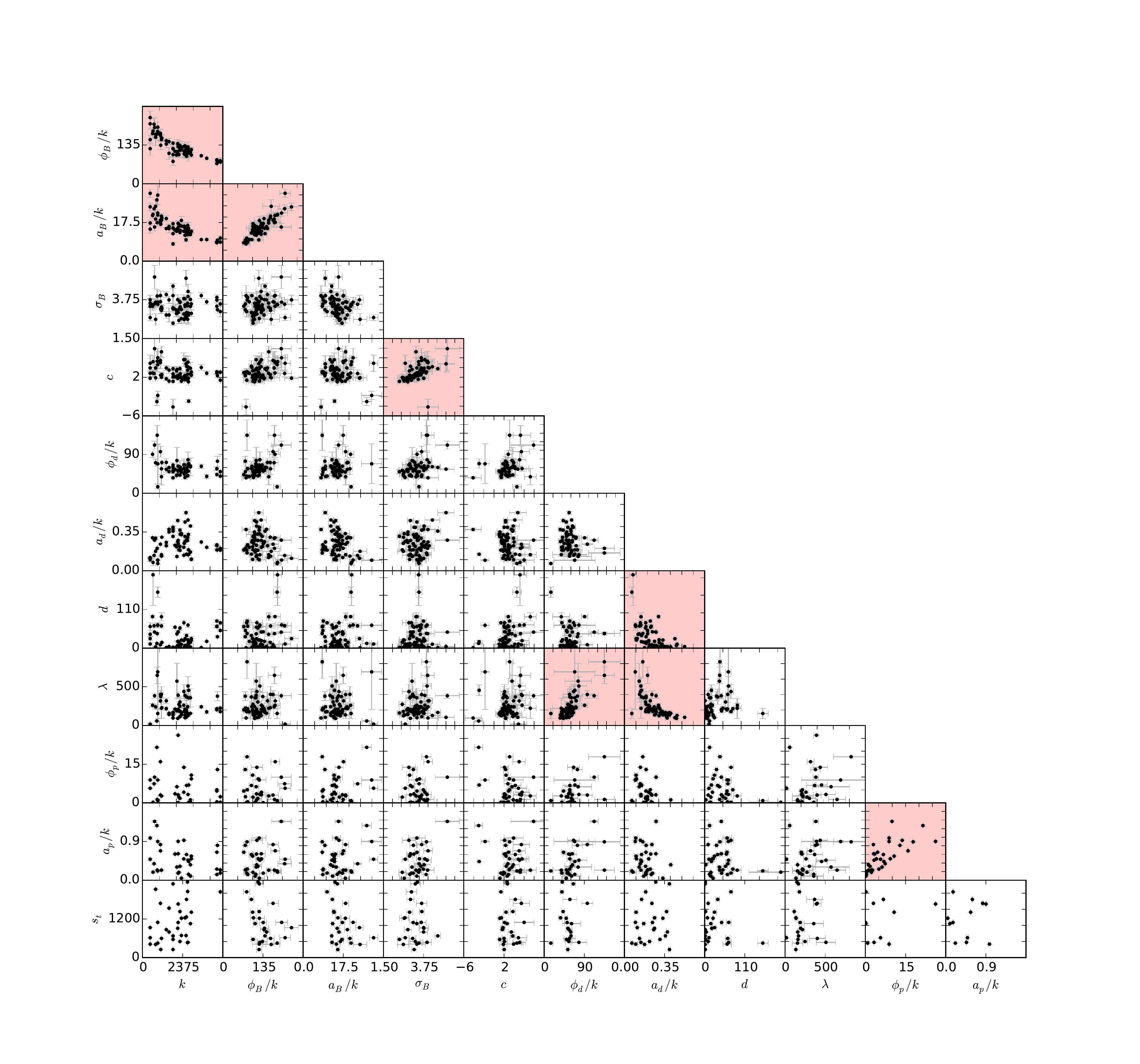}
  \caption[Covariance Matrix with a scatter plot of each pairing of the 12 normalised Normal Burst parameters listed in section \ref{sec:NormCorr}.]{\small Covariance Matrix with a scatter plot of each of the 66 pairings of the 12 Normal Burst\index{Normal burst} parameters listed in section \ref{sec:NormCorr}.  Amplitudes and fluences have been normalised by dividing by the persistent emission\index{Persistent emission} rate $k$.  Pairings which show a correlation using the Spearman Rank\index{Spearman's rank correlation coefficient} metric with a significance $\geq5\,\sigma$ are highlighted in red.}
  \label{fig:corr_n}
\end{figure}

\subsubsection{Colour Evolution}

\par To explore the spectral\index{Spectroscopy} behaviour of Normal Bursts\index{Normal burst}, Toyah Overton (\textsf{T.O.}) and I studied the evolution of the hardness\index{Colour} (the ratio between count rate in the energy bands $\sim2$--$7$ and $\sim8$--$60$\,keV energy bands) as a function of count rate during the individual bursts.  Plotting hardness-intensity diagrams\index{Hardness-intensity diagram} allow us to check for spectral evolution in a model-independent way.  We do not correct them for background\index{Background subtraction} as the count rates in both bands are very high.
\par
\begin{figure}
  \centering
  \includegraphics[width=.9\linewidth, trim={0.4cm 1cm 1.1cm 1cm},clip]{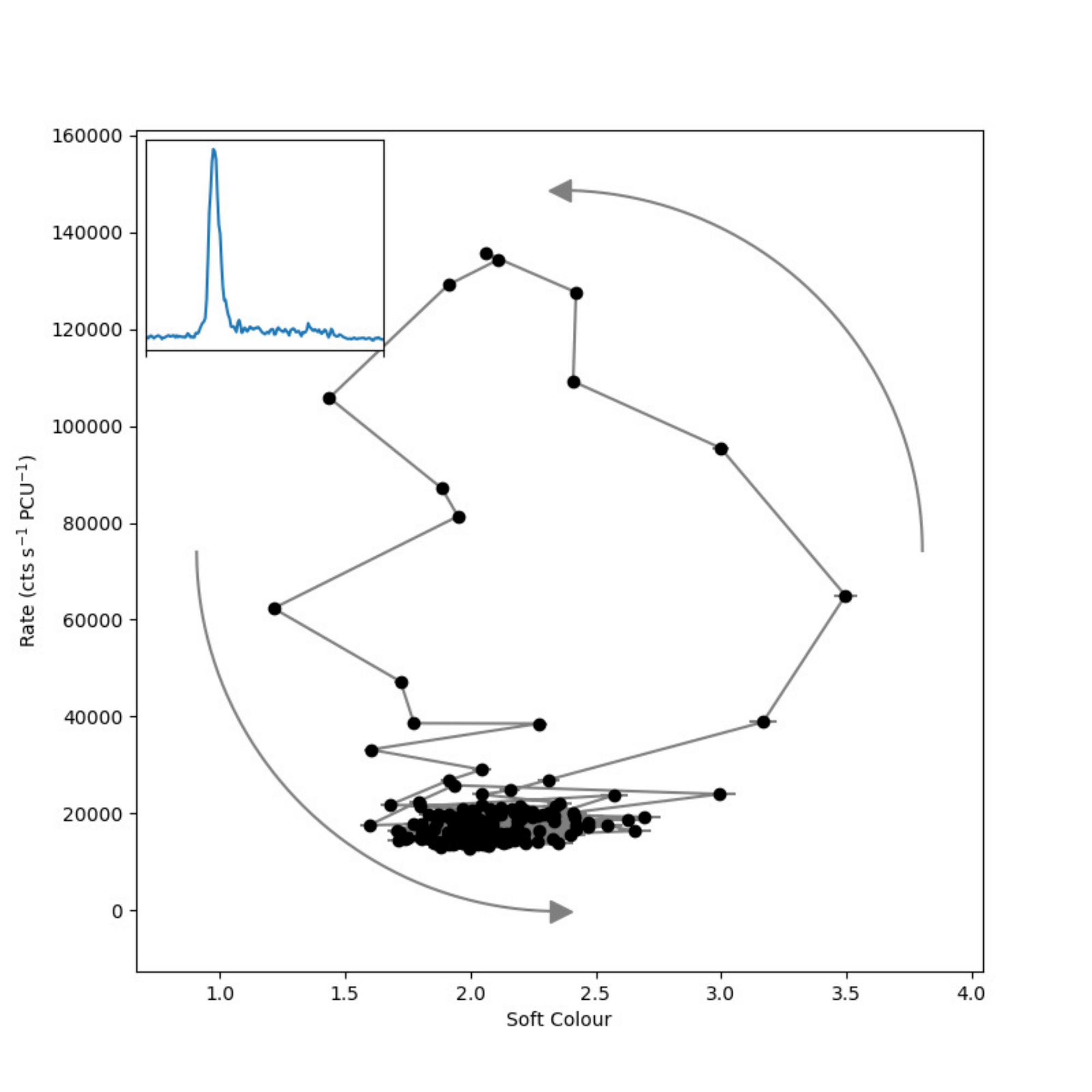}
  \caption[A hardness-intensity diagram of a typical Normal Burst.]{\small A 1\,s-binned hardness-intensity diagram\index{Hardness-intensity diagram} of a Normal Burst\index{Normal burst} from \indexpca\textit{RXTE}/PCA observation 10401-01-08-00, with an inset 2--60\,keV lightcurve\index{Lightcurve}.  Significant colour\index{Colour} evolution can be seen during the burst, taking the form of a loop\index{Hysteresis}.}
  \label{fig:loop}
\end{figure}

 \textsf{T.O.} and I find evidence of hysteretic\index{Hysteresis} loops in hardness-intensity\index{Hardness-intensity diagram} space in some, but not all, of the Normal Bursts\index{Normal burst} in my sample; see Figure \ref{fig:loop} for an example of such a loop.  The existence of such a loop suggests significant spectral\index{Spectroscopy} evolution throughout the burst.  This finding can be contrasted with results from previous studies in different energy bands (e.g. \citealp{Woods_OB2} from $\sim25$--100\,keV) which suggested no spectral evolution during Type II bursts\index{X-ray burst!Type II} in this source.

\subsection{Minibursts}

\label{sec:Minibursts}

\par I define Minibursts\index{Miniburst} as the set of all bursts\index{X-ray burst} with a peak 1\,s binned \indexpca\textit{RXTE}/PCA-equivalent count rate of $<300\%$ of the persistent rate\index{Persistent emission}.  Minibursts account for 48 out of the 190 bursts identified for this study.  They are observed during all 3 Outbursts\index{Outburst}, and occur during the same times that Normal Bursts\index{Normal burst} are present.  Minibursts occurred between MJDs 50117 and 50200 in Outburst 1, and between 50466 and 50542 in Outburst 2; during these intervals, \indexrxte\textit{RXTE} observed the source for a total of 192\,ks.  These intervals correspond to the times between the peak of each outburst and and the time that the persistent intensity falls below $\sim0.1$\,Crab.

\subsubsection{Recurrence Time}

\par There are only 10 observations with \indexrxte\textit{RXTE} which contain multiple Minibursts\index{Miniburst}.  Using these, I find minimum and maximum Miniburst recurrence\index{Recurrence time} times of 116 and 1230\,s.
\par I find 17 \indexrxte\textit{RXTE} observations which contain both a Miniburst\index{Miniburst} and a preceding Normal Burst\index{Normal burst}, and find minimum and maximum Normal Burst $\rightarrow$ Miniburst recurrence times\index{Recurrence time} of 461 and 1801\,s.

\subsubsection{Structure}

\par
\begin{figure}
  \centering
  \includegraphics[width=.9\linewidth, trim={0cm 0 0cm 0},clip]{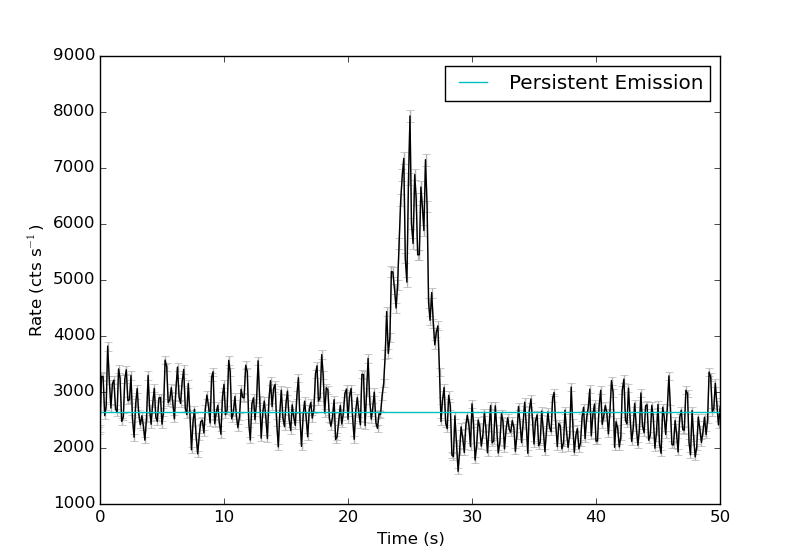}
  \caption[A representative \textit{RXTE} lightcurve of a Miniburst.]{\small  A representative \indexpca\textit{RXTE}/PCA lightcurve\index{Lightcurve} of a Miniburst\index{Miniburst} from OBSID 20077-01-03-00 in Outburst\index{Outburst} 2.}
  \label{fig:a_mini}
\end{figure}

\begin{figure}
  \centering
  \includegraphics[width=.9\linewidth, trim={0.4cm 0 1.1cm 0},clip]{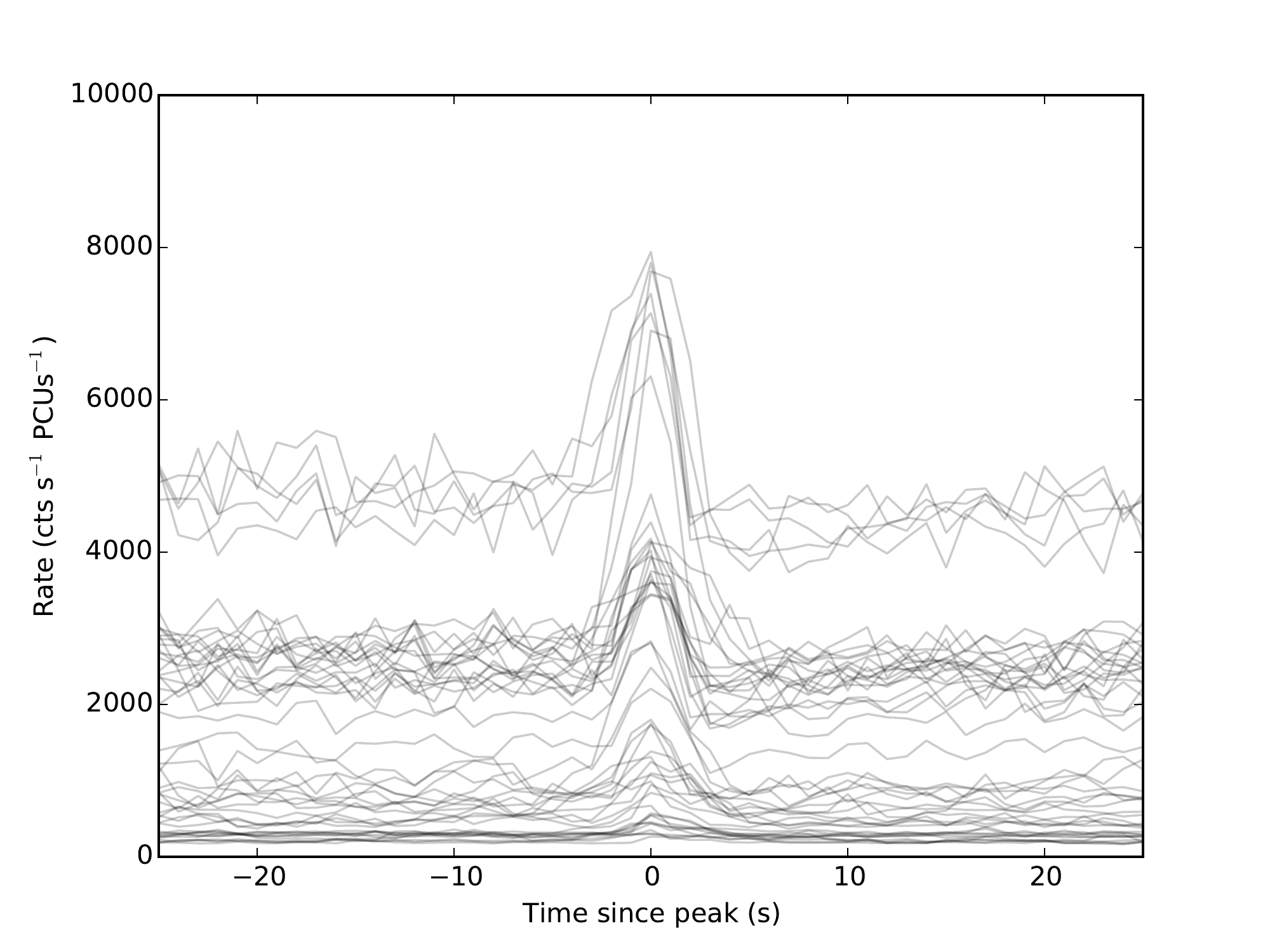}
  \includegraphics[width=.9\linewidth, trim={0.4cm 0 1.1cm 0},clip]{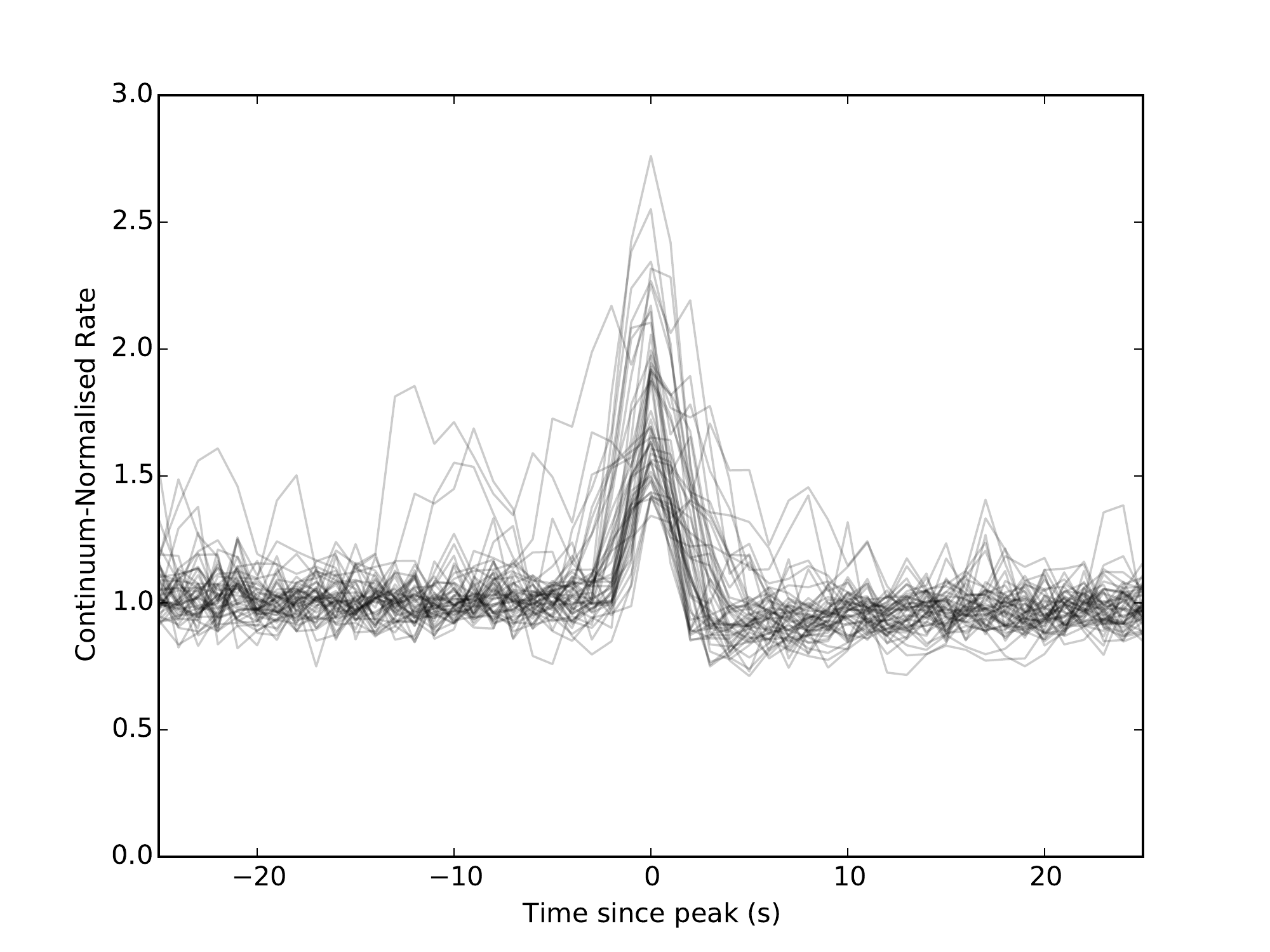}
  \caption[A plot of every Miniburst, centred by the time of its peak, overlaid on top of each other.]{\small  \textbf{Top:} a plot of every Miniburst\index{Miniburst}, centred by the time of its peak, overlaid on top of each other.  \textbf{Bottom:} a plot of every Miniburst in which count rates have been normalised by the persistent emission\index{Persistent emission} count rate during the observation from which each burst was observed.}
  \label{fig:mini_over}
\end{figure}

 In Figure \ref{fig:a_mini}, I show the lightcurve\index{Lightcurve} of a representative Miniburst\index{Miniburst}, and I show all Minibursts overplotted on each other in Figure \ref{fig:mini_over}.  These bursts are roughly Gaussian\index{Gaussian profile} in shape with a large variation in peak count rate; as can be seen in Figure \ref{fig:mini_over}, however, the persistent\index{Persistent emission}-normalised peak count rates of Minibursts are all roughly consistent with 2.


\par Minibursts\index{Miniburst} are all $\sim5$\,s in duration, and some show signs of a `dip'\index{Dip} feature similar to those seen in Normal Bursts\index{Normal burst}.  I find that the timescales of these dips are all $\lesssim10$\,s.  I estimate `missing' fluence in each dip by integrating the total persistent-rate\index{Persistent emission}-subtracted counts between the end of the burst and a point 10\,s later.  If this `missing fluence' is less than half of the standard deviation in count rate multiplied by 5\,s, which represents the smallest $<10$\,s triangle-shaped dip which would be detectable above noise in a given dataset, I treat the dip in that outburst as not being detected.
\par Due to the relatively short duration and low amplitudes of Minibursts\index{Miniburst}, I am unable to reliably discern whether they contain a single peak or multiple peaks.  For this reason I do not fit them mathematically.

\subsubsection{Parameters \& Correlations}

\label{sec:ministruc}

\par For each Miniburst\index{Miniburst}, I extract the following parameters:

\begin{itemize}
\item Total burst fluence and burst fluence divided by persistent emission\index{Persistent emission}.
\item Peak 1\,s binned rate and peak rate divided by persistent emission.
\item Rise time, fall time and total time.
\end{itemize}
The mean and standard deviation of each of these parameters, calculated from \indexrxte\textit{RXTE} data, is presented in Table \ref{tab:mini_param} for Outburst\index{Outburst} 1, Outburst 2 and the combined population of Minibursts from Outbursts 1 \& 2.  The standard deviations on the fluence and peak rates of Minibursts are very large, suggesting that these parameters are distributed broadly.

\begin{table}
\centering
\begin{tabular}{r c c c c c c}
\hline
\hline
 & \multicolumn{2}{c}{\scriptsize Outburst 1} & \multicolumn{2}{c}{\scriptsize Outburst 2} & \multicolumn{2}{c}{\scriptsize Outbursts 1\&2}  \\
 &Mean&S.D.&Mean&S.D.&Mean&S.D.\\
\hline
\scriptsize Fluence&6792&5776&4474&3307&5422&4627\\
\scriptsize Peak Rate&3501&2851&2473&1664&2902&2293\\
\scriptsize Fluence/$k$&3.67&1.13&3.58&1.47&3.61&1.34\\
\scriptsize Peak Rate/$k$&1.90&0.37&1.76&0.28&1.82&0.32\\
\scriptsize Rise Time&2.33&0.8&2.03&1.1&2.15&1.0\\
\scriptsize Fall Time&2.32&0.9&2.35&1.0&2.32&0.9\\
\scriptsize Tot. Time&4.61&1.0&4.38&01.0&4.47&1.0\\
\hline
\hline
\end{tabular}
\caption[A table showing the mean and standard deviation of 7 parameters of \textit{RXTE}-sampled Minibursts from the 1996 outburst, the 1997 outburst and both outbursts combined.]{A table showing the mean and standard deviation of 7 parameters of \indexrxte\textit{RXTE}-sampled Minibursts\index{Miniburst} from Outburst\index{Outburst} 1, Outburst 2 and both outbursts combined.  Fluence is given in cts\,PCU$^{-1}$, peak rate is given in cts\,s$^{-1}$\,PCU$^{-1}$ and rise, fall and total time are given in s.  $k$ is the persistent emission\index{Persistent emission} rate during the observation in which a given burst was detected.}
\label{tab:mini_param}
\end{table}

\par Using the Spearman's Rank\index{Spearman's rank correlation coefficient} metric, I find only two correlations above the 5$\,\sigma$ level:
\begin{itemize}
\item Fluence is correlated with peak rate ($7.3\,\sigma$).
\item Fluence divided by persistent rate\index{Persistent emission} is correlated with peak rate divided by persistent rate ($7.1\,\sigma$).
\end{itemize}
As in Normal Bursts\index{Normal burst}, a correlation between peak rate and fluence is to be expected.  However, due to the poor statstics associated with Miniburst\index{Miniburst} parameters, it is likely that other parameter pairs are also correlated.

\subsubsection{Colour Evolution}

\par
\begin{figure}
  \centering
  \includegraphics[width=.9\linewidth, trim={0.7cm 1.4cm 0.2cm 1.4cm},clip]{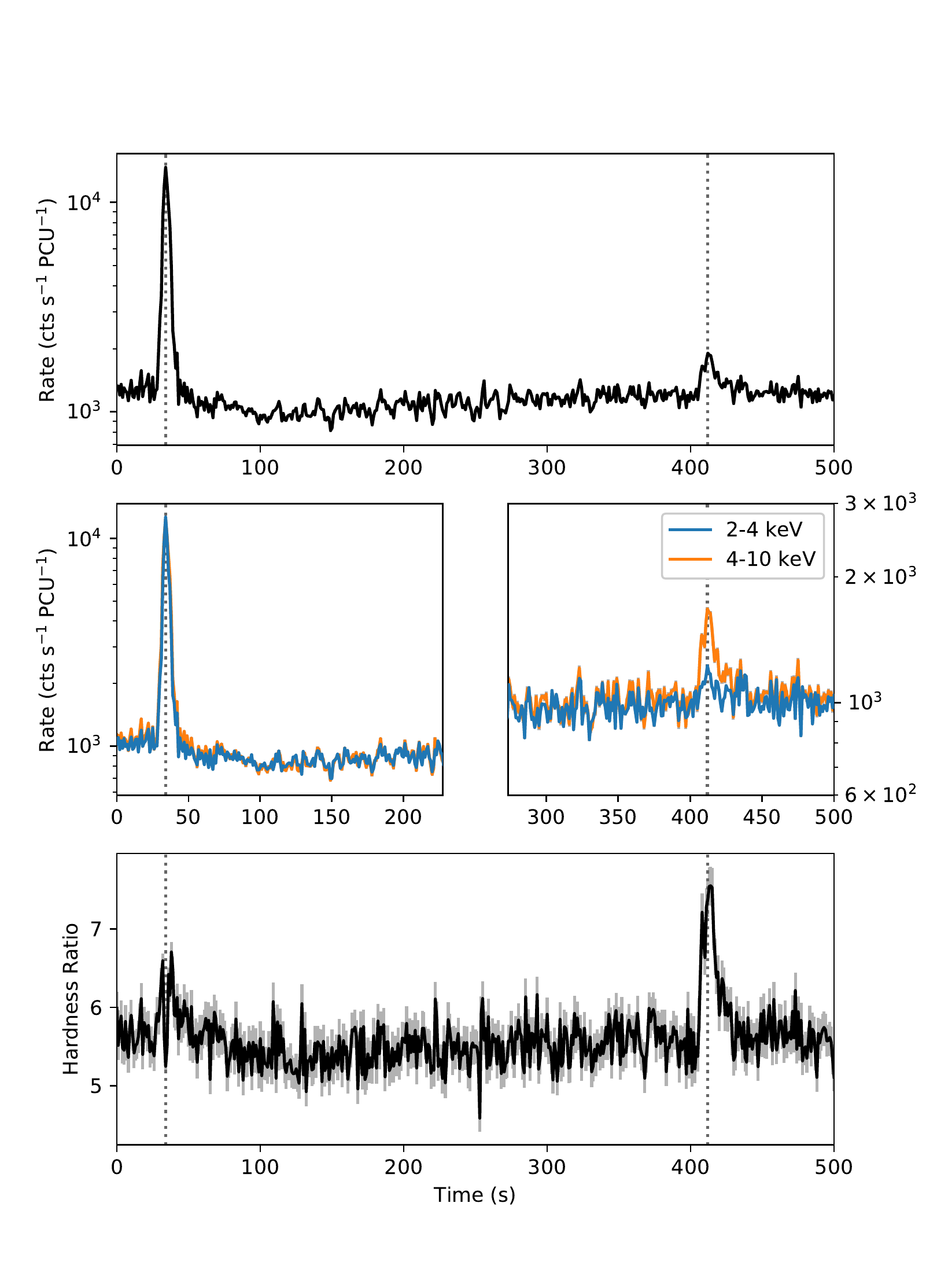}
  \caption[A portion of observation 10401-01-16-00, featuring a fNormal Burst and a Miniburst.]{\small A portion of observation 10401-01-16-00, featuring a Normal Burst\index{Normal burst} ($\sim30$\,s) and a Miniburst\index{Miniburst} ($\sim410$\,s).  The top panel shows the total 2--10\,keV lightcurve\index{Lightcurve}.  The middle panel shows lightcurves from two different energy bands; the count rates from the soft energy band have been multiplied by 5.4 so they can more easily be compared with the hard energy band.  The bottom panel shows the evolution over time of the ratio\index{Colour} between the rates in the two bands.   As can be seen in panels 2 and 3, the Miniburst has a significantly higher fractional amplitude in the 4--10\,keV energy band than in the 2--4\,keV band.}
  \label{fig:minihard}
\end{figure}

 Minibursts\index{Miniburst} show the greatest magnitude of evolution in colour\index{Colour} of all the classes of burst\index{X-ray burst}.  In Figure \ref{fig:minihard}, I show how the hardness ratio between the 4--10 and 2--4\,keV energy bands changes during an observation containing both a Miniburst and a Normal Burst\index{Normal burst}.  I find that the hardness ratio increases by $\sim50\%$ in a Miniburst, significantly more than the change in hardness during Normal or Mesobursts\index{Mesoburst}.  The statistics in minibursts were too poor to check for the presence of hysteresis\index{Hysteresis}.

\subsection{Mesobursts}

\par I define Mesobursts\index{Mesoburst} as the set of all bursts\index{X-ray burst} with a persistent-emission\index{Persistent emission}-subtracted peak 1\,s binned \indexpca\textit{RXTE}/PCA-equivalent count rate below 3000\,cts\,s$^{-1}$\,PCU$^{-1}$ in which the peak of the burst reaches at least $300\%$ of the persistent rate.  Mesobursts account for 43 out of the 190 bursts identified for this study.  They are observed in \indexrxte\textit{RXTE} data from both Outbursts\index{Outburst} 1 \& 2; in both cases they occur after the main outburst and before or during a rebrightening\index{Re-flare} event.  Mesobursts occurred between MJDs 50238 and 50248 in Outburst 1, and between 50562 and 50577 in Outburst 2; during these intervals, \textit{RXTE} observed the source for a total of 44\,ks.  As no soft X-ray instrument monitored the Bursting Pulsar during the latter stages of Outburst 3, it is unclear whether Mesobursts occurred during this outburst.  The one pointed observation of \indexnustar\textit{NuSTAR} made during this time did not detect any Mesobursts.

\subsubsection{Recurrence Time}

\par Only 6 \indexrxte\textit{RXTE} observations in Outburst\index{Outburst} 1, and 4 in Outburst 2, contain multiple Mesobursts\index{Mesoburst}.  From my limited sample I find minimum and maximum recurrence\index{Recurrence time} times of $\sim230$ and $\sim1550$\,s in Outburst 1 and minimum and maximum recurrence times of $\sim310$ and $\sim2280$\,s in Outburst 2.

\subsubsection{Structure}

\par
\begin{figure}
  \centering
  \includegraphics[width=.9\linewidth, trim={0.4cm 0 1.1cm 0},clip]{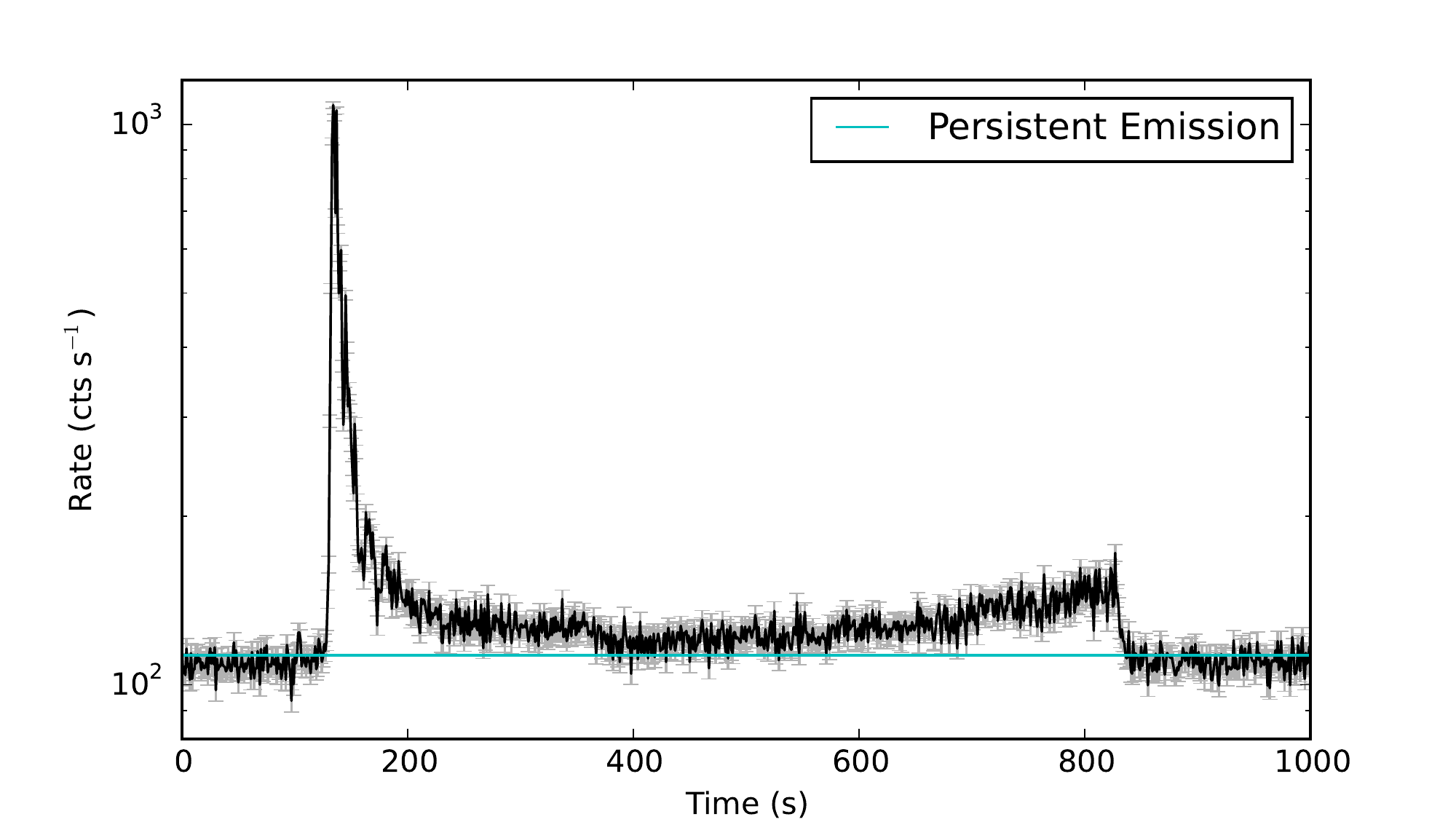}
  \caption[A lightcurve from \textit{RXTE} observation 20078-01-17-00 from Outburst 2, showing an apparent `plateau' feature after a Mesoburst.]{\small A \index{Lightcurve}lightcurve from \indexpca\textit{RXTE}/PCA observation 20078-01-17-00 from Outburst\index{Outburst} 2, showing an apparent `plateau'\index{Plateau} feature after a Mesoburst\index{Mesoburst}.}
  \label{fig:mesoplateau}
\end{figure}

 The structure of the main part of a Mesoburst\index{Mesoburst} is significantly more complex than in Normal Bursts\index{Normal burst}, consisting of a large number of secondary peaks near the main peak of the burst.  Mesobursts never show the post-burst `dip'\index{Dip} feature that we see in Normal Bursts or Minibursts\index{Miniburst}, but they can show `plateaus'\index{Plateau}.  In Figure \ref{fig:mesoplateau} I show an example of a Mesoburst with a plateau similar to those seen after Normal Bursts, suggesting a connection between the two classes.

\par
\begin{figure}
  \centering
  \includegraphics[width=.9\linewidth, trim={0.4cm 0 1.1cm 0},clip]{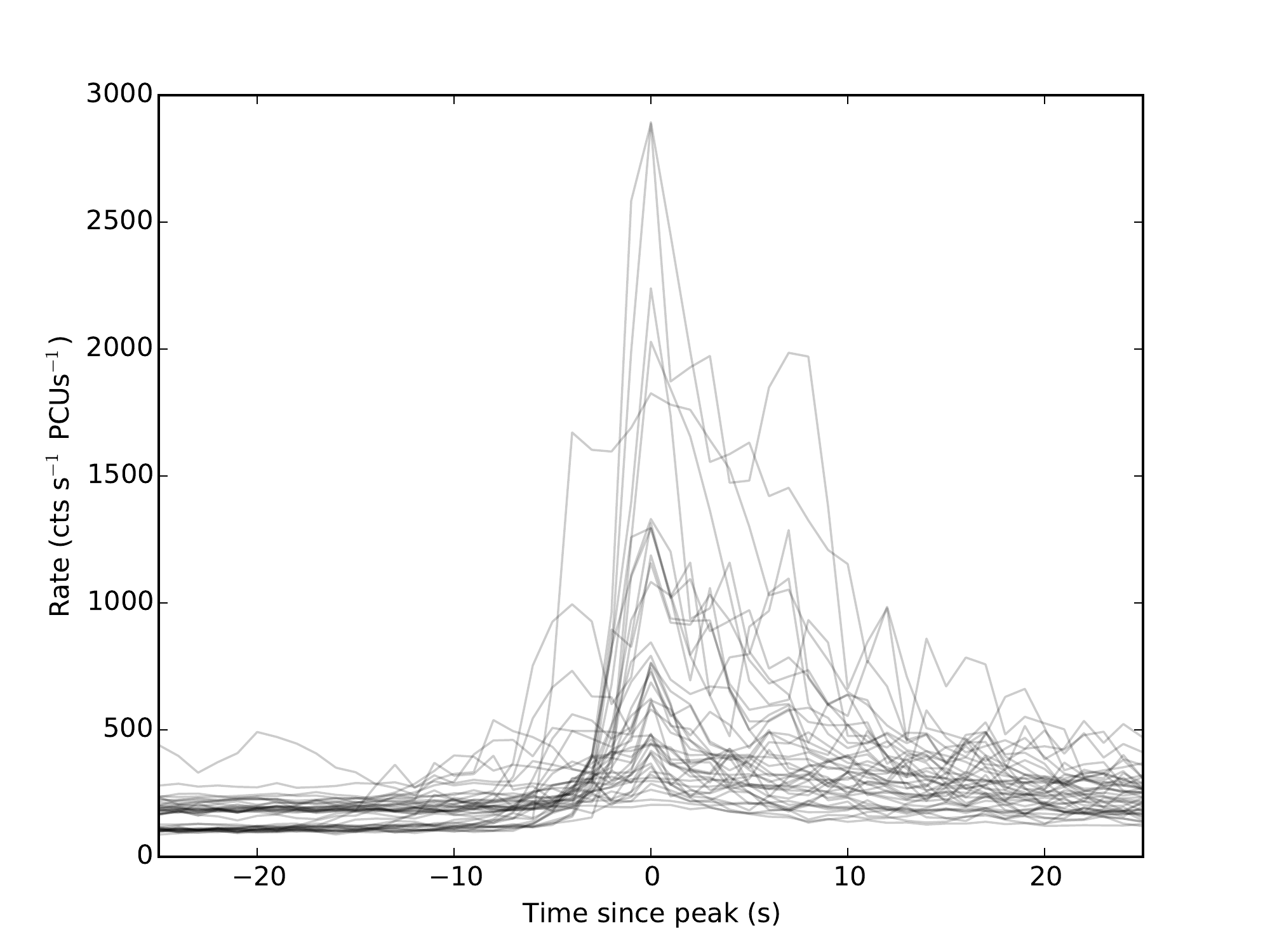}
  \includegraphics[width=.9\linewidth, trim={0.4cm 0 1.1cm 0},clip]{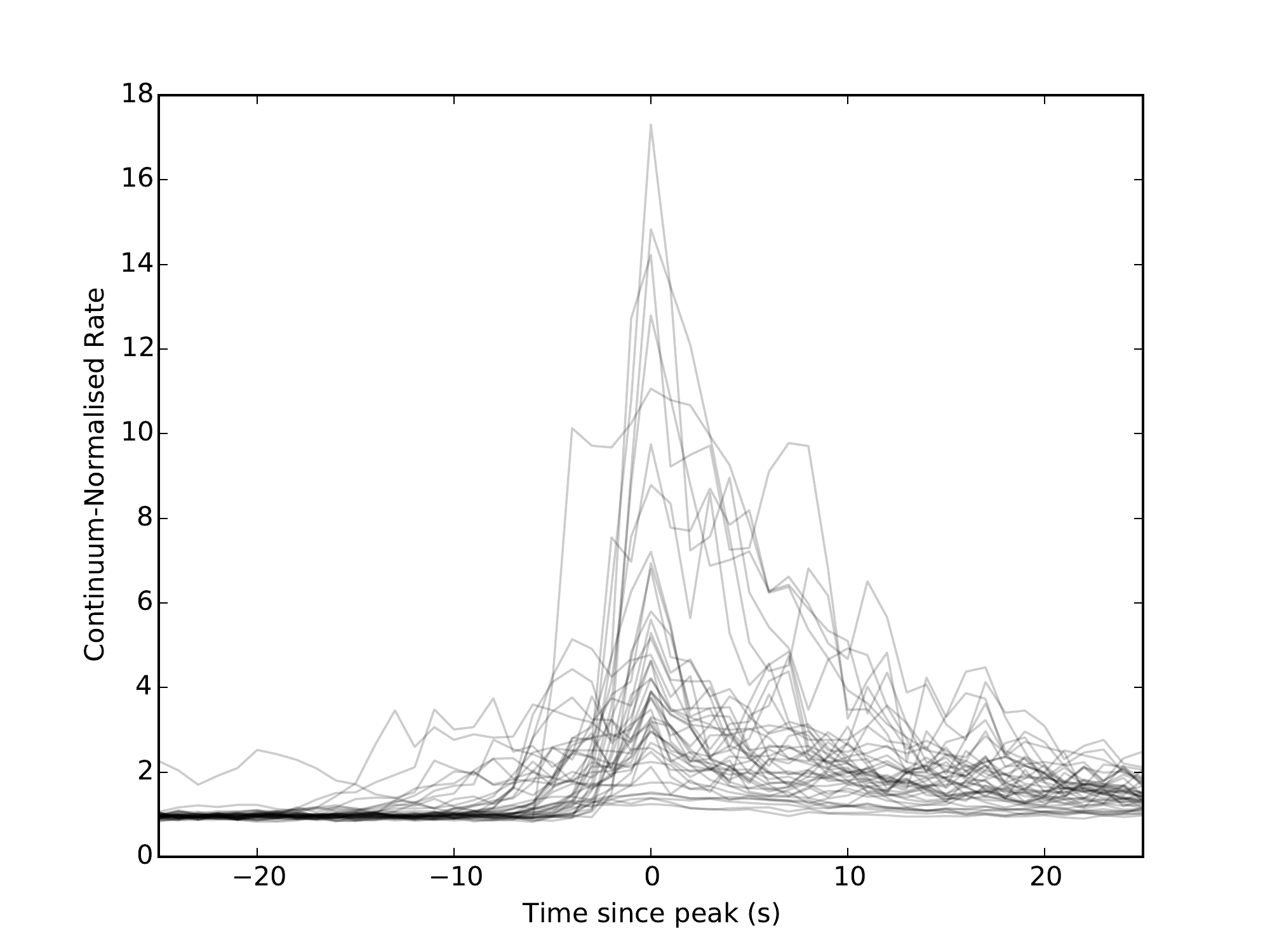}
  \caption[A plot of every Mesoburst, centred by the time of its peak, overlaid on top of each other.]{\small  \textbf{Top:} a lightcurve\index{Lightcurve} of every Mesoburst\index{Mesoburst}, centred by the time of its peak, overlaid on top of each other.  \textbf{Bottom:} a plot of every Mesoburst in which count rates have been normalised by the persistent emission\index{Persistent emission} count rate during the observation from which each burst was observed.}
  \label{fig:meso_over}
\end{figure}

 In Figure \ref{fig:meso_over} I show the lightcurves\index{Lightcurve} of all Mesobursts\index{Mesoburst} observed by \indexrxte\textit{RXTE} overlayed on top of each other before (top panel) and after (bottom panel) being renormalised by persistent emission\index{Persistent emission} rate.  It can be seen that the intensity and structure of these bursts is much more variable than in Normal Bursts\index{Normal burst} (see Figure \ref{fig:norm_overlay}).  However, each Mesoburst has a fast rise followed by a slow decay, and they occur over similar timescales of $\sim10$--$30$\,s.

\subsubsection{Parameters \& Correlations}

\label{sec:mesostruc}

\par Due to the complexity structure of Mesobursts\index{Mesoburst}, I do not fit them mathematically as I did for Normal Bursts\index{Normal burst}.  Instead I extract the same parameters as for Minibursts\index{Miniburst} (see the list in Section \ref{sec:ministruc}).  The mean and standard deviation of each of these parameters, calculated from \indexpca\textit{RXTE}/PCA data, is presented in Table \ref{tab:meso_param}.  Due to the relative low number of Mesobursts compared to Normal Bursts, I only present the results from the combined set of bursts in both Outbursts\index{Outburst} 1 \& 2.  In general, Mesobursts are longer in duration than Normal Bursts, and have significantly smaller amplitudes and fluences (compare e.g. Table \ref{tab:params_perob}).

\begin{table}
\centering
\begin{tabular}{l c c}
\hline
\hline
&Mean&Standard Deviation\\
\hline
Fluence \scriptsize(cts\,PCU$^{-1}$)&6067&6707\\
Peak Rate \scriptsize(cts\,s$^{-1}$\,PCU$^{-1}$)&665.4&658.4\\
Fluence/$k$&48.6&32.8\\
Peak Rate/$k$&5.32&4.0\\
Rise Time \scriptsize(s)&6.95&4.9\\
Fall Time \scriptsize(s)&18.28&10.8\\
Total Time \scriptsize(s)&25.88&13.3\\
\hline
\hline
\end{tabular}
\caption[A table showing the mean and standard deviation of 7 burst parameters of \textit{RXTE}-sampled Mesobursts from the 1996 \& 1997 outbursts.]{A table showing the mean and standard deviation of 7 burst parameters of \indexrxte\textit{RXTE}-sampled Mesobursts\index{Mesoburst} from Outbursts\index{Outburst} 1 \& 2.  $k$ is the persistent emission\index{Persistent emission} rate during the observation in which a given burst was detected.}
\label{tab:meso_param}
\end{table}

\par Using the Spearman's Rank metric\index{Spearman's rank correlation coefficient}, I find a number correlations above the 5$\,\sigma$ level:
\begin{itemize}
\item Fluence is correlated with peak rate ($>10\,\sigma$), peak rate divided by persistent rate\index{Persistent emission} ($6.7\,\sigma$), fall time ($6.8\,\sigma$) and total time ($6.0\,\sigma$).
\item Fluence divided by persistent rate is correlated with peak rate divided by persistent rate ($7.3\,\sigma$).
\item Peak rate is also correlated with peak rate divided by persistent rate ($7.4\,\sigma$), fall time ($5.8\,\sigma$) and persistent level ($6.2\,\sigma$).
\item Rise time correlates with total time ($5.4\,\sigma$).
\item Fall time correlates with total time ($>10\,\sigma$).
\end{itemize}

Again, the correlation between fluence and peak rate is expected, as is the correlation between peak rate and peak rate divided by persistent rate.

\subsubsection{Colour Evolution}

\par The hardness ratio\index{Colour} of the emission from the source decreases significantly during Mesobursts\index{Mesoburst}, with the PCA\indexpca\ 8--60/2--7\,keV colour decreases from $\sim0.6$ between bursts to $\sim0.2$ at the peak of a burst.  Due to the poor statistics of these features compared with Normal Bursts\index{Normal burst}, I was unable to check for evidence of hardness-intensity hysteresis\index{Hardness-intensity diagram}.

\subsection{Structured `Bursts'}

\par I define Structured Burst\index{Structured bursting} observations as observations in which the recurrence time\index{Recurrence time} between bursts\index{X-ray burst} is less than, or approximately the same as, the duration of a single burst.  Structured Bursts constitute the most complex behaviour I find in my dataset.  Unlike the other classes of burst \textsf{A.A.} and I identify, Structured Bursts are not easily described as discrete phenomena.  I find Structured Bursts in 54 observations which are listed in Appendix \ref{app:obs}.
\par
\begin{figure}
  \centering
  \includegraphics[width=.9\linewidth, trim={0.4cm 0 1.1cm 0},clip]{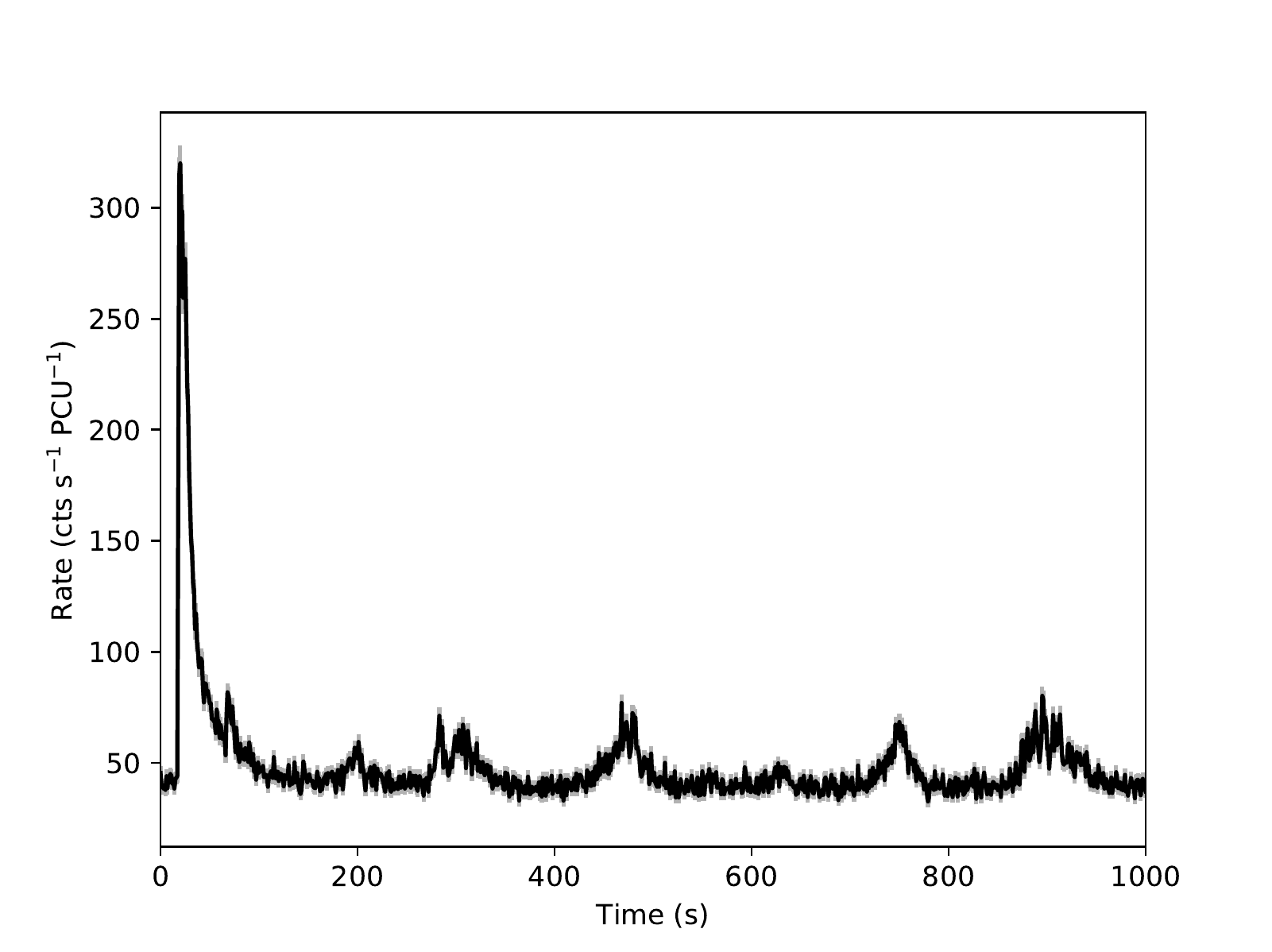}
  \caption[A lightcurve from \textit{RXTE}/PCA observation 10401-01-57-03, showing a Mesoburst occuring during a period of Structured Bursting.]{\small A lightcurve\index{Lightcurve} from \indexpca\textit{RXTE}/PCA observation 10401-01-57-03, showing a Mesoburst\index{Mesoburst} occuring during a period of Structured Bursting\index{Structured bursting}.}
  \label{fig:meso_in_struc}
\end{figure}

 In both outbursts\index{Outburst} covered by \indexrxte\textit{RXTE}, Structured Bursts occur\index{Structured bursting} in the time between the end of the main outburst and the start of a rebrightening\index{Re-flare} event.  In both cases these periods of Structured Bursts are preceded by a period populated by Mesobursts\index{Mesoburst}.  Mesobursts occurred between MJDs 50248 and 50261 in Outburst 1, and between 50577 and 50618 in Outburst 2; during these intervals, \textit{RXTE} observed the source for a total of 81\,ks.  Notably, as I show in Figure \ref{fig:meso_in_struc}, one Outburst 1 \textit{RXTE} lightcurve containing Structured Bursting also contains a bright Mesoburst.

\par In both outbursts\index{Outburst}, the amplitude of Structured Bursting\index{Structured bursting} behaviour decreases as the outburst\index{Outburst} approaches the peak of the rebrightening\index{Re-flare} event.  This amplitude continues to decrease as the Structured Burst behaviour evolves into the low-amplitude noisy variability\index{Variability} associated with the source's evolution towards the low/hard state\index{Low/Hard state}.

\subsubsection{Colour Evolution}

\par
\begin{figure}
  \centering
  \includegraphics[width=.9\linewidth, trim={0.4cm 0 1.cm 0},clip]{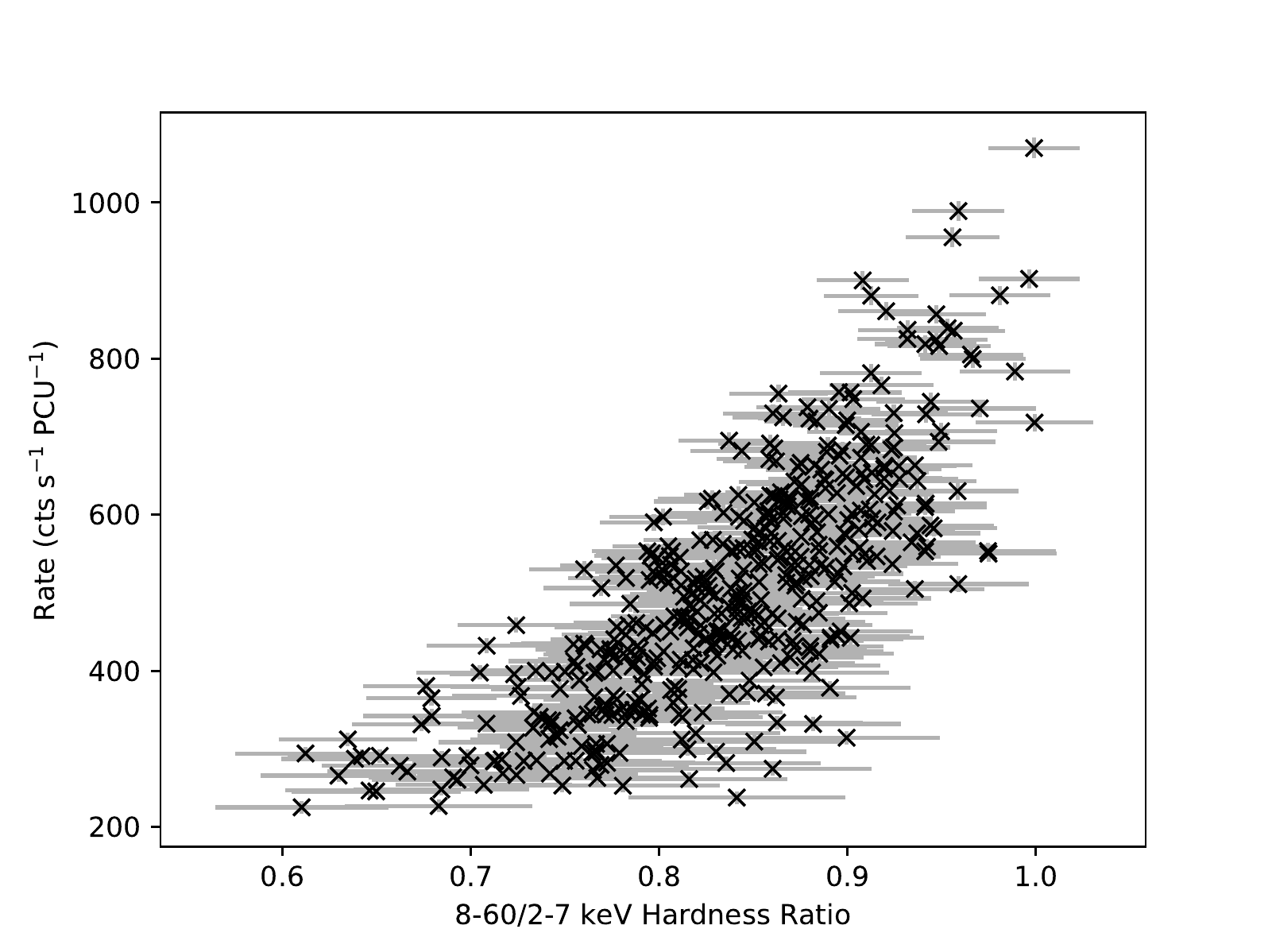}
  \caption[A hardness-intensity diagram from \textit{RXTE} observation 20078-01-23-00, showing that hardness tends to correlate with intensity during Structured Bursting.]{\small A 1\,s-binned hardness-intensity diagram\index{Hardness-intensity diagram} from \indexrxte\textit{RXTE} observation 20078-01-23-00, showing that hardness\index{Colour} tends to correlate with intensity during Structured Bursting\index{Structured bursting}.  Data are binned to 8\,s, and background\index{Background subtraction} has been estimated by subtracting mean count rates in the relevant energy bands from \textit{RXTE} OBSID 30075-01-26-00.}
  \label{fig:struc_hard}
\end{figure}

 I produce hardness-intensity diagrams\index{Hardness-intensity diagram} for a number of Structured Bursting\index{Structured bursting} observations; I show a representative example in Figure \ref{fig:struc_hard}.  I find that hardness\index{Colour} is strongly correlated with count rate during this class of bursting, but that the magnitude of the change in hardness is no greater than $\sim30\%$.  This is less than the change in hardness that I find during Normal\index{Normal burst} or Minibursts\index{Miniburst}.  I also find no evidence of hysteretic\index{Hysteresis} hardness-intensity loops from Structured Bursts.

\subsubsection{Types of Structured Bursting}
\label{sec:struc_var}

\par
\begin{figure}
  \centering
  \includegraphics[width=.9\linewidth, trim={0.8cm 0 1.5cm 0},clip]{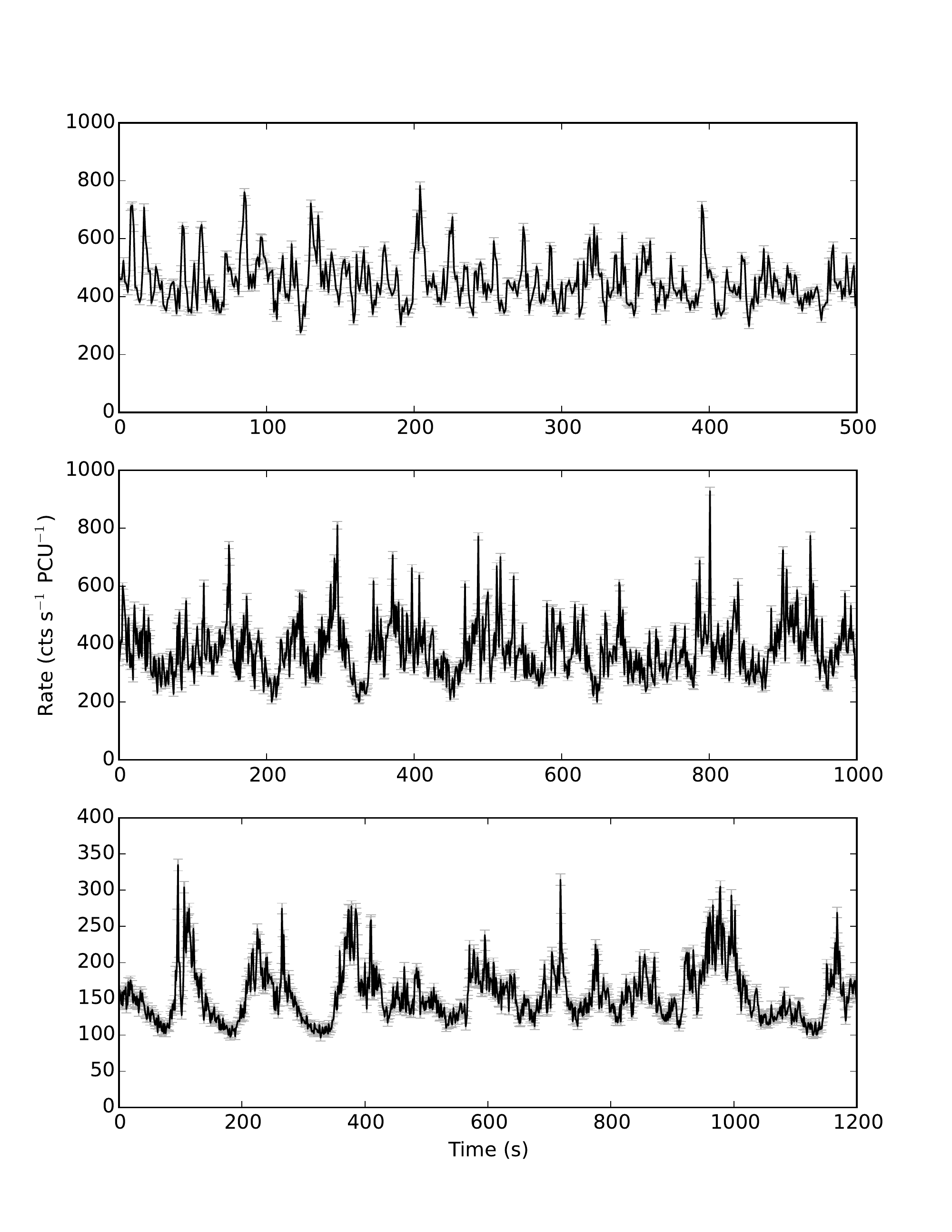}
  \caption[A selection of \textit{RXTE} lightcurves from Structured Bursting observations of the Bursting Pulsar.]{\small A selection of \indexrxte\textit{RXTE} lightcurves\index{Lightcurve} from Structured Bursting\index{Structured bursting} observations of the Bursting Pulsar\index{Bursting Pulsar}.  \textbf{Top:} a lightcurve from Outburst\index{Outburst} 1 showing flaring\index{Flare} on timescales of $\sim10$\,s.  \textbf{Middle:} a lightcurve from Outburst 1 showing the same flaring behaviour with an additional slower modulation over $\sim50$\,s.  \textbf{Bottom:} a lightcurve from Outburst 2 showing a regular sequence of flat-bottomed dips\index{Dip} and multi-peaked flaring.  These show the wide variety of variability\index{Variability} patterns that I classify as `Structured Bursting'.}
  \label{fig:Types_Struc}
\end{figure}

 In Figure \ref{fig:Types_Struc}, I present a selection of lightcurves\index{Lightcurve} which show the different types of variability\index{Variability} that can be seen during periods of Structured Bursting\index{Structured bursting}.  These consist of a variety of patterns of flares\index{Flare} and flat-bottomed dips\index{Dip}, and both \indexrxte\textit{RXTE}-observed outbursts\index{Outburst} show several of these different patterns of Structured Bursting.  As all types of Structured Bursting have similar amplitudes and occur in the same part of each outburst, I consider them to be generated by the same physical process.  I do not seperate these patterns into separate subclasses in this thesis.

\section{Discussion}

\par I analyse all available X-ray data from the first 3 outbursts\index{Outburst} of the Bursting Pulsar\index{Bursting Pulsar}.  The bursting\index{X-ray burst} behaviour evolves in a similar way during these outbursts, strongly associating them with the Bursting Pulsar and suggesting an underlying connection between the classes of burst.  I also find that both Outbursts 1 \& 2 showed `rebrightening events'\index{Re-flare} similar to those seen in a number of other LMXBs\index{X-ray binary!Low mass} (e.g. \citealp{Wijnands_1808,Patruno_Reflares2}), including IGR J17091\index{IGR J17091-3624}.
\par I find that the X-ray\index{X-ray burst} bursts from these data can be best described as belonging to four phenomenological classes: Normal Bursts\index{Normal burst}, Minibursts\index{Miniburst}, Mesobursts\index{Mesoburst} and Structured Bursts\index{Structured bursting}.  For each of these four classes, I collect a number of statistics to shed light on the physical mechanisms that generate these lightcurve\index{Lightcurve} features.
\par Normal Bursts\index{Normal burst} and Minibursts\index{Miniburst} both represent the `Type II'\index{X-ray burst!Type II} bursting behaviour which is observed most commonly from this source.   Mesobursts\index{Mesoburst} occur much later on in the outburst and show fast-rise slow-decay profiles; they are generally much fainter and more structured than Normal Bursts.  Finally, Structured Bursts\index{Structured bursting} form continuous highly structured regions of variability\index{Variability} over timescales of days.  All Normal Bursts and some Minibursts show count rate `dips'\index{Dip} after the main burst, while Mesobursts and Structured Bursts do not.  In addition to this, some Normal and Mesobursts show count rate `plateaus'\index{Plateau}; regions of roughly stable count rate above the persistent level\index{Persistent emission} which last for $\sim10$s of seconds.  These features are also sometimes seen in Mesobursts, while Minibursts and Structured Bursts never show these structures.
\par Here I discuss these results in the context of models proposed to explain Type II\index{X-ray burst!Type II} bursting.  I also compare my results with those of previous studies on bursting in both the Bursting Pulsar\index{Bursting Pulsar} and the Rapid Burster\index{Rapid Burster}.

\subsection{Evolution of Outburst and Bursting Behaviour}

\par In general, Outburst\index{Outburst} 1 was brighter than Outburst\index{Outburst} 2, with the former having a peak 2--60\,keV intensity a factor of $\sim1.7$ greater than the latter.  However, in Figure \ref{fig:global_ob} I show that both outbursts evolve in a similar way.  In both outbursts, the intensity of the Bursting Pulsar reaches a peak of order $\sim1$\,Crab before decreasing over the next $\sim100$ days to a level of a few tens of mCrab.  A few 10s of days after reaching this level, the lightcurves\index{Lightcurve} of both outbursts show a pronounced `rebrightening'\index{Re-flare} event, during which the intensity increases to $\sim100$\,mCrab for $\sim10$ days.  Outburst 1 shows a second rebrightening event $\sim50$ days after the first.  It is unclear whether any rebrightening events occurred in Outburst 3 due to a lack of late-time observations with soft X-ray telescopes.  X-ray `rebrightening' events have been seen after the outbursts of a number of other LMXBs\index{X-ray binary!Low mass} with both neutron star\index{Neutron star} and black hole\index{Black hole} primaries: including SAX J1808.4-3658\index{SAX J1808.4-3658} \citep{Wijnands_1808}, XTE J1650-500\index{XTE J1650-500} \citep{Tomsick_MiniOutbursts} and IGR J17091-3624\index{IGR J17091-3624} (see Section \ref{sec:igrobevo}).
\par As I have shown in Figures \ref{fig:ob_evo1} \& \ref{fig:ob_evo2}, the nature of bursts\index{X-ray burst} from the Bursting Pulsar\index{Bursting Pulsar} evolves in a similar way in both Outbursts\index{Outburst} 1 \& 2.  Starting from around the peak of each outburst, both Normal\index{Normal burst} and Minibursts\index{Miniburst} are observed.  The fluence of these bursts decrease over time as the X-ray intensity of the source decreases, before bursting shuts off entirely when the 2--16\,keV flux falls below $\sim0.1$\,Crab.  After a few 10s of days with no bursts, bursting switches back on in the form of Mesobursts\index{Mesoburst}; this occurs during the tail of a rebrightening\index{Re-flare} event in Outburst 1, but in the tail of the main outburst in Outburst 2.  Mesobursting continues until the 2--16\,keV source flux falls below $\sim0.03$\,Crab, at which point I observe the onset of Structured Bursting\index{Structured bursting}.  In both Outbursts, Structured Bursting stops being visible a few 10s of days later during the start of a rebrightening event.  Because this evolution is common to both of the outbursts observed by \indexrxte\textit{RXTE}, this strongly indicates that the nature of bursting in the Bursting Pulsar\index{Bursting Pulsar} is connected with the evolution of its outbursts.  Additionally, with the exceptions of Normal and Minibursts, I show that each class of burst is mostly found in a distinct part of the outburst corresponding to a different level of persistent emission\index{Persistent emission}.
\par
\begin{figure}
  \centering
  \includegraphics[width=.9\linewidth, trim={3.7cm 0cm 4.2cm 0cm},clip]{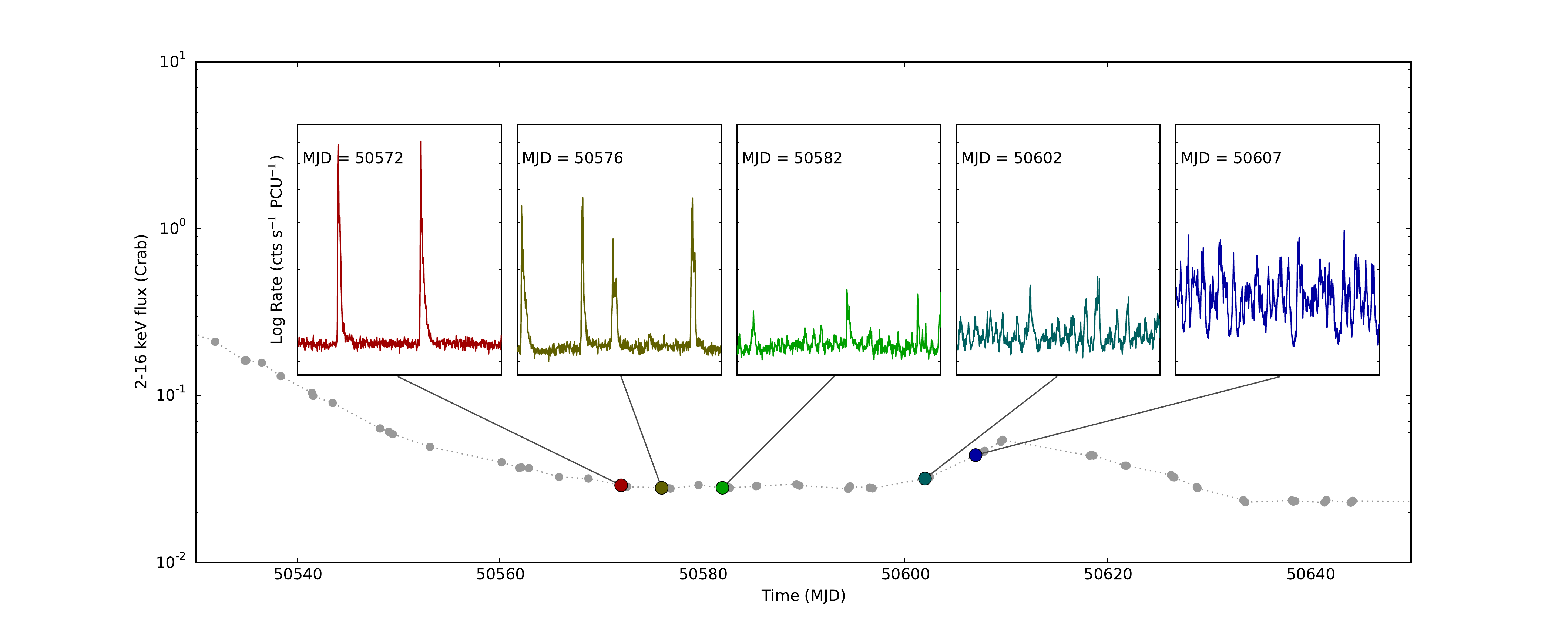}
  \caption[A series of lightcurves from \textit{RXTE}/PCA observations of Outburst 2, showing a gradual evolution from Mesobursts to Structured Bursting over a period of $\sim30$ days.]{\small A series of lightcurves\index{Lightcurve} from \indexpca\textit{RXTE}/PCA observations of Outburst\index{Outburst} 2, showing a gradual evolution from Mesobursts\index{Mesoburst} to Structured Bursting\index{Structured bursting} over a period of $\sim30$ days.  Each inset lightcurve is plotted with the same $y$-scaling, and each corresponds to 2\,ks of data.}
  \label{fig:meso_to_struc}
\end{figure}

 In Figure \ref{fig:meso_to_struc}, I show lightcurves\index{Lightcurve} from Outburst\index{Outburst} 2 taken a few days before and after the transition from Mesobursts\index{Mesoburst} to Structured Bursting\index{Structured bursting}.  We can see that, as the system approaches this transition, Mesobursts become more frequent and decrease in amplitude.  Additionally in Figure \ref{fig:meso_in_struc} I show a lightcurve which contains both a Mesoburst and Structured Bursting.  I find that, instead of a well-defined transition between these bursting classes, there is a more gradual change as Mesobursting evolves into Structured Bursting.
\par The transition between Normal Bursts\index{Normal burst} and Mesobursts\index{Mesoburst}, however, is not smooth; in both outbursts\index{Outburst} these two classes of bursting\index{X-ray burst} are separated by $\sim10$ day gaps in which no bursts of any kind were observed at all.  If all my classes of burst are caused by the same or similar processes, any model to explain them will also have to explain these periods with no bursts.

\subsection{Parameter Correlations}

\par I extracted a number of phenomenological parameters from each Normal Burst\index{Normal burst}, Miniburst\index{Miniburst} and Mesoburst\index{Mesoburst}.  For Normal Bursts, I extracted a large number of parameters by fitting a phenomenological model described in Section \ref{sec:struc}.  For Minibursts and Mesobursts I extracted recurrence times\index{Recurrence time} and persistent emission\index{Persistent emission}-subtracted peak rates; I also calculated burst\index{X-ray burst} fluences by integrating the persistent emission\index{Persistent emission}-subtracted rate over the duration of the burst.  I do not extract similar parameters for Structured Bursts\index{Structured bursting} due to their complex nature.
\par  In all three of the classes of burst\index{X-ray burst} I consider, I found that fluence and peak rate correlate strongly with persistent emission\index{Persistent emission}.  For each type of burst, the slope of these correlations is consistent with being equal during Outbursts\index{Outburst} 1 \& 2.
\par I also compared the Normal Bursts\index{Normal burst} in Outburst\index{Outburst} 1 with the Normal Bursts in Outburst 2.  The only significant statistical differences I found between these two populations\index{Population study} were in the burst peak rate and the burst fluence; both of these parameters are generally higher for Normal Bursts in Outburst 1.  As both of these parameters strongly depend on the persistent emission\index{Persistent emission}, both of these differences can be attributed to the fact that Outburst 1 was significantly brighter at peak than Outburst 2.
\par For Normal Bursts\index{Normal burst}, I found additional correlations.  Of particular note, I found that both the fall time and the recovery timescale of a `dip'\index{Dip} is proportional to its amplitude, which has implications for the possible mechanism behind these features.  I discuss this further in Section \ref{sec:mod}.
\par My findings strongly suggest that the properties of Normal\index{Normal burst}, Mini\index{Miniburst} and Mesobursts\index{Mesoburst} all depend on the persistent luminosity\index{Persistent emission} of the Bursting Pulsar\index{Bursting Pulsar}.  Assuming that this persistent luminosity is proportional to $\dot{M}$\index{Accretion rate}\index{M dot@$\dot{M}$|see {Accretion rate}}, this suggests that all classes of bursting\index{X-ray burst} are sensitive to the accretion rate of the system.  Additionally, with the exceptions of Normal and Minibursts, I find that each class of burst is mostly found in a distinct part of the outburst \index{Outburst}corresponding to a different level of persistent emission\index{Outburst}.  I suggest that Normal, Meso and Structured Bursts may in fact be manifestations of the same physical disk\index{Accretion disk} instability\index{Instability} but at different accretion rates.  This is supported by the observation of a Mesoburst during a period of Structured Bursting, which I show in the lightcurve in Figure \ref{fig:meso_in_struc}.  This shows that the conditions for both Mesobursts and Structured Bursting can be met at the same time.

\subsection{Comparison with Previous Studies}

\par In their study of bursts\index{X-ray burst} in the Bursting Pulsar\index{Bursting Pulsar}, \citet{Giles_BP} found evidence for three distinct classes of Type II\index{X-ray burst!Type II} bursts in the Bursting Pulsar:

\begin{itemize}
\item `Bursts' (hereafter G$_1$ Bursts to avoid confusion), the common Type II\index{X-ray burst!Type II} bursts\index{X-ray burst} seen from the source.
\item `Minibursts' (hereafter G$_2$  Bursts), with smaller amplitudes up to $\sim2$ times the persistent emission\index{Persistent emission} level.
\item `Microbursts' (hereafter G$_3$  Bursts), second-scale bursts with amplitudes of $\sim50$--$100\%$ of the persistent level.
\end{itemize}

We find that \citeauthor{Giles_BP}'s G$_1$ category contains the bursts that I identify as Normal Bursts\index{Normal burst}, while my Miniburst\index{Miniburst} category contains the same bursts as \citeauthor{Giles_BP}'s G$_2$ category.  \citeauthor{Giles_BP} only consider bursts\index{X-ray burst} up to MJD 50204 in their classification, and they could not classify any bursts that I identify as Mesobursts\index{Mesoburst}; under their framework, I find that Mesobursts would also be categorised as G$_1$.  I present the full mapping between \citeauthor{Giles_BP}'s classes and my classes in a schematic way in Table \ref{tab:classcomp}.

\begin{table}
\centering
\begin{tabular}{c c}
\hline
\hline
 \scriptsize My Class & \scriptsize \citeauthor{Giles_BP} Class  \\
\hline
Normal Bursts\index{Normal burst} & G$_1$ \\
Mesobursts\index{Mesoburst} & G$_1$ \\
Minibursts\index{Miniburst} & G$_2$ \\
Structured Bursts\index{Structured bursting} & - \\
 - & G$_3$ \\
\hline
\hline
\end{tabular}
\caption[A table showing how my burst classes for the Bursting Pulsar map to those described in \citet{Giles_BP}.]{A table showing how my burst\index{X-ray burst} classes map to those described in \citet{Giles_BP}.  \citeauthor{Giles_BP} do not consider the times during the outburst\index{Outburst} when Structured Bursts appear, and I consider G$_3$ bursts described by \citeauthor{Giles_BP} to be consistent with flicker noise.}
\label{tab:classcomp}
\end{table}

\par \citet{Giles_BP} note the presence of both dips\index{Dip} and plateaus\index{Plateau} in Normal Bursts.  To calculate the fluence of each main burst\index{X-ray burst} and its associated dip, \citeauthor{Giles_BP} integrate the total persistent-emission\index{Persistent emission}-subtracted counts in each feature.  They calculate that ratio between burst fluence and `missing' dip fluence ($\phi_{B}/\phi_{d}$) is between 0.26 and 0.56 in Outburst 1\index{Outburst} before correcting for dead-time\index{Dead-time} effects.  Using bursts in which my mathematical fit gave well-constrained ($>5\,\sigma$) values for both burst and dip fluence, I find that $\phi_{B}/\phi_{d}$ is between 1.3 and 2.0 in Outburst 1 and between 1.3 and 2.9 in Outburst 2.  My values differ significantly from those reported from \citeauthor{Giles_BP}; this is likely due to differing definitions of the persistent emission level\index{Persistent emission} and the start and end times of each dip, as \citeauthor{Giles_BP} do not report how they define these features.
\par My values for the ratios between burst\index{X-ray burst} and dip\index{Dip} fluences, as well as those of \citeauthor{Giles_BP}, are affected by dead-time\index{Dead-time}.  These effects cause the fluence of bursts to be under-reported, as can be inferred from Figure \ref{fig:minidips}, but the integrated counts in dips are not significantly affected \citep{Giles_BP}.  Therefore correcting for dead-time can only increase the value of $\phi_{B}/\phi_{d}$, and my result shows that the fluence of a burst is always greater than the fluence `missing' from a dip.
\par \textsf{T.O.} and I find evidence of significant colour\index{Colour}\index{Hysteresis} evolution during both Normal Bursts\index{Normal burst} and Minibursts\index{Miniburst}, which is strongly indicative of a spectral\index{Spectroscopy} evolution (see also e.g. \citealp{Woods_OB2}).  Further work on the time-resolved spectra of this source will likely allow us to better understand the underlying physics of its behaviour.
\par Using data from the KONUS\index{KONUS} experiments aboard the \textit{GGS-Wind}\index{GGS-Wind@\textit{GGS-Wind}} and \textit{Kosmos-2326}\index{Kosmos-2326@\textit{Kosmos 2326}} satellites, \citet{Aptekar_Recur} have previously found that the recurrence times\index{Recurrence time} between consecutive bursts\index{X-ray burst} in Outburst\index{Outburst} 1 are distributed with a constant mean of $\sim1776$\,s.  This is substantially longer than the value of 1209\,s that I find for Outburst 1, but my value is likely an underestimate due to a selection bias caused by the relatively short pointings of \indexrxte\textit{RXTE}.
\par Using \indexchandra\textit{Chandra} and \indexxmm\textit{XMM-Newton} data, I find a mean recurrence time\index{Recurrence time} for Outburst\index{Outburst} 3 of 1986\,s; as pointings with these instruments are significantly longer than the burst recurrence timescale, windowing effects are negligible.  As this value is close to the value that \citet{Aptekar_Recur} find for mean recurrence time, my result is consistent with the burst\index{X-ray burst} rate in all three outbursts being approximately the same.
\par Previous studies with \index{CGRO@\textit{CGRO}!BATSE}\index{CGRO@\textit{CGRO}}\textit{CGRO}/BATSE have found that the burst\index{X-ray burst} rate during the first few days of Outbursts\index{Outburst} 1 \& 2 was significantly higher than during the rest of each outburst \citep{Kouveliotou_BP,Woods_OB2}.  As \indexrxte\textit{RXTE} did not observe either of these times, I am unable to test this result with my dataset.

\subsection{Comparison with other objects}

\par Another natural comparison to the Bursting Pulsar\index{Bursting Pulsar} is the Rapid Burster\index{Rapid Burster} \citep{Lewin_RBDiscovery}, a neutron star\index{Neutron star} LMXB\index{X-ray binary!Low mass} in the globular cluster Liller I\index{Liller 1}.  This object is the only LMXB other than the Bursting Pulsar known to unambiguously exhibit Type II\index{X-ray burst!Type II} bursting behaviour during outbursts\index{Outburst}.  \citet{Rappaport_BPHistory} have previously proposed that the Bursting Pulsar, the Rapid Burster and other neutron star LMXBs form a continuum of objects with different magnetic field\index{Magnetic field} strengths.
\par I compare my study of bursts\index{X-ray burst} in the Bursting Pulsar\index{Bursting Pulsar} with studies of Type II\index{X-ray burst!Type II} bursts in the Rapid Burster\index{Rapid Burster}, particularly the detailed population study\index{Population study} performed by \citet{Bagnoli_PopStudy}.  \citet{Bagnoli_PopStudy} found that Type II bursting begins during the decay of an outburst\index{Outburst} in the Rapid Burster.  This is the same as what we see in the Bursting Pulsar, where I find Normal Bursting\index{Normal burst} behaviour starts during the outburst decay.  \citet{Bagnoli_PopStudy} found that all bursting in the Rapid Burster shuts off above an Eddington Fraction\index{Eddington limit} of $\gtrsim0.05$, whereas I find that bursting in the Bursting Pulsar shuts off \textit{below} a 2--16\,keV flux of Eddington fraction of $\sim0.1$\,Crab: assuming that the peak persistent luminosity of the Bursting Pulsar was approximately Eddington Limited (e.g. \citealp{Sazonov_BPGranat}), this value corresponds to an Eddington fraction of order $\sim0.1$.  This suggests that Type II bursting in these two objects happen in very different accretion rate\index{Accretion rate} regimes.
\par \citet{Bagnoli_PopStudy} showed that bursting\index{X-ray burst!Type II} behaviour in the Rapid Burster\index{Rapid Burster} falls into a number of `bursting modes', defined by the morphology of individual Type II bursts.  In particular, they find that Type II bursts in the Rapid Burster fall into two classes (see also \citealp{Marshall_2types}), lightcurves\index{Lightcurve} of which I reproduce in Figure \ref{fig:bagnoli_lcs}:

\begin{figure}
  \centering
  \includegraphics[width=.9\linewidth, trim={0.8cm 0 1.4cm 0},clip]{images/bagnoli_bursts.eps}
  \caption[\textit{RXTE} lightcurves of representative Long (top) and Short (bottom) Type II bursts from the Rapid Burster.]{\small \indexrxte\textit{RXTE} lightcurves\index{Lightcurve} of representative Long (top) and Short (bottom) Type II bursts\index{X-ray burst!Type II} from the Rapid Burster\index{Rapid Burster}.  These bursts were identified and classified by \citet{Bagnoli_PopStudy}.}
  \label{fig:bagnoli_lcs}
\end{figure}

\begin{itemize}
\item Short near-symmetric Bursts\index{X-ray burst!Type II} with timescales of $\sim10$s of seconds and peak rates near the Eddington Limit\index{Eddington limit}.
\item Long bursts with a fast rise, a long $\sim100$\,s plateau\index{Plateau} at peak rate followed by a fast decay.  The level of the plateau is generally at or near the Eddington Limit.
\end{itemize}

\par Short bursts\index{X-ray burst!Type II} are very similar in shape to Normal Bursts\index{Normal burst} in the Bursting Pulsar\index{Bursting Pulsar}, but I find no analogue of long bursts in my study.  \citet{Bagnoli_PopStudy} suggests that the `flat-top' profile of long bursts could be due to the effects of near-Eddington\index{Eddington limit} accretion\index{Accretion}, and they show that the intensity at the top of these bursts is close to Eddington limit.  Previous works have shown that the persistent emission\index{Persistent emission} of the Bursting Pulsar is Eddington-limited at peak, and therefore bursts\index{X-ray burst} from the Bursting Pulsar are significantly super-Eddington\index{Super-Eddington accretion} \citep{Sazonov_BPGranat}.   I suggest, therefore, that Long Bursts cannot occur in systems with a persistent rate approaching the Eddington Limit.  This could explain why Long Bursts are not seen during periods of Normal Bursting\index{Normal burst} in the Bursting Pulsar (during which the persistent emission is $\gtrsim{20}$\% of Eddington), but it remains unclear why these features are not seen later in each outburst\index{Outburst} when the Bursting Pulsar is fainter.  Alternatively, all the differences we see between bursts produced by the Rapid Burster\index{Rapid Burster} and the Bursting Pulsar could be explained if the physical mechanisms behind these bursts are indeed different between the objects.
\par \citet{Bagnoli_PopStudy} also find a number of correlations between burst\index{X-ray burst!Type II} parameters in the Rapid Burster\index{Rapid Burster}, which I can compare with my results for the Bursting Pulsar\index{Bursting Pulsar}.  I find a number of similarities between the two objects:

\begin{itemize}
\item The fluence of a burst correlates with its amplitude.
\item The duration of a burst does not correlate\footnote{We state two parameters do not correlate if their Spearman Rank score corresponds to a significance $<3\sigma$.} with the persistent emission\index{Persistent emission}.
\item The recurrence time\index{Recurrence time} between consecutive bursts does not depend on the persistent emission.
 \end{itemize}

\par There are also a number of differences between the set of correlations between burst\index{X-ray burst} parameters in these two systems:

\begin{itemize}
\item Burst duration is correlated with burst fluence in the Rapid Burster\index{Rapid Burster}, but these have not been seen to correlate in the Bursting Pulsar\index{Bursting Pulsar}.
\item Burst duration, peak rate and burst fluence are all correlated with burst recurrence time\index{Recurrence time} in the Rapid Burster.  I have not found any of these parameters to correlate with burst recurrence time in the Bursting Pulsar.
\item Peak rate and burst fluence correlate with persistent emission\index{Persistent emission} in the Bursting Pulsar, but this is not true for bursts of a given type in the Rapid Burster.
\end{itemize}

\par As the neither the fluence nor the class of a burst in the Rapid Burster\index{Rapid Burster} depend strongly on persistent emission\index{Persistent emission}, and hence $\dot{M}$\index{Accretion rate}, this suggests that the process that triggers Type-II\index{X-ray burst!Type II} bursts in this source is not strongly dependent on the global accretion rate.  However the strong correlations between persistent emission and burst peak and fluence I find in the Bursting Pulsar\index{Bursting Pulsar} show that the energetics of individual bursts\index{X-ray burst} strongly depend global accretion rate at that time.
\par It has previously been noted that consecutive Normal Bursts\index{Normal burst} in the Bursting Pulsar\index{Bursting Pulsar} do not show a strong correlation between recurrence time\index{Recurrence time} and fluence (\citealp{Taam_Evo,Lewin_BP}, however see \citealp{Aptekar_OscRel}).  This correlation would be expected if the instability\index{Instability} took the form of a relaxation oscillator\index{Relaxation oscillator}, as it does in the Rapid Burster\index{Rapid Burster} \citep{Lewin_TypeII}.  However, I also find that the arrival times of Normal Bursts from the Bursting Pulsar are not consistent with a Poisson distribution\index{Poisson distribution} with constant mean.  This implies either that bursts are also not independent events in the Bursting Pulsar, or that the frequency of these bursts is not constant throughout an outburst as reported by \citet{Aptekar_Recur}.
\par In Chapter \ref{ch:BPletter} I discuss the possibility that some of the behaviour in the Bursting Pulsar\index{Bursting Pulsar} could be due to fluctuations in the magnetospheric radius\index{Magnetospheric radius} of the system close to the co-rotation\index{Co-rotation radius} radius.  This behaviour, referred to in this thesis as `hiccup' accretion\index{Hiccup accretion}, (e.g. \citealp{Bogdanov_TMSPVar,Ferrigno_TMSPVar}) is also seen in `Transitional Millisecond Pulsars'\index{TMSP}\index{Transitional millisecond pulsar|see {TMSP}} (TMSPs): objects which alternate between appearing as X-ray pulsars\index{Pulsar} and radio pulsars (see e.g. \citealp{Archibald_Link,Papitto_Swings}).

\subsection{Comparison with Models of Type II Bursts}

\label{sec:mod}

\par
\begin{figure}
  \centering
  \includegraphics[width=.9\linewidth, trim={0cm 0 0cm 0},clip]{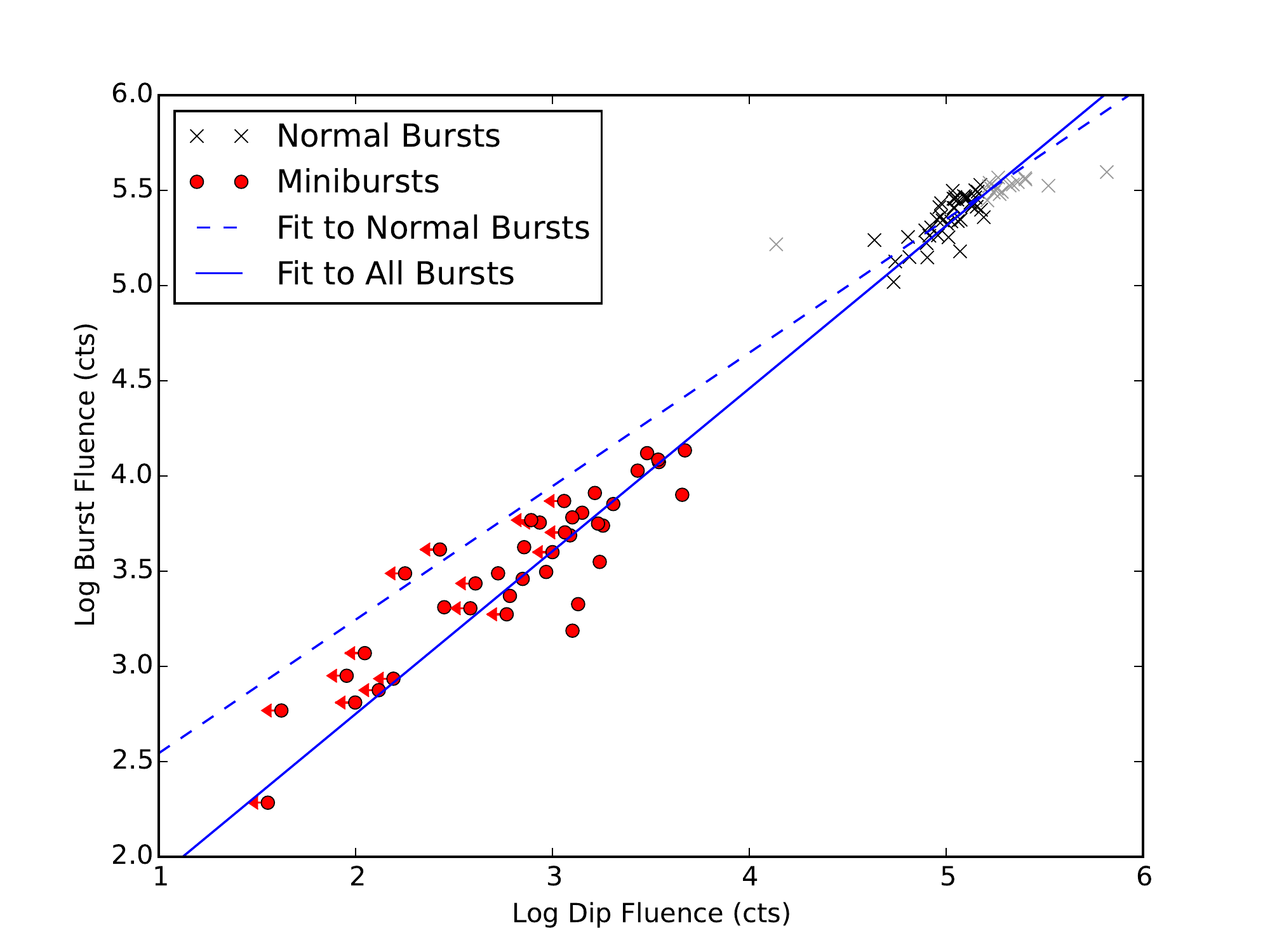}
  \caption[A scatter plot showing the relationship between burst fluence and `missing' dip fluence for Normal Bursts and Minibursts.]{\small A scatter plot showing the relationship between burst\index{X-ray burst} fluence and `missing' dip\index{Dip} fluence for Normal Bursts\index{Normal burst} (black) and Minibursts\index{Miniburst} (Red), with the best fit power law plotted in solid blue.  A power law fit to just the Normal Bursts (blue dashed line) also approaches the Minibursts.  Note that the Normal Bursts plotted in grey were not used to calculate this latter fit, as the effects of instrumental dead-time\index{Dead-time} cause high burst fluences to be under-reported.  Upper limits on Miniburst dip fluences are shown with arrows.}
  \label{fig:minidips}
\end{figure}

 All of the models of Type II\index{X-ray burst!Type II} bursting which we discuss in Section \ref{sec:TIImod} are able to reproduce some of the features we see from bursts\index{X-ray burst} in the Bursting Pulsar\index{Bursting Pulsar}.  In particular, the `dip'\index{Dip} we see after Normal Bursts\index{Normal burst} has previously been interpreted as being caused by the inner disk\index{Accretion disk}\index{Disk|see {Accretion disk}} refilling after a sudden accretion\index{Accretion} event (e.g. \citealp{Younes_Expo}).  \label{sec:Mini_Norm}As these dips are also seen after some Minibursts\index{Miniburst}, we could also interpret Minibursts as being caused by a similar cycle.  To test this idea, in Figure \ref{fig:minidips} I present a scatter plot of the burst and dip fluences for all Normal Bursts and Minibursts.  In both classes of burst, there is a strong correlation between these two parameters.  I find that a power law fit to the Normal Bursts in this parameter space also describes the Minibursts.  This suggests that the same relationship between burst fluence and missing dip fluence holds for both types of burst, although the two populations\index{Population study} are not continuous.  This suggests that Minibursts are energetically consistent with being significantly fainter versions of Normal Bursts.

\par The models of \citet{Spruit_Type2Mod} and \citet{Walker_Type2Mod} have shortcomings when used to describe the Bursting Pulsar\index{Bursting Pulsar}.  \citet{Walker_Type2Mod} state that their model only produces Type II\index{X-ray burst!Type II} bursts for a very specific set of criteria on the system parameters.   One of these criteria is an essentially non-magnetic\index{Magnetic field} ($B=0$) neutron star\index{Neutron star}.  This is inconsistent with observations of cyclotron lines\index{Cyclotron lines} from the Bursting Pulsar and the presence of a persistent pulsar\index{Pulsar}, both of which suggest a surface field strength of order 10$^{11}$\,G \citep{Doroshenko_NBFlash}.
\par Unlike models based on viscous\index{Viscosity} instability\index{Instability}, the model of \citet{Spruit_Type2Mod} does not impose a correlation between burst\index{X-ray burst} fluence and burst recurrence time\index{Recurrence time} (see e.g. the evaluation of this model in the context of the Rapid Burster\index{Rapid Burster} performed by \citealp{Bagnoli_PopStudy}).  However, it does predict a strong correlation between burst recurrence time and mean accretion rate\index{Accretion rate}, which is not consistent with my results for the Bursting Pulsar\index{Bursting Pulsar}.
\par In general, I find that models established to explain bursting\index{X-ray burst} in the Rapid Burster\index{Rapid Burster} are poor at explaining bursting in the Bursting Pulsar\index{Bursting Pulsar}.  Any model which can produce Type II\index{X-ray burst!Type II} bursting in both systems fails to explain why other systems do not also show this behaviour.  My results suggest that Type II bursts in the Rapid Burster and the Bursting Pulsar may require two separate models to be explained.

\subsubsection{Evidence of Thermonuclear Burning}

\label{sec:nuc}

\par I also consider the possibility that some of my observations could be explained by thermonuclear burning\index{Thermonuclear burning} on the Bursting Pulsar\index{Bursting Pulsar}.  A thermonuclear origin for the main part of Normal Bursts\index{Normal burst} has been ruled out by previous authors (e.g. \citealp{Lewin_BP}), but it is less clear that associated features could not be explained by this process.
\par It has been shown that, above a certain accretion rate\index{Accretion rate}, thermonuclear burning\index{Thermonuclear burning} on the surface of a neutron star\index{Neutron star} should be stable; below this rate, thermonuclear burning takes place in the form of Type I\index{X-ray burst!Type I} bursts (e.g. \citealp{Fujimoto_Shellstab,Bildsten_Regimes}).  \citet{Bildsten_Nuclear} have previously studied which form thermonuclear burning on the Bursting Pulsar\index{Bursting Pulsar} would take.  They find that the presence and profile of a thermonuclear burning event on the Bursting Pulsar would be strongly dependent on both the accretion rate $\dot{M}$ and the magnetic field\index{Magnetic field} strength $B$.  They predict that, for $B\gtrsim3\times10^{10}$\,G, burning events would take the form of a slowly propagating burning front which would result in a low-amplitude X-ray burst\index{X-ray burst} with a timescale of several minutes.  Measurements of the Bursting Pulsar taken during Outburst\index{Outburst} 3 suggest a surface field strength of $>10^{11}$\,G, in turn suggesting that the Bursting Pulsar exists in the regime in which this burning behaviour is possible.
\par The `plateau'\index{Plateau} events after Normal Bursts\index{Normal burst} are consistent with the slow burning\index{Thermonuclear burning} predicted by \citet{Bildsten_Nuclear}.  This picture is consistent with models for Type II\index{X-ray burst!Type II} X-ray bursts involving spasmodic accretion\index{Accretion} events (e.g. \citealp{Spruit_Type2Mod,Walker_Type2Mod}), as plateaus\index{Plateau} always occur after a Type II-like burst\index{X-ray burst} has deposited a large amount of ignitable material onto the neutron star\index{Neutron star} surface.  However in this picture it would be unclear why many Normal Bursts\index{Normal burst} do not show this plateau feature.  Mesobursts\index{Mesoburst} can also exhibit plateaus, and are therefore may also be products of spasmodic accretion onto the neutron star.
\par However, the interpretation of Mesobursts\index{Mesoburst} as being caused by discrete accretion\index{Accretion} events is difficult to reconcile with the fact that these features never show dips\index{Dip}.  \citet{Bildsten_Nuclear} show that, at smaller values of $\dot{M}$\index{Accretion rate}, nuclear burning\index{Thermonuclear burning} on the Bursting Pulsar\index{Bursting Pulsar} could become unstable.  Mesobursts are only seen during the latter stages of Outbursts\index{Outburst} 1 \& 2, when the accretion rate is well below 0.1 Eddington\index{Eddington limit}.  An interesting alternative possibility is that Mesobursts are a hybrid event, consisting of a flash of unstable thermonuclear X-ray burning followed by a slower quasi-stable burning of residual material in the form of a propagating burning front.
\par This picture would also be able to explain why Mesobursts\index{Mesoburst} are only seen during the latter parts of each outburst\index{Outburst}.  As the accretion rate\index{Accretion rate} onto the Bursting Pulsar\index{Bursting Pulsar} approaches Eddington\index{Eddington limit} during the peak of its outbursts, it is likely that the accretion rate is high enough that only stable burning\index{Thermonuclear burning} is permitted.  During the smaller rebrightening\index{Re-flare} events after the main part of each outburst, the accretion rate is $\sim1$--2 orders of magnitude lower, and hence the system may then be back in the regime in which Type I\index{X-ray burst!Type I} burning is possible.  Additional studies of the spectral\index{Spectroscopy} evolution of Mesobursts will be required to further explore this possibility.
\par Previous authors have discussed the possibility of a marginally stable burning\index{Thermonuclear burning} regime on the surface of neutron stars\index{Neutron star} (not to be confused with the previously mentioned quasi-stable burning).  In this regime, which occurs close to the boundary between stable and unstable burning, \citet{Heger_MargStab} showed that an oscillatory mode of burning may occur.  They associated this mode of burning with the mHz QPOs\index{Quasi-periodic oscillation} which have been observed in a number of neutron star LMXBs\index{X-ray binary!Low mass} (e.g. \citealp{Revnivtsev_MargStab,Altamirano_MargStab}).  These QPOs only occur over a narrow range of source luminosities, show a strong decrease in amplitude at higher energies, and they disappear after a Type I\index{X-ray burst!Type I} burst (e.g. \citealp{Altamirano_MargStab}).
\par Lightcurves\index{Lightcurve} of objects undergoing marginally stable burning\index{Thermonuclear burning} qualitatively resemble those of Structured Bursting\index{Structured bursting} in the Bursting Pulsar\index{Bursting Pulsar}, raising the possibility of a thermonuclear explanation for Structured Bursting.  However, as I show in Figure \ref{fig:ob_evo1}, Structured Bursting during Outburst\index{Outburst} 1 occurred during a period of time in which the Bursting Pulsar's luminosity changed by $\sim1$ order of magnitude.  In addition to this, in Figure \ref{fig:meso_in_struc} I show an example of a Mesoburst\index{Mesoburst} during a period of Structured Bursting.  If Mesobursts can be associated with Type I\index{X-ray burst!Type I} bursts, any marginally stable burning on the surface of the Bursting Pulsar should have stopped after this event.  Due to these inconsistencies with observations of marginally stable burning on other sources, it is unlikely that Structured Bursting is a manifestation of marginally stable burning on the Bursting Pulsar.
\par \citet{Linares_MargStab} observed yet another mode of thermonuclear burning\index{Thermonuclear burning} during the 2010 outburst of the LMXB\index{X-ray binary!Low mass} Terzan 5 X-2\index{Terzan 5 X-2}.  They observed a smooth evolution from discrete Type I\index{X-ray burst!Type I} bursts into a period of quasi-periodic oscillations\index{Quasi-periodic oscillation} resembling Structured Bursting\index{Structured bursting}.  This behaviour resembles the evolution I observe between Mesobursts\index{Mesoburst} and Structured Bursting in Outbursts\index{Outburst} 1 \& 2 of the Bursting Pulsar\index{Bursting Pulsar} (as shown in Figure \ref{fig:meso_to_struc}; compare with Figure 1 in \citealp{Linares_MargStab}).  However there are a number of differences between the evolutions seen in both objects.  In Terzan 5 X-2 the recurrence timescale\index{Recurrence time} of Type I bursts during the evolution is strongly related to the accretion rate\index{Accretion rate} of the source at the time, whereas there is no such strong relation between the two in Mesobursts from the Bursting Pulsar.  Additionally, the quasi-periodic oscillations in Terzan 5 X-2 evolved smoothly back into Type I bursts later in the outburst, whereas Structured Bursting does not evolve back into Mesobursts in the Bursting Pulsar.  As such, it is unclear that Mesobursts and Structured Bursting can be associated with the unusual burning mode seen on Terzan 5 X-2.

\section{Conclusions}

\par I analyse all X-ray bursts\index{X-ray burst} from the Bursting Pulsar\index{Bursting Pulsar} seen by \indexpca\textit{RXTE}/PCA during its first and second outbursts\index{Outburst}, as well as bursts seen by other missions during the third outburst of the source.  I conclude that these bursts are best described as belonging to four separate classes of burst: Normal Bursts\index{Normal burst}, Mesobursts\index{Mesoburst}, Minibursts\index{Miniburst} and Structured Bursts\index{Structured bursting}.  I find that the bursting behaviour in these four classes evolves in a similar way throughout the first two outbursts of the Bursting Pulsar.  I present a new semi-mathematical model to fit to the Normal Bursts in this object.  Using this new framework, I will be able better quantify Bursting-Pulsar-like X-ray bursts when they are observed in other objects in the future.
\par I find the bursts\index{X-ray burst} in the Rapid Burster\index{Rapid Burster} and the Bursting Pulsar\index{Bursting Pulsar} to be different in burst profile, peak Eddington\index{Eddington limit} ratio, and durations.  While the fluence of bursts in the Bursting Pulsar depend strongly on the persistent emission\index{Persistent emission} at the time, this is not the case in the Rapid Burster.  Additionally the waiting time\index{Recurrence time} between bursts in the Rapid Burster depends heavily on the fluence of the preceding burst, but I do not find this in the Bursting Pulsar.  Therefore, it would be reasonable to conclude that the bursting in these two objects is generated by two different mechanisms.
\par However, it is also important to note a number of similarities between the Bursting Pulsar\index{Bursting Pulsar} and the Rapid Burster\index{Rapid Burster}.  Bursting behaviour in both objects depends on the global accretion rate\index{Accretion rate} of the system and the evolution of its outbursts\index{Outburst}.  Additionally, the recurrence times\index{Recurrence time} of bursts do not depend on persistent emission\index{Persistent emission} in either object, and nor does the duration of an individual burst\index{X-ray burst}.  Notably while Type II\index{X-ray burst!Type II} bursts in the Rapid Burster only occur at luminosities $L\lesssim0.05L_{Edd}$\index{Eddington limit}, I find that Normal bursts\index{Normal burst} in the Bursting Pulsar only occur at $L\gtrsim0.1L_{Edd}$.  There is no overlap between the luminosity regimes, in terms of the Eddington Luminosity, at which bursting is observed in the two objects.  This leads to the alternative hypothesis that bursts in the two systems may be caused by similar processes, but that these processes take place in very different physical regimes.

\cleardoublepage

\chapter{The Bursting Pulsar GRO J1744-28: the Slowest Transitional Pulsar?}

\label{ch:BPletter}

\epigraph{\textit{I'll keep the sun behind us.  You've spent your entire life in the dark, I doubt that seeing something that bright would do you any good.}}{Lance Abell - \textit{Take the Sky}}
\vspace{1cm}

\par\noindent In Chapter \ref{ch:BPbig}, I performed a detailed analysis of all archival X-ray data of bursting\index{X-ray burst} behaviour in the Bursting Pulsar\index{Bursting Pulsar} (including \textit{RXTE}\indexrxte, \indexswift\textit{Swift}, \indexchandra\textit{Chandra}, \indexxmm\textit{XMM-Newton}, \indexsuzaku\textit{Suzaku}, \indexnustar\textit{NuStar}, and \indexintegral\textit{INTEGRAL}).  I found that the bursting phenomenology in the Bursting Pulsar is much richer than previously thought (e.g. \citealp{Giles_BP}): the characteristics of the bursts evolve with time and source luminosity. Near the end of this evolution, I observed periods of highly-structured and complex high-amplitude X-ray variability\index{Variability}.  I refer to this variability as `Structured Bursting'\index{Structured bursting}, and it is unlike what is seen in most other LMXBs\index{X-ray binary!Low mass}.
\par In Section \ref{sec:nuc}, I discuss the possibility that Structured Bursting\index{Structured bursting} is a manifestation of quasi-stable nuclear burning\index{Thermonuclear burning} on the surface of the neutron star\index{Neutron star}.  However as other types of burst\index{X-ray burst} can occur during periods of Structured Bursting without disrupting this behaviour (see e.g. Figure \ref{fig:meso_in_struc}), I consider this scenario to be unlikely.  As such, we must consider alternative explanations.  In this chapter I present the hypothesis that Structured Bursting is related to so-called `hiccup accretion'\index{Hiccup accretion}, a phenomenon seen in Transitional MilliSecond Pulsars\index{TMSP} (TMSPs).
\par \textbf{The results I present in this chapter have been published as \citet{BPletter}.}

\section{Transitional Millisecond Pulsars}

\par Millisecond Pulsars\index{Millisecond pulsar} are old radio pulsars\index{Pulsar} with spin\index{Spin} periods of order $\sim10$\,ms \citep{Backer_MSP}. They have long been believed to be the end product of systems containing a neutron star\index{Neutron star} in an LMXB\index{X-ray binary!Low mass}. In these systems, matter from a Roche-lobe\index{Roche lobe} overflowing star donates angular momentum\index{Angular momentum} to a Neutron star, spinning it up to frequencies of several 100 Hz \citep{Alpar_MSP}. A number of fast-spinning X-ray pulsars (accreting Millisecond Pulsars, or AMXPs\index{AMXP}\index{Accreting millisecond pulsar|see {AMXP}}) have been found in LMXBs (e.g. \citealp{Wijnands_XRPulsar,Altamirano_Broken,Patruno_AllAMXPs,Sanna_AMXP}), seemingly confirming this physical picture. At the end of this so-called `recycling'\index{Recycling} process, the system should transition from an accretion-powered pulsar to a rotation-powered pulsar. As such, it has long been expected that such a transition could be observed by finding a system which changes its character from an accreting Neutron star at one time to a radio pulsar at some later time. Subsequently a small family of 7 candidate objects have been discovered or proposed: these are referred to as Transitional Millisecond Pulsars (TMSPs)\index{TMSP}.
\par The first of these objects, \textbf{PSR J1023+0038}\index{PSR J1023+0038}, was identified by \citealp{Archibald_Link}. Although it appeared as a non-accreting radio pulsar\index{Pulsar} at the time of identification in 2009, previous optical studies showed that this system contained an accretion disk\index{Accretion disk} in 2002 \citep{Szkody_1023Accretion}. As such, the pulsar in this system must have switched from an accreting\index{Accretion} phase to a radio pulsar phase at some point between 2003 and 2009, confirming the identification of this system as a TMSP\index{TMSP}. The pulsar in this system has a spin\index{Spin} period of 1.69\,ms, and the companion\index{Companion star} is a star with a mass between $\sim$0.14--0.42\,M$_\odot$. \citealp{Archibald_Link} suggested that the low X-ray luminosity of PSR J1023+0038 in its accreting phase was due to accretion taking place in the propeller regime\index{Propeller effect} (see Section \ref{sec:prop}).  As previously discussed, whether a system is in the propeller regime depends on its spin\index{Spin} and its magnetic field strength\index{Magnetic field} (see also \citealp{Lewin_QPORev}). Additionally, below a certain accretion rate\index{Accretion rate}, no stable balance between ram pressure\index{Ram pressure} and radiation pressure\index{Radiation pressure} can form and any disk is ejected from the system (e.g. \citealp{Campana_NoDisk}). \citealp{Archibald_Link} suggested that the current accretion rate in PSR J1023+0038 is only slightly below this critical value, and that any small increase in accretion rate could cause accretion in this system to resume. They suggested the possibility of TMSP systems which flip back and forth between accreting and radio pulsar phases multiple times.
\par \citealp{Papitto_Swings} identified \textbf{IGR J18245-2452}\index{IGR J18245-2452} as the first known pulsar\index{Pulsar} to switch from a radio pulsar to an AMXP\index{AMXP} and back to a radio pulsar.  This source was first observed as a radio pulsar \citep{Manchester_PulsarCat}, before being observed several years later by \indexxmm\textit{XMM-Newton} \citep{Eckert_IGRJ18245} as an AMXP. Several months after the \textit{XMM-Newton} observation, \citealp{Papitto_Finding} found that the source had reactivated as a radio pulsar during X-ray quiescence\index{Quiescence}. The pulsar in this system has a period\index{Spin} of 3.93\,ms, and the companion star\index{Companion star} has a mass of $>0.17$\,M$_\odot$ \citep{Papitto_Swings}. During the 2013 outburst\index{Outburst} of IGR J18245-2452, \citealp{Ferrigno_TMSPVar} reported the presence of high-amplitude variability\index{Variability} in the X-ray lightcurve\index{Lightcurve}. They interpreted this as being due to the accretion rate\index{Accretion rate} $\dot{M}$ being very close to the critical rate at which the propeller effect\index{Propeller effect} begins to dominate the flow geometry. In this regime, small fluctuations in $\dot{M}$ cause so-called `hiccups'\index{Hiccup accretion}, in which matter alternates between being ejected by the propeller effect and being accreted onto the neutron star\index{Neutron star} poles (see our discussion of this effect in Section \ref{sec:hic}). Similar X-ray variability has subsequently been found in lightcurves from outbursts during the accreting phase of PSR J1023+0038\index{PSR J1023+0038} \citep{Bogdanov_TMSPVar}, suggesting that this variability is somehow intrinsic to TMSPs\index{TMSP} as a class of objects.
\par \textbf{1FGL J1227.9-4852}\index{1FGL J1227.9-4852} was first identified in the first \index{Fermi@\textit{Fermi}}\index{Fermi@\textit{Fermi}!LAT}\textit{Fermi}/LAT source catalogue \citep{Abdo_Catalogue}. \citealp{Hill_XSS} found that the $\gamma$-ray spectral\index{Spectroscopy} characteristics of this source are consistent with known millisecond radio pulsars\index{Pulsar}, although no radio pulsations were found. They suggested that this object could be associated with the X-ray source XSS J12270-4859\index{XSS J12270-4859|see {1FGL J1227.9-4852}}. Before 2009, XSS J12270-4859 showed optical emission lines typical of an accretion disk\index{Accretion disk} \citep{Pretorius_Optical}. \citealp{Hill_XSS} suggested that XSS J12270-4859 may also be a TMSP\index{TMSP}, which switched from an accreting \index{Accretion}phase to a radio pulsar millisecond pulsar phase between 2009 and 2011. Subsequent studies have found pulsations in both the radio \citep{Roy_12270Spin} and $\gamma$-ray \citep{Johnson_12270Spin} emissions of this source, confirming the system contains a pulsar and establishing its spin \index{Spin}period at 1.69\,ms.
\par \textbf{XMM J174457-2850.3}\index{XMM J174457-2850.3} is a neutron star\index{Neutron star} X-ray binary\index{X-ray binary}. Although no X-ray or radio pulsations\index{Pulsar} have been detected due to the faintness of the source, \citealp{Degenaar_174457} have found that the X-ray variability\index{Variability} properties of this source are similar to those seen in other TMSPs\index{TMSP}. This object also exhibits extended low-luminosity states during outbursts\index{Outburst}, which \citealp{Degenaar_174457} suggest may be symptomatic of TMSPs.
\par \textbf{3FGL J1544.6-1125}\index{3FGL J1544.6-1125}\index{1RXS J154439.4-112820|see {3FGL J1544.6-1125}} was also first identified in \index{Fermi@\textit{Fermi}}\index{Fermi@\textit{Fermi}!LAT}\textit{Fermi}/LAT data. \citealp{Bogdanov_Proxy} associated this object with the X-Ray source 1RXS J154439.4-112820. Due to the presence of $\gamma$-rays, as well as the presence of variability\index{Variability} in the X-ray lightcurve\index{Lightcurve} similar to IGR J18245-2452\index{IGR J18245-2452}, they proposed that this object is a TMSP\index{TMSP} in the accreting\index{Accretion} state. However, no pulsations\index{Pulsar} from this system have been detected in the X-ray or the radio, so the pulsar period is not known. \citealp{Bogdanov_Proxy} found a bimodality in count rate during the period of X-ray variability, suggesting that this behaviour can be explained as quick transitions between three quasi-stable accretion\index{Accretion} modes which they refer to as `low' , `high' and `flaring'. This effect has also been seen in the TMSP IGR J18245-2452\index{IGR J18245-2452} \citep{Ferrigno_TMSPVar}.
\par \citealp{Strader_6} identified the $\gamma$-ray source, \textbf{3FGL J0427.9-6704}\index{3FGL J0427.9-6704}, as a TMSP\index{TMSP}. They found that this source also displays X-ray variability\index{Variability} similar to what is seen from the other known TMSPs. Finally, \citealp{Rea_J0838} have proposed that the X-ray source \textbf{XMM J083850.4-282759}\index{XMM J083850.4-282759} may also be a TMSP. Although this source has not been detected in the gamma or the radio, the authors argued that X-ray variability coupled with X-ray flaring\index{Flare} seen from this object is reminiscent of similar behaviour seen in other TMSPs during subluminous disk\index{Accretion disk} states.
\par The phenomenology of currently known TMSPs\index{TMSP} is varied, and different methods have been used to conclude (or propose) that each individual system belongs to this class. The fact that 6 of the 7 objects show similar patterns of X-ray variability\index{Variability} during outburst\index{Outburst} suggests that this variability can be used as an indication that a system may be a TMSP.

\section{Comparison: TMSPs vs. the Bursting Pulsar}

\par \citealp{Rappaport_BPHistory} have previously suggested that the Bursting Pulsar\index{Bursting Pulsar} represents a slow X-ray pulsar\index{Pulsar} nearing the end of its accreting\index{Accretion} phase. As such it is natural to compare this system with TMSPs\index{TMSP}, which are also believed to be systems approaching this evolutionary stage. In addition to this, \citealp{Degenaar_174457} have previously noted that the Bursting Pulsar shows extended low-luminosity states during outburst\index{Outburst}, similar to those seen in the TMSP candidate XMM J174457-2850.3\index{XMM J174457-2850.3}.

\begin{figure*}
 \centering
 \resizebox{\columnwidth}{!}{\rotatebox{0}{\includegraphics[clip]{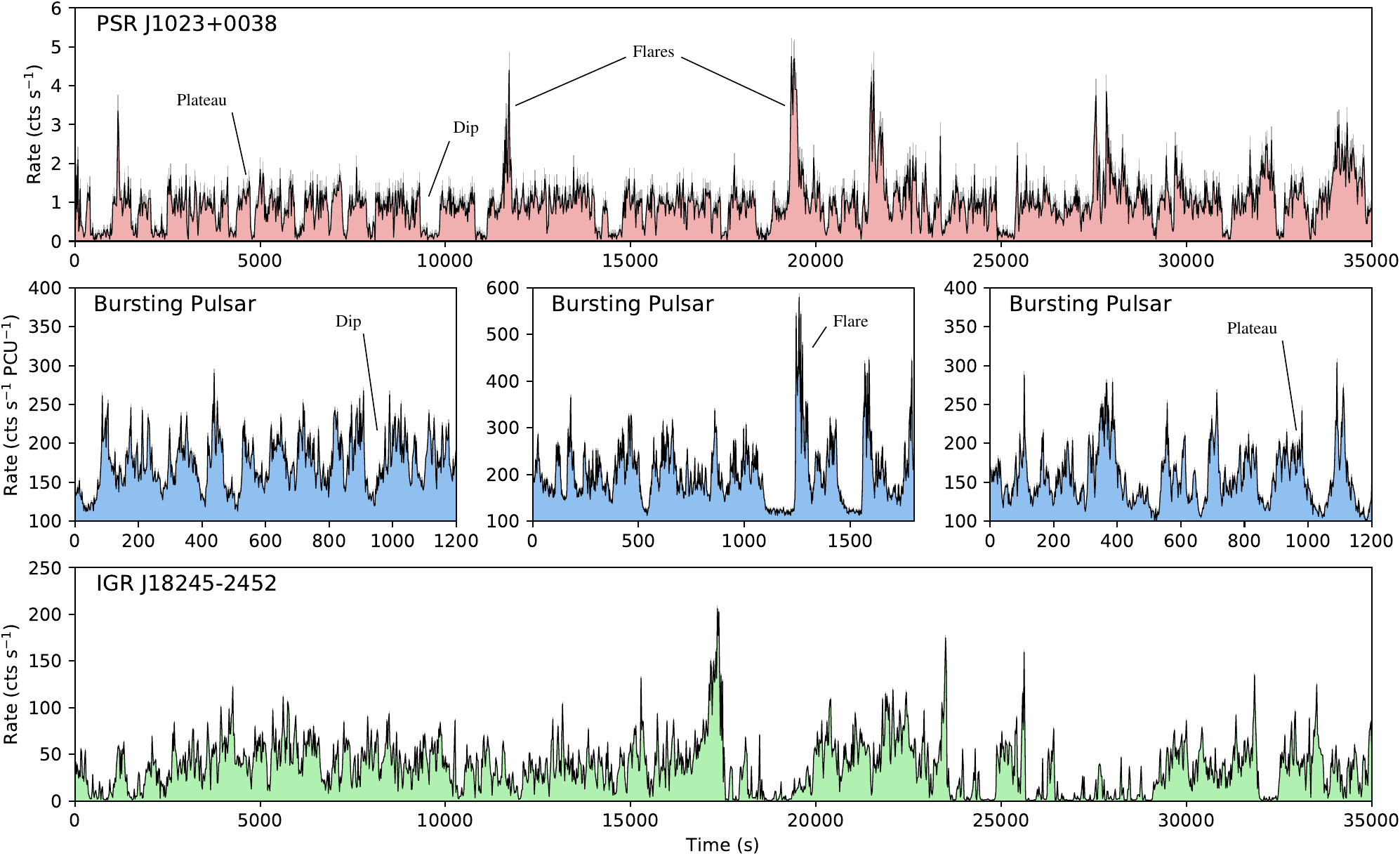}}}
 \caption[Lightcurves from the Bursting Pulsar and from two TMSPs, showing similar patterns of variability.]{\textbf{Top:} 2--15\,keV \indexxmm\textit{XMM} lightcurve\index{Lightcurve} from the TMSP\index{TMSP} PSR J1023+0038\index{PSR J1023+0038}. \textbf{Middle:} 2--60\,keV \indexrxte\textit{RXTE} lightcurves from the Bursting Pulsar during its 1996 and 1997 outbursts\index{Outburst}, showing similar variability\index{Variability} patterns to those seen in PSR J1023+0038. \textbf{Bottom:} 2--15\,keV \textit{XMM} lightcurve from the TMSP IGR J18245-2452\index{IGR J18245-2452}. \textit{XMM} lightcurves are shown from 2--15\,keV so that they can be more directly compared with \textit{RXTE}.}
 \label{fig:lcs}
\end{figure*}

\par In Figure \ref{fig:lcs}, I show \indexrxte\textit{RXTE} lightcurves\index{Lightcurve} of `Structured Bursting'\index{Structured bursting} from the Bursting Pulsar\index{Bursting Pulsar} alongside lightcurves from periods of `hiccup'\index{Hiccup accretion} variability\index{Variability} observed in the confirmed TMSPs\index{TMSP} PSR J1023+0038\index{PSR J1023+0038} and IGR J18245-2452\index{IGR J18245-2452}. All three sources show similar patterns of X-ray variability\index{Variability}:
\begin{itemize}
\item \textit{Plateaus}\index{Plateau}: periods of approximately constant count rate with high-amplitude flicker noise (all plateaus in a given observation have approximately the same mean rate),
\item \textit{Dips}\index{Dip}: Periods of low count rate ($\lesssim0.5$ of the rate in plateaus) with significantly less flicker noise, and 
\item \textit{Flares}\index{Flare}: Relatively short-lived increases of the count rate to values $\gtrsim2$ times greater than the rate during plateaus.
\end{itemize}
In TMSPs\index{TMSP}, these features are interpreted as representing three quasi-stable accretion\index{Accretion} modes: the `high', `low' and `flaring' modes respectively (e.g. \citealp{Bogdanov_TMSPVar}). The most significant difference is that, in general, the variability in the Bursting Pulsar\index{Bursting Pulsar} occurs on timescales $\sim1$ order of magnitude longer than those in TMSPs.
\par
\begin{figure}
 \centering
 \includegraphics[width=.82\linewidth, trim={1.3cm 0.1cm 1.7cm 1.1cm},clip]{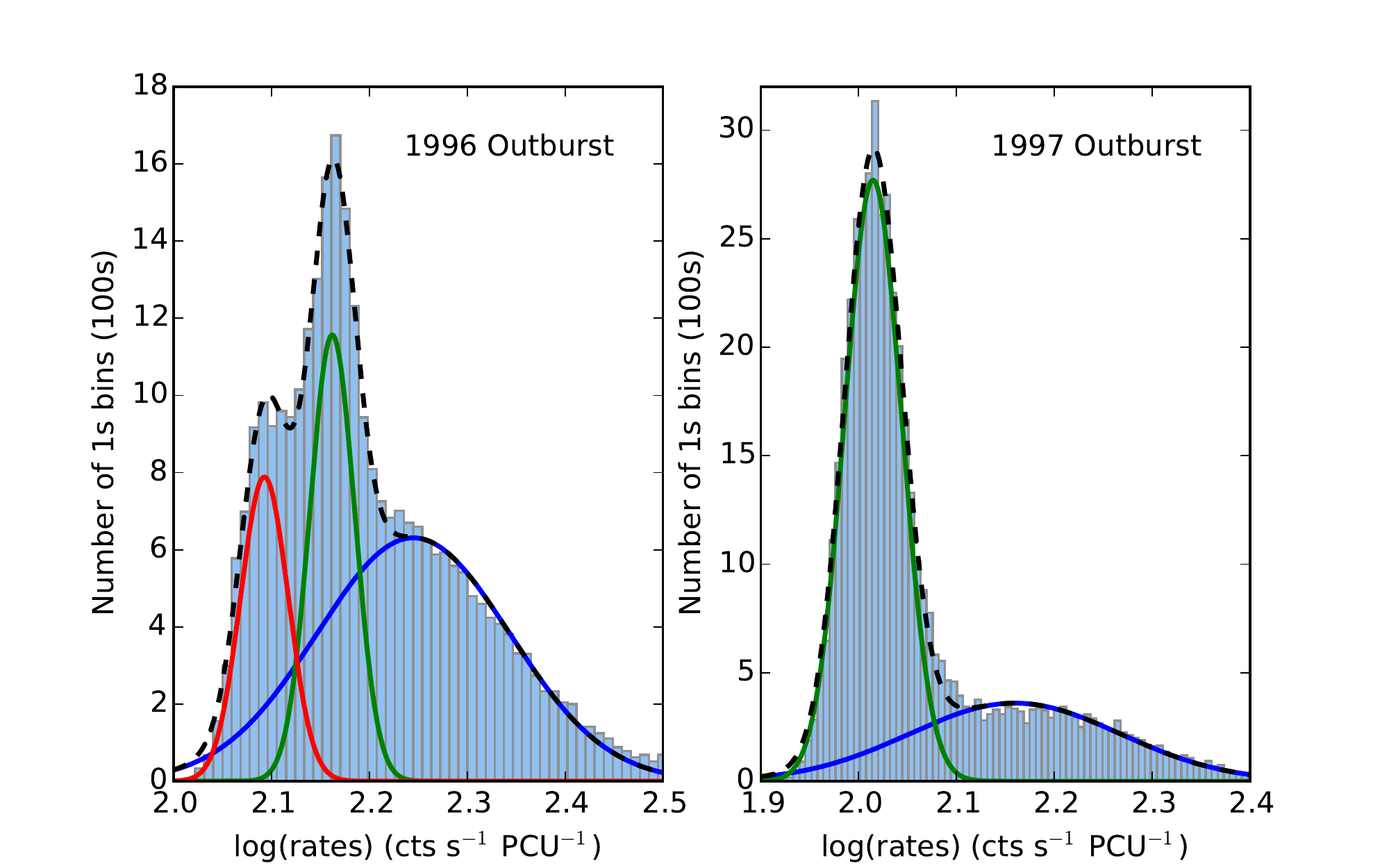}
 \caption[Histograms of the 1\,s binned count rates from all \textit{RXTE} observations of Structured Bursting in the 1996 and 1997 outbursts of the Bursting Pulsar.]{Histograms of the 1\,s binned count rates from all \indexrxte\textit{RXTE} observations of Structured Bursting\index{Structured bursting} in the 1996 (left) and 1997 (right) outbursts\index{Outburst} of the Bursting Pulsar\index{Bursting Pulsar}. For the 1996 outburst, I fit the distribution with three Gaussians, while for the 1997 outburst I fit the distribution with 2 Gaussians. The individual Gaussians are plotted in solid lines, while the combined total is plotted in a dashed line.}
 \label{fig:bimodal}
\end{figure}

 In Figure \ref{fig:bimodal} I show histograms of the 1\,s-binned count-rate from all \indexrxte\textit{RXTE} observations of Structured Bursting\index{Structured bursting} in the 1996 (left) and 1997 (right) outbursts\index{Outburst} of the Bursting Pulsar\index{Bursting Pulsar}. As is the case for TMSPs\index{TMSP}, the histograms can be described with a number of log-Normally distributed populations: 3 populations in the 1996 outburst and 2 in the 1997 outburst. It is unclear why a population would be absent from the 1997 outburst, but some TMSPs have been observed to miss the `high' mode during hiccup accretion\index{Hiccup accretion} (e.g. IGR J18245-2452\index{IGR J18245-2452}, \citealp{Ferrigno_TMSPVar}).

\par Detailed works on the low and high modes observed in the lightcurves\index{Lightcurve} of TMSPs\index{TMSP} show that X-ray pulsations\index{Pulsar} are seen during both modes. Pulsations are fractionally weaker in the low state than the high state (for example varing between $4.0\pm0.2\%$ and $16.8\pm0.2\%$ in the TMSP IGR J18245-2452\index{IGR J18245-2452}, \citealp{Ferrigno_TMSPVar}). In the case of the Bursting Pulsar\index{Bursting Pulsar}, analysis by \textsf{A.S.} detects pulsations both during the low and the high modes; much like in TMSPs, the pulsations are weaker in the low mode. For example in \indexrxte\textit{RXTE} OBSID 10401-01-59-00 (in 1996), the pulsations had amplitudes of $3.5\pm0.2\%$ and $4.9\pm0.2\%$ respectively, while in OBSID 20078-01-23-00 (in 1997), the pulsations had amplitudes of $4.5\pm0.1\%$ and $6.0\pm0.1\%$ respectively. A reduction in pulse fraction in accreting pulsars\index{Pulsar} has been interpreted as a change in accretion\index{Accretion} geometry due to a sudden decrease in the amount of matter reaching the neutron star\index{Neutron star} (e.g. \citealp{Ibragimov_PulseFrac}), and as such this result provides direct evidence that the Structured Bursting\index{Structured bursting} in the Bursting Pulsar is caused by switches between accretion\index{Accretion} and propeller\index{Propeller effect}-driven outflows.

\par TMSPs\index{TMSP} are amongst the only LMXBs\index{X-ray binary!Low mass} which are also significant $\gamma$-ray sources (e.g. \citealp{Hill_XSS}). The \index{Fermi@\textit{Fermi}}\textit{Fermi} point source 3FGL J1746.3--2851c\index{3FGL J1746.3--2851c} is spatially coincident with the Bursting Pulsar\index{Bursting Pulsar}. While the field is too crowded to unambiguously associate 3FGL J1746.3--2851c with the Bursting Pulsar, the existence of a $\gamma$-ray point source at this location is consistent with the possibility that the Bursting Pulsar and TMSPs show the same phenomenology.

\par
\begin{figure}
 \centering
 \includegraphics[width=.82\linewidth, trim={0.6cm 0.1cm 1.0cm 1.1cm},clip]{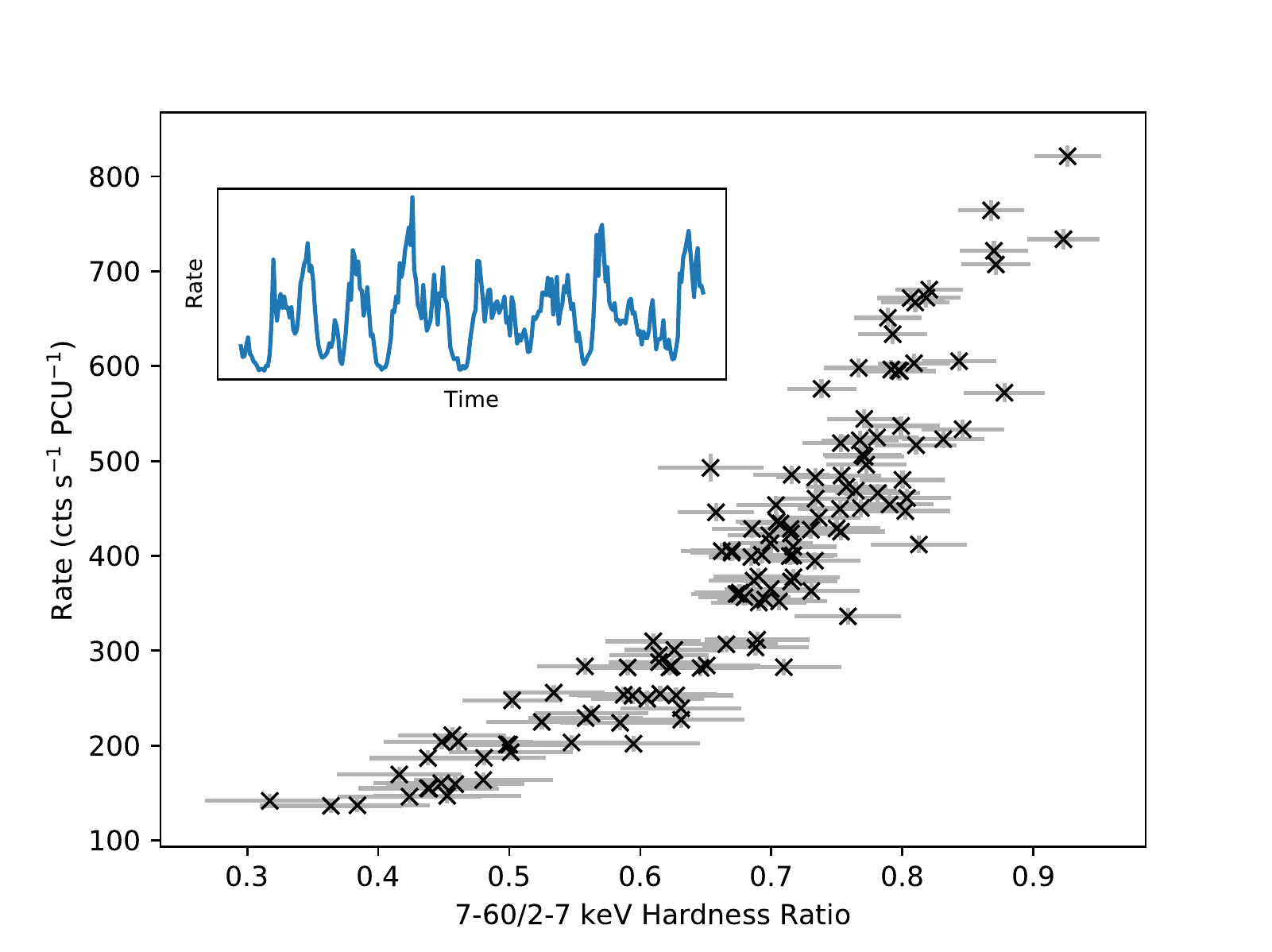}
 \caption[A 7--60/2--7\,keV hardness-intensity diagram for \textit{RXTE} observation 10401-01-59-00 of the Bursting Pulsar.]{A 7--60/2--7\,keV hardness-intensity\index{Hardness-intensity diagram} diagram for \indexrxte\textit{RXTE} observation 10401-01-59-00 of the Bursting Pulsar\index{Bursting Pulsar}; the lightcurve\index{Lightcurve} of this observation is shown in the inset. To correct for the high background\index{Background subtraction} of the region, I subtract the median count rate of \indexpca\textit{RXTE}/PCA observation 30075-01-24-00 from each band; at this time, the Bursting Pulsar was in quiescence\index{Quiescence}. I find a strong correlation between hardness\index{Colour} and count rate, with a Spearman Rank Correlation Coefficient\index{Spearman's rank correlation coefficient} of 0.93. Data for the hardness-intensity diagram are binned to 10\,s, while data for the lightcurve are binned to 5\,s.}
 \label{fig:HR}
\end{figure}

 The spectral\index{Spectroscopy} evolution of known TMSPs\index{TMSP} is varied. In PSR J1023+0038\index{PSR J1023+0038}, the low, high and flaring modes all present similar spectra \citep{Bogdanov_TMSPVar}. However in IGR J18245-2452\index{IGR J18245-2452}, \citealp{Ferrigno_TMSPVar} have found a strong correlation between spectral hardness\index{Colour} and intensity during hiccups\index{Hiccup accretion}, showing that there is spectral evolution over time in this source. In Figure \ref{fig:HR} I show the hardness-intensity diagram\index{Hardness-intensity diagram} of the Bursting Pulsar\index{Bursting Pulsar} during periods of Structured Bursting\index{Structured bursting}. I find a significant correlation, similar to what is seen in IGR J18245-2452 \citep{Ferrigno_TMSPVar}. This is in contrast with other slow accreting pulsar systems such as Vela X-1\index{Vela X-1}, which show an anticorrelation between these parameters during periods of variability\index{Variability} \citep{Kreykenbohm_Vela}.

\section{Discussion}

\par In this chapter I compare the lightcurve\index{Lightcurve}, spectral\index{Spectroscopy} and timing properties of the Bursting Pulsar\index{Bursting Pulsar} at the end of its 1996 and 1997 outbursts\index{Outburst} with those observed from Transitional Millisecond Pulsars\index{TMSP}. The data suggest that the Bursting Pulsar may have undergone `hiccup'\index{Hiccup accretion} accretion similar to that seen in TMSPs, during which matter donated to the neutron star\index{Neutron star} by the companion star\index{Companion star} alternates between being accreted\index{Accretion} onto the poles of the neutron star\index{Neutron star} and being ejected from the system by the propeller\index{Propeller effect} effect (e.g. \citealp{Ferrigno_TMSPVar}). This similarity raises the exciting prospect of studying the physics of TMSPs in a completely different regime.
\par
\begin{figure}
 \centering
 \includegraphics[width=.82\linewidth, trim={0.6cm 0.1cm 1.0cm 1.1cm},clip]{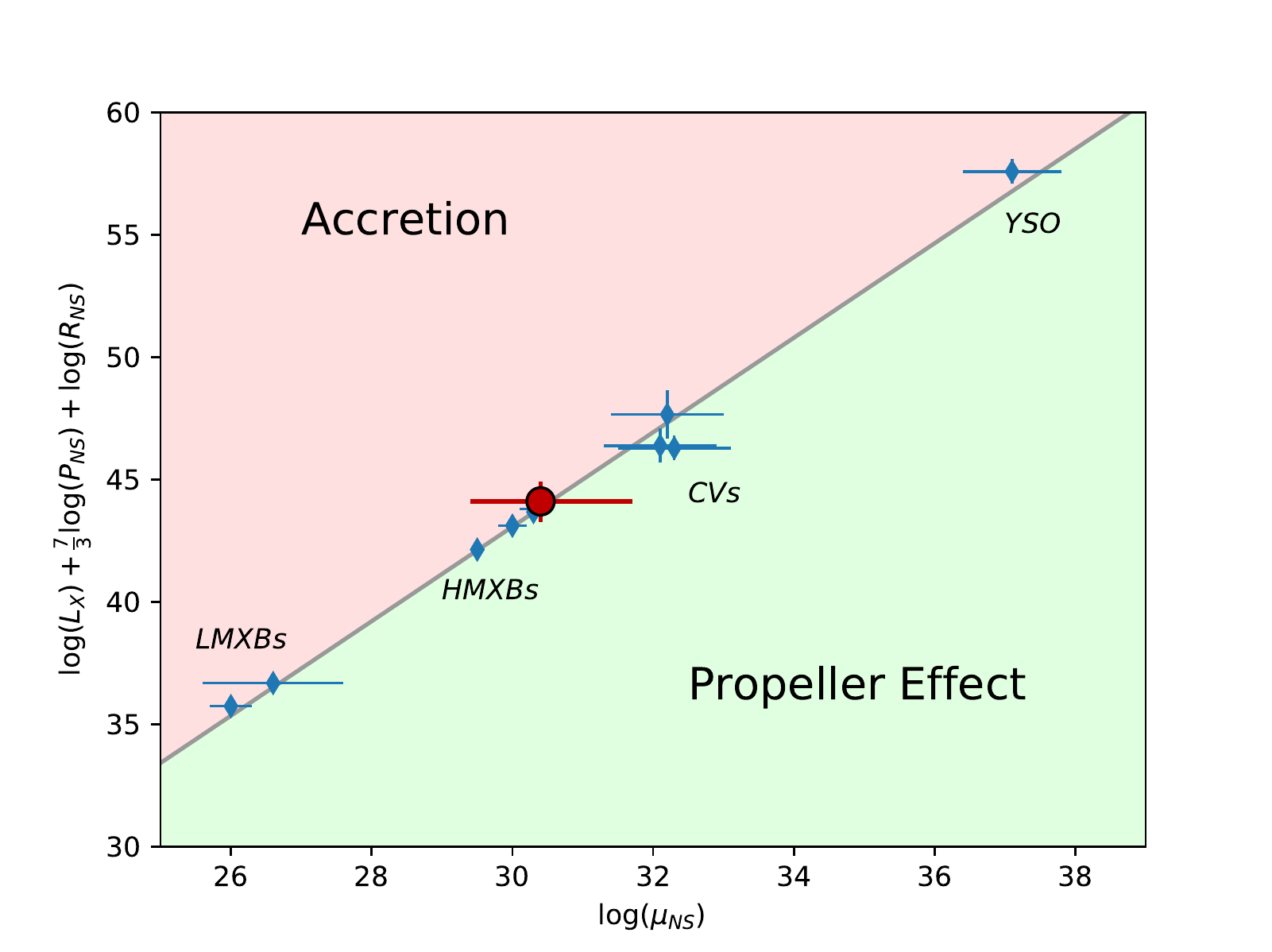}
 \caption[A plot of a number of objects ranging in scale from LMXBs and High-Mass X-ray Binaries (HMXBs) to Cataclysmic Variables (CVs) and Young Stellar Objects (YSOs). In each case, the object is plotted at the luminosity which defines its transition between propeller-mode accretion and free accretion.]{A plot of a number of objects ranging in scale from LMXBs\index{X-ray binary!Low mass} and High-Mass X-ray Binaries\index{X-ray binary!High mass} (HMXBs) to Cataclysmic Variables\index{Cataclysmic variable} (CVs) and Young Stellar Objects\index{Young stellar object}\index{YSO|see {Young stellar object}} (YSOs) (blue diamonds). In each case, the object is plotted at the luminosity which defines its transition between propeller-mode\index{Propeller effect} accretion\index{Accretion} and free accretion. \citealp{Campana_PropBorder} suggest that any object above the line of best fit accretes freely, whereas all objects below are in the propellor regime. The Bursting Pulsar\index{Bursting Pulsar} (red circle) is consistent with approaching this line during periods of Structured Bursting\index{Structured bursting}. Errorbars on the Bursting Pulsar represent the range of the reported magnetic fields\index{Magnetic field} as well as a range of stellar radii between 10--20\,km. The range in luminosity for the Bursting Pulsar is calculated using 1.5-25\,keV \indexpca\textit{RXTE}/PCA flux, assuming a distance of between 4--8\,kpc (e.g. \citealp{Kouveliotou_BP,Gosling_BPCompanion,Sanna_BP}) and a bolometric correction factor\index{Bolometric correction factor} of 1--3.  Data on the other objects taken from \citealp{Campana_PropBorder}. $L$ is the bolometric luminosity of the object in ergs\,s$^{-1}$, $P$ is the period in s, $R$ is the radius in cm and $\mu$ is the magnetic moment\index{Magnetic moment} in $Gauss\,cm^3$.}
 \label{fig:propBorder}
\end{figure}

 Recently \citealp{Campana_PropBorder} proposed a universal relation between magnetic moment\index{Magnetic moment}, spin frequency\index{Spin}, stellar radius and luminosity at the boundary between accretion\index{Accretion} and the propeller effect\index{Propeller effect}. Any object that exists on one side of this boundary should be able to accrete, whereas objects on the other side should be in the propeller phase or not accreting at all. In Figure \ref{fig:propBorder} I reproduce \citealp{Campana_PropBorder}'s results and include my estimates for the Bursting Pulsar\index{Bursting Pulsar} during the periods of Structured Bursting\index{Structured bursting}. I find that the Bursting Pulsar is consistent with lying on or near the boundary between propeller-mode and direct accretion, clustering with High Mass X-ray Binaries\index{X-ray binary!High mass} (as expected due to the Bursting Pulsar's high magnetic field\index{Magnetic field}), and supporting the link between `hiccups'\index{Hiccup accretion} and Structured Bursting.

\par
\begin{figure}
 \centering
 \includegraphics[width=.82\linewidth, trim={0cm 0cm 0cm 0cm},clip]{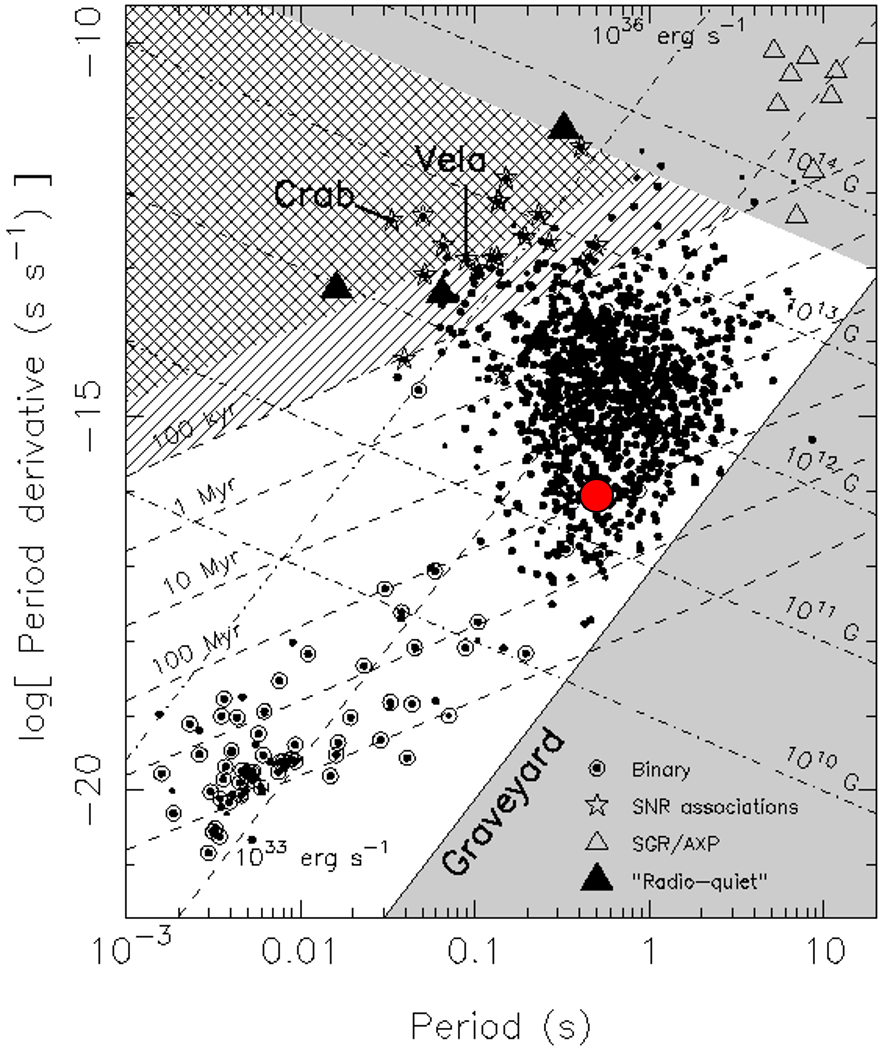}
 \caption[A plot of spin period against rate of spin-down for the population of all known radio pulsars, a so-called $P$--$\dot{P}$ diagram, showing where the Bursting Pulsar lies in this parameter space and indicating that it lies well outside of the pulsar `graveyard'.]{A plot of spin period against rate of spin-down for the population of all radio pulsars\index{Pulsar}, generally referred to as a $P$--$\dot{P}$ diagram.  Diagonal lines indicate the estimated age and surface magnetic field strength of a typical radio pulsar at any given position on the diagram.  Any pulsars to the right of the `graveyard' line on this plot are expected to be inactive in the radio, whereas objects to the left are expected to be observable as radio pulsars (e.g. \citealp{vandenHeuvel_Graveyard}).  The position of the Bursting Pulsar\index{Bursting Pulsar}, estimated from its surface field strength and spin period, is shown in red; well outside of the pulsar graveyard\index{Graveyard}.  Figure adapted from \citet{Lorimer_Handbook}, and is accurate as of the time of its first publication.}
 \label{fig:graveyard}
\end{figure}

 If the `hiccups'\index{Hiccup accretion} in the Bursting Pulsar\index{Bursting Pulsar} show that the system is transiting to a radio pulsar, then the Bursting Pulsar should not lie in the $P$-$\dot{P}$ `graveyard'\index{Graveyard} region \citep[e.g.][]{vandenHeuvel_Graveyard}. To my knowledge, there is no measurment yet of the neutron star\index{Neutron star} spin down during the Bursting Pulsar\index{Bursting Pulsar}'s X-ray quiescent\index{Quiescence} state. Under the assumption that the Bursting Pulsar becomes a radio pulsar\index{Pulsar}, and that the possible spin down during that period is due to the same mechanism as those of the known radio pulsars, I can position the Bursting Pulsar in the $P$-$\dot{P}$ diagram (the plot of pulsar spin $P$ against spin-down rate $\dot{P}$, shown in Figure \ref{fig:graveyard}) by using the orbital period and estimates of its magnetic field\index{Magnetic field}. At $B\sim2\times10^{11}$G, the Bursting Pulsar falls well outside of the pulsar graveyard. I note that \citet{Pandey-Pommier_BPRad} and \citet{Russell_BPRad} did not detect a significant radio source at the location of the Bursting Pulsar during X-ray outburst\index{Outburst}. To my knowledge, there is no report of Radio detection/non-detection during X-ray quiescence.

\subsection{Comparison with other Objects}

\par In addition to the Bursting Pulsar\index{Bursting Pulsar}, several additional sub-10\,Hz accreting X-ray pulsars\index{Pulsar} have been discovered (e.g. GX 1+4\index{GX 1+4} and 4U 1626-67\index{4U 1626-67}, \citealp{Lewin_GX1,Rappaport_4U}). The reason behind the slow spins\index{Spin} of these objects is poorly understood, but a number of these systems have been seen to undergo `torque reversal'\index{Torque reversal} events, during which $\dot{P}$ switches sign (e.g. \citealp{Chakrabarty_4U,Chakrabarty_GX14}). In some sources, the magnitude of the spin-down during an event is of the same order of magnitude as the preceding period of spin-up, resulting in little or no net spin change. Torque reversal events occur irregularly, but the recurrence timescale\index{Recurrence time} varies between objects from weeks to decades (e.g. \citealp{Bildsten_Rev}).
\par The slow accreting pulsar\index{Pulsar} Vela X-1\index{Vela X-1} has been found to show an anticorrelation between hardness\index{Colour} and intensity \citep{Kreykenbohm_Vela}, whereas I find a strong positive correlation between these parameters in the Bursting Pulsar\index{Bursting Pulsar} during periods of Structured Bursting\index{Structured bursting} (Figure \ref{fig:HR}). This significant spectral\index{Spectroscopy} difference, combined with the other phenomenological differences between these objects reinforces the idea that the Bursting Pulsar exists in a very different physical state from the other known slow accreting pulsars.
\par Given that the Bursting Pulsar\index{Bursting Pulsar} has a strongly stripped stellar companion\index{Companion star} \citep{Bildsten_Nuclear}, a high magnetic field\index{Magnetic field} and shows significant spin-up during outburst\index{Outburst} (e.g. \citealp{Finger_BP,Sanna_BP}), it is difficult to explain its low spin\index{Spin} by suggesting the system is young or that the angular momentum\index{Angular momentum} transfer is inefficient. \citealp{Rappaport_BPHistory} suggest that the magnetic field and spin could be explained if much of the mass transfer in the system occurred before the primary became a neutron star\index{Neutron star}, but they note that this scenario is inconsistent with the low mass of the donor star.
\par Torque reversal\index{Torque reversal} events in the Bursting Pulsar\index{Bursting Pulsar} (similar to those seen in other slow accreting pulsars\index{Pulsar}, e.g. \citealp{Bildsten_Rev}) could explain why the pulsar has failed to reach a spin\index{Spin} rate on par with TMSPs\index{TMSP}.  Although no torque reversal event has been reported from the Bursting Pulsar, it is feasible that the recurrence timescale\index{Recurrence time} of such an event is longer than the $\sim20$ years for which the object has been studied (this is consistent with the recurrence timescales seen in other slow accreting pulsars). The discovery of torque reversal in the Bursting Pulsar would strongly link it with the other known slow accreting pulsars.

\par The Rapid Burster\index{Rapid Burster} is often compared to the Bursting Pulsar\index{Bursting Pulsar} due to the presence of regular Type II X-ray bursts\index{X-ray burst!Type II} in both objects (e.g. \citealp{Lewin_BP}). This system also contains an accreting neutron star\index{Neutron star}. \citealp{Iaria_RB} have suggested that the vast majority of matter transferred in this system is ejected, similar to a scenario suggested by \citealp{Degenaar_BPSpec} to explain high-velocity winds\index{Wind} from the Bursting Pulsar. However it remains unclear why the Rapid Buster does not show pulsations\index{Pulsar} or display the `hiccup'\index{Hiccup accretion} behaviour seen in the Bursting Pulsar.

\section{Conclusion}

\par The Bursting Pulsar\index{Bursting Pulsar} has a spin\index{Spin} rate $\sim2$ orders of magnitude less than previously known TMSPs\index{TMSP}, and a magnetic field\index{Magnetic field} $\sim2$ orders of magnitude stronger, but it still shows lightcurve\index{Lightcurve}, timing and spectral\index{Spectroscopy} behaviour which are remarkably similar to TMSPs. This raises the exciting prospect of exploring the physics of TMSPs in a previously unexplored physical regime. If the Bursting Pulsar itself is a transitional pulsar, it should emit radio pulsations during X-ray quiescence\index{Quiescence}. Future detections of radio pulsations from this object would unambiguously confirm it as a transitional pulsar.

\cleardoublepage

\chapter{Discussion}

\label{ch:conc}

\epigraph{\textit{These thoughts are constructive criticisms. Pyramidical. I try to suppress these thoughts, but they leak out...}}{George Saden -- \textit{Zardoz}}

\vspace{1cm}

\par\noindent In Chapters \ref{ch:IGR} and \ref{ch:BPbig} I discuss new ways of classifying variability\index{Variability} in the unusual LMXB\index{X-ray binary!Low mass}s IGR J17091\index{IGR J17091-3624} and the Bursting Pulsar\index{Bursting Pulsar}.  I use the new classification frameworks I have created to compare these objects with the similar LMXBs GRS 1915\index{GRS 1915+105} and the Rapid Burster\index{Rapid Burster}.  While I find a number of similarities between these objects, I also highlight a number of differences.  For example, I find that the spectral\index{Spectroscopy} evolution during variability in IGR J17091 is very different to in GRS 1915, and the bursts\index{X-ray burst} in the Bursting Pulsar evolve in a very different way to those seen from the Rapid Burster.
\par A common theme throughout the work presented in this thesis is that the variability\index{Variability} in these unusual objects is even more complex than had previously been thought.  Application of Occam's razor\index{Occam's razor} \citep{Occam} suggests that the similar variability from these objects is generated by similar physics, but I have found that it is difficult to unify the diverse behaviours of these unusual systems.  I also compare the four objects that I focus on with other unusual XRBs\index{X-ray binary} such as Terzan 5 X-2\index{Terzan 5 X-2} and, in Chapter \ref{ch:BPletter}, TMSPs\index{TMSP}.  In this chapter I further discuss the relationships between these seemingly disparate objects, and how my findings fit in to the more general picture of accretion\index{Accretion} in these extreme and bizarre systems.

\section{General Observations}

\label{sec:disccomp}

\subsection{Variability Evolution throughout an Ouburst}

\par One feature common to at least 3\footnote{As GRS 1915\index{GRS 1915+105} has been in outburst\index{Outburst} since its discovery, it is unknown how its variability classes\index{Variability class} vary over the duration of an outburst.} of the unusual LMXBs\index{X-ray binary!Low mass} discussed in this thesis is that variability\index{Variability} changes in a predictable way over the course of an outburst.  This effect is most apparent in the results I present in Chapter \ref{ch:BPbig}, where I discuss an evolution of bursting\index{X-ray burst} which is observed in both the 1996 and 1997 outbursts of the Bursting Pulsar\index{Bursting Pulsar}:
\begin{itemize}
\item Normal Bursts\index{Normal burst} and Minibursts\index{Miniburst} begin to occur shortly after the peak of each outburst.
\item Bursting shuts off entirely when the persistent intensity\index{Persistent emission} of the source decreases below $\sim0.1$\,Crab.
\item The persistent flux of the system increases in a rebrightening\index{Re-flare} event, at which point Mesobursts\index{Mesoburst} begin to occur.
\item Mesobursts evolve into Structured Bursting\index{Structured bursting}.
\end{itemize}
\par A predictable evolution of variability\index{Variability} throughout an outburst\index{Outburst} has also been identified in the Rapid Burster\index{Rapid Burster} (e.g. \citealp{Bagnoli_PopStudy}).  In this system, Type II\index{X-ray burst!Type II} bursts near the start of each outburst are Eddington\index{Eddington limit}-Limited and persist for $\sim100$s of seconds (see e.g. the upper panel of Figure \ref{fig:bagnoli_lcs}).  As the outburst evolves, these bursts become shorter, fainter and more sharply peaked (see e.g. the lower panel of Figure \ref{fig:bagnoli_lcs}).  This evolution is qualitatively very different from the evolution of variability seen in the Bursting Pulsar\index{Bursting Pulsar}.  However, the fact that bursting in both objects changes in a predictable way throughout each outburst shows that bursting in both objects is dependent on the accretion rate\index{Accretion rate} in the system and on the state of its accretion disk\index{Accretion disk}.
\par There is also some evidence of an evolution of the variability\index{Variability} displayed by IGR J17091\index{IGR J17091-3624}.  In Figure \ref{fig:WhereCls} I show a number of lightcurves\index{Lightcurve} of the 2011 outburst\index{Outburst} of IGR J17091\index{IGR J17091-3624}, highlighting when in the outburst each of our 9 variability classes\index{Variability class} was observed.  Although the evolution between classes apparently not as strict as in the Bursting Pulsar\index{Bursting Pulsar}, it is easy to identify a number of patterns in the data, such as:
\begin{itemize}
\item Class I\indexi\footnote{See Section \ref{sec:IGRclassesintro} for a description of each class.} only occurs near the start of the outburst\index{Outburst}, within 25 days of the onset of variability\index{Variability}.
\item Class II\indexii\ only occurs during two dips\index{Dip} in the persistent flux\index{Persistent emission} to a level of $\sim20$\,mCrab.
\item Class VII\indexvii\ only occurs while the persistent flux\index{Persistent emission} of IGR J17091\index{IGR J17091-3624} is in a narrow band centred on $\sim70$\,mCrab.
\end{itemize}
It is unclear whether a similar evolution occurs during the 2016 outburst\index{Outburst} of IGR J17091\index{IGR J17091-3624}, as a variability\index{Variability} population study\index{Population study} for this outburst has not yet been performed.  However these results from the 2011 outburst suggest that variability in IGR J17091 also depends strongly on the accretion rate\index{Accretion rate} and the state of the disk\index{Accretion disk}, as it is in the Rapid Burster\index{Rapid Burster} and the Bursting Pulsar\index{Bursting Pulsar}, and therefore variability should evolve in a predictable way over the course of each outburst.

\subsection{Criteria for Exotic Variability}

\label{sec:criteria}

\par
\begin{figure}
  \centering
  \includegraphics[width=.9\linewidth, trim= 0mm 0mm 10mm 10mm,clip]{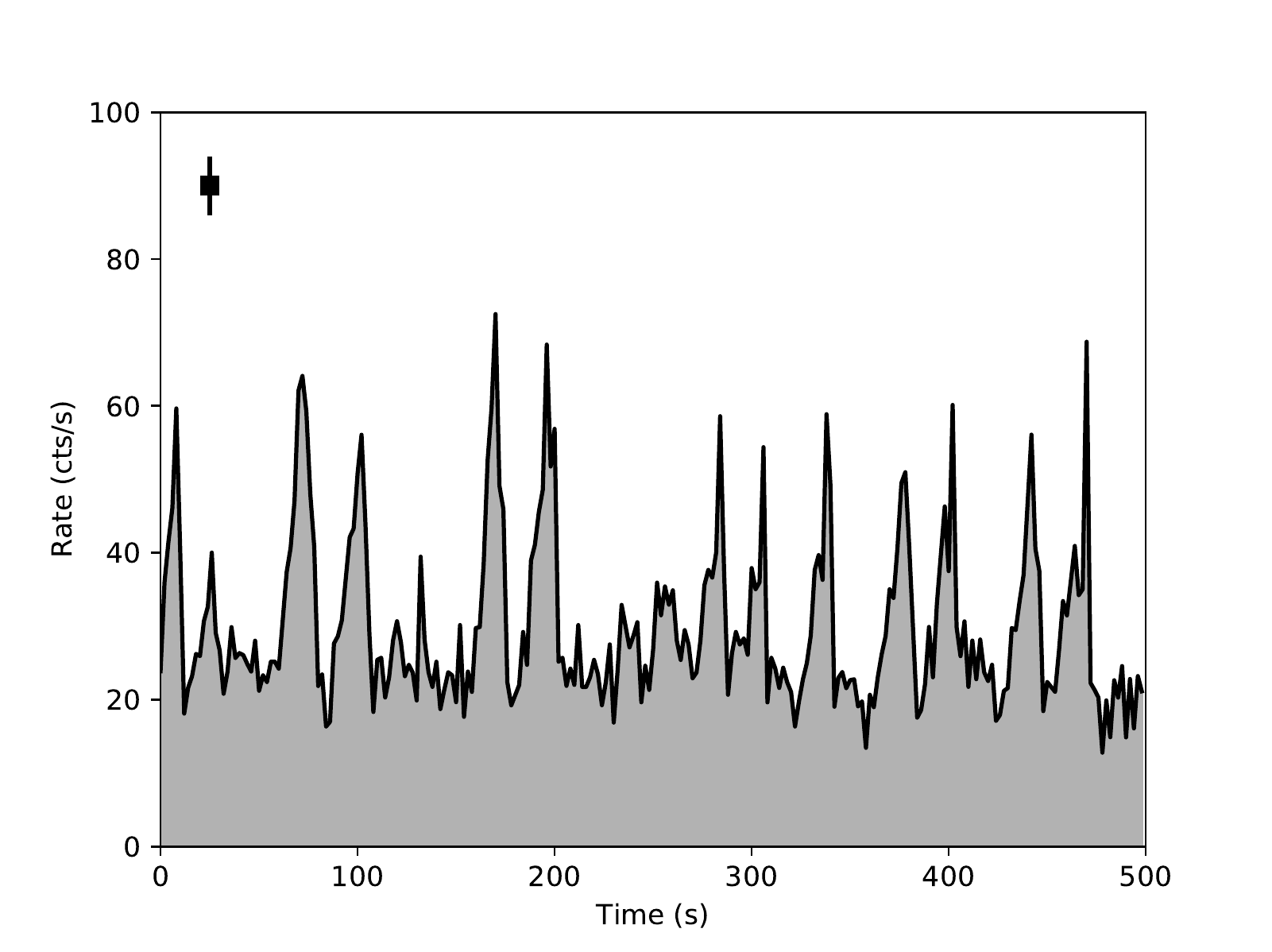}
  \caption[A \textit{Swift}/XRT lightcurve of IGR J17091-3624 during its 2016 outburst, showing Class III variability.]{0.3--10\,keV \indexxrt\textit{Swift}/XRT lightcurve\index{Lightcurve} of IGR J17091-3624\index{IGR J17091-3624} during its 2016 outburst\index{Outburst}.  The black bar shows the typical size of errors.  This lightcurve shows Class III\indexiii\ variability\index{Variability}, which I identify in the 2011 outburst of this source and describe in Section \ref{sec:classIII}.}
  \label{fig:IGR2016}
\end{figure}

 As we show in Figure \ref{fig:IGR2016}, the 2016 outburst of IGR J17091 displayed similar variability to that which it showed in 2011 (e.g. \citealp{Reynolds_2016HB}), meaning that variability was not unique to the source's 2011 outburst.  Additionally Type II bursting has been seen in the 1996, 1997 and 2014 outbursts of the Bursting Pulsar, and the Rapid Burster goes into outburst regularly every $\sim$100--200 days and always displays Type II bursts.  Therefore 3 of the 4 objects I discuss in this thesis have produced their characteristic variable behaviour during multiple separate outbursts\footnote{As GRS 1915 has been in outburst since discovery, it is not possible to tell whether its variability is repeated in subsequent outbursts.}.  This observation strongly suggests that the ability to produce such variability is a property of the system rather than of an individual outburst.  In these systems some set of parameters, which persist between outbursts, are just right to allow these exotic types of variability to occur.

\par Compact objects\index{Compact object} are relatively simple, and they can be well-defined with only a few parameters:

\begin{itemize}
\item The type of compact object\index{Compact object} (neutron star\index{Neutron star} or black hole\index{Black hole}).
\item Mass.
\item Spin\index{Spin} and rate of change of spin.
\item Radius.
\item Magnetic Field\index{Magnetic field} Strength.
\end{itemize}

Where the last two parameters only apply if the object is a neutron star\index{Neutron star}.  In addition to these, one only has to describe the companion star\index{Companion star} (mass, mass loss rate, spectral type etc.), the parameters of the system orbit (eccentricity, semi-major axis, inclination and misalignment from the spin axes of both stars) and the mass transfer rate to fully describe an LMXB\index{X-ray binary!Low mass} system.  Due to this relative simplicity, there are not many candidates for parameters which govern the existence of exotic variability\index{Variability}.
\par GRS 1915+105\index{GRS 1915+105} contains the highest mass black hole\index{Black hole} confirmed in an LMXB\index{X-ray binary!Low mass}\footnote{At least one HMXB\index{X-ray binary!High mass}, Cyg X-1\index{Cyg X-1}, is believed to contain a black hole with higher mass \citep{Orosz_CygX1}.} ($12.4\pm2.0$\,M$_\odot$, \citealp{Reid_Parallax}), although many other LMXBs are believed to contain black holes\index{Black hole} with comparable masses (e.g. V404 Cyg\index{V404 Cyg}, \citealp{Shahbaz_V404}).  The black hole in GRS 1915 also has a very high spin\index{Spin}, with a spin\index{Spin} parameter of $0.98\pm0.01$ \citep{Miller_GRSspin}, but a number of other XRBs\index{X-ray binary} are also believed to harbour near-maximal spin black holes (see e.g. \citealp{Fragos_spin}).  Therefore it seems unlikely that mass or spin alone provide the criteria for GRS 1915-like variability\index{Variability}.%  In Chapter \ref{ch:BPletter} we consider a `hiccup'-like accretion behaviour as a possible mechanism behind variability in the Bursting Pulsar; this mechanism relies on specific ranges for the values of spin and magnetic field strength, but cannot be used to explain any of the variability seen in black hole systems.
\par Lense-Thirring precession\index{Lense-Thirring precession} in the disk\index{Accretion disk}, a frame-dragging effect\index{Frame-dragging} caused by the misalignment of the orbital and spin\index{Spin} axes of the compact object\index{Compact object}, has been used to explain some of the variability\index{Variability} seen in LMXBs\index{X-ray binary!Low mass} (e.g. \citealp{Stella_LT}).  However, the timescale of the variability\index{Variability} this generates is no slower than $\sim0.1$\,s \citep{Ingram_Solid}, too fast to be linked with the $\lesssim0.1$\,Hz variability seen from GRS 1915\index{GRS 1915+105}.  Instead the other parameters on the list, or a combination thereof, must be the determining factors in whether or not an object can display exotic variability.
%\par While I find that Eddington-limited accretion is not necessary for either GRS 1915-like variability or Type II bursting, all of the objects considered in this thesis have relatively high accretion rates during the peak of their outbursts: the Rapid Burster, which is likely the least luminous of the four systems, accretes at $\sim20$\% of its Eddington rate at peak.  While a high accretion rate is obviously not sufficient for GRS 1915-like variability or Type II bursting to occur, as many systems with high accretion rates do not display either behaviour, this result suggests that a high accretion rate may be a necessary criterion.
\par GRS 1915\index{GRS 1915+105} has a very long orbital period\index{Orbital period} of $\sim30$ days \citep{Neil_GRSPeriod}, and this is believed to result in it having the largest accretion disk\index{Accretion disk} of all known X-ray binaries\index{X-ray binary}.  This likely explains how GRS 1915 has been in outburst\index{Outburst} for such a long period of time ($\gtrsim20$ years, compared to the $\lesssim2$\,year outbursts\index{Outburst} seen in most black hole\index{Black hole} LMXB\index{X-ray binary!Low mass}s).  The orbital periods of IGR J17091\index{IGR J17091-3624} and the Rapid Burster\index{Rapid Burster} are unknown, but the Bursting Pulsar\index{Bursting Pulsar} also has a relatively long orbital period of 11.8 days (e.g. \citealp{Finger_BP}), suggesting that a large disk\index{Accretion disk} may also be a factor in the generation of exotic variability\index{Variability}.
\par Another property to consider is the magnetic field\index{Magnetic field} strength in the disk\index{Accretion disk}.  Although astrophysical black holes\index{Black hole} do not have a magnetic field, magnetic fields in a black hole can still arise from one of two sources:
\begin{itemize}
\item The field of the system's donor star\index{Companion star}, including field lines advected into the disk by accretion\index{Accretion}.
\item The movement of ionised material within the disk\index{Accretion disk}.
\end{itemize}
The presence of such a magnetic field\index{Magnetic field} may be able to stabilise an accretion disk\index{Accretion disk} against the instabilities\index{Instability} described in Section \ref{sec:diskinstab} (e.g. \citealp{Sadowski_MagField}.  \citeauthor{Sadowski_MagField} calculates the minimum magnetic field strength which, when threaded through a radiation-dominated\index{Radiation pressure} disk undergoing the instability described by \citet{Shakura_Instab}, would be able to stabilise the disk.  Assuming that such a field in an LMXB\index{X-ray binary!Low mass} is provided by the companion star\index{Companion star}, they find that this minimum value in an LMXB depends mostly on the luminosity of the system and the mass of the compact object\index{Compact object}.  They estimate the value of the minimum stabilising field strength in a number of black hole X-ray binaries, and find that the field required to stabilise GRS 1915\index{GRS 1915+105} is $\sim5.7\times10^{23}$\,G\,cm$^2$; this is over twice as large as the second highest value they find for an LMXB ($1.9\times10^{23}$\,G\,cm$^2$ for XTE J1550-564\index{XTE J1550-564}).  This high value is due to the large black hole\index{Black hole} mass in GRS 1915, and the fact that this black hole accretes at a near-Eddington\index{Eddington limit} rate\index{Accretion rate} and, thus, a high luminosity.  In this picture, therefore, GRS 1915-like variability may simply be a manifestation of the \citet{Shakura_Instab} instability which is suppressed in most LMXBs.  One could test the viability of this scenario by calculating the minimum stabilising field for IGR J17091\index{IGR J17091-3624}.  Unfortunately at time of writing the companion star to IGR J17091 has not been conclusively identified, and the mass and distance (and hence luminosity) of the system remain poorly constrained.
\par It is worth noting that the scenario suggested by \citet{Sadowski_MagField} by itself is unable to account for Type II\index{X-ray burst!Type II} bursts as, in these systems, the neutron star\index{Neutron star} is able to provide more than enough magnetic\index{Magnetic field} flux to stabilise the inner disk\index{Accretion disk} in the scenario of \citeauthor{Sadowski_MagField}.  However, the scenario of \citeauthor{Sadowski_MagField} does not take into account the effects of the magnetic disruption of the inner disk, nor does it account for any effects caused by the disk's interaction with the rapidly spinning neutron star magnetic field.

\subsection{Evidence of System Memory}

\par Another notable observation from these objects is that variability\index{Variability} in both GRS 1915\index{GRS 1915+105} and IGR J17091\index{IGR J17091-3624} falls into a discrete set of variability classes\index{Variability class}.  In both objects, a variability class could be observed on one day, not be observed for weeks and then reappear in a later observation.  Somehow, the physics that governs variability in these systems only permit the system to occupy one of a discrete set of variability classes.  In Figure \ref{fig:IGR2016} I show evidence that at least one of the variability classes I identified in the 2011 outburst\index{Outburst} of IGR J17091 occurred again in 2016.  This suggests that GRS 1915-like systems are somehow able to `remember' which variability classes they can occupy, and that this memory persists between outbursts.  This in turn means that, in addition to determining whether or not GRS 1915-like variability can occur, the simple parameters that define a black hole\index{Black hole} LMXB\index{X-ray binary!Low mass} also determine \textit{which} classes of GRS 1915-like variability can occur.
\par In Section \ref{sec:IGRcomp}, I report that two of the classes\index{Variability class} of variability I find in IGR J17091\index{IGR J17091-3624} (Classes VII\indexvii\ and VIII\indexviii) are unlike anything which has ever been seen in GRS 1915\index{GRS 1915+105} (see also Table \ref{tab:class_assign}).  This leads to one of three possibilities:
\begin{itemize}
\item Class VII\indexvii\ and VIII\indexviii-like variability has occurred in GRS 1915\index{GRS 1915+105} since its discovery, but coincidentally we have never observed it.  As GRS 1915 has been observed extensively during its ongoing $\gtrsim20$ year-long outburst\index{Outburst}, this is unlikely to be true.
\item Class VII and VIII variability preferentially occurs during a specific disk\index{Accretion disk} configuration which occurs at a specific point in the evolution of an outburst, and GRS 1915 is not currently in this state.
\item Some set of system parameters in IGR J17091\index{IGR J17091-3624} are different from in GRS 1915 in such a way that many variability classes\index{Variability class} can appear in both objects, but a small set of different variability classes are permitted in each object.
\end{itemize}
\par The final possibility is of great interest.  Future observations of IGR J17091\index{IGR J17091-3624} will aim to better constrain the parameters that define the system.  If these parameters are then compared to the already well-constrained parameters of GRS 1915\index{GRS 1915+105}, then we may be able to learn exactly which physical properties of the system govern which set of variability classes\index{Variability class} can be displayed.
\par There is also evidence of some degree of system memory in the Type II\index{X-ray burst!Type II} bursting systems.  The near-identical evolutions of variability\index{Variability} in the 1996 and 1997 outbursts\index{Outburst} of the Bursting Pulsar\index{Bursting Pulsar} indicate that the factors governing this evolution persist between outbursts in this object as well.
\par The Normal Bursts\index{Normal burst} and Minibursts\index{Miniburst} of the Bursting Pulsar\index{Bursting Pulsar}, which I describe in Sections \ref{sec:Normal_Bursts} and \ref{sec:Minibursts} respectively, also provide evidence of system memory in the Bursting Pulsar.  Both types of burst\index{X-ray burst} display a number of similar features and they occur interchangeably during the same period of each outburst\index{Outburst}, leading to the possibility that they are generated by the same physical instability\index{Instability}.  As I show in Figure \ref{fig:minidips}, Minibursts and Normal Bursts fall into two clear populations\index{Population study} when plotted by their amplitudes: we find no Normal or Minibursts with fluences between $\sim10^4$ and $\sim10^5$ 2--60\,keV PCA\indexpca\ counts.  Due to the large number of Minibursts and Normal Bursts observed in my study, it is highly likely that this gap is real.  Observations of future outbursts of the Bursting Pulsar will allow us to see whether this fluence gap always spans the same range, or whether this gap changes between outbursts as a function of the longer-term evolution of the system.

\section{IGR J17091 vs. the Bursting Pulsar: A Comparison}

\par It is clear that there are a number of significant similarities between objects which display Type II\index{X-ray burst!Type II} bursts and GRS 1915\index{GRS 1915+105}-like variability\index{Variability}.  What remains unclear is how, if at all, the physics of GRS 1915-like variability and Type II bursting are related to each other.  Many of the models proposed to explain GRS 1915-like variability and Type II bursts rely on similar viscous\index{Viscosity} disk instabilities\index{Instability} (see Sections \ref{sec:models_GRS} and \ref{sec:TIImod}), and the discovery of GRS 1915-like patterns in lightcurves\index{Lightcurve} of the Rapid Burster\index{Rapid Burster} strongly suggests a link between these two classes of object \citep{Bagnoli_RB}.
\par Historically, the Bursting Pulsar\index{Bursting Pulsar} has been considered a `twin' system to the Rapid Burster\index{Rapid Burster} (however, see \citealp{Lewin_BP}), but the many differences I find between these object calls this comparison into doubt.  In Chapter \ref{ch:BPletter} I consider the possibility that some of the bursting\index{X-ray burst} seen in the Bursting Pulsar is a result of the object being similar to TMSPs\index{TMSP}.  In this section, I consider the alternative possibility that bursting in the Bursting Pulsar is instead a manifestation of GRS 1915-like variability.

\subsection{Variability Classes and Burst Classes}

\par A number of disparate features in data from the Bursting Pulsar\index{Bursting Pulsar} are at least superficially similar to behaviours I identify in IGR J17091-3624\index{IGR J17091-3624}.  In Figures \ref{fig:BP_with_IGR1} and \ref{fig:BP_with_IGR2} I identify features in IGR J17091\index{IGR J17091-3624} data which resemble Normal Bursts\index{Normal burst} and Structured Bursting\index{Structured bursting} in the Bursting Pulsar.  As discussed in Section \ref{sec:disccomp}, there are a number of system similarities between IGR J17091 and the Bursting Pulsar, so perhaps the similarities in their lightcurves\index{Lightcurve} should not be surprising.  However, any attempt to compare Bursting classes in the Bursting Pulsar with variability classes in IGR J17091 encounters a number of difficulties:

\begin{figure}
  \centering
  \includegraphics[width=.9\linewidth, trim= 0mm 0mm 0mm 0mm,clip]{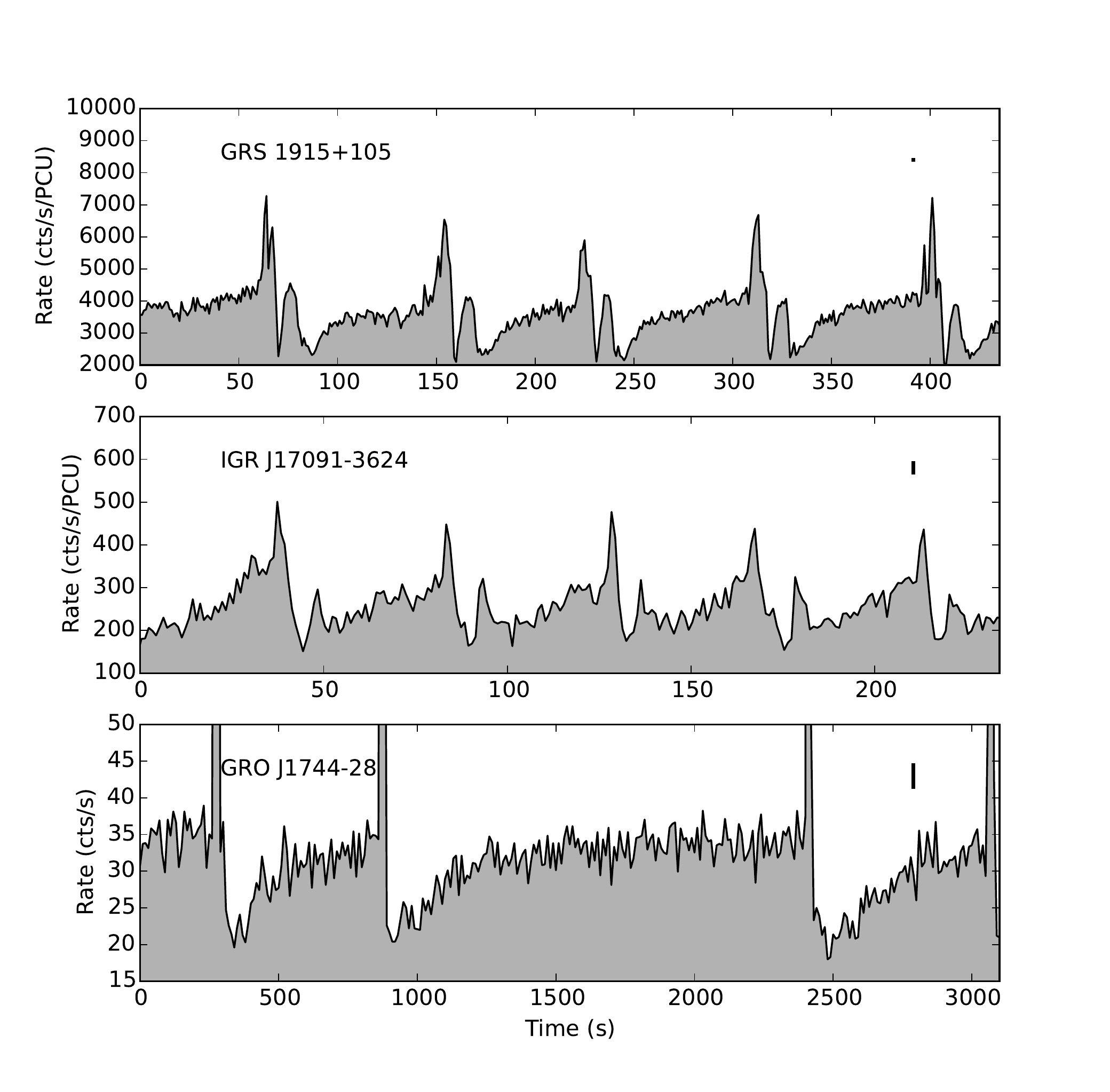}
  \caption[Lightcurves from GRS 1915, IGR J17091 and the Bursting Pulsar, showing lightcurves with Normal Burst-like behaviour for each.]{Lightcurves\index{Lightcurve} from GRS 1915\index{GRS 1915+105}, IGR J17091\index{IGR J17091-3624} and the Bursting Pulsar\index{Bursting Pulsar}, each showing heartbeat\indexrho-like variability\index{Variability} over timescales of 10s to 1000s of seconds.  Black bars indicate average error in each case.  GRS 1915 and IGR J17091 data taken from \indexpca\textit{RXTE}/PCA, Bursting Pulsar data taken from \indexchandra\textit{Chandra}.}
  \label{fig:BP_with_IGR1}
\end{figure}

\begin{figure}
  \centering
  \includegraphics[width=.9\linewidth, trim= 0mm 0mm 0mm 0mm,clip]{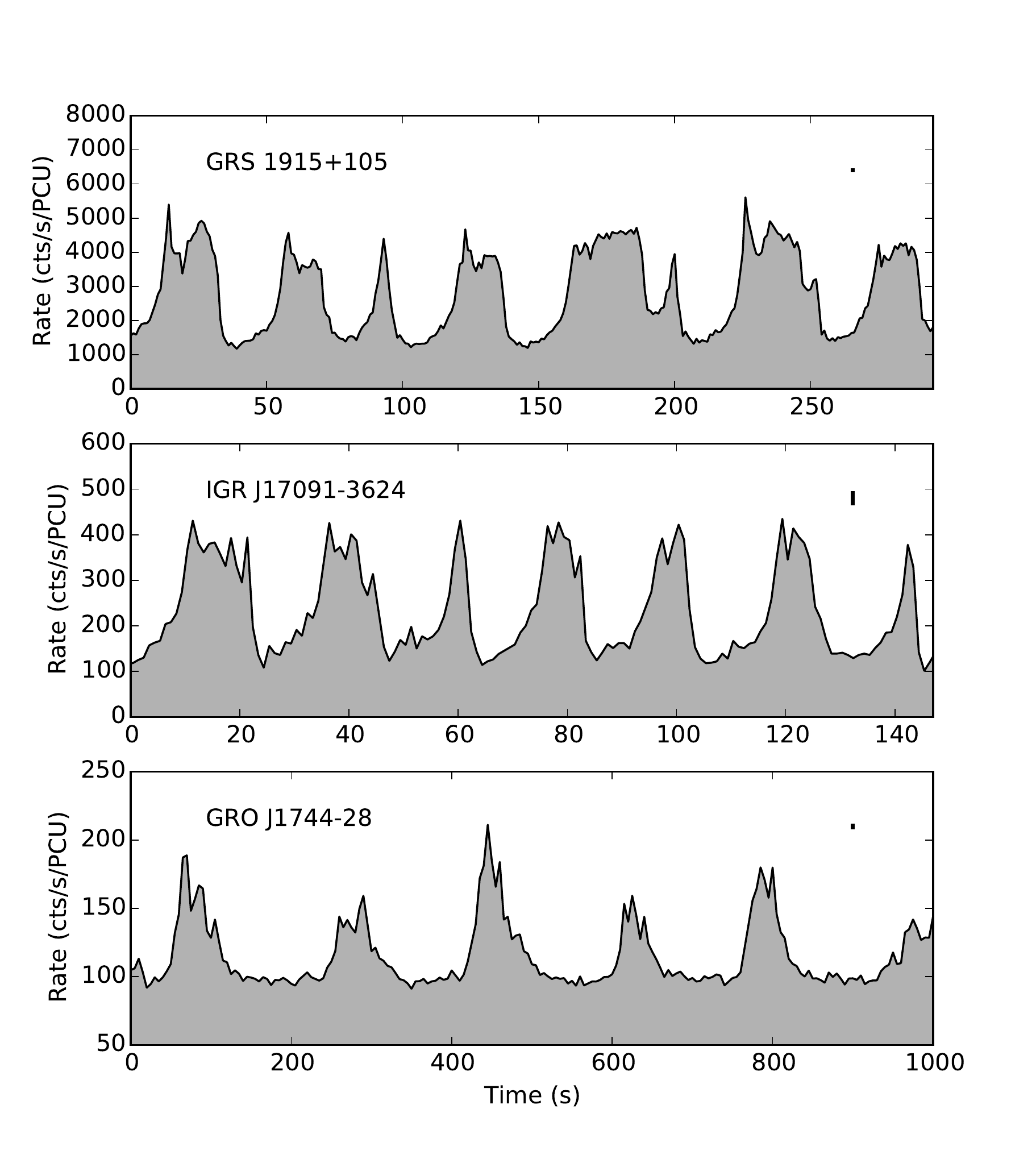}
  \caption[Lightcurves from GRS 1915, IGR J17091 and the Bursting Pulsar, showing lightcurves with Structured Bursting-like behaviour for each.]{Lightcurves\index{Lightcurve} from GRS 1915\index{GRS 1915+105}, IGR J17091\index{IGR J17091-3624} and the Bursting Pulsar\index{Bursting Pulsar}, each showing a Structured Bursting\index{Structured bursting}-like variability over timescales of 10s to 100s of seconds.  Black bars indicate average error in each case.  Data taken from \indexpca\textit{RXTE}/PCA.}
  \label{fig:BP_with_IGR2}
\end{figure}

\begin{itemize}
\item Variability\index{Variability} in IGR J17091\index{IGR J17091-3624} is likely caused by some instability\index{Instability} in the inner, radiation-dominated\index{Radiation pressure} part of its accretion disk\index{Accretion disk}.  The inner region of the disk in the Bursting Pulsar\index{Bursting Pulsar} is dominated by magnetic pressure\index{Magnetic pressure} rather than radiation pressure, leading to different possible instabilities.
\item IGR J17091 can evolve from one variability class\index{Variability class} to another quickly (over timescales of $\lesssim1$ day), whereas Normal Bursts\index{Normal burst} and Minibursts\index{Miniburst} occur continuously in the Bursting Pulsar for many weeks during each outburst\index{Outburst}.
\item IGR J17091 shows complex variability from the peak of each outburst until the time it enters the low/hard state\index{Low/Hard state}, whereas the Bursting Pulsar shows a large gap with no bursts between the end of Normal Bursts and the onset of Mesobursts\index{Mesoburst}.
\item All complex variability in IGR J17091 occurs over a relatively narrow range of luminosities (a factor of $\sim3$, see e.g. Figure \ref{fig:WhereCls}).  Bursting in the Bursting Pulsar occurs at luminosities spanning more than an order of magnitude.
\item All complex variability in IGR J17091 occurs during the main high-soft portion\index{High/Soft state} of its outbursts, whereas Mesobursts and Structured Bursting are seen during rebrightening\index{Re-flare} events in the Bursting Pulsar.
\end{itemize}

Because of these differences, it is unlikely that Burst Classes in the Bursting Pulsar\index{Bursting Pulsar} can be considered as being generated by the exact same phenomenon as variability classes\index{Variability class} in IGR J17091\index{IGR J17091-3624}.  Some of the apparent similarities between the phenomena could instead be explained by phenomenological limit cycles common to both.  For example, if both Class IV\indexiv\ variability and Normal Bursts\index{Normal burst} involve the filling and depletion of a portion of the inner part of the accretion disk\index{Accretion disk}, then it is to be expected that the flares\index{Flare} in both types of variability\index{Variability} have similar morphologies.

\subsection{Structured Bursting}

\par Structured Bursting\index{Structured bursting} in the Bursting Pulsar\index{Bursting Pulsar} on its own can also be compared with the variability classes\index{Variability class} observed in IGR J17091\index{IGR J17091-3624}.  As discussed in Section \ref{sec:struc_var}, and shown in Figure \ref{fig:Types_Struc}, Structured Bursting is a highly variable phenomenon.  Like variability in IGR J17091, Structured Bursting in the Bursting Pulsar consists of flares\index{Flare}, flat-bottomed dips\index{Dip} in flux and periods of seemingly unstructured noise.  As such, an alternative hypothesis to the `hiccup'\index{Hiccup accretion} scenario presented in Chapter \ref{ch:BPletter} is that Structured Bursting is an example of GRS 1915\index{GRS 1915+105}-like variability manifesting in a neutron star\index{Neutron star} LMXB\index{X-ray binary!Low mass}.
\par There are a number of problems with simply equating Structured Bursting\index{Structured bursting} with GRS 1915\index{GRS 1915+105}-like variability classes\index{Variability class}.  Variability in GRS 1915 and IGR J17091\index{IGR J17091-3624} shows hysteresis\index{Hysteresis} in hardness-intensity diagrams\index{Hardness-intensity diagram}, indicating a finite lag\index{Hard lag} between hard and soft emission from the source.  However no such hysteresis exists in Structured Bursting from the Bursting Pulsar\index{Bursting Pulsar}: as we show in Figure \ref{fig:HR}, hardness\index{Colour} and intensity simply correlate during periods of Structured Bursting.  In addition to this, the source intensities involved in GRS 1915-like variability and Structured Bursting are very different; GRS 1915 is a near-Eddington\index{Eddington limit} source, but Structured Bursting in the Bursting Pulsar occurs at a luminosity no greater than 0.5\% of its Eddington Luminosity.
\par If however Structured Bursting\index{Structured bursting} and GRS 1915\index{GRS 1915+105}-like variability\index{Variability} are the same phenomenon, then these issues may be resolved in a number of ways.  While the hard lag\index{Hard lag} in GRS 1915 is positive in every variability class\index{Variability class}, I find that its sign can vary in different variability classes in IGR J17091.  Therefore, it is feasible to imagine a GRS 1915-like system in which this lag is always close to zero, resulting in a simple correlation between rate and hardness in a HID\index{Hardness-intensity diagram} rather than a hysteretic\index{Hysteresis} loop.  Notably, the GRS 1915-like lightcurves\index{Lightcurve} reported from the Rapid Burster\index{Rapid Burster} \citep{Bagnoli_RB} also show no hysteretic loops.
\par The apparent different luminosity regimes of GRS 1915\index{GRS 1915+105} and the Bursting Pulsar\index{Bursting Pulsar} can be resolved if GRS 1915\index{GRS 1915+105}-like variability\index{Variability} does not require near-Eddington\index{Eddington limit} accretion.  I discuss this possibility in Section \ref{sec:criteria}.  The Rapid Burster\index{Rapid Burster} is known to accrete\index{Accretion rate} at $\sim20$\% of its Eddington Limit, and I find that IGR J17091\index{IGR J17091-3624} likely accretes at 5--33\% of its Eddington Limit (Section \ref{sec:newmass}), so these systems set possible precedents for GRS 1915-like variability in systems which are not near-Eddington limited.
\par To further investigate the similarities between GRS 1915-like\index{GRS 1915+105} variability\index{Variability} and Structured Bursting\index{Structured bursting}, the natural next step would be to perform phase-resolved spectroscopy\index{Spectroscopy!Phase-resolved} on Structured Bursting data.  Archival data of Structured Bursting from the Bursting Pulsar\index{Bursting Pulsar} only exists from the 1996 and 1997 outbursts\index{Outburst} of the source, as no observations were taken during the latter stages of its outbursts in 2014 or 2017.  The Bursting Pulsar is a faint source during periods of Structured Bursting, and is in a crowded region of the sky populated by many other X-ray sources, and such a study is difficult to perform on data from instruments launched before 1996.  As such, it remains unclear whether Structured Bursting\index{Structured bursting} is a neutron star\index{Neutron star} manifestation of GRS 1915\index{GRS 1915+105}-like variability\index{Variability} or whether it is a manifestation of `hiccup'\index{Hiccup accretion} accretion as I suggest in Chapter \ref{ch:BPletter}.

\section{Future Research}

\par A number of recently-launched and planned satellites have sensitivities, collecting areas and energy resolutions which exceed those of the instruments I use in this study.  For example, the currently operational \textit{Neutron Star Interior Composition Explorer} (\textit{NICER}, \citealp{Gendreau_Nicer}\index{NICER@\textit{NICER}}) has a 0.5--10\,keV X-ray sensitivity around 30 times greater than that of \textit{RXTE}\indexrxte , with an energy resolution comparable to \textit{XMM-Newton} and \textit{Chandra}\indexxmm\indexchandra.  The European Advanced Telescope for High Energy Astrophysics (\textit{ATHENA}, \citealp{Johnson_Athena}\index{ATHENA@\textit{ATHENA}}), planned for launch in the 2030s, is expected to have a sensitivity around 2 orders of magnitude greater than \textit{XMM-Newton} or \textit{Chandra}.  This generation of highly sensitive instruments will allow us to perform phase-resolved spectroscopy\index{Spectroscopy!Phase-resolved} of variability\index{Variability} from fainter objects, such as IGR J17091\index{IGR J17091-3624} and the Bursting Pulsar\index{Bursting Pulsar} during periods of Structured Bursting\index{Structured bursting}.
\par A phase-reolved spectroscopic\index{Spectroscopy!Phase-resolved} study of the variability classes\index{Variability class} in IGR J17091\index{IGR J17091-3624} will allow us to identify the physical changes in the accretion disk\index{Accretion disk} that occur during each class.  This study will be able to be compared to the phase-resolved spectral study of GRS 1915\index{GRS 1915+105} by \citet{Neilsen_GRSModel}, allowing us to further understand the similarities and differences between these two systems.
\par Phase-resolved spectral studies will also be possible to perform on the fainter classes of bursting\index{Burst} seen in the Bursting Pulsar\index{Bursting Pulsar}.  These will allow us to understand the physical mechanisms underlying each class of burst, in particular identifying which bursts if any are a result of thermonuclear burning\index{Thermonuclear burning} on the surface of the neutron star\index{Neutron star}.  This information will allow us to better understand whether bursting in the Bursting Pulsar is a manifestation of the same instabilities\index{Instability} seen in GRS 1915\index{GRS 1915+105} and IGR J17091\index{IGR J17091-3624}, and will allow us to understand where all of these systems fit in a picture of accretion disk\index{Accretion disk} instability as a whole.

\cleardoublepage

\chapter{Conclusions}

\epigraph{\textit{And this goes on and on and back and forth for 90 or so minutes, until it just sort of... ends.}}{Dennis Reynolds -- \textit{It's Always Sunny in Philadelphia}}

\vspace{1cm}

\par\noindent In this thesis, I have presented the results of phenomenological studies of the X-ray variability\index{Variability} seen in two unusual LMXBs\index{X-ray binary!Low mass}: IGR J17091-3624\index{IGR J17091-3624} and GRO J1744-28\index{Bursting Pulsar} (the ``Bursting Pulsar'').  I have analysed these results in the context of previous studies of variability in GRS 1915+105\index{GRS 1915+105} and MXB 1730-335\index{Rapid Burster} (the ``Rapid Burster''), systems which are often compared to IGR J17091 and the Bursting Pulsar respectively.  In doing so I have discovered a number of new similarities and differences between these objects.  On the back of this analysis, I have evaluated the physical models and scenarios which have been proposed to explain the variability in these objects.  In doing so, I have brought us closer to an understanding of the accretion\index{Accretion} physics that underlies this exotic behaviour.
\par In Chapter \ref{ch:IGR}, I have presented a new set of variability classes\index{Variability class} to describe IGR J17091\index{IGR J17091-3624}: these classes are analogous to, but independent from, the classes presented by \citet{Belloni_GRS_MI} to describe GRS 1915\index{GRS 1915+105}.  Comparing my set of variability classes to those of \citeauthor{Belloni_GRS_MI}, I found a number of variability classes which are only seen in one of the two objects, as well as a number of types of variability which are seen in both.  When studying the spectral\index{Spectroscopy} timing properties of both objects, I found another significant difference: while hard photons lag\index{Hard lag} soft photons in every variability class of GRS 1915, the sign of this lag varies from class to class in IGR J17091.  This finding rules out any physical picture in which the hard lag is caused by a corona\index{Corona} reacting to changes in the flux from the disk\index{Accretion disk}, instead suggesting that the hard lag is generated by a spectral change in the emission from the disk.
\par In Chapter \ref{ch:BPbig}, I have presented a new set of classifications for Type II-like\index{X-ray burst!Type II} X-ray bursts\index{X-ray burst} from the Bursting Pulsar\index{Bursting Pulsar}.  In doing so, I have discovered previously unreported bursting behaviour in the late stages of outbursts\index{Outburst} of the Bursting Pulsar, namely `Mesobursts'\index{Mesoburst} and `Structured Bursting'\index{Structured bursting}.  I find that Mesobursts may be a manifestation of quasi-stable thermonuclear burning\index{Thermonuclear burning} on the surface of the neutron star\index{Neutron star}; a phenomenon which has long been predicted to occur on the Bursting Pulsar but which has never been conclusively identified (e.g. \citealp{Bildsten_Nuclear}).  There are similarities between lightcurves\index{Lightcurve} of Structured Bursting and the lightcurves of another form of quasi-stable nuclear burning predicted by \citet{Heger_MargStab}.  However I find that at least one Mesoburst occured during a period of Structured Bursting without disrupting it, suggesting that Structured Bursting is non-nuclear in nature.  Instead, in Chapter \ref{ch:BPletter} I have identified similarities between Structured Bursting and variability\index{Variability} seen in Transitional Millisecond Pulsars\index{TMSP}.  This raises the possibility that Structured Bursting is a manifestation of `hiccup' accretion\index{Hiccup accretion}; spasmodic accretion onto a neutron star caused by small perturbations of its magnetospheric radius\index{Magnetospheric radius} near the boundary of the propeller regime\index{Propeller effect}.
\par From a phenomenological standpoint, I find that the variability\index{Variability} seen in IGR J17091\index{IGR J17091-3624} and the Bursting Pulsar\index{Bursting Pulsar} is generally even more complex than previously thought.  IGR J17091 and the Bursting Pulsar have often been considered `twin systems' of GRS 1915\index{GRS 1915+105} and the Rapid Burster\index{Rapid Burster} respectively, but I find a number of differences between each pair of twins which makes such a simple picture seem unlikely.  Variability in GRS 1915 has traditionally been thought to be tied to its near-Eddington\index{Eddington limit} accretion rate\index{Accretion rate}; however, I find that IGR J17091 likely accretes at $\lesssim33$\% of its Eddington Limit.  The Rapid Burster shows Type II\index{X-ray burst!Type II} bursts which transition smoothly between 2 classes over the course of an outburst\index{Outburst}, whereas I find that bursts in the Bursting Pulsar can be described in no less than 4 classes which take place at different periods of each outburst.  Rather than suggesting that these pairs of objects are unrelated, I suggest that further study of their differences will lead to better understanding of the physics behind the instabilities\index{Instability} that they present.
\par In Chapter \ref{ch:conc} I discuss the relationship between GRS 1915-like\index{GRS 1915+105}\index{Variability} variability and Type II X-ray bursts\index{X-ray burst!Type II}.  While there are a number of problems with assuming that these two types of variability are the same, I find a number of similarities between them that suggest at least some of the physics underlying these phenomena are similar.  Finally, I suggest that phase-resolved spectral\index{Spectroscopy!Phase-resolved} studies by the next generation of space telescopes will allow us to fully understand the relationships, or lack thereof, between these four enigmatic objects.
\par I also present a number of results unconnected to the variability\index{Variability} seen in these objects.  In Chapter \ref{ch:IGR} I provide new constraints on the distance of IGR J17091-3624\index{IGR J17091-3624} and the mass of its black hole\index{Black hole}, and I also present the first \textit{INTEGRAL}\indexintegral\ detection of the object above 150\,keV.  Additionally in Chapters \ref{ch:IGR} and \ref{ch:BPbig} I report the discovery of `re-flares'\index{Re-flare} in the tails of outbursts\index{Outburst} in both IGR J17091 and the Bursting Pulsar\index{Bursting Pulsar}.  I have also created a number of algorithms to identify bursts\index{X-ray burst} or flares\index{Flare}, and to `fold'\index{Folding} datasets which show repeating variability\index{Variability} with a non-constant frequency (Chapter \ref{ch:methods}).  These are encoded as part of my own suite of computational tools to analyse X-ray data (\texttt{PANTHEON}\index{PANTHEON@\texttt{PANTHEON}}, see Appendix \ref{app:PAN}).
\par In conclusion, the work I present in this thesis provides a comprehensive framework for future study of variability\index{Variability} in IGR J17091\index{IGR J17091-3624} and the Bursting Pulsar\index{Bursting Pulsar}.  Using this framework, I have been able to rule out a number of models and physical scenarios which have been proposed to explain the behaviour seen in these systems.  Further studies of the key properties of these systems will allow us to better understand the exotic instabilities\index{Instability} which can be present in accretion disks\index{Accretion disk} and, as such, improve our knowledge of the physics of accretion\index{Accretion} in general.

\cleardoublepage

\begin{appendices}
\chapter{Model-Independent Classification of each Observation of IGR J17091-3624}
\label{app:Obsids}

\par In Table \ref{tab:obsids}, I present observation IDs, and orbit IDs, for every \indexrxte\textit{RXTE} observation and observation segment that was used in my study of variability\index{Variability} in IGR J17091-3624\index{IGR J17091-3624} (Chapter \ref{ch:IGR}).  Note that not all of every observation was used; in many cases, large spikes caused by PCA\indexpca\ PCUs switching off or on rendered $\sim100$\,s unusable.  As these often occurred very close to the beginning or end of an observation segment, small sections of data before or after these spikes was also sometimes discarded.  Every observation segment is presented along with the variability class\index{Variability class} assigned to it by this study.

\begin{table*}
\caption[The variability class assigned to each \textit{RXTE} observation of IGR J17091-3624 considered in this thesis.]{Here is listed the Observation IDs for every \indexrxte\textit{RXTE} observation that was used in my analysis of variability\index{Variability} in IGR J17091-3624\index{IGR J17091-3624}, along with the variability class\index{Variability class} which has been assigned to it.  \textit{Orb.} is the orbit ID (starting at 0) of each observation segment, \textit{Exp.} is the exposure time in seconds and \textbf{X} is the prefix 96420-01.  This table is continued overleaf in Tables \ref{tab:obsids2}-\ref{tab:obsids4}.}
\label{tab:obsids}
\begin{tabular}{llllrllllr}
\hline
\hline
MJD&OBSID&\textit{Orb.}&Class&\textit{Exp.}&MJD&OBSID&\textit{Orb.}&Class&\textit{Exp.}\\
\hline
55622&\textbf{X}-01-00&0&I\indexi&1840&55643&\textbf{X}-04-01&0&III\indexiii&1190\\
55622&\textbf{X}-01-000&0&I&3480&55644&\textbf{X}-04-03&0&III&2903\\
55622&\textbf{X}-01-000&1&I&1656&55645&\textbf{X}-05-02&0&I&3578\\
55622&\textbf{X}-01-000&2&I&3384&55647&\textbf{X}-05-00&0&IV\indexiv&2872\\
55622&\textbf{X}-01-000&3&I&3400&55647&\textbf{X}-05-000&0&IV&3472\\
55622&\textbf{X}-01-000&4&I&3384&55647&\textbf{X}-05-000&1&IV&3520\\
55623&\textbf{X}-01-01&0&I&1240&55647&\textbf{X}-05-000&2&IV&3512\\
55623&\textbf{X}-01-01&1&I&752&55647&\textbf{X}-05-000&3&IV&3520\\
55623&\textbf{X}-01-01&2&I&992&55647&\textbf{X}-05-000&4&IV&3512\\
55623&\textbf{X}-01-01&3&I&1184&55647&\textbf{X}-05-000&5&IV&648\\
55623&\textbf{X}-01-01&4&I&1056&55649&\textbf{X}-05-03&0&IV&2409\\
55623&\textbf{X}-01-010&0&I&2080&55650&\textbf{X}-05-01&0&IV&1473\\
55623&\textbf{X}-01-010&1&I&1832&55651&\textbf{X}-05-04&0&IV&2954\\
55623&\textbf{X}-01-010&2&I&1648&55653&\textbf{X}-06-00&0&IV&2723\\
55623&\textbf{X}-01-010&4&I&1424&55654&\textbf{X}-06-01&0&IV&3388\\
55623&\textbf{X}-01-010&5&I&400&55656&\textbf{X}-06-02&0&IV&2908\\
55623&\textbf{X}-01-02&0&I&3056&55657&\textbf{X}-06-03&0&V\indexv&1842\\
55623&\textbf{X}-01-02&1&I&2792&55661&\textbf{X}-07-00&0&V&1754\\
55623&\textbf{X}-01-02&2&I&2432&55662&\textbf{X}-07-01&0&V&3365\\
55623&\textbf{X}-01-020&0&I&3456&55663&\textbf{X}-07-02&0&V&3373\\
55623&\textbf{X}-01-020&1&I&3464&55666&\textbf{X}-08-00&0&V&3338\\
55623&\textbf{X}-01-020&2&I&3512&55669&\textbf{X}-08-01&0&V&3368\\
55623&\textbf{X}-01-020&3&I&3520&55670&\textbf{X}-08-03&0&VI\indexvi&2489\\
55623&\textbf{X}-01-020&4&I&3512&55671&\textbf{X}-08-02&0&VI&2609\\
55623&\textbf{X}-01-020&5&I&464&55673&\textbf{X}-09-03&0&VI&1011\\
55624&\textbf{X}-02-00&0&I&1758&55674&\textbf{X}-09-00&0&VI&1386\\
55626&\textbf{X}-02-01&0&I&1380&55675&\textbf{X}-09-05&0&IX\indexix&1148\\
55628&\textbf{X}-02-02&0&I&3305&55676&\textbf{X}-09-06&0&VI&3540\\
55630&\textbf{X}-02-03&0&I&1876&55677&\textbf{X}-09-01&0&V&1676\\
55632&\textbf{X}-03-00&0&I&1712&55678&\textbf{X}-09-04&0&V&2090\\
55634&\textbf{X}-03-01&0&III&3590&55679&\textbf{X}-09-02&0&V&2306\\
55639&\textbf{X}-04-00&0&IV&3099&55680&\textbf{X}-10-02&0&V&952\\
55642&\textbf{X}-04-02&0&IV&2972&55681&\textbf{X}-10-00&0&V&3725\\
\hline
\hline
\end{tabular}
\end{table*}

\begin{table*}
\caption[]{A continuation of Table \ref{tab:obsids}.  This table is continued overleaf in Tables \ref{tab:obsids3} and \ref{tab:obsids4}.}
\label{tab:obsids2}
\begin{tabular}{llllrllllr}
\hline
\hline
MJD&OBSID&\textit{Orb.}&Class&\textit{Exp.}&MJD&OBSID&\textit{Orb.}&Class&\textit{Exp.}\\
\hline
55682&\textbf{X}-10-03&0&V\indexv&1157&55720&\textbf{X}-15-04&0&IV\indexiv&1486\\
55684&\textbf{X}-10-01&0&III\indexiii&1504&55721&\textbf{X}-15-05&0&IV&1500\\
55686&\textbf{X}-10-04&0&III&1127&55722&\textbf{X}-16-00&0&IV&900\\
55686&\textbf{X}-10-05&0&II\indexii&2179&55723&\textbf{X}-16-01&0&III&1004\\
55687&\textbf{X}-11-00&0&II&3537&55724&\textbf{X}-16-02&0&II&1923\\
55688&\textbf{X}-11-01&0&II&1153&55725&\textbf{X}-16-03&0&II&1919\\
55690&\textbf{X}-11-02&0&II&1408&55726&\textbf{X}-16-04&0&III&1935\\
55691&\textbf{X}-11-03&0&II&886&55727&\textbf{X}-16-05&0&II&730\\
55692&\textbf{X}-11-04&0&II&3566&55728&\textbf{X}-16-06&0&II&1953\\
55693&\textbf{X}-11-05&0&II&1817&55729&\textbf{X}-17-00&0&II&2735\\
55694&\textbf{X}-12-00&0&II&2761&55730&\textbf{X}-17-01&0&II&3556\\
55695&\textbf{X}-12-01&0&II&1374&55731&\textbf{X}-17-02&0&II&3605\\
55695&\textbf{X}-12-02&0&II&2041&55732&\textbf{X}-17-03&0&II&1647\\
55696&\textbf{X}-12-03&0&II&1456&55733&\textbf{X}-17-04&0&II&1459\\
55698&\textbf{X}-12-04&0&II&1916&55734&\textbf{X}-17-05&0&III&1736\\
55698&\textbf{X}-12-05&0&II&3139&55735&\textbf{X}-17-06&0&III&3653\\
55700&\textbf{X}-12-06&0&II&1189&55736&\textbf{X}-18-00&0&III&2317\\
55701&\textbf{X}-13-00&0&II&1214&55737&\textbf{X}-18-01&0&IV&1387\\
55702&\textbf{X}-13-01&0&II&980&55738&\textbf{X}-18-02&0&V&1291\\
55704&\textbf{X}-13-02&0&II&732&55739&\textbf{X}-18-03&0&V&2178\\
55705&\textbf{X}-13-03&0&III&1217&55740&\textbf{X}-18-04&0&V&1478\\
55706&\textbf{X}-13-04&0&III&1161&55741&\textbf{X}-18-05&0&VII\indexvii&782\\
55707&\textbf{X}-13-05&0&IV&2763&55743&\textbf{X}-19-00&0&VII&1412\\
55708&\textbf{X}-14-00&0&IV&1188&55744&\textbf{X}-19-01&0&VIII\indexviii&1938\\
55709&\textbf{X}-14-01&0&IV&3342&55745&\textbf{X}-19-02&0&VII&2172\\
55710&\textbf{X}-14-02&0&IV&1094&55747&\textbf{X}-19-03&0&VIII&1691\\
55712&\textbf{X}-14-03&0&IV&1404&55748&\textbf{X}-19-04&0&VI\indexvi&1283\\
55713&\textbf{X}-14-04&0&V&871&55749&\textbf{X}-19-05&0&VIII&1417\\
55714&\textbf{X}-14-05&0&V&1311&55751&\textbf{X}-20-05&0&VI&1726\\
55715&\textbf{X}-15-00&0&IV&1241&55752&\textbf{X}-20-01&0&VIII&1079\\
55716&\textbf{X}-15-01&0&IV&1262&55753&\textbf{X}-20-02&0&VIII&1433\\
55717&\textbf{X}-15-02&0&III&1557&55754&\textbf{X}-20-03&0&VII&1122\\
55718&\textbf{X}-15-03&0&III&1334&55756&\textbf{X}-20-04&0&VIII&1486\\
\hline
\hline
\end{tabular}
\end{table*}

\begin{table*}
\caption[]{A continuation of Table \ref{tab:obsids}.  This table is continued overleaf in Table \ref{tab:obsids4}.}
\label{tab:obsids3}
\begin{tabular}{llllrllllr}
\hline
\hline
MJD&OBSID&\textit{Orb.}&Class&\textit{Exp.}&MJD&OBSID&\textit{Orb.}&Class&\textit{Exp.}\\
\hline
55757&\textbf{X}-21-00&0&VIII\indexviii&3372&55790&\textbf{X}-25-05&0&V\indexv&1473\\
55758&\textbf{X}-21-01&0&VIII&3383&55791&\textbf{X}-25-06&0&V&922\\
55759&\textbf{X}-21-02&0&VI\indexvi&1938&55792&\textbf{X}-26-00&0&V&2336\\
55761&\textbf{X}-21-04&0&VII\indexvii&1497&55794&\textbf{X}-26-01&0&V&1385\\
55762&\textbf{X}-21-05&0&VII&1548&55795&\textbf{X}-26-02&0&VIII&1458\\
55763&\textbf{X}-21-06&0&VII&2202&55796&\textbf{X}-26-03&0&VI&1325\\
55764&\textbf{X}-22-00&0&VII&1682&55798&\textbf{X}-26-04&0&VI&2075\\
55765&\textbf{X}-22-01&0&VII&1221&55799&\textbf{X}-27-00&0&VI&1396\\
55766&\textbf{X}-22-02&0&V&720&55800&\textbf{X}-27-01&0&VI&2684\\
55767&\textbf{X}-22-03&0&V&1801&55801&\textbf{X}-27-02&0&VI&1016\\
55768&\textbf{X}-22-04&0&VIII&1983&55802&\textbf{X}-27-03&0&VI&1179\\
55769&\textbf{X}-22-05&0&VIII&999&55803&\textbf{X}-27-04&0&VI&1304\\
55770&\textbf{X}-22-06&0&VIII&667&55805&\textbf{X}-27-05&0&VI&1663\\
55771&\textbf{X}-23-00&0&VIII&2075&55806&\textbf{X}-28-00&0&VI&1456\\
55772&\textbf{X}-23-01&0&VII&3385&55808&\textbf{X}-28-01&0&VIII&577\\
55773&\textbf{X}-23-02&0&VII&2218&55810&\textbf{X}-28-02&0&VI&1251\\
55774&\textbf{X}-23-03&0&V&1811&55811&\textbf{X}-28-03&0&VI&2000\\
55775&\textbf{X}-23-04&0&V&3356&55813&\textbf{X}-29-00&0&VIII&1309\\
55776&\textbf{X}-23-05&0&V&2603&55819&\textbf{X}-29-04&0&VIII&1686\\
55777&\textbf{X}-23-06&0&IV\indexiv&912&55820&\textbf{X}-30-00&0&VI&1488\\
55777&\textbf{X}-23-06&1&IV&1544&55821&\textbf{X}-30-01&0&VI&1503\\
55778&\textbf{X}-24-00&0&IV&1309&55822&\textbf{X}-30-02&0&VI&1417\\
55779&\textbf{X}-24-01&0&IV&3599&55823&\textbf{X}-30-03&0&VI&1290\\
55779&\textbf{X}-24-02&0&IV&2013&55824&\textbf{X}-30-04&0&VI&1489\\
55782&\textbf{X}-24-03&0&V&1761&55825&\textbf{X}-30-05&0&VI&2581\\
55782&\textbf{X}-24-04&0&V&1725&55826&\textbf{X}-30-06&0&VI&2747\\
55784&\textbf{X}-24-05&0&V&3144&55827&\textbf{X}-31-00&0&VI&1559\\
55784&\textbf{X}-24-06&0&V&2591&55828&\textbf{X}-31-01&0&VI&2954\\
55785&\textbf{X}-25-00&0&V&2366&55829&\textbf{X}-31-02&0&IX\indexix&3005\\
55786&\textbf{X}-25-01&0&V&1804&55830&\textbf{X}-31-03&0&IX&1472\\
55787&\textbf{X}-25-02&0&V&1951&55830&\textbf{X}-31-03&1&IX&288\\
55788&\textbf{X}-25-03&0&V&1619&55831&\textbf{X}-31-04&0&IX&1586\\
55789&\textbf{X}-25-04&0&V&2601&55832&\textbf{X}-31-05&0&VI&3812\\
\hline
\hline
\end{tabular}
\end{table*}

\begin{table*}
\caption[]{A continuation of Table \ref{tab:obsids}.}
\label{tab:obsids4}
\begin{tabular}{llllrllllr}
\hline
\hline
MJD&OBSID&\textit{Orb.}&Class&\textit{Exp.}&MJD&OBSID&\textit{Orb.}&Class&\textit{Exp.}\\
\hline
55833&\textbf{X}-31-06&0&IX\indexix&3675&55867&\textbf{X}-36-05&0&IX&1732\\
55834&\textbf{X}-32-00&0&IX&1217&55868&\textbf{X}-36-06&0&IX&1657\\
55835&\textbf{X}-32-01&0&IX&1445&55871&\textbf{X}-37-00&0&IX&815\\
55836&\textbf{X}-32-02&0&IX&1591&55871&\textbf{X}-37-02&0&IX&1460\\
55837&\textbf{X}-32-03&0&IX&2155&55872&\textbf{X}-37-03&0&IX&1683\\
55838&\textbf{X}-32-04&0&IX&2641&55873&\textbf{X}-37-04&0&IX&1402\\
55838&\textbf{X}-32-05&0&IX&2077&55874&\textbf{X}-37-05G&0&IX&1536\\
55840&\textbf{X}-32-06&0&IX&3392&55875&\textbf{X}-37-06&0&IX&1536\\
55840&\textbf{X}-32-06&1&IX&3512&55876&\textbf{X}-38-00&0&IX&1497\\
55840&\textbf{X}-32-06&2&IX&3934&55877&\textbf{X}-38-01&0&IX&1134\\
55840&\textbf{X}-32-06&3&IX&3880&55878&\textbf{X}-38-02&0&IX&1289\\
55840&\textbf{X}-32-06&4&IX&1896&55879&\textbf{X}-38-03&0&IX&1433\\
55841&\textbf{X}-33-00&0&IX&1188&&&&&\\
55842&\textbf{X}-33-01&0&IX&855&&&&&\\
55843&\textbf{X}-33-02&0&IX&1156&&&&&\\
55845&\textbf{X}-33-04&0&IX&1713&&&&&\\
55846&\textbf{X}-33-05&0&IX&934&&&&&\\
55847&\textbf{X}-33-06&0&IX&717&&&&&\\
55848&\textbf{X}-34-00&0&IX&1159&&&&&\\
55849&\textbf{X}-34-01&0&IX&973&&&&&\\
55851&\textbf{X}-34-02&0&IX&2261&&&&&\\
55852&\textbf{X}-34-03&0&IX&1092&&&&&\\
55853&\textbf{X}-34-04&0&IX&741&&&&&\\
55856&\textbf{X}-35-00&0&IX&797&&&&&\\
55857&\textbf{X}-35-01&0&IX&1912&&&&&\\
55859&\textbf{X}-35-02&0&IX&200&&&&&\\
55859&\textbf{X}-35-02&1&IX&1296&&&&&\\
55860&\textbf{X}-35-03&0&IX&1372&&&&&\\
55861&\textbf{X}-35-04&0&IX&836&&&&&\\
55862&\textbf{X}-36-00&0&IX&1145&&&&&\\
55863&\textbf{X}-36-01&0&IX&1322&&&&&\\
55865&\textbf{X}-36-03&0&IX&1485&&&&&\\
55866&\textbf{X}-36-04&0&IX&1795&&&&&\\
\hline
\hline
\end{tabular}
\end{table*}
\cleardoublepage
\chapter{List of \textit{RXTE} Observations of the Bursting Pulsar}
\label{app:obs}

\par In Table \ref{tab:obslist} we present a table of all \indexrxte\textit{RXTE} observations used in our study of burst\index{X-ray burst} evolution in the Bursting Pulsar\index{Bursting Pulsar}.  The prefixes \textbf{A}, \textbf{B}, \textbf{C}, \textbf{D} and \textbf{E} correspond to OBSIDs beginning with 10401-01, 20077-01, 20078-01, 20401-01 and 30075-01 respectively.

\begin{table*}
\centering
\begin{tabular}{lllllllll}
\hline
\hline
\scriptsize Obsid&\scriptsize Exp.&\scriptsize Date&\scriptsize Obsid&\scriptsize Exp.&\scriptsize Date&\scriptsize Obsid&\scriptsize Exp.&\scriptsize Date\\
\hline
\textbf{A}-01-00&3105&119&\textbf{A}-34-00&1831&213&\textbf{A}-59-00&1152&257\\
\textbf{A}-02-00&1655&117&\textbf{A}-35-00&2563&216&\textbf{A}-59-01&2203&257\\
\textbf{A}-03-00&6724&122&\textbf{A}-36-00&3683&219&\textbf{A}-59-02&768&257\\
\textbf{A}-03-000&2372&122&\textbf{A}-37-00&3446&215&\textbf{A}-60-00&1907&260\\
\textbf{A}-03-01&768&122&\textbf{A}-38-00&1536&217&\textbf{A}-60-01&3376&260\\
\textbf{A}-04-00&639&128&\textbf{A}-39-00&2317&218&\textbf{A}-60-02&1783&260\\
\textbf{A}-05-00&1990&129&\textbf{A}-40-00&1239&220&\textbf{A}-60-03&1559&260\\
\textbf{A}-06-00&1280&134&\textbf{A}-41-00&1363&221&\textbf{A}-61-00&3292&262\\
\textbf{A}-08-00&2431&142&\textbf{A}-42-00&2728&224&\textbf{A}-61-01&3035&262\\
\textbf{A}-09-00&640&138&\textbf{A}-43-00&2079&225&\textbf{A}-61-02&2013&262\\
\textbf{A}-10-00&2470&143&\textbf{A}-44-00&2076&226&\textbf{A}-62-00&2390&264\\
\textbf{A}-11-00&2381&148&\textbf{A}-45-00&2050&228&\textbf{A}-62-01&1703&264\\
\textbf{A}-12-00&3352&151&\textbf{A}-47-00&2687&232&\textbf{A}-62-02&2719&264\\
\textbf{A}-13-00&3480&155&\textbf{A}-48-00&2267&234&\textbf{A}-63-00&517&266\\
\textbf{A}-14-00&1839&158&\textbf{A}-49-00&35&236&\textbf{A}-63-01&3077&266\\
\textbf{A}-15-00&1595&161&\textbf{A}-50-00&3719&238&\textbf{A}-64-00&2381&268\\
\textbf{A}-16-00&3470&156&\textbf{A}-51-00&3590&240&\textbf{A}-64-01&3110&268\\
\textbf{A}-17-00&4481&164&\textbf{A}-52-00&2518&241&\textbf{A}-65-00&2003&270\\
\textbf{A}-18-00&384&171&\textbf{A}-53-00&3063&243&\textbf{A}-65-01&2744&270\\
\textbf{A}-19-00&128&172&\textbf{A}-55-00&3328&245&\textbf{A}-65-02&4331&270\\
\textbf{A}-20-00&2087&178&\textbf{A}-55-01&3395&245&\textbf{A}-66-00&2203&272\\
\textbf{A}-21-00&2711&181&\textbf{A}-55-02&2667&245&\textbf{A}-66-01&1723&272\\
\textbf{A}-22-00&2816&183&\textbf{A}-56-00&512&250&\textbf{A}-66-02&2533&272\\
\textbf{A}-22-01&2911&185&\textbf{A}-56-01&1280&250&\textbf{A}-67-00&395&274\\
\textbf{A}-23-00&1678&187&\textbf{A}-56-02&1664&250&\textbf{A}-67-01&3533&274\\
\textbf{A}-24-00&2509&189&\textbf{A}-56-03&1920&250&\textbf{A}-67-02&3466&274\\
\textbf{A}-25-00&2846&192&\textbf{A}-57-00&2432&250&\textbf{A}-68-00&1841&276\\
\textbf{A}-26-00&768&194&\textbf{A}-57-01&894&253&\textbf{A}-69-00&3659&278\\
\textbf{A}-27-00&2923&196&\textbf{A}-57-02&1408&253&\textbf{A}-70-00&2022&280\\
\textbf{A}-28-00&6839&199&\textbf{A}-57-03&1792&253&\textbf{A}-71-00&3474&283\\
\textbf{A}-29-00&3478&201&\textbf{A}-58-00&1024&255&\textbf{A}-72-00&5687&285\\
\textbf{A}-30-00&5906&203&\textbf{A}-58-01&1401&255&\textbf{A}-73-00&3109&287\\
\textbf{A}-31-00&6170&206&\textbf{A}-58-02&1679&255&\textbf{A}-74-00&1659&289\\
\textbf{A}-32-00&2712&209&\textbf{A}-58-03&1683&255&\textbf{A}-75-00&1798&291\\
\hline
\hline
\end{tabular}
\caption[A list of all \textit{RXTE} observations of the Bursting Pulsar used in this thesis.]{A list of all \indexrxte\textit{RXTE} observations of the Bursting Pulsar used in this study.  Exposure is given in seconds, and date is given in days from MJD 50000.  The prefixes \textbf{A}, \textbf{B}, \textbf{C}, \textbf{D} and \textbf{E} correspond to OBSIDs beginning with 10401-01, 20077-01, 20078-01, 20401-01 and 30075-01 respectively.  This table is continued in Tables \ref{tab:obslist2}-\ref{tab:obslist3}}
\label{tab:obslist}
\end{table*}

\begin{table*}
\centering
\begin{tabular}{lllllllll}
\hline
\hline
\scriptsize Obsid&\scriptsize Exp.&\scriptsize Date&\scriptsize Obsid&\scriptsize Exp.&\scriptsize Date&\scriptsize Obsid&\scriptsize Exp.&\scriptsize Date\\
\hline
\textbf{A}-76-00&1558&293&\textbf{B}-10-00&1009&482&\textbf{C}-11-00&2330&527\\
\textbf{A}-77-00&1738&295&\textbf{B}-11-00&2864&487&\textbf{C}-11-01&290&527\\
\textbf{A}-78-00&463&297&\textbf{B}-12-00&1847&489&\textbf{C}-11-02&2399&527\\
\textbf{A}-79-00&1024&299&\textbf{B}-13-00&2805&497&\textbf{C}-12-00&3345&534\\
\textbf{A}-80-00&5818&301&\textbf{B}-14-00&3741&499&\textbf{C}-12-01&2048&534\\
\textbf{A}-81-00&6898&303&\textbf{B}-15-00&384&501&\textbf{C}-13-00&1735&541\\
\textbf{A}-82-00&3537&306&\textbf{B}-16-00&768&503&\textbf{C}-13-01&1691&541\\
\textbf{A}-83-00&512&308&\textbf{B}-17-00&2399&509&\textbf{C}-14-00&3579&549\\
\textbf{A}-84-00&6361&310&\textbf{B}-18-00&2306&511&\textbf{C}-14-01&2785&549\\
\textbf{A}-85-00&10391&312&\textbf{B}-19-00&3477&516&\textbf{C}-15-00&7494&579\\
\textbf{A}-86-00&9232&314&\textbf{B}-20-00&1922&520&\textbf{C}-16-00&4941&562\\
\textbf{A}-87-00&3109&316&\textbf{C}-01-00&8200&389&\textbf{C}-16-01&671&561\\
\textbf{A}-88-00&6630&318&\textbf{C}-02-00&1408&400&\textbf{C}-16-02&1159&562\\
\textbf{A}-89-00&2569&320&\textbf{C}-02-01&896&401&\textbf{C}-17-00&3537&568\\
\textbf{A}-90-00&2209&323&\textbf{C}-02-02&512&401&\textbf{C}-18-00&2981&576\\
\textbf{A}-91-00&2317&325&\textbf{C}-03-00&3409&465&\textbf{C}-18-01&3103&576\\
\textbf{A}-92-00&2199&327&\textbf{C}-03-01&2635&466&\textbf{C}-19-00&3286&582\\
\textbf{A}-93-00&3720&331&\textbf{C}-03-02&2645&466&\textbf{C}-19-01&2893&582\\
\textbf{A}-94-00&3216&332&\textbf{C}-04-00&2620&478&\textbf{C}-19-02&470&582\\
\textbf{A}-95-00&9487&333&\textbf{C}-04-01&2956&477&\textbf{C}-20-00&3460&589\\
\textbf{A}-96-00&2627&337&\textbf{C}-04-02&2515&476&\textbf{C}-20-01&1126&589\\
\textbf{A}-97-00&3341&340&\textbf{C}-05-00&1421&484&\textbf{C}-21-00&3659&596\\
\textbf{A}-98-00&99&343&\textbf{C}-05-01&1995&484&\textbf{C}-21-01&2907&596\\
\textbf{A}-99-00&2783&345&\textbf{C}-05-02&2505&485&\textbf{C}-21-02&1086&596\\
\textbf{A}-99-01&1001&344&\textbf{C}-06-00&2770&492&\textbf{C}-22-00&1967&602\\
\textbf{B}-01-00&1664&467&\textbf{C}-06-01&2375&492&\textbf{C}-22-01&3086&602\\
\textbf{B}-02-00&1920&468&\textbf{C}-06-02&2203&492&\textbf{C}-22-02&1024&602\\
\textbf{B}-03-00&2982&469&\textbf{C}-07-00&1258&494&\textbf{C}-23-00&3697&607\\
\textbf{B}-04-00&3530&470&\textbf{C}-08-00&3305&505&\textbf{C}-23-01&3091&607\\
\textbf{B}-05-00&2025&472&\textbf{C}-08-01&777&505&\textbf{C}-24-00&1152&618\\
\textbf{B}-06-00&2677&473&\textbf{C}-09-00&1377&513&\textbf{C}-24-01&2300&618\\
\textbf{B}-07-00&3365&473&\textbf{C}-09-01&1536&513&\textbf{C}-24-02&1386&618\\
\textbf{B}-08-00&3113&475&\textbf{C}-10-00&1664&517&\textbf{C}-25-00&4069&626\\
\textbf{B}-09-00&2868&480&\textbf{C}-10-01&3796&518&\textbf{C}-25-01&1920&626\\
\hline
\hline
\end{tabular}
\caption[]{A continuation of Table \ref{tab:obslist}.  This table is further continued in Table \ref{tab:obslist3}.}
\label{tab:obslist2}
\end{table*}

\begin{table*}
\centering
\begin{tabular}{lllllllll}
\hline
\hline
\scriptsize Obsid&\scriptsize Exp.&\scriptsize Date&\scriptsize Obsid&\scriptsize Exp.&\scriptsize Date&\scriptsize Obsid&\scriptsize Exp.&\scriptsize Date\\
\hline
\textbf{C}-25-02&768&626&\textbf{C}-39-01&4690&723&\textbf{D}-19-00&3158&650\\
\textbf{C}-26-00&2071&633&\textbf{C}-40-00&3419&730&\textbf{D}-20-00&751&672\\
\textbf{C}-26-01&4043&633&\textbf{C}-40-01&3419&730&\textbf{E}-01-00&512&831\\
\textbf{C}-27-00&1792&638&\textbf{C}-40-02&896&764&\textbf{E}-02-00&1836&845\\
\textbf{C}-27-01&2495&638&\textbf{C}-41-00&5255&735&\textbf{E}-03-00&1871&859\\
\textbf{C}-27-02&3082&638&\textbf{C}-41-01&2387&735&\textbf{E}-04-00&1927&873\\
\textbf{C}-28-00&3454&644&\textbf{C}-41-02&1141&744&\textbf{E}-05-00&2088&889\\
\textbf{C}-28-01&1359&644&\textbf{C}-42-00&1476&744&\textbf{E}-06-00&2003&901\\
\textbf{C}-28-02&756&644&\textbf{C}-43-00&5277&764&\textbf{E}-07-00&1536&914\\
\textbf{C}-29-00&1535&652&\textbf{C}-44-00&6712&769&\textbf{E}-08-00&967&935\\
\textbf{C}-30-01&3435&658&\textbf{D}-01-00&2688&523&\textbf{E}-09-00&1598&949\\
\textbf{C}-31-00&1920&662&\textbf{D}-02-00&3469&525&\textbf{E}-10-00&1835&961\\
\textbf{C}-31-01&1152&662&\textbf{D}-03-00&3026&528&\textbf{E}-11-00&1741&975\\
\textbf{C}-31-02&1012&657&\textbf{D}-04-00&3050&531&\textbf{E}-12-00&1032&991\\
\textbf{C}-32-00&4646&678&\textbf{D}-05-00&3485&536&\textbf{E}-13-00&1231&1001\\
\textbf{C}-32-01&2803&678&\textbf{D}-06-00&1367&538&\textbf{E}-14-00&1608&1016\\
\textbf{C}-33-00&4334&747&\textbf{D}-07-00&3196&543&\textbf{E}-15-00&1712&1030\\
\textbf{C}-33-01&3534&748&\textbf{D}-08-00&2617&548&\textbf{E}-16-00&1440&1045\\
\textbf{C}-33-02&2957&748&\textbf{D}-09-00&2598&553&\textbf{E}-17-00&1888&1057\\
\textbf{C}-34-00&3477&687&\textbf{D}-10-00&4069&560&\textbf{E}-18-00&1847&1071\\
\textbf{C}-34-01&1008&687&\textbf{D}-11-00&2686&572&\textbf{E}-19-00&1792&1086\\
\textbf{C}-34-02&2831&687&\textbf{D}-12-00&2867&565&\textbf{E}-20-00&1904&1101\\
\textbf{C}-35-00&1497&756&\textbf{D}-13-00&2021&585&\textbf{E}-21-00&1921&1115\\
\textbf{C}-35-01&1959&755&\textbf{D}-13-01&765&585&\textbf{E}-22-00&1769&1129\\
\textbf{C}-35-02&2023&755&\textbf{D}-14-00&2640&594&\textbf{E}-23-00&1892&1135\\
\textbf{C}-36-00&2825&702&\textbf{D}-14-01&1719&594&\textbf{E}-24-00&1943&1197\\
\textbf{C}-36-01&1592&702&\textbf{D}-15-00&3226&621&\textbf{E}-25-00&2237&1210\\
\textbf{C}-37-00&2092&709&\textbf{D}-15-01&1373&621&\textbf{E}-26-00&1396&1224\\
\textbf{C}-37-01&384&710&\textbf{D}-16-00&2432&609&&&\\
\textbf{C}-38-00&1752&716&\textbf{D}-16-01&1562&609&&&\\
\textbf{C}-38-01&1536&716&\textbf{D}-17-00&1790&628&&&\\
\textbf{C}-38-02&1144&716&\textbf{D}-17-01&1291&628&&&\\
\textbf{C}-38-03&338&717&\textbf{D}-18-00&1959&641&&&\\
\textbf{C}-39-00&2756&723&\textbf{D}-18-01&2614&641&&&\\
\hline
\hline
\end{tabular}
\caption[]{A continuation of Table \ref{tab:obslist}.}
\label{tab:obslist3}
\end{table*}

\cleardoublepage
\chapter{Normal Burst Histograms}
\label{app:hists}

\par
\begin{figure}
  \centering
  \includegraphics[width=.9\linewidth, trim={0cm 0 0cm 0},clip]{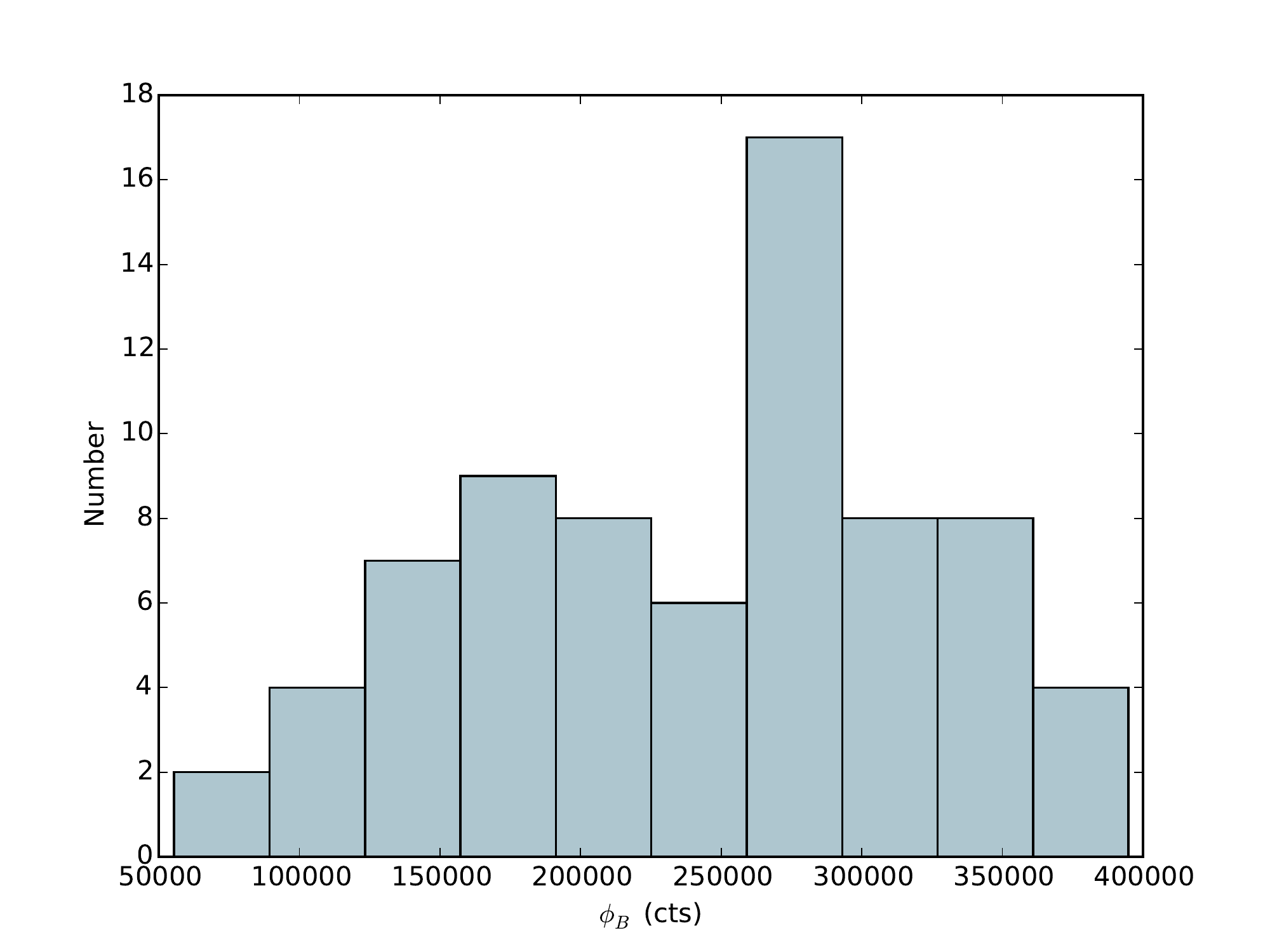}
  \caption[Histogram showing the distribution of $\phi_B$ amongst Normal Bursts.]{A histogram showing the distribution of burst fluence $\phi_B$ amongst our sample of Normal Bursts.\index{Normal burst}}
  \label{fig:app_hist_phib}
\end{figure}

\begin{figure}
  \centering
  \includegraphics[width=.9\linewidth, trim={0cm 0 0cm 0},clip]{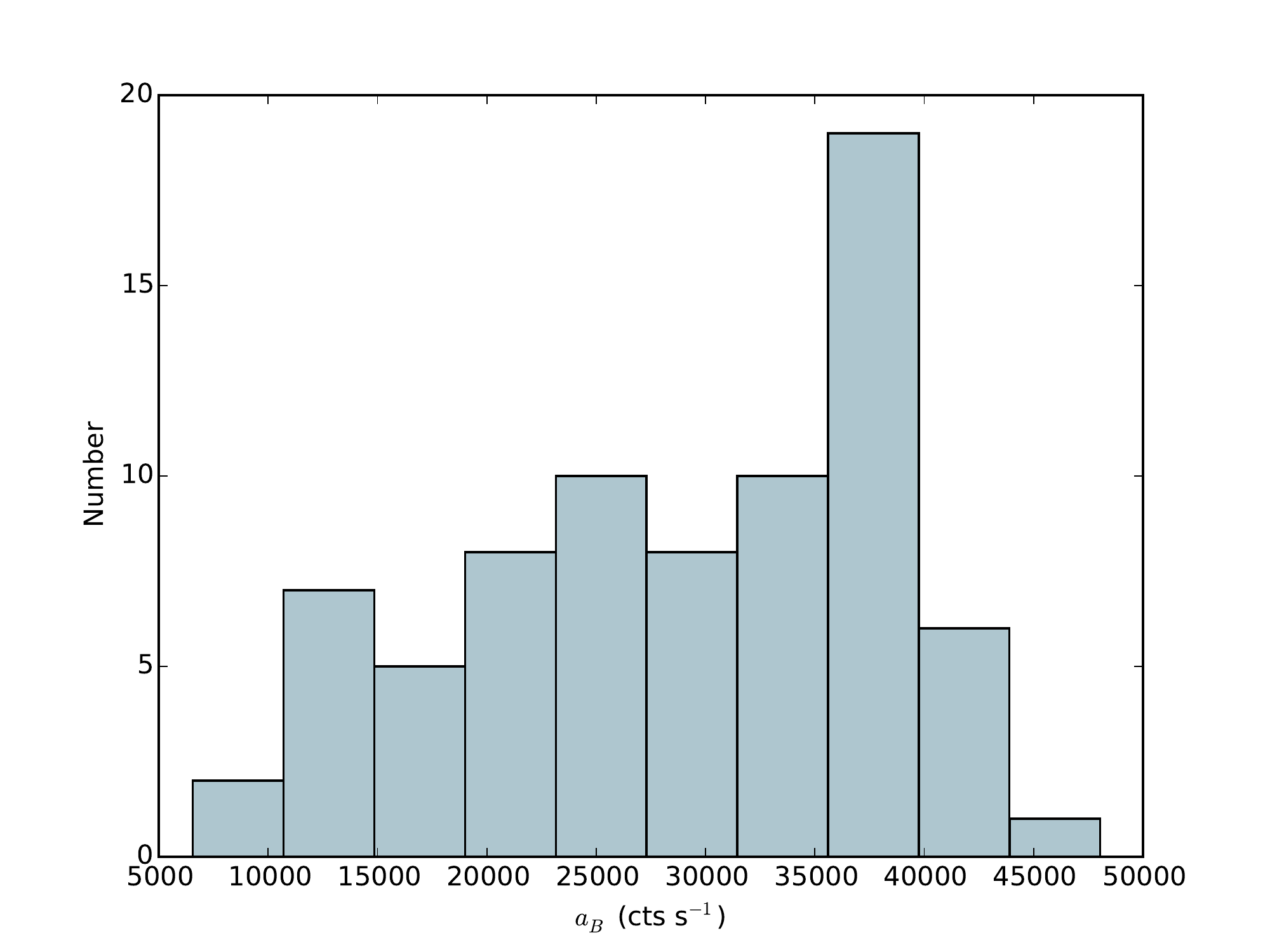}
  \caption[Histogram showing the distribution of $a_B$ amongst Normal Bursts.]{A histogram showing the distribution of burst amplitude $a_B$ amongst our sample of Normal Bursts.\index{Normal burst}}
  \label{fig:app_hist_ab}
\end{figure}

\begin{figure}
  \centering
  \includegraphics[width=.9\linewidth, trim={0cm 0 0cm 0},clip]{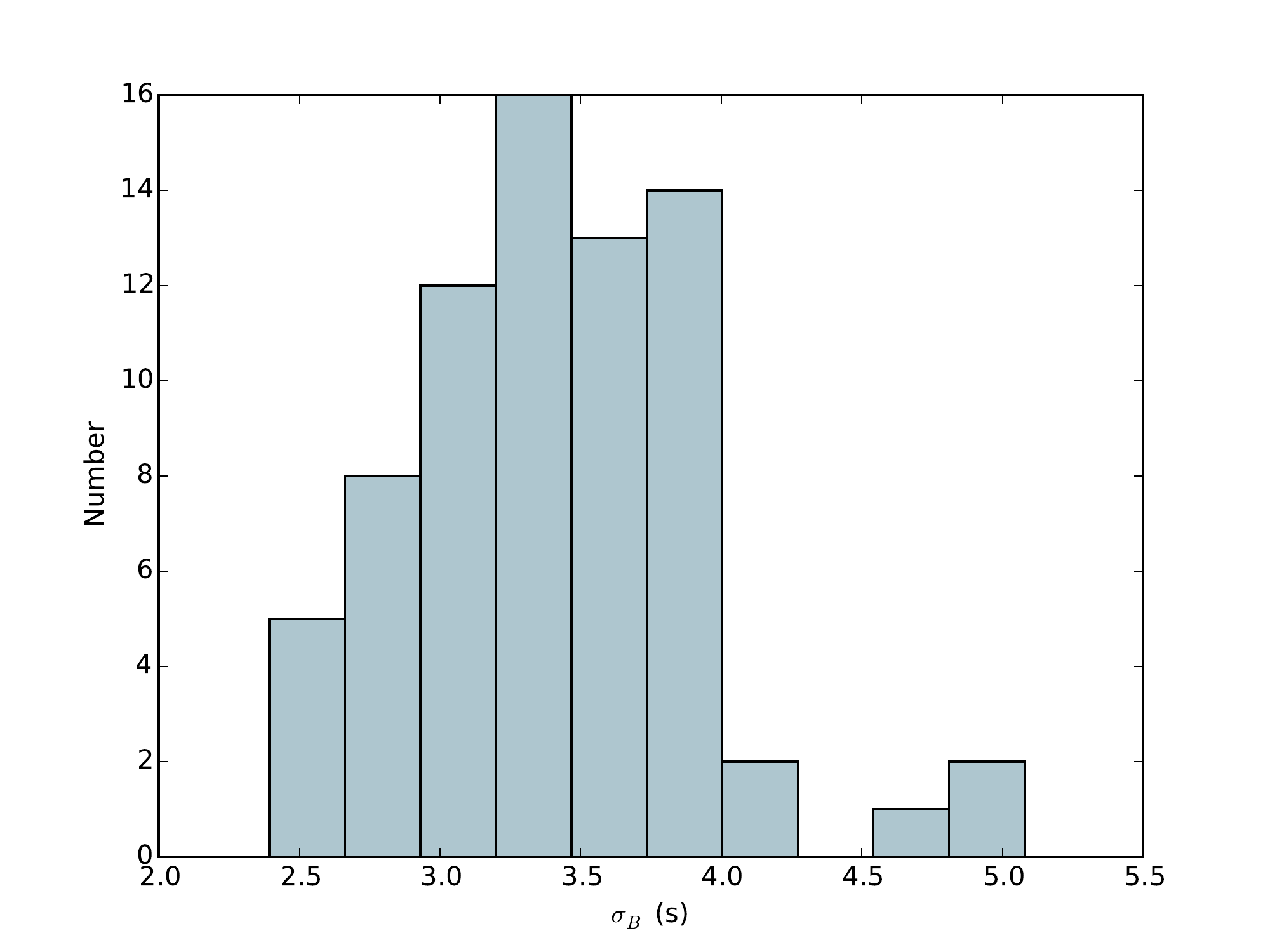}
  \caption[Histogram showing the distribution of $\sigma_B$ amongst Normal Bursts.]{A histogram showing the distribution of burst width $\sigma_B$ amongst our sample of Normal Bursts.\index{Normal burst}}
  \label{fig:app_hist_sigb}
\end{figure}

\begin{figure}
  \centering
  \includegraphics[width=.9\linewidth, trim={0cm 0 0cm 0},clip]{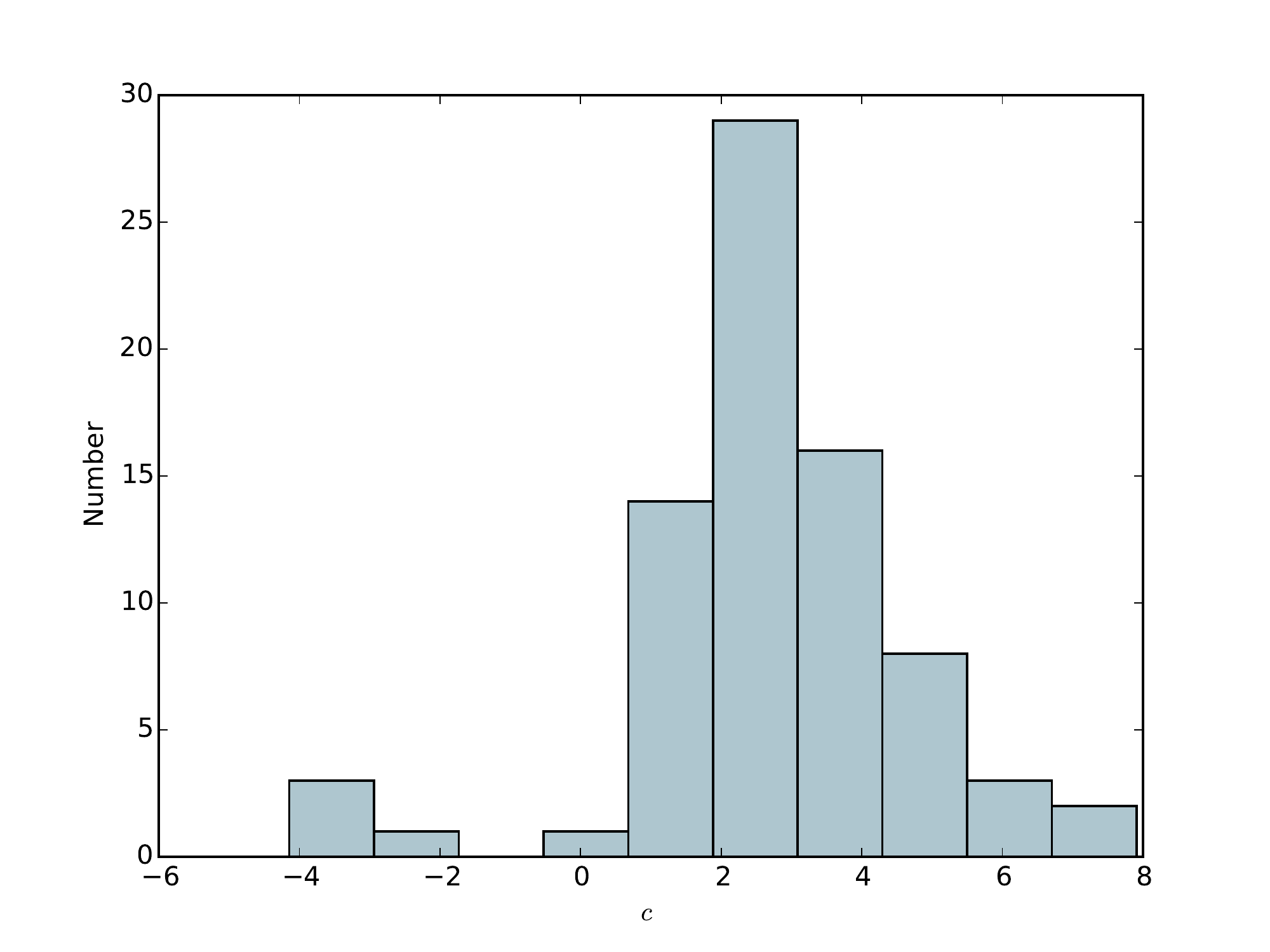}
  \caption[Histogram showing the distribution of $c$ amongst Normal Bursts.]{A histogram showing the distribution of burst skewness $c$ amongst our sample of Normal Bursts.\index{Normal burst} }
  \label{fig:app_hist_c}
\end{figure}

\begin{figure}
  \centering
  \includegraphics[width=.9\linewidth, trim={0cm 0 0cm 0},clip]{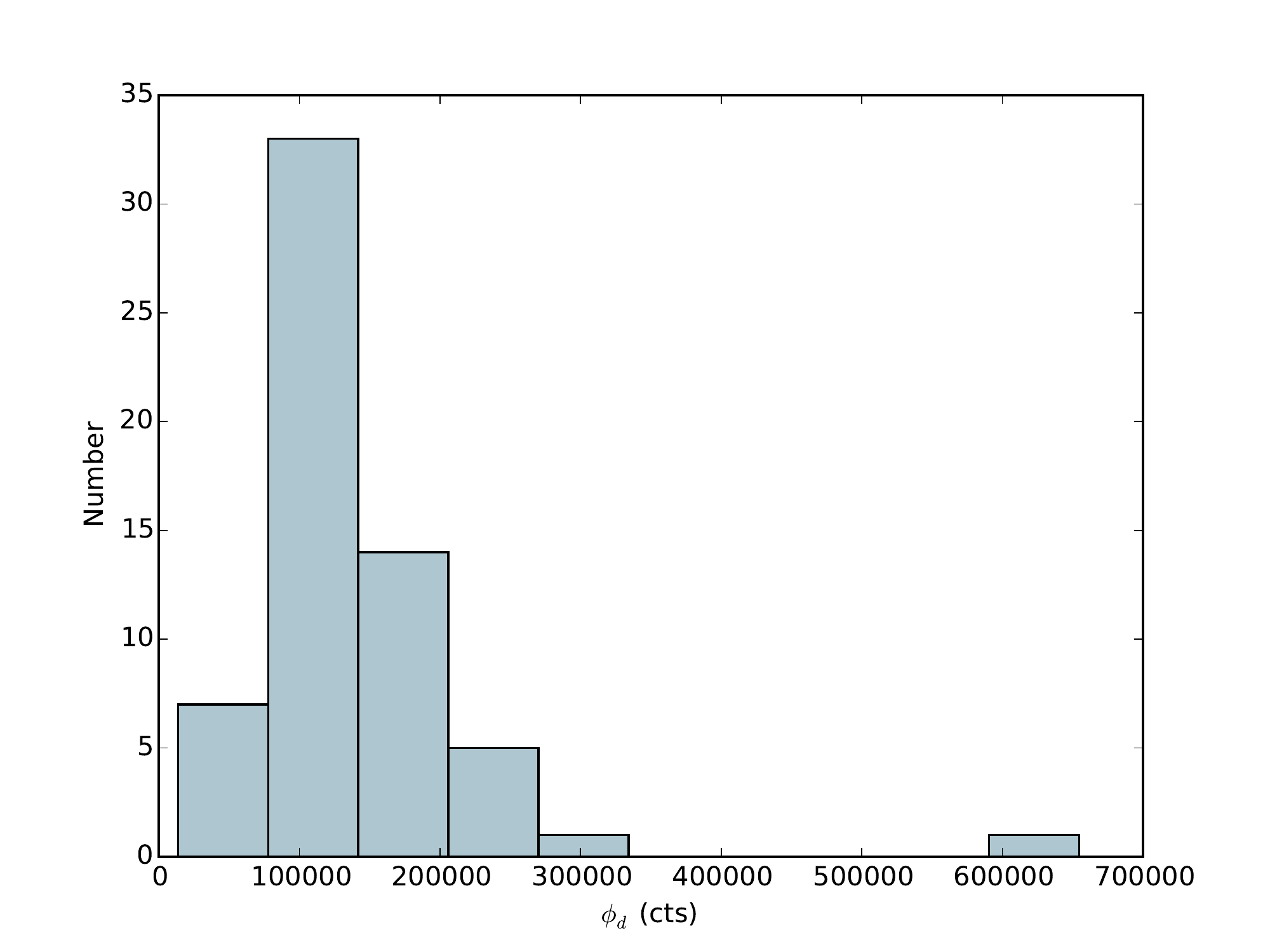}
  \caption[Histogram showing the distribution of $\phi_d$ amongst Normal Bursts.]{A histogram showing the distribution of dip fluence $\phi_d$ amongst our sample of Normal Bursts.\index{Normal burst}}
  \label{fig:app_hist_phid}
\end{figure}

\begin{figure}
  \centering
  \includegraphics[width=.9\linewidth, trim={0cm 0 0cm 0},clip]{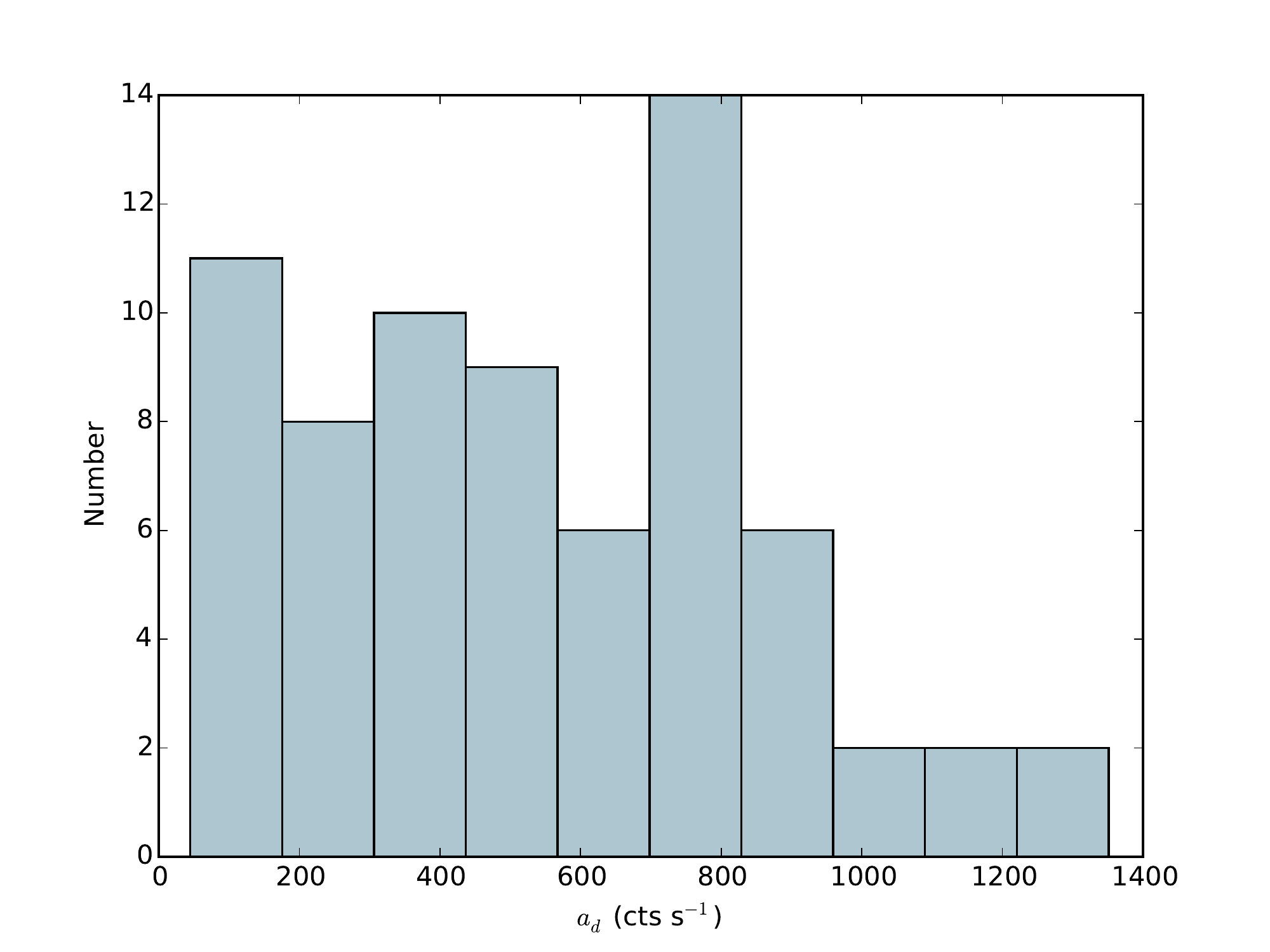}
  \caption[Histogram showing the distribution of $a_d$ amongst Normal Bursts.]{A histogram showing the distribution of dip amplitude $a_d$ amongst our sample of Normal Bursts.\index{Normal burst}}
  \label{fig:app_hist_ad}
\end{figure}

\begin{figure}
  \centering
  \includegraphics[width=.9\linewidth, trim={0cm 0 0cm 0},clip]{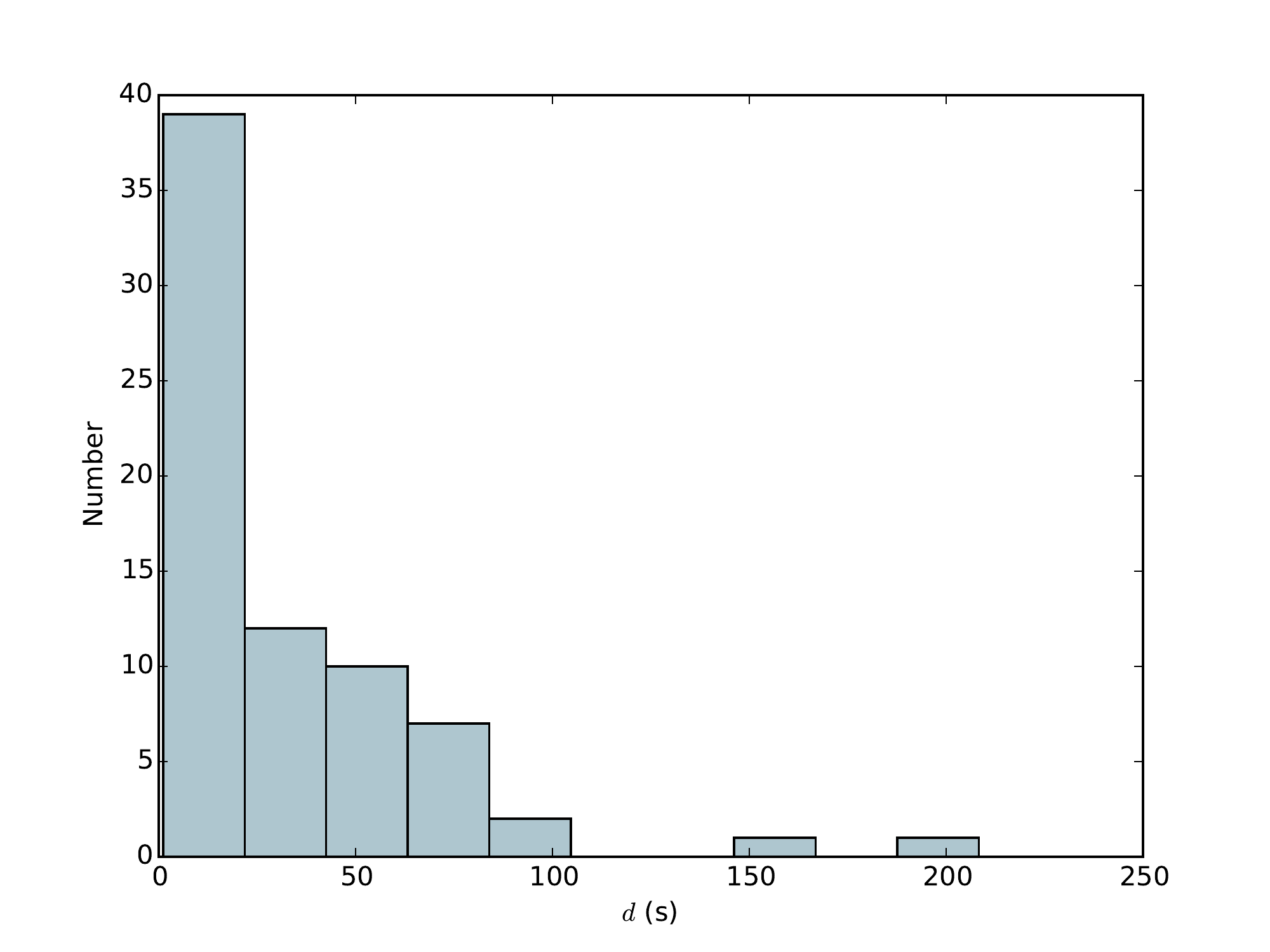}
  \caption[Histogram showing the distribution of $d$ amongst Normal Bursts.]{A histogram showing the distribution of dip fall-time $d$ amongst our sample of Normal Bursts.\index{Normal burst}}
  \label{fig:app_hist_d}
\end{figure}

\begin{figure}
  \centering
  \includegraphics[width=.9\linewidth, trim={0cm 0 0cm 0},clip]{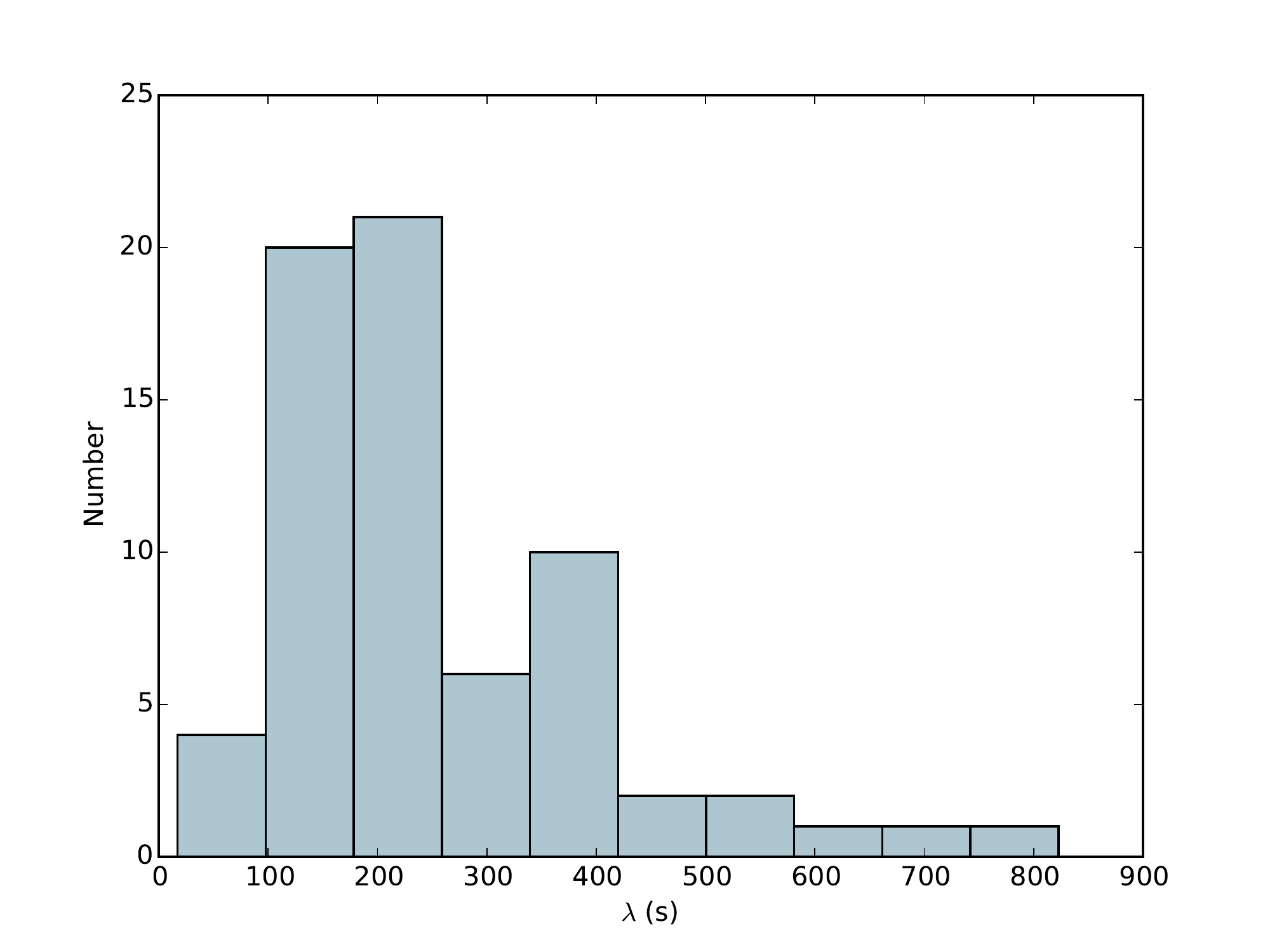}
  \caption[Histogram showing the distribution of $\lambda$ amongst Normal Bursts.]{A histogram showing the distribution of dip recovery timescale $\lambda$ amongst our sample of Normal Bursts.\index{Normal burst}}
  \label{fig:app_hist_lamb}
\end{figure}

\begin{figure}
  \centering
  \includegraphics[width=.9\linewidth, trim={0cm 0 0cm 0},clip]{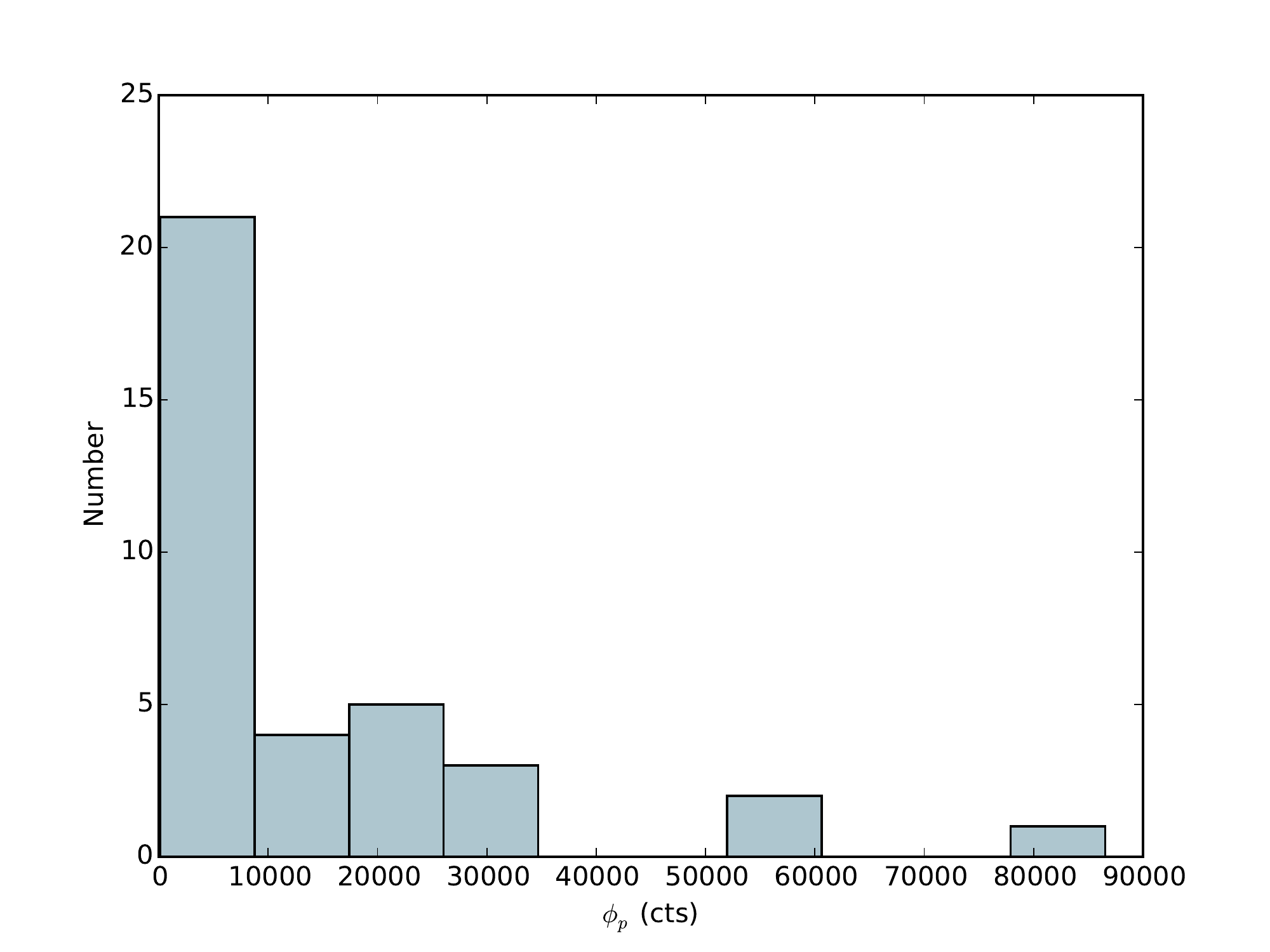}
  \caption[Histogram showing the distribution of $\phi_p$ amongst Normal Bursts.]{A histogram showing the distribution of plateau fluence $\phi_p$ amongst our sample of Normal Bursts.\index{Normal burst}}
  \label{fig:app_hist_phip}
\end{figure}

\begin{figure}
  \centering
  \includegraphics[width=.9\linewidth, trim={0cm 0 0cm 0},clip]{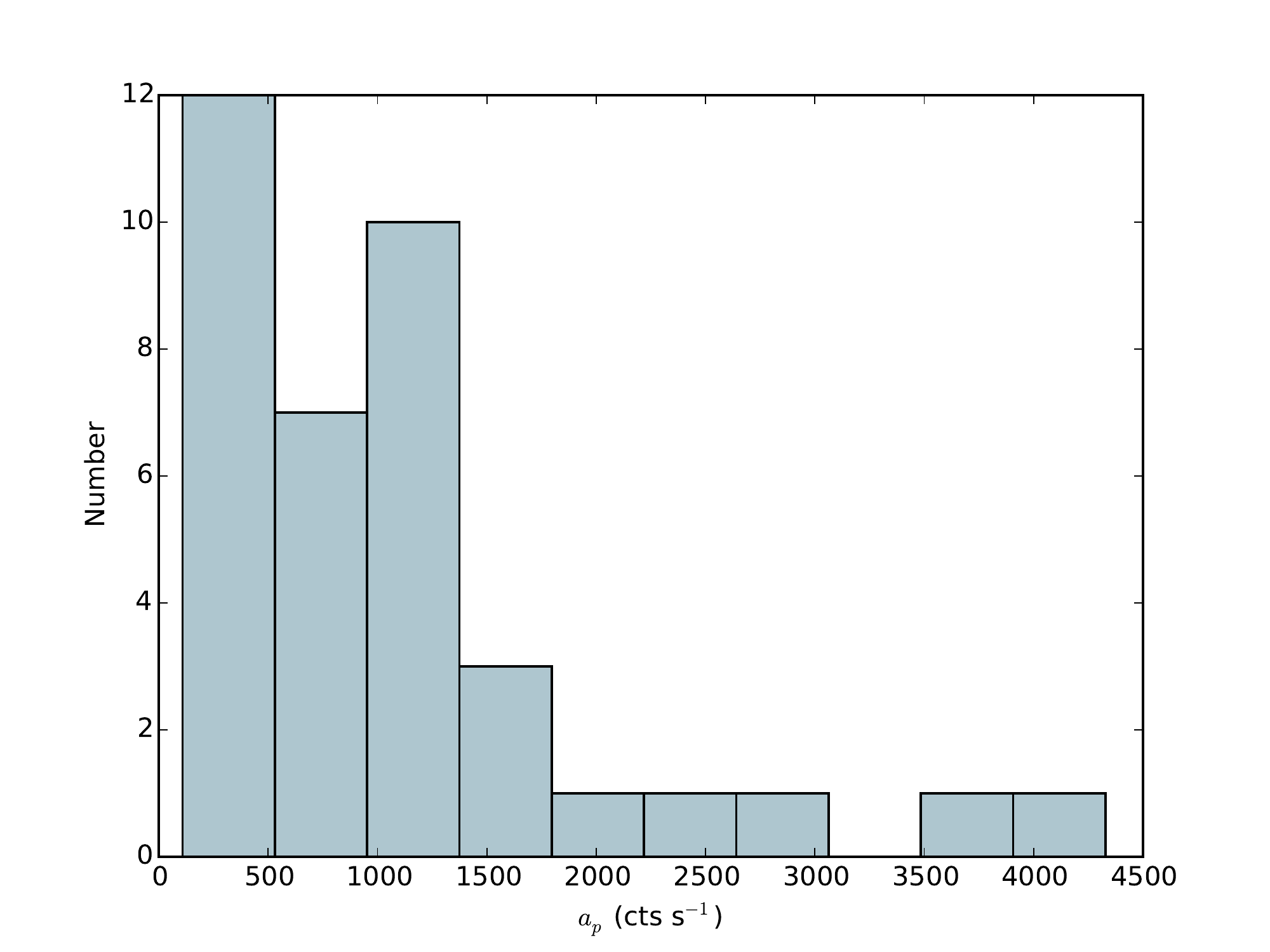}
  \caption[Histogram showing the distribution of $a_p$ amongst Normal Bursts.]{A histogram showing the distribution of plateau amplitude $a_p$ amongst our sample of Normal Bursts.\index{Normal burst}}
  \label{fig:app_hist_ap}
\end{figure}

\begin{figure}
  \centering
  \includegraphics[width=.9\linewidth, trim={0cm 0 0cm 0},clip]{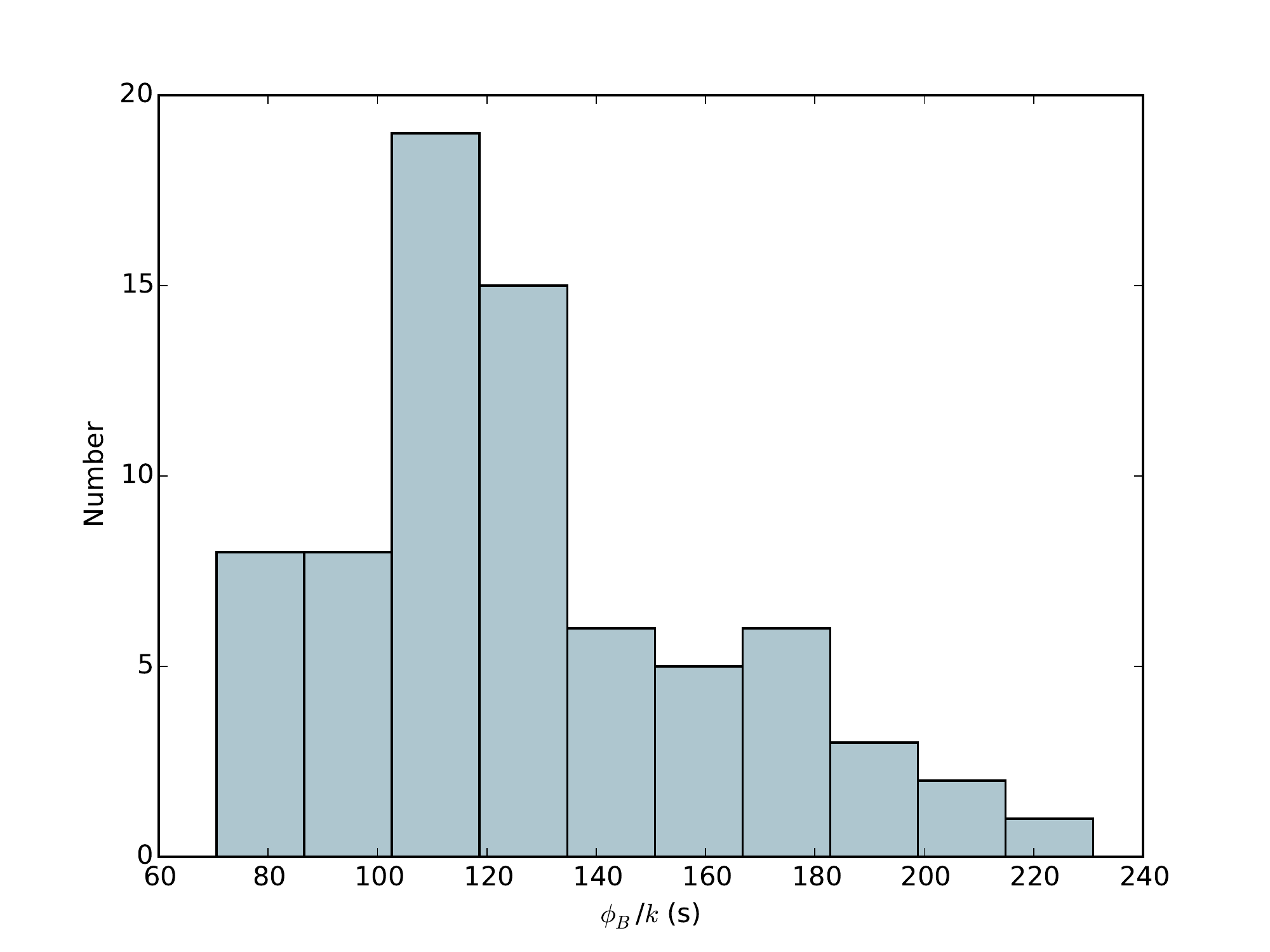}
  \caption[Histogram showing the distribution of $\phi_B/k$ amongst Normal Bursts.]{A histogram showing the distribution of persistent-emission-normalised burst fluence $\phi_B/k$ amongst our sample of Normal Bursts.\index{Normal burst}}
  \label{fig:app_hist_phib_n}
\end{figure}

\begin{figure}
  \centering
  \includegraphics[width=.9\linewidth, trim={0cm 0 0cm 0},clip]{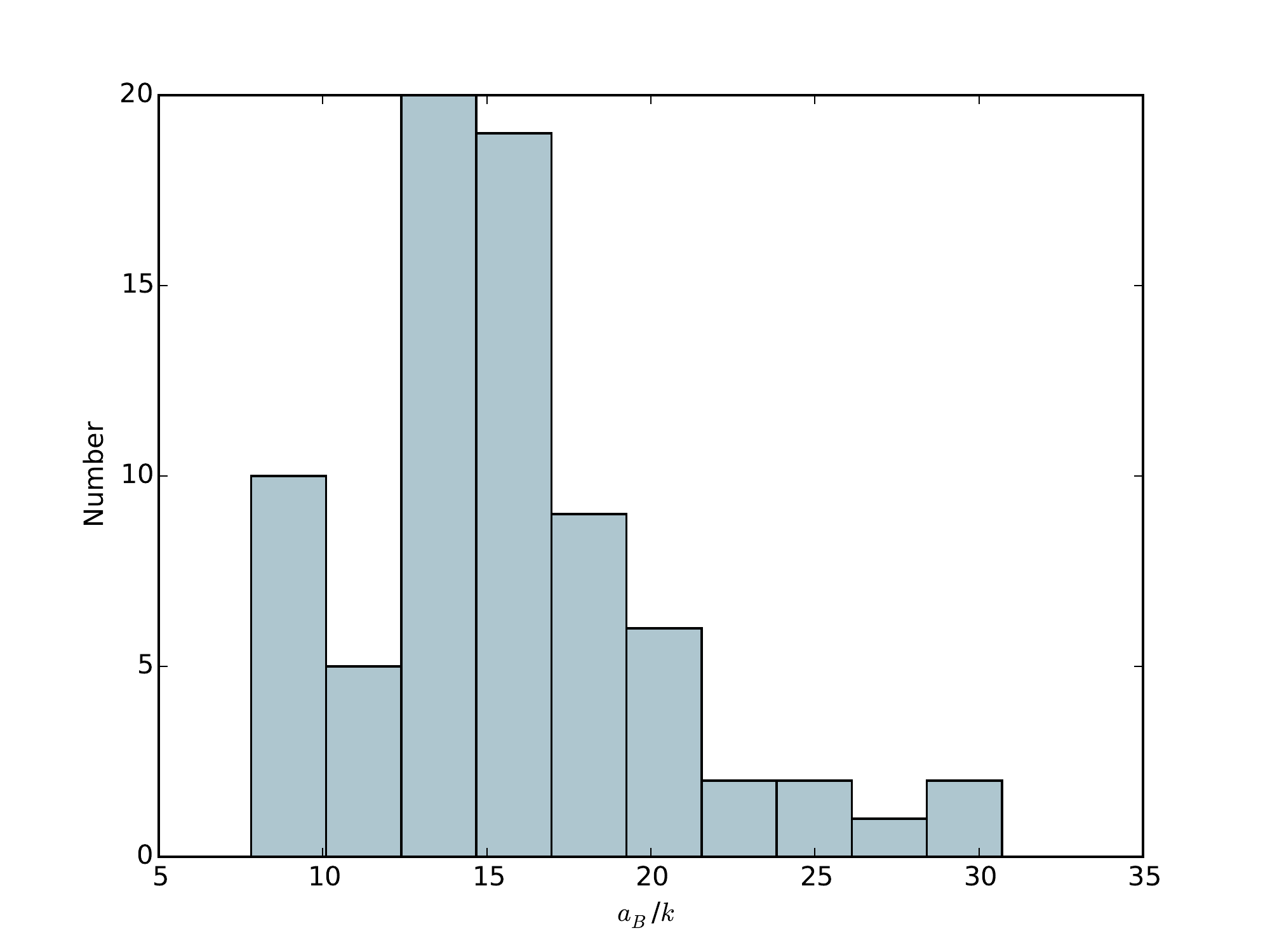}
  \caption[Histogram showing the distribution of $a_B/k$ amongst Normal Bursts.]{A histogram showing the distribution of persistent-emission-normalised burst amplitude $a_B/k$ amongst our sample of Normal Bursts.\index{Normal burst}}
  \label{fig:app_hist_ab_n}
\end{figure}

\begin{figure}
  \centering
  \includegraphics[width=.9\linewidth, trim={0cm 0 0cm 0},clip]{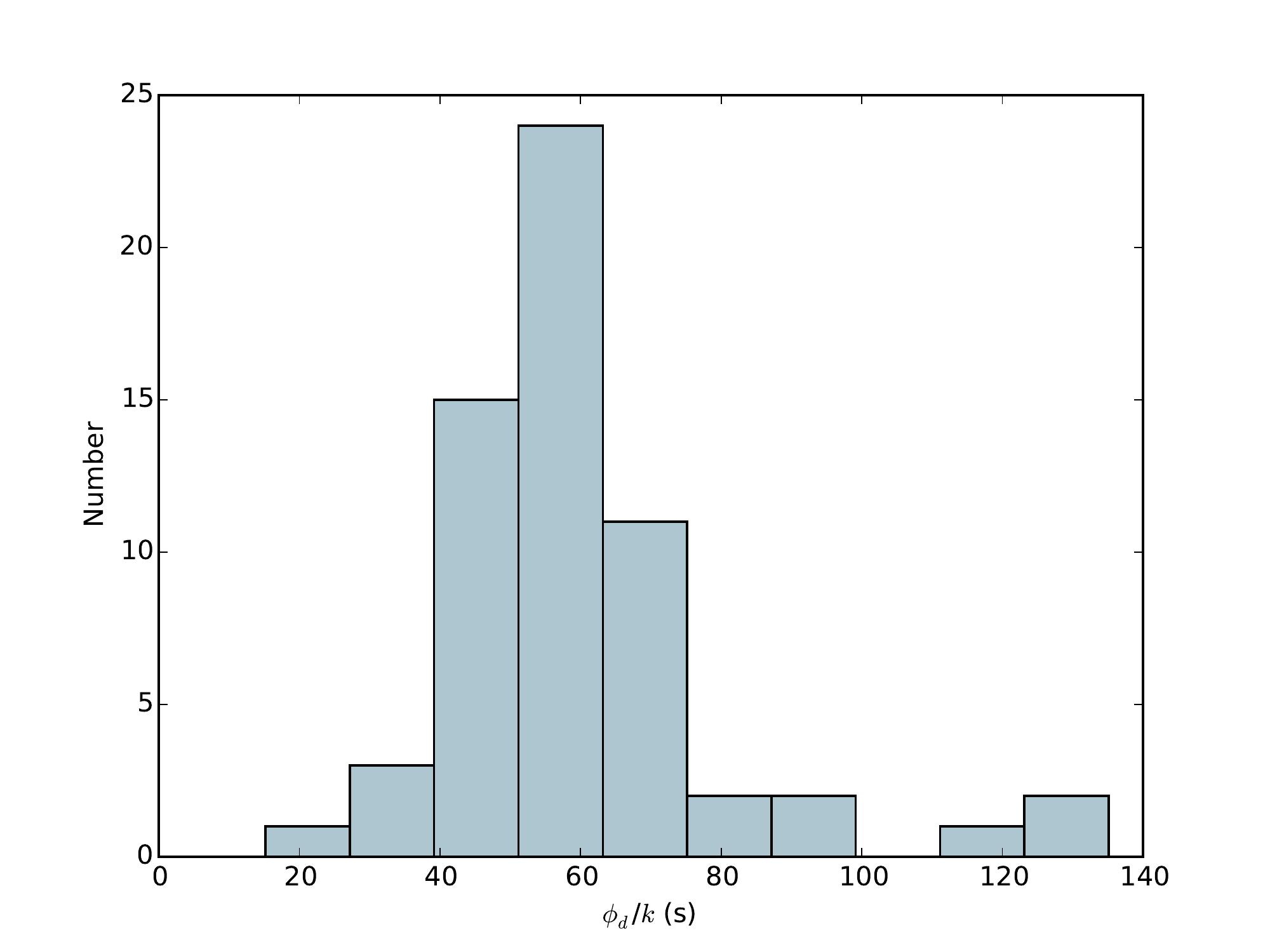}
  \caption[Histogram showing the distribution of $\phi_d/k$ amongst Normal Bursts.]{A histogram showing the distribution of persistent-emission-normalised dip fluence $\phi_d/k$ amongst our sample of Normal Bursts.\index{Normal burst}}
  \label{fig:app_hist_phid_n}
\end{figure}

\begin{figure}
  \centering
  \includegraphics[width=.9\linewidth, trim={0cm 0 0cm 0},clip]{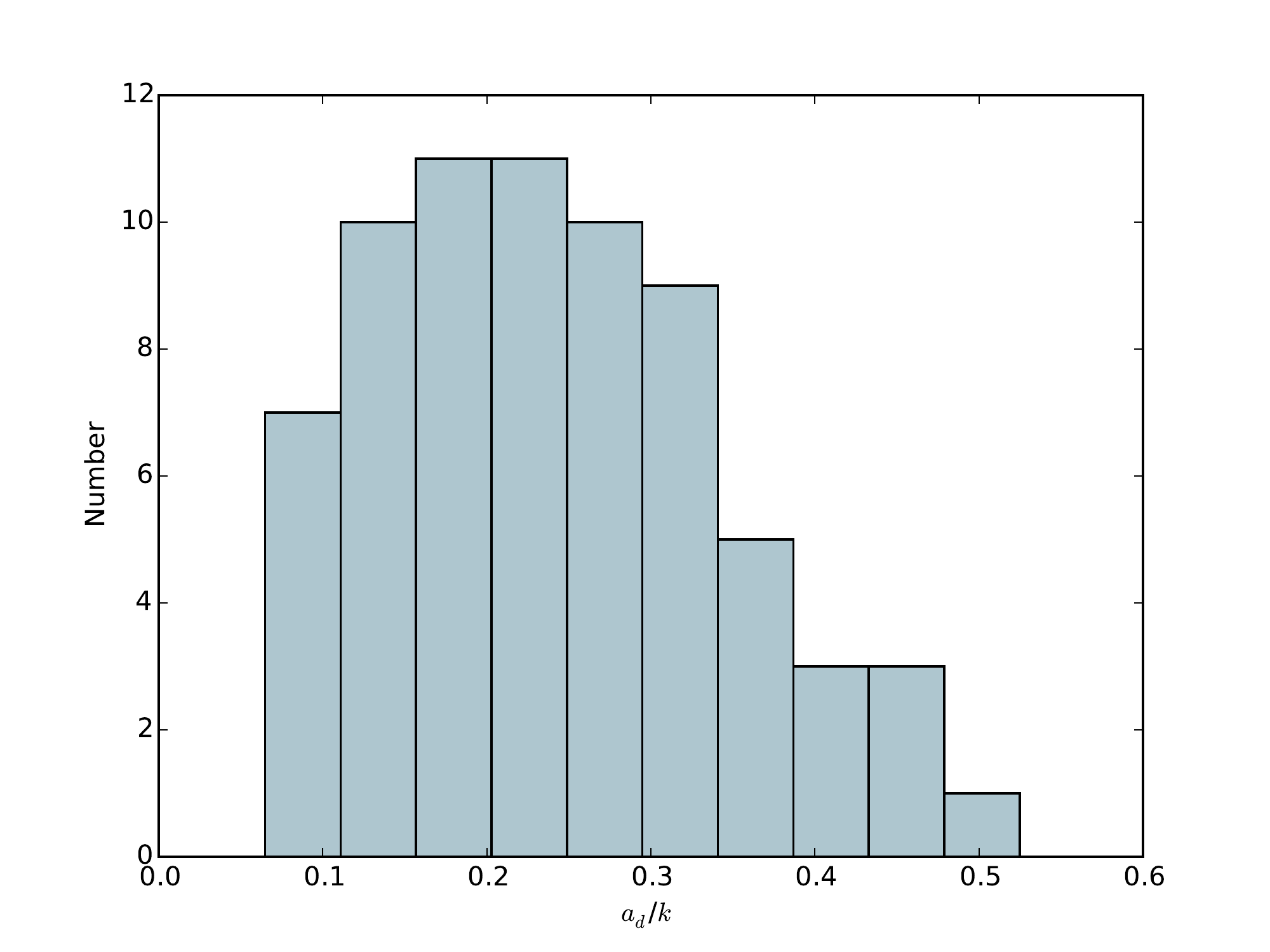}
  \caption[Histogram showing the distribution of $a_d/k$ amongst Normal Bursts.]{A histogram showing the distribution of persistent-emission-normalised dip amplitude $a_d/k$ amongst our sample of Normal Bursts.\index{Normal burst}}
  \label{fig:app_hist_ad_n}
\end{figure}

\begin{figure}
  \centering
  \includegraphics[width=.9\linewidth, trim={0cm 0 0cm 0},clip]{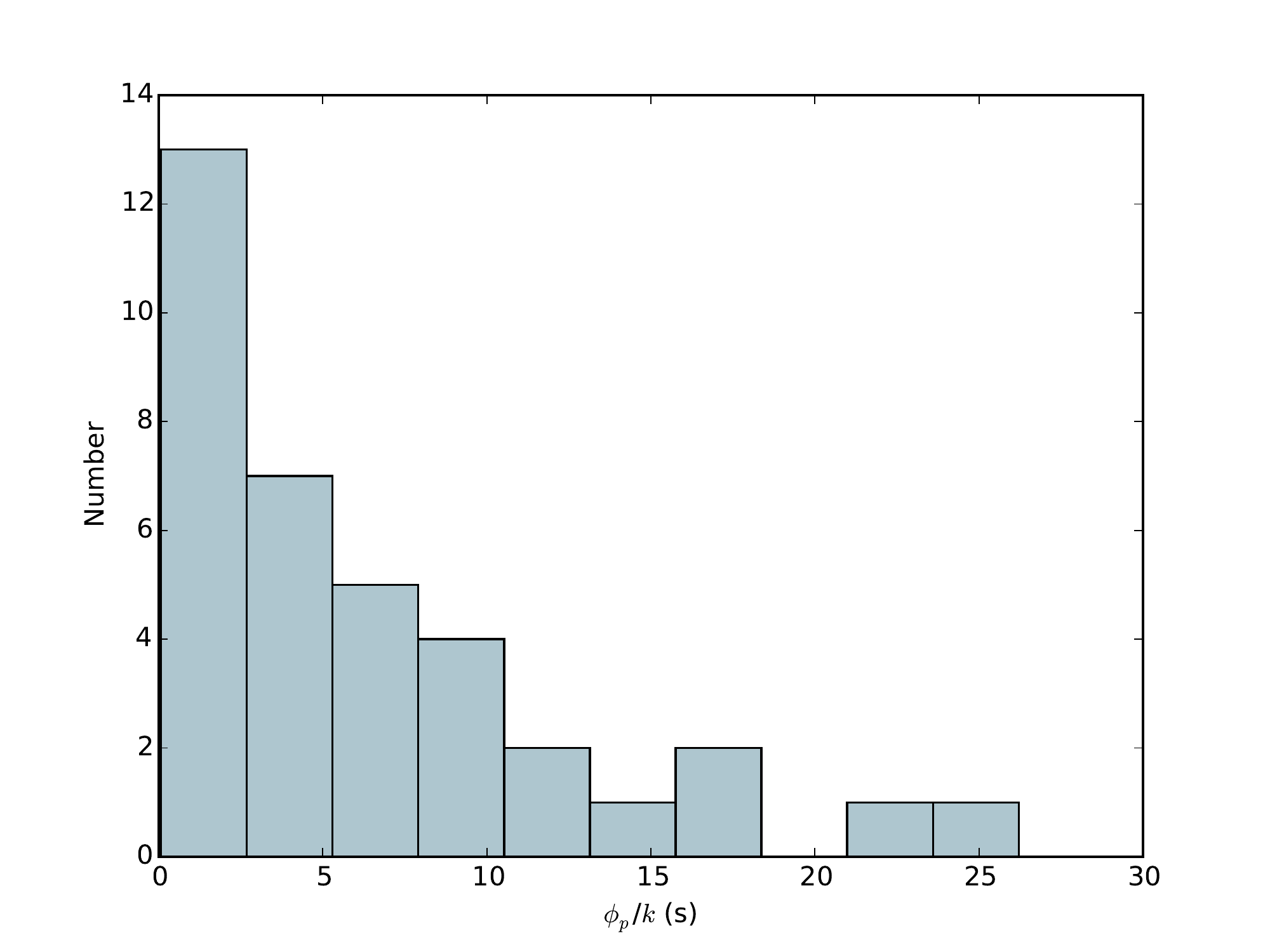}
  \caption[Histogram showing the distribution of $\phi_p/k$ amongst Normal Bursts.]{A histogram showing the distribution of persistent-emission-normalised plateau fluence $\phi_p/k$ amongst our sample of Normal Bursts.\index{Normal burst}}
  \label{fig:app_hist_phip_n}
\end{figure}

\begin{figure}
  \centering
  \includegraphics[width=.9\linewidth, trim={0cm 0 0cm 0},clip]{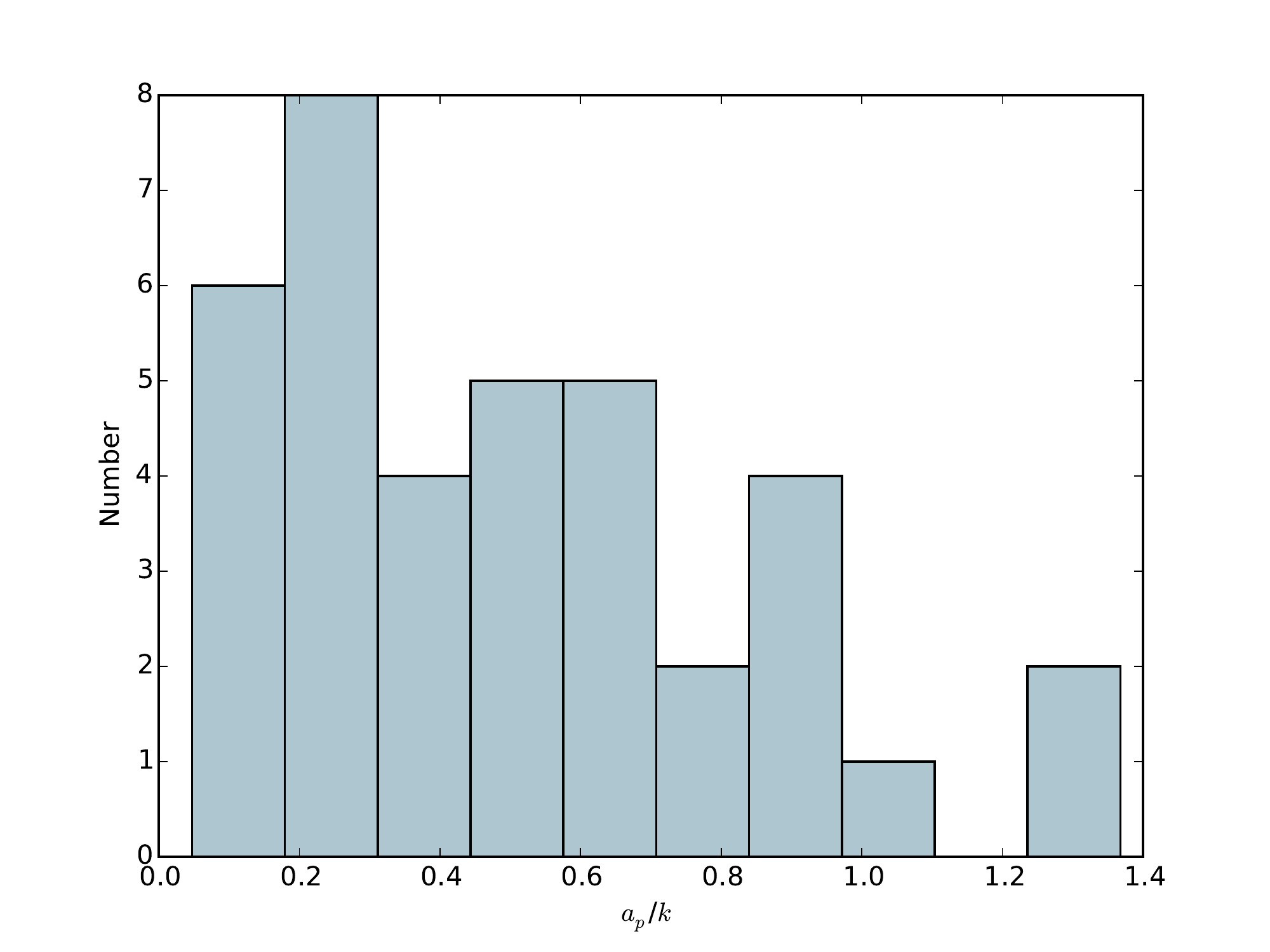}
  \caption[Histogram showing the distribution of $a_p/k$ amongst Normal Bursts.]{A histogram showing the distribution of persistent-emission-normalised plateau amplitude $a_p/k$ amongst our sample of Normal Bursts.\index{Normal burst}}
  \label{fig:app_hist_ap_n}
\end{figure}

 In Figures \ref{fig:app_hist_phib}--\ref{fig:app_hist_ap}, we present histograms showing the distributions of $\phi_B$, $a_B$, $\sigma_B$, $c$, $\phi_d$, $a_d$, $d$, $\lambda$, $\phi_p$ and $a_p$ we find in our population study\index{Population study} of Normal Bursts\index{Normal burst} in the Bursting Pulsar\index{Bursting Pulsar}.  Each of these is a parameter we used to fit the Normal Bursts in our sample: see Section \ref{sec:struc} for a full explanation of these parameters.  In Figures \ref{fig:app_hist_phib_n}--\ref{fig:app_hist_ap_n} we show the distributions of $\phi_B$, $a_B$, $\phi_d$, $a_d$, $\phi_p$ and $a_p$ after being normalised by the persistent emission rate $k$ at the time of each burst.

%-----------------

\cleardoublepage
\include{paramcorr}
\cleardoublepage
\chapter{PANTHEON suite}

\label{app:PAN}

\par In this section I present the headers and internal command lists of every piece of code in \textit{PANTHEON}\index{PANTHEON@\texttt{PANTHEON}} (Python ANalytical Tools for High-energy Event data manipulatiON), that suite of tools that I created and used during the work presented in this thesis.  The full code is available at \url{https://github.com/jmcourt/PANTHEON}.  PANTHEON makes use of the Astropy \citep{Astropy}, Matplotlib \citep{Hunter_MatPlotLib}, Numpy, Scipy \citep{NumPy} and Numba \citep{Numba}.

\section{FITS Genie}

\par \texttt{FITS Genie} is a script that allows the user to extract data from raw \texttt{FITS}\index{FITS@\texttt{FITS}} files.  The script was designed to interface with \indexrxte\textit{RXTE} data, but there is also limited implementation with \indexsuzaku\textit{Suzaku}.  The script produces \texttt{.plotd} and \texttt{.speca} files, which can be further processed with \textit{Plot Demon} and \textit{Spec Angel}.

\begin{minted}[fontsize=\scriptsize]{python}
#! /usr/bin/env python

# |----------------------------------------------------------------------|
# |-----------------------------FITS GENIE-------------------------------|
# |----------------------------------------------------------------------|

# Call as ./fitsgenie.py FILE1 PROD_REQ [LCHAN] [HCHAN] [BINNING] [FOURIER RES] 
#   [FOURIER SEP] [BGEST] [FLAVOUR]
#
# Takes 1 FITS Event file and produces .speca and .plotd formatted products to be
#   analysed by plotdemon and specangel.
#
# Arguments:
#
#  FILE1
#   The absolute path to the file to be used.
#
#  PROD_REQ
#   The products requested by the user.  The following inputs are valid:
#      'spec','speca','s' will cause FITSGenie to produce only a .speca file as output
#      'plot','plotd','p' will cause FITSGenie to produce only a .plotd file as output
#      'both','all','b','a','sp','ps' will cause both files to be output
#
#  [LCHAN]
#   Optional: The lowest channel on the PCA instrument on RXTE which will be used to
#     populate the data.  Default of 0 (minimum).
#
#  [HCHAN]
#   Optional: The highest channel on the PCA instrument on RXTE which will be used to
#     populate the data.  Default of 255 (maximum).
#
#  [BINNING]
#   Optional: The size, in seconds, of bins into which data will be sorted.  Takes the
#     value of the time resolution of the data if not specified by the user.  Default
#     of 2^-15s
#
#  [FOURIER RES]
#   Optional: The size of the individual time windows in which the data is to be split.
#   Fourier spectra will be made of each of these windows.  Default of 128s.
#
#  [FOURIER SEP]
#   Optional: The separation of the startpoints of individual time windows in which the
#     data is to be split.  Fourier spectra will be made of each of these windows.
#     Default of 128s.
#
#  [BGEST]
#   Optional: The approximate average background count rate during the observation in
#     cts/s.  Default of 30cts/s.
#
#  [FLAVOUR]
#   Optional: A useful bit of text to put on plots to help identify them later on.
\end{minted}

\section{Plot Demon}

\par \texttt{Plot Demon} is a script for manipulating the \texttt{.plotd} files output by \texttt{Fits Genie}, as well as time series stored in comma-separated variable files (CSVs).  The script allows the user to produce lightcurves\index{Lightcurve}, hardness-intensity diagrams\index{Hardness-intensity diagram} and colour-colour diagrams\index{Colour-colour diagram}, among other products.  The script also allows the user to modify the data by rebinning, clipping or folding\index{Folding} it using the algorithms I detail in Section \ref{sec:vfold}.

\begin{minted}[fontsize=\scriptsize]{python}
#! /usr/bin/env python

# |----------------------------------------------------------------------|
# |------------------------------PLOT DEMON------------------------------|
# |----------------------------------------------------------------------|

# Call as ./plotdemon.py FILE1 [FILE2] [FILE3] BINNING
#
# Takes 1-3 .plotd files and plots relevant astrometric plots
#
# Arguments:
#
#  FILE1
#   The absolute path to the first file to be used (generally the lowest energy band)
#
#  [FILE2]
#   The absolute path to the second file to be used
#
#  [FILE3]
#   The absolute path to the third file to be used (generally the highest energy band)
#
#  [BINNING]
#   Optional: the size, in seconds, of bins into which data will be sorted.

def give_inst():                                      # Define printing this list of
                                                      # instructions as a function
   print 'COMMANDS: Enter a command to manipulate data.'
   print ''
   print 'DATA:'
   print '* "rebin" to reset the data and load it with a different binning.'
   print '* "clip" to clip the data.'
   print '* "norm time" to renormalise the times by the start time of the data'
   print '* "mask" to remove a range of data.'
   print '* "rms" to return the fractional rms of the data.'
   print '* "fold" to fold data over a period of your choosing'+(' (requires PyAstron'+
         'omy module!)' if not module_pyastro else '')+'.'
   print '* "autofold" to automatically seek a period over which to fold data'+(' (re'+
         'quires PyAstronomy module!)' if not module_pyastro else '')+'.'
   print '* "varifold" to fold over a non-constant period using an algorithm optimise'+
         'd for high-amplitude quasi-periodic flares.'
   print '* "plot bursts" to plot the results of the peak-finding algorithm used in v'+
         'arifold.'
   print ''
   print '1+ DATASET PLOTS:'
   print '* "lc" to plot a simple graph of flux over time.'
   print '* "bg" to plot background over time, if background has been estimated for t'+
         'hese files.'
   print '* "animate" to create an animation of the lightcurve as the binning is incr'+
         'eased.'
   print '* "circanim" to create an animation of the lightcurve circularly folded as '+
         'the period is increased.'
   print '* "lombscargle" to create a Lomb-Scargle periodogram of the lightcurve.'
   print '* "autocor" to plot the auto-correlation function.'
   print '* "rmsflux" to plot the rms-flux relationship of the data.'
   if nfiles>1:                                       # Only display 2-data-set inst-
                                                      # -ructions if 2+ datasets given
      print ''
      print '2+ DATASET PLOTS:'
      print '* "hardness21" to plot a hardness/time diagram of file2/file1 colour ove'+
            'r time.'
      print '* "hardness12" to plot a hardness/time diagram of file1/file2 colour ove'+
            'r time.'
      print '* "hid21" to plot a hardness-intensity diagram of file2/file1 colour aga'+
            'inst total flux.'
      print '* "hid12" to plot a hardness-intensity diagram of file1/file2 colour aga'+
            'inst total flux.'
      print '* "calcloop21" to return the probability of a null hysteresis in the 12 '+
            'HID.'
      print '* "col21" to plot file2/file1 colour against time.'
      print '* "col12" to plot file1/file2 colour against time.'
      print '* "band" to plot the lightcurve of a single energy band.'
      print '* "bands" to plot lightcurves of all bands on adjacent axes.'
      print '* "xbands" to plot lightcurves of all bands on the same axes.'
      print '* "compbands21" to plot lightcurves of bands 2 and 1 against each other.'
      print '* "crosscor21" to plot the cross-correlation function of band 1 with ban'+
            'd 2.'
      print '* "timeres crosscor21" to plot the time-resolved cross-correlation funct'+
            'ion of band 1 with band 2' 
      print '* "all" to plot all available data products.'
   if nfiles==3:                                      # Only display 3-data-set inst-
                                                      # -ructions if 3 datasets given
      print ''
      print '3 DATASET PLOTS:'
      print '* "hardness32" to plot a hardness/time diagram of file3/file2 colour ove'+
            'r time.'
      print '* "hardness23" to plot a hardness/time diagram of file2/file3 colour ove'+
            'r time.'
      print '* "hardness31" to plot a hardness/time diagram of file3/file1 colour ove'+
            'r time.'
      print '* "hardness13" to plot a hardness/time diagram of file1/file3 colour ove'+
            'r time.'
      print '* "hid32" to plot a hardness-intensity diagram of file3/file2 colour aga'+
            'inst total flux.'
      print '* "hid23" to plot a hardness-intensity diagram of file2/file3 colour aga'+
            'inst total flux.'
      print '* "calcloop32" to return the probability of a null hysteresis in the 32 '+
            'HID.'
      print '* "hid31" to plot a hardness-intensity diagram of file3/file1 colour aga'+
            'inst total flux.'
      print '* "hid13" to plot a hardness-intensity diagram of file1/file3 colour aga'+
            'inst total flux.'
      print '* "calcloop31" to return the probability of a null hysteresis in the 31 '+
            'HID.'
      print '* "col32" to plot file3/file2 colour against time.'
      print '* "col23" to plot file2/file3 colour against time.'
      print '* "col31" to plot file3/file1 colour against time.'
      print '* "col13" to plot file1/file3 colour against time.'
      print '* "compbands31" to plot lightcurves of bands 3 and 1 against each other.'
      print '* "compbands32" to plot lightcurves of bands 3 and 2 against each other.'
      print '* "ccd" to plot a colour-colour diagram (3/1 colour against 2/1 colour).'
      print '* "timeres crosscor31" to plot the time-resolved cross-correlation funct'+
            'ion of band 3 with band 1'
      print '* "timeres crosscor32" to plot the time-resolved cross-correlation funct'+
            'ion of band 3 with band 2'
      print '* "crosscor31" to plot the cross-correlation function of band 3 with ban'+
            'd 1.'
      print '* "crosscor32" to plot the cross-correlation function of band 3 with ban'+
            'd 2.'
   print ''
   print 'BURST ANALYSIS:'
   print '* "burst get" to interactively extract burst data for analysis.'
   print '* "burst peaks" for a histogram of peak heights of extracted bursts.'
   print '* "burst risetimes" for a histogram of rise times of extracted bursts.'
   print '* "burst falltimes" for a histogram of fall times of extracted bursts.'
   print '* "burst lengths" for a histogram of durations of extracted bursts.'
   print '* "burst help" for further information on burst analysis.'
   print ''
   print 'SAVING DATA TO ASCII:'
   print '* "export" to dump the lightcurve and colour data into an ASCII file.'
   print '* "bgdump" to export background lightcurve to an ASCII file.'
   print '* "timenorm" to toggle absolute or relative time values on x-axis.'
   print ''
   print 'TOGGLE OPTIONS:'
   print '* "errors" to toggle whether to display errors in plots.'
   print '* "lines" to toggle lines joining points in graphs.'
   print '* "ckey" to toggle colour key (red-blue) for the first five points in all p'+
         'lots.'
   print '* "save" to save to disk any plots which would otherwise be shown.'
   print ''
   print 'ADVANCED OPTIONS:'
   print '* "burstalg" to select algorithm for finding pulse peaks in lightcurve.'
   print ''
   print 'OTHER COMMANDS:'
   print '* "info" to display a list of facts and figures about the current PlotDemon'+
         ' session.'
   print '* "reflav" to rewrite the flavour text used for graph titles.'
   print '* "help" or "?" to display this list of instructions again.'
   print '* "quit" to quit.'

give_inst()                                          # Print the list of instructions
\end{minted}

\section{Spec Angel}

\par \texttt{Spec Angel} is a script to allow users to produce power spectra\index{Fourier analysis} from \texttt{.speca} files output by \texttt{Fits Genie}.  These power spectra can be linearly or logarithmically binned, and can be normalised in a number of different ways.

\begin{minted}[fontsize=\scriptsize]{python}

#! /usr/bin/env python

# |----------------------------------------------------------------------|
# |------------------------------SPEC ANGEL------------------------------|
# |----------------------------------------------------------------------|

# Call as ./specangel.py FILE1 [LBINNING]

# Takes 1 RXTE FITS Event file and produces an interactive spectrogram
#
# Arguments:
#
#  FILE1
#   The absolute path to the file to be used.
#
#  [LBINNING]
#   Optional- the logarithmic binning factor 'x'; frequency data will be binned into 
#   bins which have their lefthand edges defined by the formula 10**(ix) for integer i.
#

def give_inst():                                      # Define printing this list of
                                                      # instructions as a function
   print 'COMMANDS: Enter a command to manipulate data.'
   print ''
   print 'DATA:'
   print '* "rebin" to reset the data and load it with a different normalisation and '+
         ' binning.'
   print '* "clip" to clip the range of data.'
   print '* "reset" to reset data.'
   print ''
   print 'SPECTROGRAM:'
   print '* "sg plot" to plot the spectrogram currently being worked on.'
   print '* "sg floor" to set a minimum value for the spectrogram'+"'"+'s z-axis colo'+
         'ur key.'
   print '* "sg ceil" to set a maximum value for the spectrogram'+"'"+'s z-axis colou'+
         'r key.'
   print '* "sg auto" to automatically set colour floor and ceiling.'
   print '* "sg log" to toggle logarithmic spectrogram plotting.'
   print ''
   print 'POWER SPECTRA:'
   print '* "aspec" to plot the average spectrum and return the frequency of its high'+
         'est peak.'
   print '* "gspec" to get an individual spectrum at any time and plot it.'
   print '* "peaks" to plot a graph of the frequency of the strongest oscillation aga'+
         'inst time.'
   print '* "rates" to get a simple lightcurve of the data.'
   print '* "fqflux" to plot "peaks" against "rates".'
   print ''
   print 'TOGGLE OPTIONS:'
   print '* "errors" to toggle errorbars on power spectra plots.'
   print '* "save" to save to disk any plots which would otherwise be shown.'
   print ''
   print 'OTHER COMMANDS:'
   print '* "info" to display a list of facts and figures about the current SpecAngel'+
         ' session.'
   print '* "reflav" to rewrite the flavour text used for graph titles.'
   print '* "export" to create an ASCII file of the average power density spectrum.'
   print '* "help" or "?" to display this list of instructions again.'
   print '* "quit" to Quit'

give_inst()                                          # Print the list of instructions
\end{minted}

\section{Back Hydra}

\par \texttt{Back Hydra} is a script to subtract the background\index{Background subtraction} from the data in a \texttt{.plotd} file.  It requires a background lightcurve created by the \texttt{pcabackest} tool available from NASA's \texttt{FTOOLS}\index{FTOOLS@\texttt{FTOOLS}} suite.

\begin{minted}[fontsize=\scriptsize]{python}
#! /usr/bin/env python

# |----------------------------------------------------------------------|
# |-----------------------------BACK HYDRA-------------------------------|
# |----------------------------------------------------------------------|

# Call as ./bckghydra.py DATA_FILE BACK_FILE SAVE_FILE

# Takes a .plotd file and a background file created with PCABACKEST and returns
#
# Arguments:
#
#  DATA_FILE
#   The absolute path to the file to be used as data.
#
#  BACK_FILE
#   The file to be used as background; does not need to be the same binning as File 1.
#   suggest using pcabackest from FTOOLS to produce this file.
#   FTOOLS can be found at http://heasarc.gsfc.nasa.gov/ftools/
#
#  SAVE_FILE
#   The location to save the resultant background-subtracted file
#
\end{minted}

\section{PAN Lib}

\par \texttt{PAN Lib} is a library of functions.  I use this library extensively in the other scripts in this suite.

\begin{minted}[fontsize=\scriptsize]{python}
#! /usr/bin/env python

# |----------------------------------------------------------------------|
# |-------------------------------PAN_LIB--------------------------------|
# |----------------------------------------------------------------------|

# A selection of useful functions which are placed here to reduce clutter in the other
#  files of

# PANTHEON.
#
# Contents:
#
# ARGCHECK   - compares the list of arguments against a value given as the minimum
#              allowed number of arguments.  If the list of arguments is too short,
#              throw a warning and kill the script.
#
#  BINIFY    - takes a x-series with its associated y-axis data and y-axis errors.
#              Rebins the data into larger linear bins with a width of the user's
#              choosing, and returns the tuple x,y,y_error.
#
#  BOOLVAL   - takes a list of Boolean values and, interpreting it as binary, returns 
#              its integer value.
#
#  EQRANGE   -
#
#  EVAL_BURST-
#
#  FILENAMECHECK  - checks to see whether a proposed input file has the correct file
#                   extension.
#
#  FOLDIFY   - takes a time series with its associated y-axis data and y-axis errors. 
#              Folds this data over a time period of the user's choosing, and returns
#              them as the tuple x,y,y_error.
#
#  FOLD_BURSTS - uses GET_BURSTS to obtain burst locations then interpolates to
#                populate phase information for all other points
#
#  GET_BURSTS- takes an array of data, looks for bursts and returns an array of tuples
#              containing the start and end points of these bursts.
#
#  GET_DIP   - returns the index of the lowest point between two user-defined flags in
#              a dataset.
#
#  GTIMASK   - returns a data mask when given a time series and a GTI object
#
#  LBINIFY   - takes a linearly binned x-series with associated y-axis data and y-axis
#              errors and rebins them into bins of a constant width in logx space.
#              In places where the logarithmic bins would be finer than the linear
#              bins, the linear bins are retained.
#
#  LEAHYN    - takes the raw power spectrum output from the scipy FFT algorithm and
#              normalises it using Leahy normalisation.
#
#  LH2RMS    - takes a Leahy-normalised power spectrum and converts it to an
#              (RMS/Mean)^2-normalised power spectrum.
#
#  LHCONST   - returns the normalisation of the white noise component in a Leahy-
#              normalised power spectrum with no features in the range 1.5kHz - 4kHz.
# 
#
#  MXREBIN   - takes a 2-dimensional set of data and corresponding errors linearly
#              binned on the x-axis and rebins them by an integer binning factor of
#              the user's choice.
#
#  NONES     - like np.zeros, but with None.
#
#  PDCOLEX   - extracts colours from a set of 2 or 3 lightcurves
#
#  PLOTDLD   - load and unpickle a .plotd file and extract its data.
#
#  PLOTDSV   - collect a selection of data products as a library, pickle it and save
#              as a .plotd file.
#
#  RMS_N     - takes the raw power spectrum output from the scipy FFT algorithm and
#              normalises it using (RMS/Mean)^2 normalisation.
#
#  SAFE_DIV  - Divides two arrays by each other, replacing NaNs that would be caused
#              by div 0 errors with zeroes.
#
#  SIGNOFF   - prints an dividing line with some space.  That's all it does.
#
#  SINFROMCOS- calculates the sines of an array of values when also passed their co-
#              sines.  If both sines and cosines of the array are required, this method
#              is faster than calling both trig functions.
#              Also contains function COSFROMSIN.
#
#  SLPLOT    - plots an x-y line plot of two sets of data, and then below plots the
#              same data on another set of axes in log-log space.
#
#  SPECALD   - load and unpickle a .speca file and extract its data.
# 
#  SPECASV   - collect a selection of data products as a library, pickle it and save
#              as a .speca file.
#
#  SRINR     - calculates whether a value given by a user is within an existant evenly
#              spaced array and, if it is, returns the index value of the closest
#              match to this value within the array.
#              Intended for validating subranges specified by user.
#
#  TNORM     - takes a list of times, and subtracts the lowest value from each entry
#              such that a new list starting with 0 is produced.  Large number
#              subtraction errors are avoided by checking that every entry is an
#              integer number of time-resolution steps from zero.
#
#  UNIQFNAME - checks if a proposed filename is currently in use and, if so, proposes
#              an alternative filename to prevent overwrite.
#
#  XTRFILLOC - takes a filepath and outputs the file name and its absolute(ish)
#              location
#
\end{minted}
\end{appendices}

\cleardoublepage

\nocite{*}
\addcontentsline{toc}{chapter}{Bibliography}
\bibliography{refs.bib}
\bibliographystyle{apalike}

\cleardoublepage

\addcontentsline{toc}{chapter}{Index}
\printindex

\end{document}